\documentclass[12pt,titlepage]{report}
\usepackage{babel,german,a4}
\usepackage{epsf}
\usepackage{cite}  
\usepackage[dvips]{epsfig}
\usepackage{graphicx}
\usepackage{axodraw}
\usepackage{longtable}
\usepackage{ifthen_mod}
\usepackage{rotating_mod} 
\usepackage{longtable}
\begin{document}
\thispagestyle{empty}
\vspace*{-2cm}
\begin{flushright}
{\tt
DESY-THESIS-2000-030
\\
August 2000
\\
hep-ph/0009068
}
\end{flushright}

\bigskip

\begin{center} 
\LARGE
Semi-analytical calculation 
of QED radiative corrections to $e^{+}e^{-}\to \bar{f} f$ 
with special emphasis on kinematical cuts to the final state

\vspace*{1cm}

\LARGE
{\bf Dissertation}

\vspace*{1cm}

\large
zur Erlangung des akademischen Grades
\\
doctor rerum naturalium
\\
(Dr. rer. nat.)
\\
im Fach Physik

\vspace*{1cm}

\large
eingereicht an der
\\ 
Mathematisch-Naturwissenschaftlichen 
Fakult\"at I
\\
der Humboldt-Universit\"at zu Berlin

\vspace*{1cm}

\large 
von 

\bigskip

Mark Alexander Jack

\end{center}

\newpage

%---------------------
\selectlanguage{english}
\abstract
In this dissertation a complete calculation of 
QED radiative corrections is presented for total cross sections
and forward-backward asymmetries for
$s$-channel fermion pair production in $e^+e^-$ annihilation 
with kinematical cuts to the final state.
This includes cuts on the maximal acollinearity angle $\theta_{\rm acol}$
and on the minimal energies $E_{\min}$ of the final state fermion pair and
on the cosine of the scattering angle of one fermion $\cos\vartheta$. 
The applied cuts pose a realistic alternative
for leptonic final states to cuts on the 
invariant mass squared $s'$ of the fermion 
pair and on $\cos\vartheta$.

The derived QED flux functions $\rho(s'/s)$, with $s$ as 
center-of-mass (c.m.) energy squared, are convoluted over $s'$ 
in an improved Born approximation and were implemented into the 
semi-analytical Fortran program {\tt ZFITTER}. It is used 
at the LEP and SLC experiments for precision tests to the 
Standard model and for searches of new particle physics 
phenomena. This calculation had become necessary due to 
the much higher experimental precision obtained now  
and the insufficient accuracy of the approximate, 
earlier coding.

The hard photon flux functions partly correct 
unpublished earlier results or constitute otherwise new,
general formulae which contain known results 
as special cases. Very compact expressions are obtained
when omitting a cut on $\cos\vartheta$. 

An analysis of the updated code
together with other numerical two-fermion codes
yields for the acollinearity cut option 
an agreement of cross sections 
and asymmetries of better than 0.1 per mil on the 
$Z$ boson resonance, and of better than 0.3 to 1 per mil 
at the wings ($\sqrt{s}=M_Z\pm 3\,\mbox{GeV}$). 
Deviations between the 
codes at LEP~2 energies up to several per cent in case 
of $Z$ radiative return events are now understood due 
to an approximation for higher order QED effects 
with an acollinearity cut in {\tt ZFITTER}.
At c.m.~energies up to roughly $800\,\mbox{GeV}$,
{\tt ZFITTER} and other codes deviate not more than 0.5 
to 1 per cent for different $s'$ cuts and different 
higher order QED corrections.

The new formulae and the analysis presented in this thesis 
for high energies and luminosities form 
an essential building block for an upgrading of 
two-fermion codes like {\tt ZFITTER} 
for a future $e^+e^-$ Linear Collider. 
\selectlanguage{german}
\abstract
In dieser Disser\-tation wird eine voll\-st\"an\-dige Berechnung der
QED-Strah\-lungs\-kor\-rek\-turen f\"ur totale 
Wirkungs\-querschnitte und Vor\-w\"arts-R\"uck\-w\"arts\-asym\-met\-rien 
f\"ur $s$-Kanal-Fermion\-paar\-produk\-tion in $e^+e^-$-Annihilation dargestellt, 
einschlie{\ss}lich kinema\-tischer Schnitte am End\-zustand.
Diese umfassen Schnitte an dem maxi\-malen Akol\-linea\-rit\"ats\-winkel 
$\theta_{\rm acol}$ und den mini\-malen Energien $E_{\min}$ des 
Fermion\-paares im Endzustand und an dem Kosinus 
des Streu\-winkels eines Fermions $\cos\vartheta$. 
F\"ur lep\-tonische End\-zust\"ande bilden diese Schnitte 
eine realis\-tische Alter\-native zu Schnit\-ten an der 
in\-varian\-ten Masse $s'$ des Fermion\-paares und an $\cos\vartheta$.
Die berechneten QED-Radiator\-funk\-tionen $\rho(s'/s)$ mit $s$ als 
quadrierter Schwer\-punkts\-energie werden \"uber $s'$ in einer effektiven 
Born\-appro\-xi\-mation gefalten und wurden in das 
semi-ana\-ly\-tische For\-tran\-pro\-gramm {\tt ZFITTER} 
implementiert, welches f\"ur Pr\"a\-zisions\-tests zum
Stan\-dard\-modell und f\"ur die Suche nach neuen Physik\-ph\"anomenen
bei den LEP- und SLC-Ex\-peri\-menten eingesetzt wird.
Diese Berechnung war durch die jetzt h\"ohere ex\-peri\-men\-telle Pr\"a\-zision 
und die un\-zu\-reichende Genauig\-keit der ap\-pro\-xi\-ma\-tiven fr\"uheren 
Pro\-gram\-mierung notwendig geworden.
Die Ra\-diator\-funk\-tionen zur harten 
Brems\-strahlung kor\-rigieren zum Teil \"altere, un\-pub\-lizierte Re\-sul\-tate
oder bilden ansonsten neue, allgemeine 
Formeln, welche bekannte Resul\-tate 
als Spezial\-f\"alle enthalten. Die Ver\-nach\-l\"assigung des Schnittes 
an $\cos\vartheta$ liefert sehr kompakte Ausdr\"ucke. 
Eine Analyse des erneuerten Pro\-gramms mit anderen Zwei\-fermion\-programmen 
ergibt jetzt auf der $Z$-Boson\-resonanz f\"ur Wir\-kungs\-quer\-schnitte 
und Asym\-metrien eine \"Uberein\-stimmung um besser als 0.1 Permille 
und um besser als 0.3 bis 1 Permille an den Fl\"ugel\-regionen
($\sqrt{s}=M_Z\pm 3\,\mbox{GeV}$). Mehrere Prozent gro{\ss}e 
Ab\-wei\-chungen zwischen den Pro\-grammen bei LEP~2-Energien im 
Falle einer radiativen R\"uckkehr zur $Z$-Resonanz 
sind jetzt verstanden und k\"onnen auf eine N\"aherung der
h\"oheren QED-Kor\-rek\-turen mit 
Akol\-lineari\-t\"ats\-schnitt in {\tt ZFITTER} zu\-r\"uck\-ge\-f\"uhrt werden. 
F\"ur Schwer\-punkts\-energien bis un\-gef\"ahr $800\,\mbox{GeV}$
weichen {\tt ZFITTER} und andere Pro\-gramme um nicht mehr als 
0.5 bis 1 Prozent ab, f\"ur ver\-schiedene Schnitte an $s'$
und f\"ur ver\-schie\-dene QED-Kor\-rek\-turen h\"oherer Ordnung.
Die in dieser Thesis dar\-gestell\-ten neuen Formeln 
und die Analyse im Falle hoher Energien und 
Lumi\-nosi\-t\"aten bilden eine essen\-tielle Vor\-arbeit 
f\"ur eine Auf\-wertung von Zwei\-fermion\-pro\-grammen
wie {\tt ZFITTER} f\"ur einen zu\-k\"unf\-tigen $e^+e^-$-Li\-near\-be\-schleu\-niger.
%
%_____________________
\newpage
\thispagestyle{empty}
%---------------------
\newpage
\selectlanguage{english}
\pagenumbering{roman}
\tableofcontents
\listoffigures
\listoftables
\newpage

%\setlength{\textwidth}{16.0cm}
%\setlength{\textheight}{23.5cm}
%\setlength{\evensidemargin}{0.0cm}
%\setlength{\oddsidemargin}{0.0cm}
%\setlength{\topmargin}{-1.0cm}
%
%\setcounter{secnumdepth}{3}
%\setcounter{tocdepth}{2}
%
%\hyphenation{brems-strah-lung}
%
\newcommand{\LEPI}{LEP~I}
\newcommand {\zf}{{\tt ZFITTER}}
%------------------------------------------------------------------------------
\newcommand{\stm}{\cal SM}
\newcommand{\Order}[1]{{\cal O}($#1$)}
\newcommand{\ta}{{\tilde{a}}}
\newcommand{\ee}{$e^+e^-$}
\newcommand{\st}{$\sigma_{T}\:$}
\newcommand{\oalf}{\mbox{${\cal O}(\alpha)$}}
\newcommand{\oalff}{\mbox{${\cal O}(\alpha^2) \:$}}
\newcommand{\oalfs}{\mbox{${\cal O}(\alpha \alpha_s) \:$}}
\newcommand{\oalss}{\mbox{${\cal O}(\alpha_s^2) \:$}}
\newcommand{\ost}{\mbox{${\cal O}(\alpha \alpha_s m_t^2) \:$}}
\newcommand{\ostw}{\mbox{${\cal O}(\alpha \alpha_s m_t^2/M_W^2) \:$}}
\newcommand{\oaa}{\mbox{${\cal O}(\alpha^2 m_t^4) \:$}}
\newcommand{\oaaw}{\mbox{${\cal O}(\alpha^2 m_t^4/M_W^4) \:$}}
\newcommand{\dsdc}{$\frac{d\sigma}{dcos\vartheta}$}
\newcommand{\Sw}{$\sin^2\theta_W$}
\newcommand{\MZ}{\mbox{$M_{_Z}$}}
\newcommand{\mz}{\mbox{$M_{_Z}$}}
\newcommand{\MW}{\mbox{$M_W$}}
\newcommand{\MH}{\mbox{$M_{_H}$}}
\newcommand{\GAMZ}{\mbox{$\Gamma_{_Z}$}}
\newcommand{\GAMe}{$\Gamma_{initial}$}
\newcommand{\GAMf}{$\Gamma_{final}$}
\newcommand{\R}{$\rho$}
\newcommand{\RS}{$\sqrt{s}$}
\newcommand{\BB}{$b\bar{b}$}
\newcommand{\FF}{$f\bar{f}$}
\newcommand{\LL}{$l\bar{l}$}
\newcommand{\QQ}{$q\bar{q}$}
\newcommand{\MM}{$\mu^+\mu^-$}
\newcommand{\MwS}{\mbox{$M^2_{_W}$}}
\newcommand{\MzS}{\mbox{$M^2_{_Z}$}}
\newcommand{\mq }{\mbox{$m_q  $}}
\newcommand{\mqS}{\mbox{$m^2_q$}}
\newcommand{\ms }{\mbox{$m_s  $}}
\newcommand{\msS}{\mbox{$m^2_s$}}
\newcommand{\mc }{\mbox{$m_c  $}}
\newcommand{\mcS}{\mbox{$m^2_c$}}
\newcommand{\mb }{\mbox{$m_b  $}}
\newcommand{\mbS}{\mbox{$m^2_b$}}
\newcommand{\MSB}{\mbox{$\overline{MS}$}}
\newcommand{\alem}{\mbox{$\alpha_{em}$}}
\newcommand{\alsS}{\mbox{$\alpha^2_{_S}$}}
\newcommand {\vai}  {\mbox{$(v_e+a_e \gamma_5)$}}
%------------------------------------------------------------------------------
\def\mev{{\hbox{MeV}}}
\def\gev{{\hbox{GeV}}}
\def\tev{{\hbox{TeV}}}
\def\msb{{\overline{MS}}}
\def\als{\alpha_{_S}}
\def\afb{\mbox{$A_{_{FB}}$}}
\def\gv{g_{_V}}
\def\ga{g_{_A}}
\def\barf{\overline f}
\def\barq{\overline q}
\def\barb{\overline b}
\def\bart{\overline t}
\def\barc{\overline c}
\def\gvf{g^f_{_{V}}}
\def\gaf{g^f_{_{A}}}
\def\gvl{g^l_{_{V}}}
\def\gal{g^l_{_{A}}}
\def\ste{\sin\theta}
\def\stes{\sin^2\theta}
\def\dr{\Delta r}
\def\Pgg{\Pi_{\gamma\gamma}}
\def\Pf{\Pi_{_F}}
\def\chig{\chi_{\gamma}}
\def\chiz{\chi_{_Z}}
\def\nn{\nonumber}
\def\ra{\rightarrow}
\def\lra{\leftrightarrow}
\def\ipi{\frac{i}{16\pi^2}}
\def\alpi{\frac{\alpha}{4\pi}}
\def\api{\frac{\alpha}{\pi}}
\def\alfs{\frac{\alpha^2}{4s}}
\def\alspi{\frac{\alpha_s}{\pi}}
\def\apw{\frac{\alpha}{2\pi}}
\def\Gmu{G_{\mu}}
\def\gamu{\gamma_{\mu}}
\def\gimu{\gamma^{\mu}}
\def\ganu{\gamma_{\nu}}
\def\ginu{\gamma^{\nu}}
\def\giro{\gamma^{\rho}}
\def\gisi{\gamma^{\sigma}}
\def\garo{\gamma_{\rho}}
\def\gasi{\gamma_{\sigma}}
\def\gafi{\gamma_5}
\def\Gu{\Gamma_{\mu}}
\def\noi{\noindent}
\def\epmf{e^+e^- \rightarrow f\bar{f}}
\def\epmm{e^+e^- \rightarrow \mu^+\mu^-}
\def\epmb{e^+e^- \rightarrow b\bar{b}}
\def\epm{e^+e^-}
\def\vaee{v_e a_e}
\def\vaff{v_f a_f}
\def\vvef{v_e v_f}
\def\aaef{a_e a_f}
\def\ve{(v_e^2+a_e^2)}
\def\vf{(v_f^2+a_f^2)}
\def\vff{(v_f^2+a_f^2-4\mu_f a_f^2)}
\def\vam{(v_e^2-a_e^2)}
\def\rchi{\mbox{Re}\,\chi_Z(s)}
\def\chizh{\mid\chi_Z(s) \mid^2}
\def\rc{radiative corrections }
\def\su{SU(2)$\times$U(1) }
\def\siw{\sin^2\theta_W}
\def\cow{\cos^2\theta_W}
\def\sinw{\sin\theta_W}
\def\cosw{\cos\theta_W}
\def\Dr{\Delta r}
\def\Drb{\overline{\Delta r}}
\def\iprop{s-M_Z^2+i\displaystyle \frac{s}{M_Z}\Gamma_Z}
\def\prop{\displaystyle \frac{1}{s-M_Z^2
 +i \displaystyle \frac{s}{M_Z}\Gamma_Z}}
\def\eps{\epsilon}
\def\veps{\varepsilon}
\def\alp{\frac{\alpha}{2\pi}}
\def\dlog{\mbox{Li}_2}
\def\oal{O(\alpha^2)}
\def\adu{\overline{u}}
\def\add{\overline{d}}
\def\text{\textstyle}

\def\bea{\begin{eqnarray}}
\def\eea{\end{eqnarray}}
\newcommand{\bq}{\begin{equation}}
\newcommand{\eq}{\end{equation}}
\newcommand{\beq}{\begin{eqnarray}}
\newcommand{\eeq}{\end{eqnarray}}
\newcommand{\ba}{\begin{eqnarray}}
\newcommand{\ea}{\end{eqnarray}}
%
% Added by Mark Jack 19.01.98:
%
\newcommand{\ct}{\cos\vartheta}
\newcommand{\ctt}{\cos^2\vartheta}
\newcommand{\sint}{\sin\vartheta}
\newcommand{\sintt}{\sin^2\vartheta}
\newcommand{\ctg}{\cos\theta_\gamma}
\newcommand{\ctgg}{\cos^2\theta_\gamma}
\newcommand{\stg}{\sin\theta_\gamma}
\newcommand{\stgg}{\sin^2\theta_\gamma}
\newcommand{\cxi}{\cos\xi}
\newcommand{\ctxi}{\cos^2\xi}
\newcommand{\cxih}{\cos{\frac{\xi}{2}}}
\newcommand{\ctxih}{\cos^2{\frac{\xi}{2}}}
\newcommand{\cxihb}{\cos{\frac{\bar{\xi}}{2}}}
\newcommand{\ctxihb}{\cos^2{\frac{\bar{\xi}}{2}}}
\newcommand{\sxi}{\sin{\frac{\xi}{2}}}
\newcommand{\stxi}{\sin^2{\frac{\xi}{2}}}
\newcommand{\sxib}{\sin{\frac{\bar{\xi}}{2}}}
\newcommand{\stxib}{\sin^2{\frac{\bar{\xi}}{2}}}
\newcommand{\sxim}{\sin{\frac{\xi^{\max}}{2}}}
\newcommand{\stxim}{\sin^2{\frac{\xi^{\max}}{2}}}
\newcommand{\Rp}{R_E^{\bar{f}}}
\newcommand{\Rm}{R_E^f}
\newcommand{\Rpm}{R_E^{f,\bar{f}}}
\newcommand{\Ep}{\bar{E}_f}
\newcommand{\Em}{\bar{E}_{\bar{f}}}
\newcommand{\Epm}{\bar{E}_{f,\bar{f}}}
\newcommand{\Rtp}{\bar{R}_E^{\bar{f}}}
\newcommand{\Rtm}{\bar{R}_E^f}
\newcommand{\Rtpm}{\bar{R}_E^{f,\bar{f}}}
\newcommand{\Rxip}{R_\xi^{\bar{f}}}
\newcommand{\Rxim}{R_\xi^f}
\newcommand{\Rxipm}{R_\xi^{f,\bar{f}}}
%
%Added by Penka   22.01.1998
%
\newcommand{\val } {\mbox{$(v_{\ell} + a_{\ell} \gamma_5)$}}
\def\epml{e^+e^- \rightarrow \ell \bar{\ell}}
\newcommand{\me }{\mbox{$m_e  $}}
\newcommand{\meS}{\mbox{$m^2_e$}}
\newcommand{\meQ}{\mbox{$m^4_e$}}
\newcommand{\ml }{\mbox{$m_{\ell}  $}}
\newcommand{\mlS}{\mbox{$m^2_{\ell}$}}
\def\qel{\mid Q_e\mid\,\mid Q_{\ell}\mid}
\def\QeS{Q^2_e }
\def\QlS{Q^2_{\ell}}
\def\moe{\mid \vec k_1 \mid}
\def\mop{\mid \vec k_2 \mid}
\def\mok{\mid \vec k \mid}
\def\mol{\mid \vec p_1 \mid}
\def\mola{\mid \vec p_2 \mid}
\def\dsdcpl{\frac{d\sigma}{dcos\theta}(h_-=+1)}
\def\dsdcmi{\frac{d\sigma}{dcos\theta}(h_-=-1)}
\def\dmin{\dsdcpl-\dsdcmi}
\def\dplu{\dsdcpl+\dsdcmi}
%
%Added by Mark Jack 29.01.1998
%
\newcommand{\ang}{{\scriptstyle\prec\,\!\!\!\!)}\,}
\newcommand{\DS}{\displaystyle}
\newcommand{\phig}{\varphi_\gamma}
\newcommand{\theg}{\theta_\gamma}
\newcommand{\costg}{\cos\theta_\gamma}
\newcommand{\sintg}{\sin\theta_\gamma}
\newcommand{\cost}{\cos\theta}
\newcommand{\sinth}{\sin\theta}
\newcommand{\sa}{\sqrt{a_1}}
\newcommand{\scn}{\sqrt{c^0_1}}
\newcommand{\be}{\beta_0}
\newcommand{\arsinh}{\mbox{arsinh}}
\newcommand{\go}{\gamma_1}
\newcommand{\gt}{\gamma_2}
\newcommand{\got}{\gamma_{1,2}}
\newcommand{\Cr}{C_1^{\frac{3}{2}}}
\newcommand{\Rr}{R^{\frac{3}{2}}}
\def\gabe{\gamma_{\beta}}
\def\gaal{\gamma^{\alpha}}
\def\slash{\!\!\!/}
\def\sslash{\!\!\!\!\!\!\:/}
\def\LS{\,{L}\slash\,}
\def\qs{\,{q}\slash\,}
\def\qso{\,{q_1}\sslash\,}
\def\qst{\,{q_2}\sslash\,}
\def\ps{\,{p}\slash\,}
\def\pso{\,{p_1}\sslash\,}
\def\pst{\,{p_2}\sslash\,}
\def\ks{\,{k}\slash\,}
\def\kso{\,{k_1}\sslash\,}
\def\kst{\,{k_2}\sslash\,}
\def\ksi{\,{k_i}\sslash\,}
\def\ksf{\,{k_f}\sslash\,}
\newcommand{\vae } {\mbox{$(v_e + a_e \gamma_5)$}}
\newcommand{\vaf } {\mbox{$(v_f + a_f \gamma_5)$}}

% A useful Journal macro
\def\Journal#1#2#3#4{{#1} {\bf #2}, #3 (#4)}

% Some useful journal names
\def\NCA{\em Nuovo Cimento}
\def\NIM{\em Nucl. Instrum. Methods}
\def\NIMA{{\em Nucl. Instrum. Methods} A}
\def\NPB{{\em Nucl. Phys.} B}
\def\PLB{{\em Phys. Lett.}  B}
\def\PRL{\em Phys. Rev. Lett.}
\def\PRD{{\em Phys. Rev.} D}
\def\ZPC{{\em Z. Phys.} C}
% Some other macros used in the sample text
\def\st{\scriptstyle}
\def\sst{\scriptscriptstyle}
\def\mco{\multicolumn}
\def\epp{\epsilon^{\prime}}
\def\vep{\varepsilon}

\def\CPbar{\hbox{{\rm CP}\hskip-1.80em{/}}}%temp replacement due to no font

\renewcommand{\d}{{\sf d}} 
\newcommand{\EE}{\EPL\EMI} 
\newcommand{\EPL}{e^+} 
\newcommand{\EMI}{e^-} 
\newcommand{\sm}{{\tt SM}} 
\newcommand{\STWO}{ s_{2}} 
\newcommand{\hf}{\hfill} 

\def\nl{\nonumber\\}
\def\dz{\lambda_Z}
\def\theequation{\arabic{section}.\arabic{equation}}
\newcommand{\ezero}{\setcounter{equation}{0}}
\newcommand{\litwo}{\mbox{Li}_2}
\newcommand{\lsim}{\raisebox{-0.13cm}{~\shortstack{$<$ \\[-0.07cm] $\sim$}}~}
\newcommand{\gsim}{\raisebox{-0.13cm}{~\shortstack{$>$ \\[-0.07cm] $\sim$}}~}
\setlength{\parskip}{1ex plus 0.5ex minus 0.2ex}

% LEP1
%\newcommand{\mz}{M_{_Z}}
\newcommand{\afba}[1]{A^{#1}_{_{\rm FB}}}

% LEP2
%For clones of tables of PCP report: 
\newcommand{\tbn}[1]{Table~\ref{#1}}
\newcommand{\tbns}[2]{Tables~\ref{#1}--\ref{#2}}
\newcommand{\tbnsc}[2]{Tables~\ref{#1},~\ref{#2}}
\newcommand{\nc}{\newcommand}
\nc{\GeV}{\,\mbox{GeV}}
\nc{\MeV}{\,\mbox{MeV}}
\nc{\pb}{\,\mbox{pb}}
\nc{\nb}{\,\mbox{nb}}
\nc{\btu}{\bigtriangleup}
\nc{\DD}{\displaystyle}

% Appendix: Initial State Radiation
\newcommand{\RAp}{R_A^{+}}
\newcommand{\RAm}{R_A^{-}}
\newcommand{\RApb}{\bar{R}_A^{+}}
\newcommand{\RAmb}{\bar{R}_A^{-}}
\newcommand{\RAcp}{R_{Ac}^{+}}
\newcommand{\RAcm}{R_{Ac}^{-}}

%--------------------------------------------------
% Appendix: Final state radiation
\newcommand{\req}[1]{(\ref{#1})}
\newcommand {\mf}  {\mbox{$m^2_f     $}}
\newcommand {\Qe}  {\mbox{${\tt Q}^2_e $}}
\newcommand {\Qf}  {\mbox{${\tt Q}^2_f $}}
\newcommand {\Lf}  {\mbox{${ln{\frac{s}{\mf} }} $}}
\newcommand{\nll}{{\nonumber\\}}

%--------------------------------------------------
% Appendix: Initial state virtual corrections
% Indices
%--------
\newcommand{\gQ}{\ph{\scriptscriptstyle{Q}}}
\newcommand{\gL}{\ph{\scriptscriptstyle{L}}}
\newcommand{\gZQ}{\ph{\scriptscriptstyle{(Z)Q}}}
\newcommand{\gZL}{\ph{\scriptscriptstyle{(Z)L}}}
\newcommand{\ZQ}{\scriptscriptstyle{ZQ}}
\newcommand{\ZL}{\scriptscriptstyle{ZL}}
\newcommand{\QL}{\scriptscriptstyle{QL}}
\newcommand{\LQ}{\scriptscriptstyle{LQ}}
\newcommand{\zg}{{\scriptscriptstyle{Z}}\ph}
\newcommand{\sss}[1]{\scriptscriptstyle{#1}}
\newcommand{\Zi}{\sss{Z}}
\newcommand{\gf}{G_{\mu}}
\newcommand{\gfs}{G^2_{\mu}}
%   Fields
\newcommand{\ph}{\gamma}
\newcommand{\ab}{A}
\newcommand{\zb}{Z}
\newcommand{\wb}{W}
\newcommand{\hb}{H}
%  Fermions
\newcommand{\fe}{e}
\newcommand{\fbe}{{\bar{e}}}
\newcommand{\ff}{f}
\newcommand{\ffp}{f'}
\newcommand{\fep}{e^{+}}
\newcommand{\fem}{e^{-}}
\newcommand{\fnue}{\nu_e}
\newcommand{\barl}{\overline{l}}
\newcommand{\fu}{u}
\newcommand{\fd}{d}
\newcommand{\fc}{c}
\newcommand{\fs}{s}
\newcommand{\ft}{t}
\newcommand{\fb}{b}
\newcommand{\ffb}{b}
\newcommand{\fl}{l}
\newcommand{\fq}{q}
\newcommand{\flm}{\mu}
\newcommand{\flmp}{\mu^{+}}
\newcommand{\flmm}{\mu^{-}}
\newcommand{\flt}{\tau}
\newcommand{\gap}{\lpar 1+\gamma_5\rpar}
\newcommand{\gadi}[1]{\gamma_{#1}}
\newcommand{\cff}[5]{C_{#1}\lpar #2;#3,#4,#5\rpar}    
%   Masses
\newcommand{\mws}{M^2_{\sss{W}}}
\newcommand{\mzs}{M^2_{\sss{Z}}}
\newcommand{\mzc}{M^3_{\sss{Z}}}
\newcommand{\mhs}{M^2_{\sss{H}}}
\newcommand{\mts}{m^2_{t}}
\newcommand{\mvs}{M^2_{_V}}
\newcommand{\mV }{M^2_{_V}}
\newcommand{\mfp}{m^2_{f'}}
\newcommand{\mfh}{m^2_{h}}
\newcommand{\mt }{m^2_t}
\newcommand{\mes}{m^2_e}
\newcommand{\mfpq}{m^4_{f'}}
\newcommand{\mwq }{M^4_{\sss{W}}}
\newcommand{\mzq }{M^4_{\sss{Z}}}
\newcommand{\mhq }{M^4_{\sss{H}}}
\newcommand{\mtq }{m^4_{t}}
\newcommand{\mhl }{M_{\sss{H}}}
\newcommand{\mVl }{M_{_V}}
\newcommand{\mwl }{M_{\sss{W}}}
\newcommand{\mzl }{M_{\sss{Z}}}
\newcommand{\mfl }{m_f}
\newcommand{\mtl }{m_t}
\newcommand{\mel }{m_e}
\newcommand{\mfpl}{m_{f'}}
\newcommand{\mfhl}{m_{h}}
\newcommand{\mml }{m_{\mu}}
\newcommand{\mfs }{m^2_f}
\newcommand{\mls }{m^2_l}
\newcommand{\uml}{ m   _{t}}
\newcommand{\um }{ m^2_{t} }
\newcommand{\umf}{ m^4_{t} }
\newcommand{\wml}{ M  _{\sss{W}}}
\newcommand{\zml}{ M  _{\sss{Z}}}
\newcommand{\dml}{ m   {_f}}
\newcommand{\dms}{ m^2_{_f}}
\newcommand{\hml}{ M  _{\sss{H}}}
\newcommand{\rtc}{ r^3_  t }

\pagenumbering{arabic}
%
%---------------------------------------------------------------

%======================================================================
\chapter*{Introduction 
\label{ch_intro}
}
%======================================================================
\addcontentsline{toc}{chapter}{Introduction}
%\addtocontents{toc}{\protect\thispagestyle}
\pagestyle{myheadings}
\markright{\it INTRODUCTION}
%\pagenumbering{arabic}

%-----------------------------------------------
\section*{Motivation
\label{motiv}
}
%-----------------------------------------------
\addcontentsline{toc}{section}{Motivation}
Up to now the phenomenology of particles and their interactions 
has been so successfully described by the {\it Standard Model}
\cite{Glashow:1961ez,Weinberg:1967pk,Salam:1968rm,Fritzsch:1973pi,Politzer:1973fx,Gross:1973id,Weinberg:1973un}.
The fundamental particles are classified as fermions into 
three families consisting of leptons and quarks.
Electromagnetic, weak, and strong interactions are 
mediated between the fermions in a quantum field theoretical 
picture through the exchange of spin-1 vector bosons.
The underlying symmetries of these particles and 
interactions can be described by the semi-simple gauge
group $SU_C(3)\times SU_L(2)\times U_Y(1)$, i.e.~the
Lagrangian ${\cal L}$ of the theory has to be invariant under 
these local symmetry transformations applied to the 
fermionic and bosonic fields.
Leptons and quarks are combined into left-handed doublets 
or described as right-handed singlets with respect to 
the $SU_L(2)$ gauge group, while an extra quantum number
called {\it color} is assorted to each quark making it
to a triplet of the fundamental representation of $SU_C(3)$.
A key characteristic of electroweak interactions is that
flavor changing neutral current transitions are not 
observed, in the Standard model explained by 
the {\it GIM mechanism} \cite{Glashow:1970st}. 

While {\it quantum chromodynamics} can be treated 
as an exact, non-abelian gauge symmetry $SU_C(3)$
with massless, self-interacting gauge bosons, the {\it gluons},
the unified description of electromagnetic and weak 
forces in nature is not exactly given by an 
$SU_L(2)\times U_Y(1)$ gauge symmetry, but 
the symmetry is `broken': the weak charged 
and neutral vector bosons were discovered to be massive 
with different masses 
\cite{Arnison:1983rp,Banner:1983jy,Arnison:1983mk,Bagnaia:1983zx}, 
while the photon is massless with an 
exact $U_{em}(1)$ as gauge symmetry.

The {\it Higgs-Kibble mechanism}, which introduces 
an extra complex scalar doublet to the theory
with a non-vanishing vacuum expectation value,
shows a possible way out to describe the electroweak 
`symmetry breakdown' in a gauge-invariant way
\cite{Higgs:1964a,Kibble:1967e}. 
In principal the underlying 
symmetry of ${\cal L}$ is not really broken, but
hidden: We obtain a multiplet of vacuum states which do not 
possess the symmetry of the Lagrangian ${\cal L}$, therefore when 
choosing one, the underlying gauge symmetry of ${\cal L}$  
is not apparent anymore in this vacuum state.
Choosing a specific gauge, the {\it unitary gauge}, transforms 
the unphysical degrees of freedom of the scalar fields into 
longitudinal components of the weak vector bosons, thus giving 
them masses while the photon remains massless. From the scalar 
fields one massive scalar boson, the {\it Higgs boson}, remains
which is so extensively searched for at existing high-energy 
colliders as last missing building block of the mass generation 
mechanism in the electroweak sector. 
%This generalization of the 
%Goldstone theorem \cite{Goldstone}, when including gauge
%fields, naturally also produces one massive scalar boson,
%the {\it Higgs boson}, which is so extensively searched 
%for at existing high-energy colliders as last missing 
%building block of the mass generation mechanism in the 
%electroweak sector.

At the high-energy $e^+e^-$ colliders LEP~1 
\cite{Grunewald:1998kw,Grunewald:1999wn,Barate:1999ce,Abbiendi:1999eh,Abreu:1998kh,Acciarri:2000ai,Abbaneo:2000aa}
and SLC \cite{Abe:2000dq,Abe:2000ey}, 
for example, the main objective is to determine 
the neutral current properties of the electroweak theory:
the mass $M_Z$ and the total and partial decay widths 
of the $Z$ boson into fermion pairs $\Gamma_Z$ and $\Gamma_f$ 
and the neutral current vector and axial-vector couplings 
$v_f$ and $a_f$. For example, from the precise measurement of 
the shape of the $Z$ boson resonance curve, i.e the fermionic 
peak cross sections, $M_Z$ and $\Gamma_Z$ were determined 
at LEP~1 and SLC with relative errors of roughly $2\times 10^{-5}$, 
or $10^{-3}$ respectively \cite{Grunewald:1998kw,Grunewald:1999wn}.

The calculation of quantum effects to observables in 
the Standard Model is in this context of course absolutely
mandatory. For a perturbative expansion of the 
$S$ matrix of scattering processes in the small coupling 
constants of the theory it was shown
that all divergences arising during 
the calculation can be completely removed by a redefinition 
of bare fields and parameters in the Lagrangian 
\cite{Veltman:1968ki,'tHooft:1971rn,'tHooft:1972fi,'tHooft:1972ue}.
This can be done once and for all by adding a finite number 
of counter terms removing all singularities to all orders 
of perturbation theory. A basic, finite set of parameters 
fixed by experiment at a certain energy scale suffices as 
input to calculate all other {\tt SM} observables.
Thus, the Standard Model ({\tt SM}) as gauge-invariant 
field theory is renormalizable and yields finite and 
therefore physically meaningful results.

{From} virtual radiative corrections
one can, for example, also determine indirectly the top quark 
mass $m_t$ at LEP or SLC as test for the Tevatron results where  
the top quark can be directly produced \cite{Abe:1995hr,Abachi:1995iq}. 
Even indirect upper bounds on the mass $M_H$ of the 
{\tt SM} Higgs boson \cite{Veltman:1977kh}, which has still escaped 
direct observation, can be obtained in this way. But finally, searching 
for {\it physics beyond the {\tt SM}} at existing or future colliders 
is probably the main motivation for any high-energy physicist to ask 
for higher and higher luminosities and energies.
Either one studies again virtual effects of massive new particles 
coupling to the {\tt SM} ones, predicted for example by supersymmetric, 
grand-unifying, or string-inspired models, or one tries and produces 
them directly at high energies. 

{For} this, radiative corrections to cross section observables    
have to be accurately dealt with. In the {\tt SM} we have 
pure QED, electroweak, and QCD corrections
which all influence observables like fermion pair cross 
sections and asymmetries. On the $Z$ boson resonance the 
QED corrections dominate, but also the inclusion of the other 
corrections is ultimately important to correctly 
reproduce the experimental observations.

In this dissertation the effects of QED radiative 
corrections to cross sections and asymmetries are examined
for the $Z$ boson resonance region, for LEP~2 energies, 
and for higher energies like at a future $e^+e^-$ Linear
Collider. Real soft and hard photon emission 
from the initial and final state fermions is considered 
together with the QED interference and the virtual 
photonic corrections. This is done for the semi-analytical
Fortran program {\tt ZFITTER} \cite{Bardin:1987hva,Bardin:1989cw,Bardin:1989di,Bardin:1991de,Bardin:1991fu,Bardin:1992jc2,Christova:1999cc,Bardin:1999yd-orig}
in comparison with other numerical codes for two-fermion production 
in $e^+e^-$ annihilation. The program {\tt ZFITTER} is used for example 
together with other codes e.g. at LEP in data-fitting routines.
New results for hard photon radiation where derived 
with realistic cuts for totally integrated and differential 
cross sections. They constitute general analytical formulae which 
contain earlier results in the literature
with kinematically simpler cuts as special case 
\cite{Bardin:1991de,Bardin:1991fu} and yield 
very compact expressions when omitting one angular cut. The applied
energy and angular cuts are experimentally especially interesting 
for leptonic final states like $\bar{\mu}\mu$, where they pose
an alternative to a kinematically simpler cut on the final state
invariant mass squared \cite{Acciarri:2000ai}.

In order to illustrate the general importance of QED bremsstrahlung,
a brief discussion of its influence on cross sections at the 
$Z$ boson resonance shall be given in the remaining part of this Introduction.
The modifications to cross sections at the $Z$ peak
by QED radiative corrections are {\it universal}: 
They arise from multiple soft or virtual photonic corrections, 
i.e.~finite contributions due to real photon emission or virtual 
photon exchange for vanishing photon momenta. 
In first approximation they do not depend on the details 
of the final state phase space like kinematical cuts or final 
state masses. The main corrections develop from the photonic 
corrections to the initial state fermion pair. 
Exactly on the $Z$ peak, the emission of hard photons is 
strongly suppressed and will therefore be neglected 
during most of the calculation in our introductory 
approximate description of cross sections and asymmetries
around the $Z$ peak. 
With the per mil, and partly better than per mil precision 
experimentally available for observables in the $Z$ boson 
resonance region the hard QED bremsstrahlung will, however, 
have to be included for {\tt SM} precision tests.
This latter point will be treated in detail in Chapter \ref{ch_lep1slc}.

Away from the $Z$ boson resonance and for
higher center-of-mass energies the importance of hard photon
emission grows and its dependence on the kinematical cuts 
applied has to be correctly taken into account. Events with 
a {\it radiative return to the $Z$ boson}, where the effective 
center-of-mass energy after initial state hard photon emission 
is shifted onto the $Z$ boson mass in case of no or only loose 
kinematical cuts, produce a strong cross section enhancement. 
Also higher order QED corrections will start to play a larger 
role. Both issues will be discussed for center-of-mass energies 
typical at LEP~2 or at a future $e^+e^-$ Linear 
Collider (LC) in Chapters \ref{ch_lep2} and \ref{ch_linac}.

%-----------------------------------------------
\section*{The $Z$ line shape around the $Z$ boson resonance
\label{peakcross}
}
%-----------------------------------------------
\addcontentsline{toc}{section}{The $Z$ line shape around the $Z$ boson resonance}
The three major effects to the $Z$ line shape in the 
$Z$ boson resonance region, i.e. to cross sections 
at center-of-mass energies of roughly 
$88\,\mbox{GeV} < \sqrt{s} < 95\,\mbox{GeV}$,
can be summarized through following three `rules of thumb':
\begin{itemize} 
\item[1.] The peak cross section $\sigma_{\max}$ is lowered with respect 
          to the Born case $\sigma^0_{\max}$ approximately by a factor 
          $(\Gamma_Z/M_Z)^{\beta_e}$ \cite{Greco:1967,Etim:1967}, 
          with $M_Z$ and $\Gamma_Z$ as mass and total width of the $Z$ boson 
          \cite{Wetzel:1983mh,Akhundov:1986fc,Jegerlehner:1986vs,Beenakker:1988pv,Bernabeu:1988me,Lynn:1990hd}:         
          \ba
          \label{peakheight}
          \sigma_{\max} &=& \sigma(\sqrt{s}\approx M_Z) = 
          \sigma^0_{\max}\, 
          \left(\frac{\Gamma_Z}{M_Z}\right)^{\beta_e}\, (1+\bar{S}),
          \\
          \label{betae}
	  {\beta_e} &=& \frac{2\alpha}{\pi}\left[L_e(M_Z^2)-1\right],
          \quad L_e(M_Z^2) = \ln\frac{M_Z^2}{m_e^2},
          \quad 
          (\Gamma_Z/M_Z)^{\beta_e} \approx 0.7.
          \nonumber\\
          \ea
          The term $\bar{S}$ stands for the finite corrections from soft and
          virtual photons.
          
\item[2.] The peak position $\sqrt{s_{\max}}$ is shifted with respect 
          to the $Z$ boson mass $M_Z$ to a slightly higher value:
          \ba          
          \label{peakpos}
          \sqrt{s_{\max}} = M_Z\left[1+\frac{\pi\beta_e}{4}\gamma
          -\frac{\gamma^2}{4}\right],\quad \gamma=\frac{\Gamma_Z}{M_Z},
          \ea
          where the first term $\pi\beta_e\gamma/4$ 
          \cite{Cahn:1987qf,Barroso:1987ae,Berends:1988bg,Bardin:1989qr,Beenakker:1990ec}, is due 
          to the main bremsstrahlung corrections to 
          the $Z$ boson resonance described by a 
          Breit-Wigner resonance formula,
          while the second term proportional to $\gamma^2$ 
          arises also taking into account the $\gamma Z$ 
          interference contribution and an $s$-dependence 
          of the $Z$ boson width in the $Z$ boson
          propagator of the squared matrix element, 
          $\Gamma_Z(s)\equiv s/M_Z^2\cdot\Gamma_Z$ 
          \cite{Akhundov:1986fc,Bardin:1988xt,Bardin:1989di,Riemann:1997kt,Riemann:1997tj}.

\item[3.] The hard photon emission enhances the cross section
          values $\sigma(s)$ with respect to the Born values 
          $\sigma^0(s)$ for center-of-mass energies $\sqrt{s}$ 
          above the $Z$ boson resonance region roughly by a factor 
          $C\cdot M_Z/\Gamma_Z$, with $C$ being a factor 
          of $O(1)$ \cite{Riemann:1997kt}.
\end{itemize} 
Corresponding effects to the forward-backward asymmetries 
$A_{FB}(s)$ at or above the $Z$ boson resonance
\cite{Bohm:1989pb,Riemann:1992gv,Riemann:1997kt,Riemann:1997tj}
will be briefly mentioned at the end of 
our presentation for total cross sections there 
\cite{Barroso:1987ae,Berends:1988bg,Bardin:1989qr,Beenakker:1990ec,Riemann:1997kt,Riemann:1997tj}.

The dominant QED corrections to the total cross section $\sigma(s)$ 
by initial state bremsstrahlung are given by following approximate 
formula:
\ba 
\label{crossini}
\sigma(s) &=& 
\int^{s}_{s_{\min}}
\,\frac{d{s'}}{s}\,\sigma^0(s') R(v) 
\\
&=&
\label{crossini2}
\int^{s}_{s_{\min}}
\,\frac{d{s'}}{s}\,\sigma^0(s')
\cdot\left[(1+\bar{S})\beta_e v^{\beta_e-1}+\bar{H}(v)\right],
\\
&& \mbox{with}\quad v\equiv 1-\frac{s'}{s}.
\ea

The term $\sigma^0(s')$ is the effective Born cross section. 
The variable $s'$ is the effective invariant mass squared after 
initial state radiation and $v$ is equivalent to the 
normalized photon energy.
The {\it improved Born term} $\sigma^0(s')$ contains the 
remaining electroweak and QCD corrections. At the $Z$ 
boson resonance they decouple from the pure QED radiative 
corrections and can be described accurately through effective 
couplings, form factors, and an $s$-dependent $Z$ boson width
in the effective Born expressions
\cite{Akhundov:1986fc,Bardin:1988xt,Bardin:1989di}. 
%For the following discussion, details of the electroweak
%and QCD part are not important here, but will be addressed
%again in the main part of the thesis Chapter (\ref{ch_lep1slc}).

For the QED description, $\sigma^0(s')$ is convoluted over 
$s'$ with a flux function $R(v)$ for the QED bremsstrahlung. 
The integration reaches from a minimum value $s_{\min}$ 
to the maximal value $s$. Finite corrections from soft and virtual photons
or from hard photons are contained in the terms $\bar{S}$ 
and $\bar{H}$, respectively. The factor $\beta_e v^{\beta_e-1}$ 
denoted in (\ref{crossini}) combines all leading logarithmic soft 
and virtual photonic corrections in a {\it soft and virtual photon exponentiation}
\cite{Yennie:1961ad,Greco:1975rm,Greco:1980}.
%
%Kinematical cuts to the hard photon phase space, with 
%exception of a trivial cut on the maximal energy of the radiated 
%photon, are not of relevance for this introductory 
%discussion of the peak cross section. They will, however, play a 
%major role in the rest of this work, when having in mind sub-per-mil
%level cross section predictions at the $Z$ boson resonance 
%and better than per cent predictions at LEP~2 or LC energies. 

In order to demonstrate the QED effects on the  
cross section maximum at the $Z$ boson resonance (peak cross sections), 
it is instructive in a first approach to use a Breit-Wigner
resonance formula with a constant width $\Gamma_Z$, where the 
$\gamma Z$ interference term $\sigma^0_{\gamma Z}$ and the pure 
QED part $\sigma^0_{\gamma}$ (see \ref{simple_peak1} to \ref{simple_qed}) 
have been omitted for simplicity.
Their corrections to the peak cross sections will then be 
discussed afterwards. 

%-----------------------------------------------
\subsection*{\large{Simple approach}
\label{sub_simple}}
%-----------------------------------------------
%
So, starting with the effective Born cross section $\sigma^0(s)$
of our convolution integral (\ref{crossini}) for the corrected 
peak cross section, we have
\ba
\label{simple_peak1}
\sigma^0(s) &=& \sigma^0_{Z}(s) + \sigma^0_{\gamma Z}(s)
                 + \sigma^0_{\gamma}(s)\approx \sigma^0_{Z}(s),
             \\
\sigma^0_{Z}(s) &=&
             \sigma^0_{\max}\cdot\frac{s\Gamma_Z^2}
             {(s-M_Z^2)^2+ M_Z^2\Gamma_Z^2},
\label{simple_res}
             \\
\sigma^0_{\gamma Z}(s) &=&
             \frac{4\pi\alpha^2}{3}J_f\frac{s-M_Z^2}
             {(s-M_Z^2)^2+ M_Z^2\Gamma_Z^2},
\label{simple_gZ}
             \\
\sigma^0_{\gamma}(s) &=&
             \frac{4\pi\alpha^2}{3 s} Q_e^2 Q_f^2 {N_c}_f,
\label{simple_qed}
\ea
with the factor $J_f$ given by {\tt SM} couplings 
and the {\tt SM} maximal Born cross section $\sigma^0_{\max}$ by
\ba
\label{born_peak}
\sigma^0_{\max} &=& \frac{12\pi\Gamma_e\Gamma_f}
{M_Z^2\,\Gamma_Z^2}.
\ea

In this simple approach \cite{Barroso:1987ae,Bardin:1989qr,Beenakker:1990ec},
we just consider the Breit-Wigner resonant part $\sigma^0_{Z}(s)$
of $\sigma^0(s)$ in (\ref{crossini}) and omit the hard photon 
term $\bar{H}(v)$. For the integration of the soft photon part 
over the variable $v\equiv 1-R$ only a negligible numerical error
is introduced when one extends the integration region from $[0;1]$ to $[0;\infty]$.
\ba
\label{simple_peak2}
\sigma(s) &=& 
\int^{s}_{s_{\min}}
\,\frac{d{s'}}{s}\,
\sigma^0(s')
\,(1+\bar{S})\,\beta_e v^{\beta_e-1}
\\
&\approx&
 \sigma^0_{\max}(1+\bar{S})\,\frac{M_Z^2\Gamma_Z^2}{s}\, 
\beta_e
\int\limits^{\infty}_{0}
{d{v}}\,
v^{\beta_e-1}
\,\frac{1-v}{v^2+2\eta\cos\zeta v+\eta^2} 
\label{simple_peak2b}
\\
&\approx&
 \sigma^0_{\max}(1+\bar{S})\,\frac{M_Z^2\Gamma_Z^2}{s}\cdot 
\left\{
{\cal J}_{\beta_e}\left[\eta\left(\frac{M_Z}{s},\frac{\Gamma_Z}{s}\right),
\zeta\left(\frac{M_Z}{s},\frac{\Gamma_Z}{s}\right)\right]
\right.
\nonumber\\
&&
\hspace*{2cm}-\left.
\frac{\beta_e}{\beta_e+1}
{\cal J}_{\beta_e+1}\left[\eta\left(\frac{M_Z}{s},\frac{\Gamma_Z}{s}\right),
\zeta\left(\frac{M_Z}{s},\frac{\Gamma_Z}{s}\right)\right]
\right\},
\label{simple_peak2c}
\ea
with 
\ba
\label{simple_J}
{\cal J}_{\beta_e}(\eta,\zeta) &=& \beta_e 
\int^\infty_{0}\,\frac{d{x}}{x}\, x^{\beta_e}
\cdot\frac{1}{x^2+2\eta\cos\zeta\cdot x+\eta^2}
\\
\nonumber\\
\label{simple_J2}
&=& \eta^{\beta_e-2}\cdot \Phi(\cos\zeta,\beta_e),
\\
\nonumber\\
\label{simple_J3}
\Phi(\cos\zeta,\beta_e) &=& 
\frac{\pi\beta_e\sin\left[(1-\beta_e)\zeta\right]}
{\sin(\pi\beta_e)\sin\zeta},  
\ea
and
\ba
\label{simple_para}
\eta^2 &=& a^2 + b^2,
\quad\cos\zeta = \frac{a}{\eta},\quad \sin\zeta = \frac{b}{\eta},
\quad a = M_Z^2/s-1,\quad b = (M_Z\Gamma_Z)/s.
\nonumber\\ 
\ea
The above equation (\ref{simple_peak2c}) then leads us to:

\ba
\label{simple_peak3}
\sigma_{\max}
&=& \sigma(\sqrt{s_{\max}})
= 
\sigma^0_{\max}\, (1+\bar{S})\, \frac{M_Z\Gamma_Z}{s} 
\,
\eta^{\beta_e-1}
\left\{ 
\frac{\pi\beta_e\sin\left[(1-\beta_e)\zeta\right]}
{\sin(\pi\beta_e)}
\right.
\nonumber\\
&&
\hspace*{2.5cm}-\left.
\eta\,\frac{\beta_e}{\beta_e+1}\,
\frac{\pi(\beta_e+1)\sin\left[(1-(\beta_e+1))\zeta\right]}
{\sin(\pi(\beta_e+1))}
\right\}.  
\ea
For the evaluation of the peak height $\sigma_{\max}$, 
we can safely neglect the second term proportional to $\eta$ 
in (\ref{simple_peak3}) with $\eta=\Gamma_Z/M_Z \ll 1$ and $\zeta=\pi/2$:

\ba
\label{simple_peak4}
\sigma_{\max}
&=& 
\sigma^0_{\max}\,(1+\bar{S})\,
\frac{\pi\beta_e\sin\left[(1-\beta_e)(\pi/2)\right]}
{\sin(\pi\beta_e)}
\cdot \left(\frac{\Gamma_Z}{M_Z}\right)^{\beta_e}
\\
&=& 
\sigma^0_{\max}\,
\frac{\pi\beta_e/2}{\sin(\pi\beta_e/2)}\,(1+\bar{S})
\cdot \left(\frac{\Gamma_Z}{M_Z}\right)^{\beta_e}.
\ea
With $\frac{\pi\beta_e/2}{\sin(\pi\beta_e/2)}\approx 1$, 
we therefore observe the decrease of the Born peak
cross section by a factor $(\Gamma_Z/M_Z)^{\beta_e}$ 
\cite{Greco:1967,Etim:1967} with a small correction 
term $(1+\bar{S})$ \cite{Bardin:1989qr,Beenakker:1990ec}. 
This effect at the $Z$ peak is solely produced 
by the multiple soft and virtual photon corrections 
and independent of the small hard photon emission
there. It is therefore a universal, i.e.~completely process 
and cut-independent phenomenon.
 
In order to see the shift of the peak position $\sqrt{s_{\max}}$, 
we introduce the variable $y:=(s-M_Z^2)/(\Gamma_Z M_Z)$
and use the fact that the shift $\sqrt{s_{\max}}$ will be 
negligibly small compared to the $Z$ boson width $\Gamma_Z$. 
With $y\ll 1$ and $\gamma\equiv \Gamma_Z/M_Z\ll 1$, we finally get,
omitting terms of $O(y^3)$ and $O(y\gamma)$:
%
%\ba
%\label{simple_app1}
%&&\eta = \frac{M\Gamma}{s}\left(1+\frac{1}{2}y^2+O(y^2)\right),
%\quad
%\eta^{\beta_e-1} = (\frac{M\Gamma}{s})^{\beta_e-1}
%\left(1+\frac{\beta_e-1}{2}y^2+O(y^2)\right),
%\nonumber\\
%\\
%&&\sin\left[(1-\beta_e)\zeta\right]
%=
%\sin\left[(1-\beta_e)\left(\frac{\pi}{2}-y+O(y^3)\right)\right]
%\nonumber\\
%&=&
%\cos(\frac{\beta_e\pi}{2}) + (1-\beta_e)\sin(\frac{\beta_e\pi}{2})y
%- \frac{(1-\beta_e)^2 y^2}{2}\cos(\frac{\beta_e\pi}{2})y^2+O(y^3).
%\ea
%
%Similarly, we can expand the overall factor 
%$(M\Gamma/s)^{\beta_e}$ in (\ref{simple_peak3}) 
%for $\Gamma/M\ll 1$, neglecting all terms 
%of $O(\Gamma/M\cdot y)$. 
%
%\ba
%\label{simple_app2}
%(\frac{M\Gamma}{s})^{\beta_e} &\approx& 
%(\frac{\Gamma}{M})^{\beta_e}\left[
%1-\beta_e\frac{\Gamma}{M}y+\frac{1}{2}\beta_e(\beta_e+1)
%\left(\frac{\Gamma}{M} y\right)^2+O(y^3)
%\right] 
%\ea
%
%The logarithmic term $\beta_e = \beta_e(s)\approx \beta_e(M_Z^2)$ 
%can safely be treated as constant for the small deviations
%$y$, which we are considering. With these approximations 
%and inserting (\ref{simple_app1}) into (\ref{simple_peak3}), 
%omitting all terms of $O(y^n)$ for $n\geq 3$, we get:
%
\ba
\label{simple_peak5}
\sigma(s(y))
&=&
\sigma^0_{\max}\,\frac{\pi\beta_e}{\sin(\pi\beta_e)}\, 
(1+\bar{S})\,\left(\frac{\Gamma}{M}\right)^{\beta_e}\cdot
\\
&&
\left[
\cos\left(\frac{\beta_e\pi}{2}\right) 
+ (1-\beta_e)\,\sin\left(\frac{\beta_e\pi}{2}\right)y
- \frac{(1-\beta_e)^2}{2}\cos\left(\frac{\beta_e\pi}{2}\right)y^2
\right].
\nonumber
\ea
The logarithmic term $\beta_e=\beta_e(s\approx M_Z^2)$ 
from (\ref{betae}) can safely be treated as constant 
for the small deviations $y$ which are considered here. 
Setting ${d{\sigma}}/{d{s}}(s=s_{\max})=0$,
%$\frac{d{\sigma}}{d{s}}|_{s=s_{\max}}=
%1/(M\Gamma)\cdot\left.\frac{d{\sigma}}{d{y}}\right|_{y=y(s_{\max})} = 0$,
we finally obtain the following result for the peak position 
$\sqrt{s}_{\max}$ which is 
slightly shifted with respect to $M_Z$ through the QED effects 
\cite{Barroso:1987ae,Berends:1988bg,Bardin:1989qr,Beenakker:1990ec}:
\ba
\label{simple_peak6}
&& y = \frac{\tan(\pi\beta_e/2)}{2-\beta_e} 
\approx \frac{\pi\beta_e}{4},
\\
\label{peak_pos2}
\longrightarrow&&
\sqrt{s}_{\max} \approx M_Z\left(1+\frac{\pi\beta_e}{8}\gamma\right)
\quad\mbox{with}\quad \gamma\equiv \frac{\Gamma_Z}{M_Z}.
\ea

%-----------------------------------------------
\subsection*{\large{Effects by an $s$ dependent width $\Gamma_Z(s)$}
\label{sub_swidth}
}
%-----------------------------------------------
%
The total $Z$ boson decay width $\Gamma_Z$ can be calculated from 
perturbation theory and is basically given by the imaginary part of 
the self-energy correction $\Sigma_Z$ to the $Z$ boson propagator 
\cite{Bardin:1988xt,Berends:1988bg}:
\ba
\label{more_peak0}
M_Z\Gamma_Z &=& \frac{{\Im m}\left(\Sigma_Z(M_Z^2)\right)}
{1+\Pi_Z\left(M_Z^2\right)},
\quad 
\Pi_Z(s) = \frac{\partial}{\partial{s}}
{\Re e}\left(\Sigma_Z(s)\right). 
\ea
The value $\Gamma_Z$ can be calculated in different renormalization schemes, 
for example, with $M_Z$, $G_\mu$, and $\alpha_{em}(M_Z^2)$ as input values.
$G_\mu$ is the muon decay constant and can be 
written at tree-level as $G_\mu=\pi\alpha/(\sqrt{2}\sin^2\theta_W M_W^2)$,
while $\alpha_{\rm em}(M_Z^2)$ is the running 
electromagnetic coupling constant derived at $s=M_Z^2$.
This introduces an $s$-dependency to the width $\Gamma_Z$
\cite{Akhundov:1986fc,Jegerlehner:1986vs,Beenakker:1988pv}:
\ba
\label{more_peak1}
\Gamma_Z(s) &=& \frac{\sqrt{2} G_\mu M_Z s}{3\pi} 
\sum\limits_f {N_c}_f (v_f^2+a_f^2)\equiv \frac{s}{M_Z^2} \Gamma_Z,
\ea
with ${N_c}_f$ as fermionic color factor, $v_f$ and $a_f$ 
as weak neutral current couplings, and the constant 
$\Gamma_Z$ extracted as a factor in the way 
given in (\ref{more_peak1}). 

In a more realistic approach, we therefore have to 
replace the constant-width resonance curve in (\ref{simple_peak1}) 
with the following Breit-Wigner ansatz, which provides a much better 
description of the actual case:
\ba
\label{simple_peak1s}
\sigma^0(s') &=&
             \sigma^0_{\max}\cdot\frac{s'\Gamma_Z^2}
             {(s'-M_Z^2)^2+ {s'}^2(\Gamma_Z/M_Z)^2}.
\ea

It is now straightforward to show that the $s$-dependency
of the width can be removed in the denominator 
by the {\it $Z$ boson transformation} 
\cite{Bardin:1988xt,Bardin:1989di,Leike:1991pq,Riemann:1997tj}:
\ba
\label{more_peak2}
M_Z &=& \frac{\bar{M}_Z}{\sqrt{1+\Gamma_Z^2}},
\\
\Gamma_Z &=& \frac{\bar{\Gamma}_Z}{\sqrt{1+\Gamma_Z^2}},
\\
G_\mu &=& \frac{\bar{G}_\mu}{1-i\gamma}
\quad \mbox{with}\quad 
\gamma\equiv\frac{\Gamma_Z}{M_Z}=\frac{\bar{\Gamma}_Z}{\bar{M}_Z}.
\ea
This leads us from (\ref{simple_peak1s}) to following Born 
expression which is to be convoluted into (\ref{simple_peak2})
with the QED radiator there:
\ba
\label{simple_peak2s}
\sigma^0(s') &=&
             \sigma^0_{\max}\cdot\frac{s'\bar{\Gamma}_Z^2}
             {(s'-\bar{M}_Z^2)^2 + (\bar{M}_Z\bar{\Gamma}_Z)^2}.
\ea

This $Z$ boson transformation in (\ref{more_peak2}) 
produces an effective $Z$ boson mass $\bar{M}_Z$ and a 
new effective constant width $\bar{\Gamma}_Z$.\footnote{
The back transformation $\bar{M}_Z,\bar{\Gamma}_Z \to M_Z,\Gamma_Z$ 
is absolutely symmetric with 
$\bar{M}_Z = M_Z/\sqrt{1+\Gamma_Z^2}$,
$\bar{\Gamma}_Z = \Gamma_Z/\sqrt{1+\Gamma_Z^2}$,
and 
$\gamma\equiv \frac{\Gamma_Z}{M_Z}$.
}
The next term in the expansion yields \cite{Bardin:1989qr,Beenakker:1990ec}:
\ba
\label{peak_pos3}
\sqrt{s}_{\max} \approx 
\bar{M}_Z\left(1+\frac{\pi\beta_e}{8}\gamma
+\frac{1}{4}\gamma^2\right).
\ea

The peak position is now given in terms of the effective values 
$\bar{M}_Z$ and $\bar{\Gamma}_Z$ and after transformation 
can be re-expressed in terms of the $Z$ boson mass and width, 
$M_Z$ and $\Gamma_Z$, which naturally produces a further negative shift
\cite{Bardin:1988xt,Riemann:1997kt,Riemann:1997tj}:
\ba
\label{peak_pos4}
\sqrt{s}_{\max} \approx M_Z\left(1+\frac{\pi\beta_e}{8}\gamma
-\frac{1}{4}\gamma^2\right).
\ea

The Born level shift $+\frac{1}{4}\gamma^2$ 
in (\ref{peak_pos3})
of the peak position with respect to $M_Z$
is about $+17\,\mbox{MeV}$.
The additional shift $\Delta\sqrt{s}_{\max}$ of the 
peak position in (\ref{peak_pos4}) 
then amounts to about $+112\,\mbox{MeV}$ \cite{Bardin:1989qr} 
for the leading logarithmic approximation of soft and virtual photons
and to about $+128\,\mbox{MeV}$ taking into account
also the $O(\alpha^2)$ QED corrections with a soft and 
virtual photon resummation \cite{Bardin:1988xt}.
Finally a further negative shift of $-34\,\mbox{MeV}$ arises from 
the $s$-dependence of the width \cite{Bardin:1988xt}. 
The shift agrees nicely within a few $\mbox{MeV}$
with the experimental observation \cite{Bardin:1989qr}.
%The peak height $\sigma_{\max}$ is naturally not 
%affected by the introduction of an $s$-dependent width 
%$\Gamma_Z(s)$ with $\Gamma_Z(s=M_Z^2)\equiv\Gamma_Z$.

Further small corrections to the height and position 
of the peak cross section are introduced by the 
$\gamma Z$ interference and pure QED terms 
$\sigma^0_{\gamma Z}(s')$ and $\sigma^0_{\gamma}(s')$   
in the effective Born term (\ref{simple_peak1}). 
%While $\sigma^0_{\gamma Z}(s')$ disappears at the $Z$ boson 
%resonance and has no effect on the peak position, 
The QED part $\sigma^0_{\gamma}(s')$ especially has an 
effect for the muon pair final state due to logarithmic 
corrections proportional to $\ln(M_Z^2/m_{\mu}^2)$.\footnote{
To see this, one has to replace the lower integration 
limit in (\ref{simple_peak2}) by $4 m_{\mu}^2/s$.
}
This slightly increases $\sigma_{\max}$ for muon pairs
by a few per cent with respect to the hadronic case.
The QED term does not change the peak position. 

The peak position is however modified by $\sigma^0_{\gamma Z}(s')$,
which can also be measured with the available experimental precisions.
The $\gamma^2$-dependent term in (\ref{peak_pos4}) is 
replaced by \cite{Riemann:1997kt,Riemann:1997tj}:
\ba
\label{peak_pos5}
+\frac{1}{4}\Gamma_Z^2\left(1+\frac{{\cal J}}{{\cal R}}\right)
-\frac{1}{2}\Gamma_Z^2,
\ea
with the factor ${\cal J/R}$ parameterizing the relative 
weight of the coupling factors to $\sigma^0_{Z}(s')$
and $\sigma^0_{\gamma Z}(s')$.

These effects can finally also be discussed in the context of 
$ Z Z'$ mixing when searching for extra heavy neutral 
gauge bosons $Z'$ predicted in different extensions to
the {\tt SM}. This has been treated in a 
model-independent approach in \cite{Leike:1992uf}
and produces a further shift of $M_Z$ depending on 
the $Z Z'$ mass splitting and neutral current couplings
to $Z$ and $Z'$ which can be experimentally tested. 

\subsection*{\large{The radiative tail from hard photon emission}
\label{sub_tail}
}
%-----------------------------------------------
%
The third important effect of QED radiative corrections
to the peak cross section is the radiative tail, 
i.e.~the cross section enhancement compared to the 
Born situation, observed at energies 
above the $Z$ boson resonance. 
%
%This happens 
%if no or only weak kinematical cuts are applied 
%to the final state, i.e.~if hard photon emission is 
%kinematically allowed. We can correctly derive the dominant 
%hard photon effects in our previous analytic approach
%inserting the first-order hard photonic radiator $\bar{H}(v)$ 
%of (\ref{crossini}), (\ref{softvirthard}) and (\ref{firstorder})
%into (\ref{simple_peak2}) and using one of the discussed 
%effective Born terms $\sigma^0(s')$. For the detailed 
%analytic calculation, we would like to refer to \cite{Cahn:1987qf}.

\begin{figure}[htp] 
\begin{flushleft}
\vspace*{-0.5cm}
\begin{tabular}{ll}
\hspace*{-0.5cm}
\mbox{
 \epsfig{file=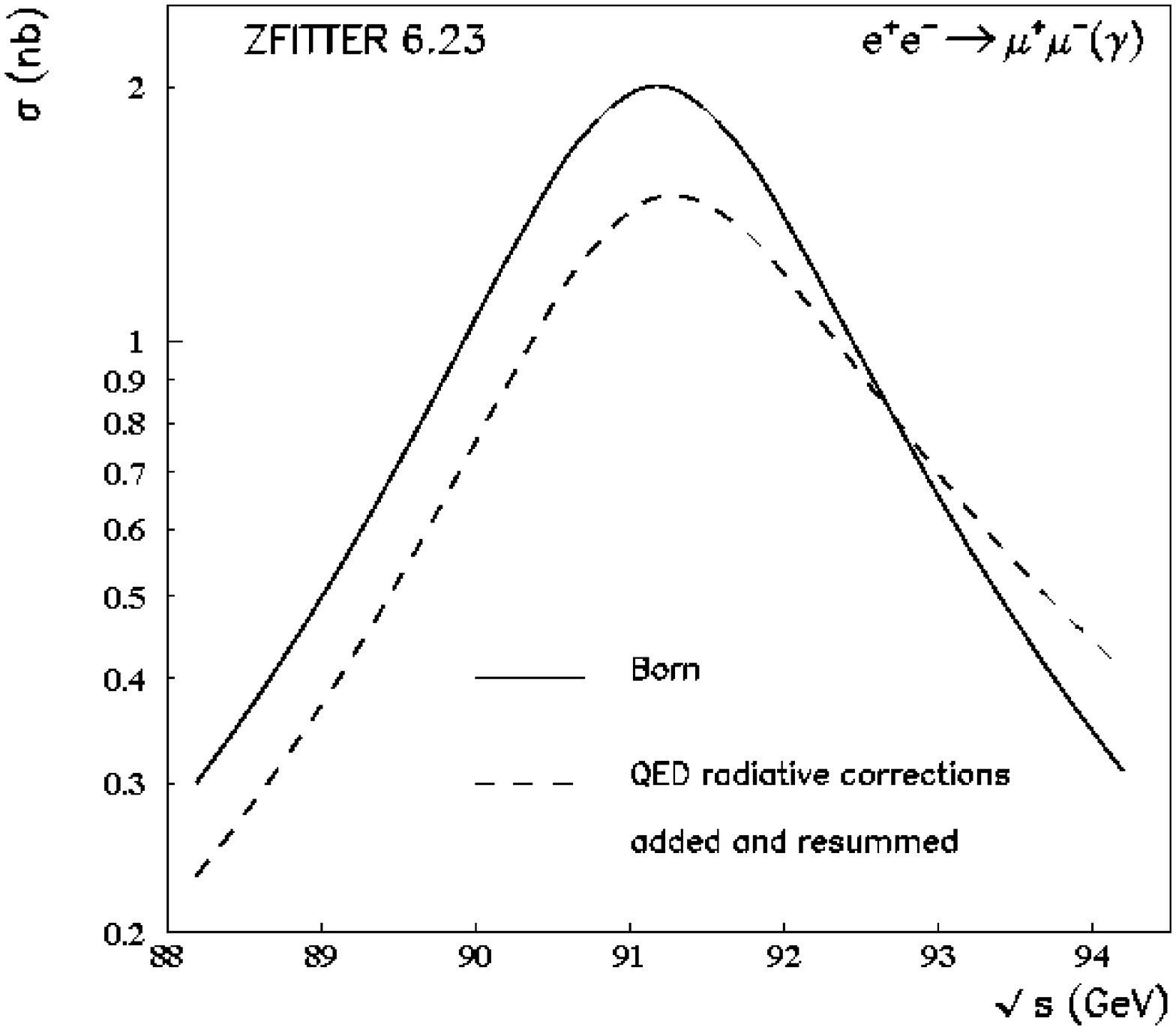,width=7.25cm}}
\vspace*{-0.25cm}
&
\hspace*{-0.8cm}
\mbox{
 \epsfig{file=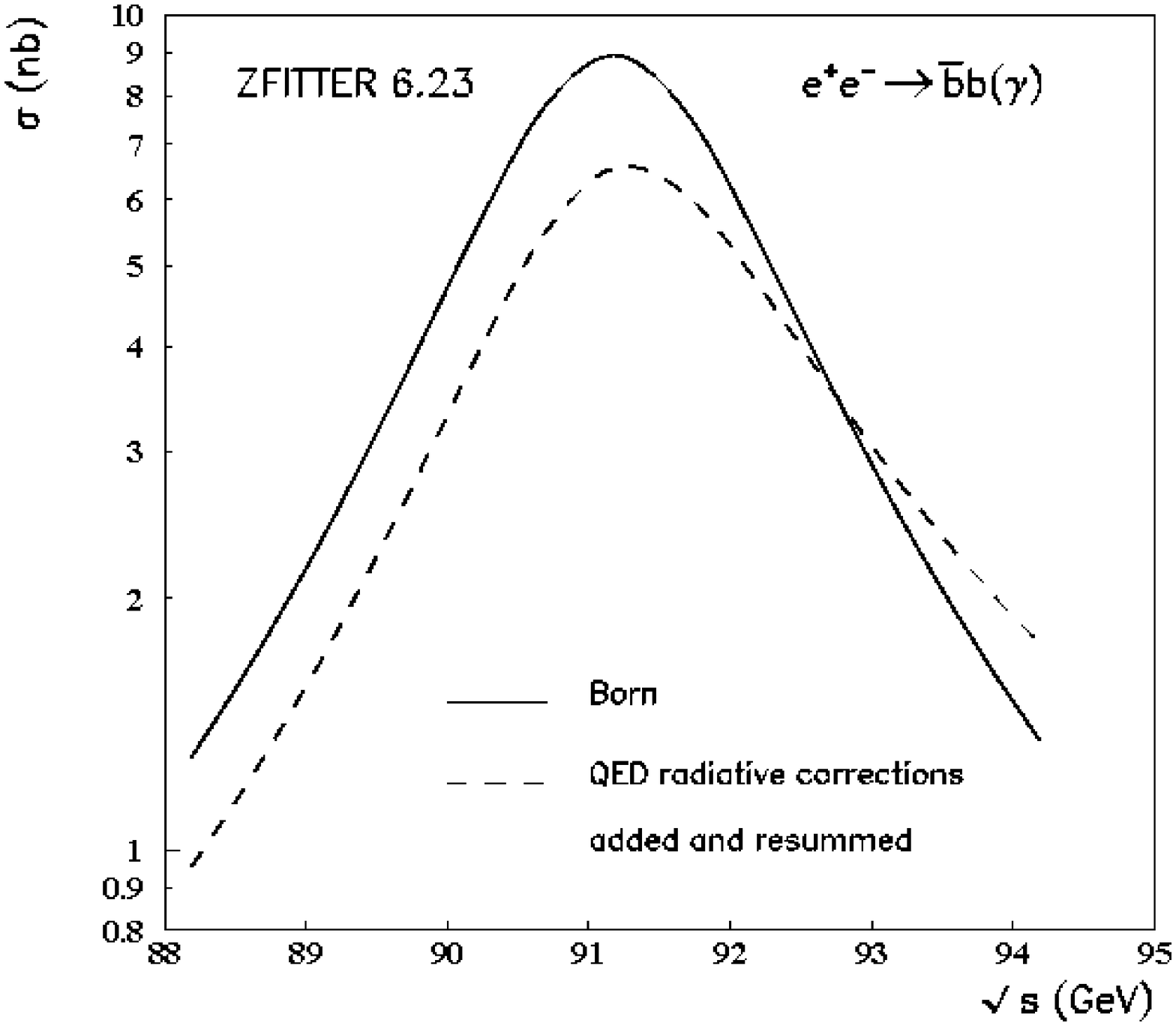,width=7.25cm}}
\vspace*{-0.25cm}
\\
\hspace*{-0.5cm}
\mbox{
 \epsfig{file=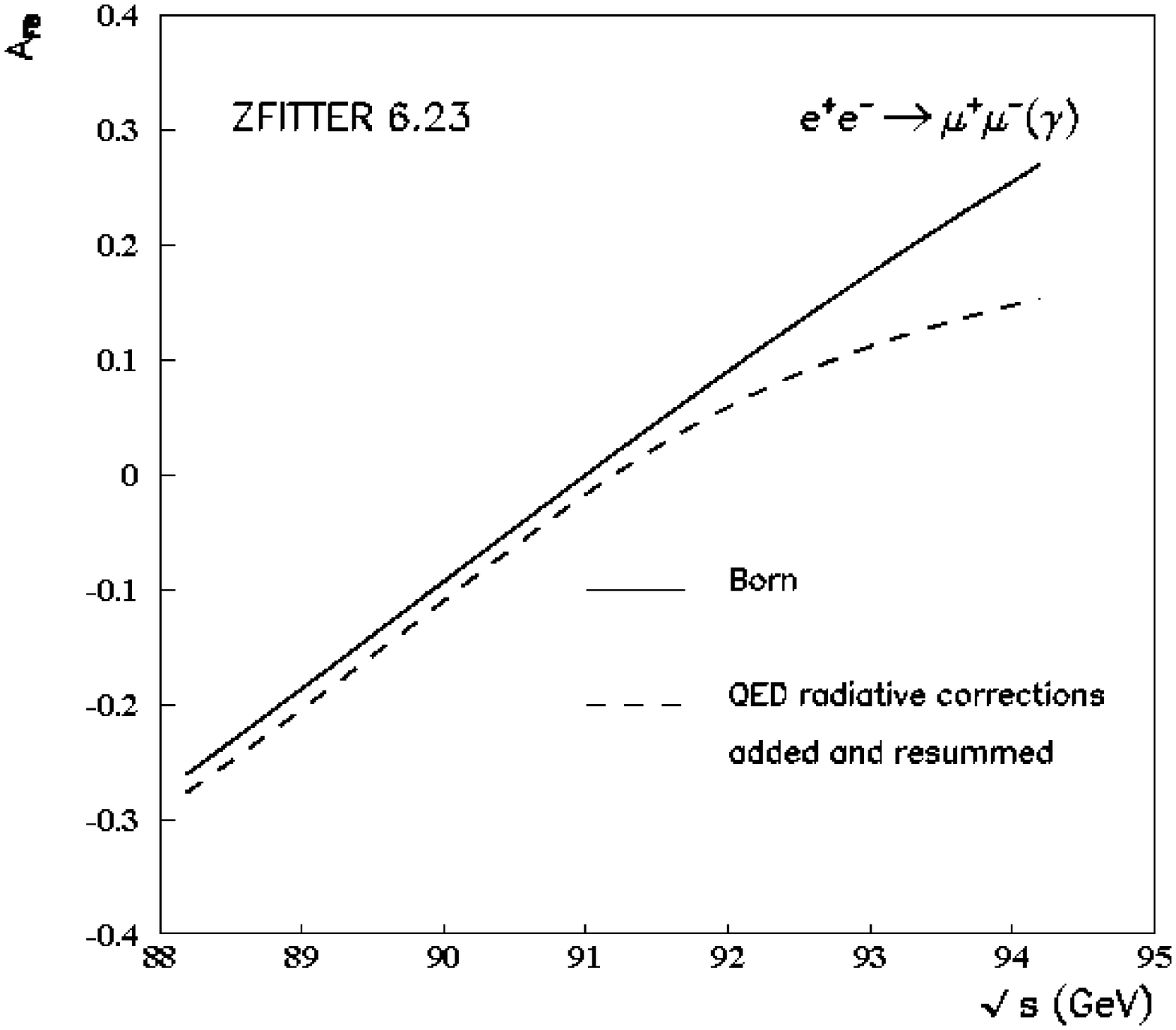,width=7.25cm}}
&
\hspace*{-0.8cm}
\mbox{
 \epsfig{file=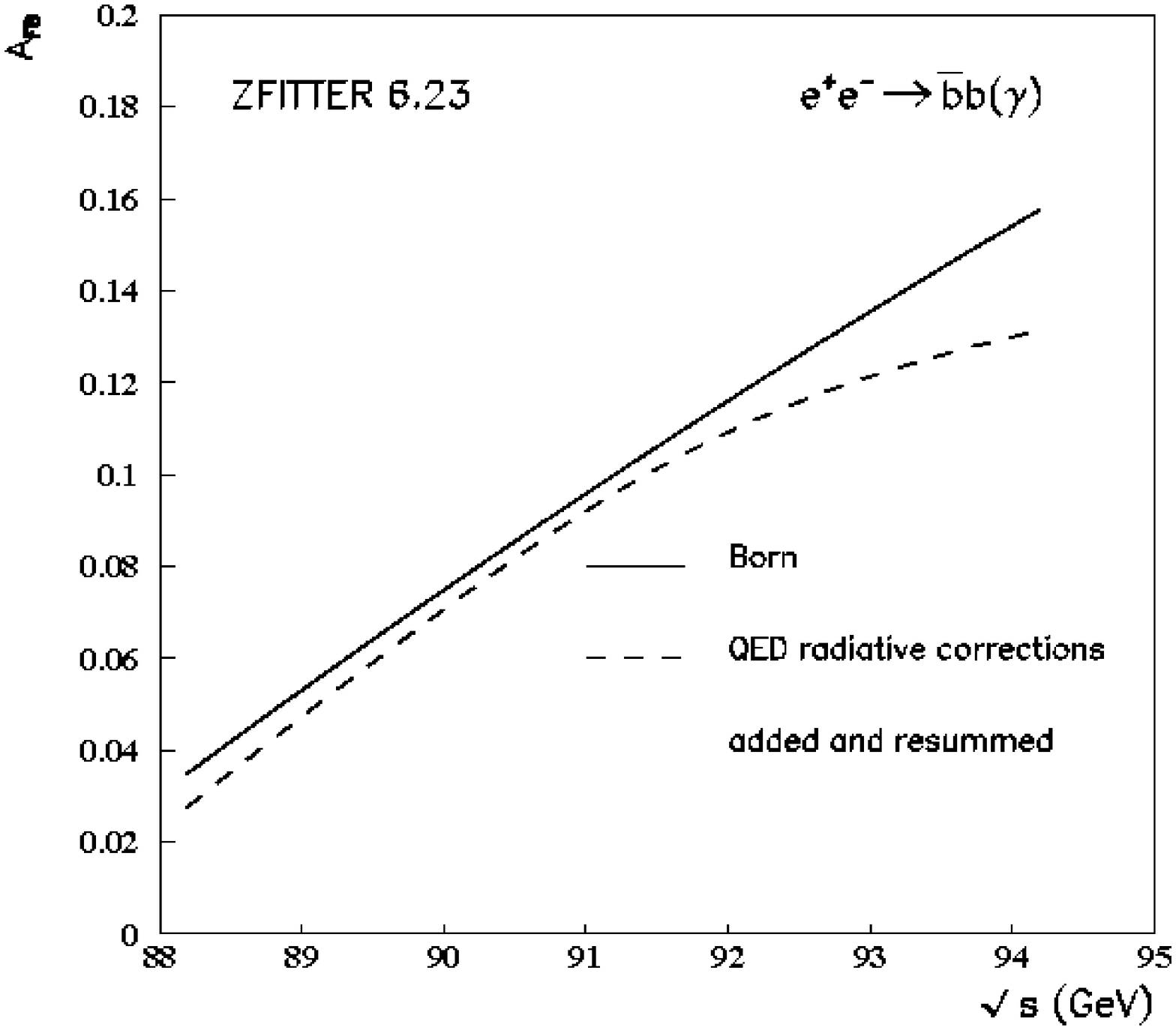,width=7.25cm}}
\end{tabular}
\end{flushleft}
\vspace*{-1cm}
\caption[Peak cross sections and forward-backward asymmetries]
{\sf
Born and QED corrected 
peak cross sections and forward-backward asymmetries
for muon and $b$ quark pair production calculated 
with {\tt ZFITTER} v.6.23 
\cite{Bardin:1992jc2,Bardin:1999yd-orig}.
QED corrections include full $O(\alpha^2)$ and 
leading logarithmic $O(\alpha^3)$ corrections
with soft  and virtual photon exponentiation 
for initial state bremsstrahlung. 
\label{sig_afb_peak} 
} 
\end{figure}

We want to give a quick ad-hoc computation 
of the tail effect. First, for the Born case,
the resonance part in (\ref{simple_peak1s}) 
develops for $\sqrt{s}> M_Z$ roughly like 
\ba
\label{tail1}
\sigma^0(s) = \frac{C}{M_Z^2}\cdot 
\frac{s}{(s-M_Z^2)^2+\Gamma_Z^2 M_Z^2}
\approx \frac{C}{M_Z^2}\cdot \frac{1}{s},
\ea
with a constant factor $C\approx O(1)$.
For the QED convoluted case we can write 
generically for $\sigma(s)$: 

\ba
\label{tail2}
\sigma(s) &=& C\cdot\int^{s}_{s_{\min}}
\frac{d{s'}}{s}\, f\left(\frac{s'}{s}\right)\cdot 
\frac{1}{|s'-M_Z^2 + i M_Z\Gamma_Z|^2},
\ea
with $f\left(s'/s\right)$ being the QED radiator and 
$1/|s-M_Z^2 + iM_Z\Gamma_Z|^2$ the squared 
$Z$ boson propagator. (\ref{tail2}) can then be 
reexpressed in terms of the imaginary part of 
the propagator:

\ba
\label{tail3}
\sigma(s) &=& \frac{C}{M_Z\Gamma_Z}
\cdot\int^{s}_{s_{\min}}
\frac{d{s'}}{s}\, f\left(\frac{s'}{s}\right)\cdot 
{\Im m}\left(\frac{1}{s'-M_Z^2 + i M_Z\Gamma_Z}\right).
\ea
We first consider the case, $s_{\min} < M_Z < s$,
taking into account QED bremsstrahlung through the 
flux function $f\left(s'/s\right)$.
The function $f\left(s'/s\right)$ evolves slowly over most 
of the integration region in (\ref{tail2}) and we can 
replace $f\left(s'/s\right)$ by some medium value
$f\left(s_0/s\right)$. Its detailed structure is 
not important here. We have therefore following approximation:

\ba
\label{tail4}
\sigma(s) &\approx& -\frac{C}{M_Z\Gamma_Z}\cdot f\left(\frac{s_0}{s}\right) 
\cdot\int^{s}_{s_{\min}}
\frac{d{s'}}{s}\,  
{\Im m}\left(\frac{1}{s'- M_Z^2 + i M_Z\Gamma_Z}\right)
\\
&=&
-\frac{C'}{M_Z\Gamma_Z}
\cdot\int^{s}_{s_{\min}}
\frac{d{s'}}{s}\,  
{\Im m}\left(\frac{1}{s'- M_Z^2 + i M_Z\Gamma_Z}\right).
\ea
With the general relation 
$\ln(-|R|\pm i\varepsilon)\approx \ln|R|\pm i\pi$,
we can obtain the following approximation for $\sigma(s)$:

\ba
\label{tail5}
&&\int^{s}_{s_{\min}}\frac{d{s'}}{s}\,  
{\Im m}\left(\frac{1}{s'- M_Z^2 + i M_Z\Gamma_Z}\right)
=
\frac{1}{s}{\Im m}\left\{\ln\left(-\frac{s- M_Z^2 + i M_Z\Gamma_Z}
{M_Z^2-s_{\min}- i M_Z\Gamma_Z}\right)\right\} 
\nonumber\\
&=&
\frac{1}{s}{\Im m}\left\{\ln\left(-\frac{s - M_Z^2}{M_Z^2-s_{\min}} 
- i M_Z\Gamma_Z 
\frac{s-s_{\min}}{(s - M_Z^2)(M_Z^2-s_{\min})}\right)\right\} 
\approx -\pi\cdot\frac{1}{s},
\nonumber\\
\\
&\rightarrow &\qquad\qquad
\sigma(s) = {C'''}\frac{\pi}{M_Z\Gamma_Z}\cdot\frac{1}{s}.
\ea
For the last step in (\ref{tail5}) $\Gamma_Z/M_Z\ll 1$ was used.
The factors $C$, $C'$, $C''$, and $C'''$ are all terms of $O(1)$.
We finally observe a substantial cross section enhancement
by roughly a factor $M_Z/\Gamma_Z$ \cite{Riemann:1997kt}, 
comparing (\ref{tail5}) with the Born case (\ref{tail1}). 
The radiative tail effects above the $Z$ peak 
can also be seen from Fig.~\ref{sig_afb_peak}.

If we had considered the case, 
$M_Z < s_{\min} < s$, through a sufficiently strong cut $s_{\min}$
on $s'$ against hard photon emission, we would have had instead:
$\ln(|R|\pm i\varepsilon)\approx\ln|R|\pm i\varepsilon$
and with this,
\ba
\label{tail6}
&&\int^{s}_{s_{\min}}\frac{d{s'}}{s}\,  
{\Im m}\left(\frac{1}{s'- M_Z^2 + i M_Z\Gamma_Z}\right)
=
\frac{1}{s}{\Im m}\left\{\ln\left(\frac{s - M_Z^2 + i M_Z\Gamma_Z}
{s_{\min}- M_Z^2 + i M_Z\Gamma_Z}\right)\right\}  
\nonumber\\
&=&
\frac{1}{s}{\Im m}\left\{
\ln\left(\frac{s - M_Z^2}{s_{\min}-M_Z^2} 
- i \frac{\Gamma_Z}{M_Z}\cdot 
\frac{M_Z(s - s_{\min})}{(s_{\min}-M_Z^2)^2}
\right)\right\} 
= -O\left(\frac{\Gamma_Z}{M_Z}\right), 
\ea
and therefore no radiative tail is developed.
These general considerations do not change for the more 
realistic case with an $s$-dependent width, which can be 
easily seen using the $Z$ boson transformation.
The changes of peak height and position have been 
depicted for muon and $b$ quark pair cross sections
in Fig.~\ref{sig_afb_peak}.
%
%--------------------------------------------------------------------------------
%--------------------------------------------------------------------------------

The above discussion can be similarly done
for the shift of the zero position of $A_{FB}$ with respect
to $M_Z$ \cite{Riemann:1992gv,Riemann:1997kt,Riemann:1997tj}.
The figure also shows the changes to forward-backward asymmetries 
$A_{FB}$ at the $Z$ peak and the shifts of peak positions 
and $A_{FB}$ zero positions.

An overall correction factor arises from hard photons 
compared to the Born asymmetries $A^0_{FB}$ which lowers the asymmetry 
values for growing $s>M_Z^2$ (see Fig.~\ref{sig_afb_peak}). 
In order to derive this, it is crucial 
to take into account the $\gamma Z$ interference term 
$\sigma^0_{FB,\gamma Z}(s')$ in the Born asymmetry 
$\sigma^0_{FB}(s')$, which is added to the resonant part 
$\sigma^0_{FB,Z}(s')$ and convoluted with the QED flux function 
in analogy to (\ref{simple_peak1}). 
%The complete analytic discussion can also be 
%found in more detail for cross sections and 
%asymmetries in \cite{Cahn:1987qf,Bardin:1989qr,Bohm:1989pb}. 
Further numerical analysis is also given in 
\cite{Berends:1988ab,Montagna:1989at,Bardin:1989di}.
%\cite{Berends:1988ab,Nicrosini:1988sw,Montagna:1989at,Bardin:1989di}.
More recently some discussion has also started in this context  
on the importance
of electroweak two-loop corrections in self-energy corrections
to the $Z$ boson propagator and on the definition of masses and 
widths of massive bosons like the $Z$ and $Higgs$ boson 
\cite{Willenbrock:1991hu,Sirlin:1991rt,Sirlin:1991fd,Kniehl:1998vy,Bohm:2000jw}.

Summarizing, we have seen that QED radiative corrections
have profound effects on cross sections
and asymmetries at the $Z$ peak. 
Theoretically, this can be described in 
a semi-analytical approach, convoluting QED bremsstrahlung 
flux functions with improved Born observables. The modifications to 
peak cross sections and asymmetries by QED are not influenced 
by the remaining electroweak and QCD corrections, which can be
included correctly in effective paramaters in the 
improved Born approach.
For cross section predictions at the per mil level or better 
by theory at the $Z$ boson resonance we will see that the exact 
treatment of hard photon bremsstrahlung with kinematical cuts 
will be absolutely mandatory. 
%-----------------------------------------------
%\section{Lepton flavor violation in $Z$ decays
%\label{fcnc}
%}
%-----------------------------------------------
%Neutrino physics at future high-energy and high-luminosity 
%$e^+e^-$ colliders is of course only one very hot topic 
%in the field of {\it New Physics} searches. Other examples 
%are searches for extra heavy gauge bosons 
%predicted from grand-unification models or supersymmetric
%particles, or even effects from extra spatial
%dimensions in low-scale quantum gravity models like 
%from the exchange of spin-2 bosons  
%in angular cross section distributions, just to name a
%few.
 
%-----------------------------------------------
\section*{Outline of this Thesis
\label{outline}
}
%-----------------------------------------------
\addcontentsline{toc}{section}{Outline of this Thesis}
The evaluation of radiative corrections to fermion pair observables
now also forms the main task pursued in this dissertation:
\\
We shall present in this work new analytical formulae for total cross
sections and forward-backward asymmetries
for $s$-channel fermion pair production $e^+e^-\to \bar{f}f$. 
We will focus on the dominant radiative corrections from QED bremsstrahlung
and apply realistic kinematical cuts to the final state.
These calculations are given for first order flux functions for 
the complete hard photon corrections with kinematical 
cuts on the acollinearity angle and the energies of the final state 
fermion pairs, and, optionally, on the scattering angle of one fermion.
The remaining electroweak and QCD corrections can be described 
in effective Born observables convoluted with our derived QED 
flux functions, which gives a well-established and -justified 
approximate description of realistic observables. 
These results 
%%for initial state bremsstrahlung and the 
%%QED interference with the final state radiation 
partly correct or replace so-far unpublished 
older results \cite{Bilenkii:1989zg,MBilenky:1989ab} 
or are completely new and constitute especially
for leptonic final states an alternative to corresponding 
formulae with a kinematically simpler cut on the final state 
invariant mass squared $s'$ \cite{Bardin:1991de,Bardin:1991fu}. 
Corrections from soft and virtual photons or higher order 
QED effects can be straightforwardly 
included in our flux functions' description, 
in order to have physically complete and finite predictions.
%
%These results are obtained also in the context of the semi-analytical 
%program {\tt ZFITTER} 
%\cite{Bardin:1999yd-orig} 
%which calculates fermion pair production
%observables in $e^+e^-$ annihilation including all relevant 
%electroweak, QCD and QED radiative corrections.
%The code was updated with the new analytical formulae obtained,
%numerically checked, and compared with other existing numerical 
%programs on fermion pair production at LEP and LC energies. 
%%The code poses a quick, experimentally useful tool for data-fitting
%routines, due to only one necessary numerical integration over 
%the kinematical variable $s'$,
%and can be seen as complimentary approach to other numerical 
%programs, like Monte-Carlo programs or event generators,  
%for the evaluation of different fermion pair observables.

The dissertation is structured as follows:
In Chapter \ref{ch_lep1slc} a quick 
guide is given through the gauge theory description of the 
electroweak sector of the Standard Model. 
We will then discuss in Chapter \ref{ch_lep1slc}
fermion pair precision observables for LEP~1 
and SLC applications with radiative corrections.
%We want to focus on the dominant corrections by QED bremsstrahlung to 
%$s$-channel 2-fermion processes, $e^+e^-\to \bar{f}f$,
%$f\neq e,\nu_e$ and briefly illustrate the 
%effects and the treatment of the remaining electroweak 
%and QCD corrections with precision observables. 
This is done together with a description 
of the implementation of fermion pair observables 
in the semi-analytical Fortran program {\tt ZFITTER} 
\cite{Bardin:1987hva,Bardin:1989cw,Bardin:1989di,Bardin:1991de,Bardin:1991fu,Bardin:1992jc2,Christova:1999cc,Bardin:1999yd-orig}.
In this context, we will discuss the importance 
of kinematical cuts to the hard photon phase space.
%with the main focus on cuts to the maximal acollinearity 
%angle and the minimal energies of the final state fermion
%pair, together with a possible cut to one fermion's 
%scattering angle $\cos\vartheta$ \cite{Passarino:1982,Bilenkii:1989zg}.

As one of the main analytical results of this thesis we will 
present very compact formulae for the initial state, final state, 
and initial-final state interference flux functions to 
total cross sections and forward-backward asymmetries with 
the above mentioned cuts, but omitting a cut on the scattering angle.
These results have been published in \cite{Christova:1999cc} 
and implemented in the code {\tt ZFITTER} recently \cite{Bardin:1999yd-orig}. 
{For} the initial state case also one example will be presented 
for the general hard photon flux functions with all angular cuts. 
The numerical effects at LEP~1 energies are
analyzed for the program {\tt ZFITTER} in comparison with other 
available numerical codes. We then expand this analysis in 
Chapters \ref{ch_lep2} and \ref{ch_linac} to the LEP~2 
case and the especial situation at a future $e^+e^-$ Linear 
Collider. As an instructive example, we will look at 
muon pair production and take into account higher order 
QED corrections with different kinematical cuts.
In the Summary, the presented results of this 
dissertation are briefly reviewed and an outlook on  
possible future applications of the program {\tt ZFITTER} is given.

The Appendix derives a suitable parameterization of the 
hard photon phase space for the applied cuts and then a 
complete calculation of all hard photon radiator functions 
with all mentioned cuts for total cross sections and asymmetries.
The Appendix also contains for completeness a collection 
of all formulae for the remaining soft and virtual photonic 
flux functions. 
%A resumming procedure for the higher-order, 
%leading logarithmic initial- and final state soft 
%and virtual contributions then ends this presentation. 
%
%Mass terms have been neglected where possible in order
%to analytically integrate over the phase space,
%which also delivers compact and numerically precise
%cross section predictions at LEP energies when integrating
%over the final state invariant mass squared, $s'$.
%Finally, all other 2-fermion observables like e.g.
%$A_{LR}$, $A_{LR,pol}$, or $P_\tau$ 
%can be constructed straightforwardly from these results.
%Our formulae are a direct generalization of earlier results 
%given in the literature on QED radiative corrections with 
%kinematically simpler cuts on $s'$, and optionally 
%$\cos\vartheta$. 

%
\pagestyle{headings}
%-----------------------------------------------------------
\def\thechapter{\arabic{chapter}}
\setcounter{chapter}{0}
\setcounter{section}{0}
\setcounter{subsection}{0}
\def\theequation{\arabic{chapter}.\arabic{equation}}
\def\thetable{\arabic{chapter}.\arabic{table}}
\def\thefigure{\arabic{chapter}.\arabic{figure}}
\setcounter{equation}{0}
\setcounter{table}{0}
\setcounter{figure}{0}
%-----------------------------------------------------------
%

%-----------------------------------------------------
\chapter{The Electroweak Standard Model 
\label{ch_sm}
}
%-----------------------------------------------------

The observed gauge symmetry of electroweak interactions 
is described in the Standard Model ({\tt SM}) 
by the semi-simple gauge group
\cite{Glashow:1961ez,Weinberg:1967pk,Salam:1968rm}
\ba
\label{gaugegroup}
G\equiv SU_L(2) \times U_Y(1).
\ea

The subscript $L$ in (\ref{gaugegroup}) indicates 
that the unitary transformations of the
weak isospin under $SU_L(2)$ only apply to left-handed doublet 
fields, while $U_Y(1)$ is the abelian gauge group of weak 
hypercharge. One constructs left-handed
lepton or quark doublet fields $\Psi_1(x)$ and 
right-handed singlets $\Psi_{2}(x)$ under $SU_L(2)$.
For the first particle generation this is:
\ba
\label{doubsing}
\Psi_1(x) =  
\left(
\begin{array}{l}
\nu_e  
\\
e^-
\end{array}
\right)_L,
\quad 
\left(
\begin{array}{l}
u  
\\
d
\end{array}
\right)_L,
\quad
\mbox{and}
\quad 
\Psi_2(x) = (\nu_e)_R, e_R^{-},\, u_R,\, d_R. 
\ea
The free Lagrangian
\ba
\label{freeL0}
{\cal L}_0 = \sum\limits_{j=1}^{3} 
i\bar{\Psi}_j(x)\,\gamma_\mu\,\partial_\mu\,{\Psi}_j(x), 
\ea
is invariant under global $SU_L(2)\times U_Y(1)$
transformations of the fermion fields:

\ba
\label{transf}
 {\Psi}_j(x) \longrightarrow {\Psi}_j'(x)\equiv 
exp\left(i\frac{\tau_k}{2}\alpha_k\right) exp(i Y_j \beta) {\Psi}_j(x).
\ea
The $\tau_k$, $k=1,2,3$, are the Pauli-matrices 
with the commutation relation
\ba
\label{paulimat}
[\tau_i,\tau_j] = 2\,i\varepsilon_{ijk}\tau_k.
\ea
The index $k$ in (\ref{transf}) is summed.
The value $Y_j$ denotes the weak hypercharge value of the field 
${\Psi}_j(x)$, with the operator $Y$ acting on left- and 
right-handed fields.
For the further discussion it suffices to consider
infinitesimal transformations: 
\ba
\label{transf2}
 {\Psi}_j(x) \longrightarrow 
{\Psi}_j'(x)
\equiv 
\left(1 + i \frac{\tau_k}{2}\alpha_k + i Y_j \beta\right) 
{\Psi}_j(x).
\ea
The invariance of ${\cal L}_0$ under 
local $SU_L(2)\times U_Y(1)$ transformations, 
with $\alpha_k(x)$ and $\beta(x)$ now depending on $x$,
can be achieved by the {\it minimal substitution}:
\ba
\label{minisub}
\partial_\mu{\Psi}_j(x) \longrightarrow {D_\mu}{\Psi}_j(x)
\equiv \left(\partial_\mu - i g \frac{\tau_k}{2} W^k_\mu(x) 
-i g' Y_j B_\mu(x)\right) {\Psi}_j(x),
\ea
with ${D_\mu}{\Psi}_j(x)$ now transforming like ${\Psi}_j(x)$. 
This naturally introduces the gauge fields 
$W^k_\mu$ and $B_\mu$ which have to transform infinitesimally 
like 
\ba
\label{transgauge}
B_\mu(x) &\longrightarrow& {B_\mu}'(x)
\equiv B_\mu(x)+\frac{1}{g'} \partial_\mu \beta(x),
\label{bmu}
\\
W^i_\mu(x) &\longrightarrow& {W^i_\mu}'(x)
\equiv W^i_\mu(x)+\frac{1}{g} \partial_\mu \alpha_k(x) 
- \varepsilon_{ijk}\, \alpha_j(x)\, W^k_\mu(x),   
\label{wkmu}
\ea
in order to keep ${\cal L}_0$ invariant.
The parameters $g$ and $g'$ are the couplings of the 
fermion-gauge field interactions.
The complete Lagrangian ${\cal L}$ of course
also has to contain the free kinetic terms of the gauge fields. 
Introducing the field strength tensors  
\ba
\label{fieldst1}
\qquad B_{\mu\nu} = \left[D_\mu\, B_\nu, D_\nu\, B_\mu\right],
&&
W_{\mu\nu} = \left[D_\mu\, W_\nu, D_\nu\, W_\mu\right],
\\
\nonumber\\
\label{fieldst2}
\longrightarrow B_{\mu\nu} = \partial_\mu B_\nu - \partial_\nu B_\mu, 
&&
W^i_{\mu\nu} = \partial_\mu W^i_\nu - \partial_\nu W^i_\mu 
+ g \varepsilon_{ijk} W^j_\mu W^k_\nu, 
\ea
with $W^\mu =\sum\limits_i W^\mu_i\tau_i$, 
we can construct the electroweak Lagrangian without mass terms:
\ba
\label{lagrange}
{\cal L} = 
\sum\limits_{j=1}^{3} 
i\bar{\Psi}_j(x) {D_\mu} {\Psi}_j(x) - \frac{1}{4}B_{\mu\nu} B^{\mu\nu}
- \frac{1}{4} W^k_{\mu\nu} W_k^{\mu\nu}.
\ea
So, (\ref{lagrange}) produces with (\ref{fieldst2})
for the $SU_L(2)$ gauge fields $W^k_\mu$ 
self-interaction terms which are trilinear and quadrilinear   
in the gauge fields. Such gauge-field self-interactions 
are a characteristic feature
of a non-abelian gauge theory. The coupling of the gauge
fields is provided by the same coupling constant $g$ 
as for the fermion-gauge field interactions.

The charged-current interaction term can now be easily 
deduced from (\ref{lagrange}): 
\ba
\label{chargedL}
{\cal L}_{CC} = \frac{g}{2\sqrt{2}}\left\{
W^{\dagger}_\mu \cdot J^\mu + W^\mu\cdot J^{\dagger}_\mu
\right\},
\ea
with 
\ba
\label{charcurr}
W^{(\dagger)}_\mu = 
\frac{1}{\sqrt{2}}\left(W^1_\mu \pm i W^2_\mu\right)  
\quad
\mbox{and}
\quad
J^\mu \equiv \bar{u}\gamma^\mu \frac{1-\gamma_5}{2} d
+\bar{\nu}_e\gamma^\mu \frac{1-\gamma_5}{2} e.
\ea

For the neutral current case we have to consider that the 
two neutral gauge fields $W^3_\mu$ and $B^\mu$, connected
to the two diagonal generators $I_3=\tau_3/2$ and 
$Y$, mix in order to produce the photon field $A_\mu$
and the weak neutral gauge field $Z_\mu$. 
Parameterizing this transition as a rotation 
of the neutral fields $W^3_\mu$ and $B_\mu$ 
via the weak mixing angle $\theta_W$,  
\ba
\label{neutralcurr}
\left(
\begin{array}{l}
W^3_\mu  
\\
B_\mu   
\end{array}
\right)
&=&
\left(
\begin{array}{ll}
\quad\cos\theta_W  & \sin\theta_W
\\
-\sin\theta_W & \cos\theta_W
\end{array}
\right)
\left(
\begin{array}{l}
Z_\mu  
\\
A_\mu   
\end{array}
\right),
\ea
we obtain for the neutral current Lagrangian:  
\ba
\label{neutralL}
{\cal L}_{NC} = \sum\limits_j 
\bar{\Psi}_j\gamma^\mu\left\{
A_\mu\left[
\frac{g}{2}\tau_3\sin\theta_W + g' Y_j \cos\theta_W
\right]
\right.
\nonumber\\
\left. 
\hspace*{4cm} 
+\, Z_\mu\left[
\frac{g}{2}\tau_3\cos\theta_W - g' Y_j \sin\theta_W
\right]
\right\}
{\Psi}_j.
\ea
{From} (\ref{neutralL}) we can immediately derive a relation between the 
couplings and the weak mixing angle if we demand, 
as is observed in nature, that $A_\mu$ only 
couples to particles with electric charges:  
\ba
\label{couplweak}
g\sin\theta_W = g'\cos\theta_W = e.
\ea
For (\ref{couplweak}) we also had to impose that 
the electromagnetic charge $Q$, the weak
isospin $I_3$, and the weak hypercharge $Y$
fulfill the relation:  
\ba
\label{generators}
Q = I_3 + Y.
\ea
The neutral current part ${\cal L}_{NC}$ 
of the Lagrangian  
can also be written in terms of currents:
\ba
\label{neutralL2}
{\cal L}_{NC} &=& {\cal L}_{QED} + {\cal L}^Z_{NC} 
=
e A_\mu J^\mu_{em} + \frac{e}{2\sin\theta_W\cos\theta_W}
Z_\mu J^\mu_{Z},
\ea
with
\ba
\label{currents}
J^\mu_{Z} &=& J^\mu_{3} - 2\sin^2\theta_W  J^\mu_{em},
\label{jmuz}
\\
\nonumber\\
J^\mu_{em} &=& 
\sum\limits_j\bar{\Psi}_j(x)\gamma^\mu Q_j {\Psi}_j(x),
\label{jmuem3}
\qquad
J^\mu_{3} = 
\sum\limits_j\bar{\Psi}_j(x)\gamma^\mu\tau_3{\Psi}_j(x),
\ea
or equivalently in terms of the vector and axial-vector couplings 
$v_f$ and $a_f$:
\ba
\label{neutralL3}
{\cal L}_{NC} &=& 
e A^\mu 
\sum\limits_f\, Q_f\, \bar{f}\, \gamma^\mu\, f
\,+\,
\frac{e}{2\sin\theta_W\cos\theta_W}
Z^\mu 
\sum\limits_f\, \bar{f}\, \gamma^\mu\, (v_f - a_f \gamma_5)\, f,
\nonumber\\
\\
%\label{neucoupl}
%v_f &=& I_3^f-2 Q_f \sin^2{\theta}_W,\quad a_f = I_3^f,
%\\
%&& Q_e = -1, \quad a_e = -\frac{1}{2}.
%
v_e &=& -\frac{1}{2} + 2 \siw,
\qquad a_e = \frac{1}{2},
\qquad Q_e = -1,
\label{veae}
\\ 
v_f &=& I_3^f-2 Q_f \sin^2{\theta}_W,\quad a_f = I_3^f.
\label{vfaf}
\ea
%
%From (\ref{neutralL3}) one can also see that 
%flavor changing neutral current transitions 
%do not exist in the Standard Model at the tree level
%\cite{Glashow:1970st}. 
%
Inserting (\ref{neutralcurr}) into (\ref{lagrange})
and applying (\ref{couplweak}) and (\ref{generators}), 
one can also derive the gauge boson interaction terms 
between the physically observed 
fields $W^{\pm}_\mu$, $Z_\mu$, and $A_\mu$.

The mere addition 
of mass terms of the form $M_W^2 W_\mu^{+} {W^\mu}^{-}$ 
or $m\bar{\Psi}\Psi$ to the Lagrangian ${\cal L}$
would now introduce combinations of right-handed 
singlet fields with left-handed doublet fields,
which is not invariant under $SU_L(2)$ transformations
and therefore forbidden. A nice possibility to generate 
masses for both the weak gauge bosons and the fermions
is the {\it Higgs-Kibble mechanism} 
%{\it Higgs-Englert-Brout-Kibble mechanism}
%{\it Higgs-mechanism}, 
%\cite{Higgs:1964a,Higgs:1964b,Englert:1964c,Higgs:1966d,Kibble:1967e}, 
\cite{Higgs:1964a,Kibble:1967e}, 
which can be regarded as a generalization of the 
{\it Goldstone mechanism} \cite{Goldstone:1961} to gauge theories. 
The weak gauge bosons acquire masses after a spontaneous breakdown 
of the $SU_L(2)\times U_Y(1)$ symmetry through a coupling to a complex 
scalar doublet $\Phi(x)$ introduced to the theory with $Y_\Phi=1/2$: 
\ba
\label{doublet}
\Phi(x)
&\equiv&
\left(
\begin{array}{l}
\phi^{(+)}(x) 
\\
\phi^{(0)}(x)
\end{array} 
\right).
\ea
The couplings to the gauge fields are again constructed 
via a minimal substitution like in (\ref{minisub})
respecting the underlying gauge symmetry.
\ba
\label{lagrangeH}
{\cal L}_H  &=& 
\left({D_\mu}\Phi\right)^{\dagger}
{D^\mu}\Phi -\mu^2\,\Phi^{\dagger}\Phi 
-h\left(\Phi^{\dagger}\Phi\right),
\\
{D_\mu}{\Phi}(x)
&\equiv& \left(\partial_\mu - i g \frac{\tau_k}{2} W^k_\mu(x) 
-i g' Y_\Phi B_\mu(x)\right) {\Phi}(x).
\ea
With both $\mu^2$ and $h>0$, the introduced complex doublet
$\Phi(x)$ develops a non-vanishing vacuum-expectation value $v$:
\ba
\label{lagrangeH2}
|<0|\phi^{(0)}|0>| =\frac{-\mu^2}{2 h}
\equiv \frac{v}{\sqrt{2}}.
\ea
That is, while the Lagrangian ${\cal L}_H$ is invariant 
under $SU_L(2)\times U_Y(1)$ transformations, its vacuum 
ground state $\phi^{(0)}(x)$ is not. To be exact, one obtains 
a multiplet of degenerate ground states which can be transformed 
into each other via $SU_L(2)\times U_Y(1)$ rotations.
Due to the presence of gauge fields these massless degrees
of freedom can be absorbed in the longitudinal components of 
the weak gauge fields by choosing a suitable gauge, the 
{\it physical or unitary gauge}.
\ba
\label{doubbreak}
\Phi(x)
&\rightarrow&
\frac{1}{\sqrt{2}}
\left(
\begin{array}{c}
0
\\
v + H(x)
\end{array}
\right).
\ea
At the same time this generates the masses of the
three weak gauge bosons $W^\pm$ and $Z^0$ and of 
one scalar boson, the Higgs boson $H$:
\ba
 \label{lagrmass}
{\cal L}_{H} &=& 
\frac{1}{2} \partial_\mu H \partial^\mu H + ( v + H )^2
\left\{
\frac{g^2}{4} W^{\dagger}_\mu W^\mu
+ \frac{g^2}{8\cos^2\theta_W} Z_\mu Z^\mu 
\right\} 
+ 
\ldots ,
\ea
with
\ba
\label{gmasses}
M_Z\cos\theta_W = M_W = \frac{1}{2} g v.
\ea
The doublet $\Phi(x)$ with $Y_\Phi=1/2$ in (\ref{doublet}) 
has been chosen in such a way that the photon $A$
necessarily remains massless: Only one real, charge-conserving
component $\phi^0(x)$ of $\Phi(x)$ is still present after 
rotation. One can easily verify that the number of degrees of freedom
remains the same before and after the symmetry breakdown as
it should: Four massless gauge bosons with 8 and a complex  
scalar doublet with 4 degrees of freedom are transformed 
into three massive gauge fields $W^\pm$ and $Z$ with 9 degrees 
of freedom, by acquiring extra longitudinal components, one massless
photon $A$ with 2 possible spin orientations, and one massive 
scalar particle providing the missing degree of freedom.

For the fermionic case a similar mass generation can be obtained 
adding the extra complex scalar doublet $\Phi'(x)$ to the Lagrangian,
\ba
\label{doublet2}
\Phi'(x)
&\equiv&
\left(i\Phi(x)\tau_2
\right)^{\dagger}
=
\left(
\begin{array}{l}
\phi^{(0)}(x) 
\\
\phi^{(-)}(x)
\end{array}
\right),
\ea
which transforms like $\Phi(x)$ but 
with hypercarge $Y'=-1/2$. For the fermions one can construct Yukawa-type 
terms invariant under $SU_L(2)\times U_Y(1)$ 
of the form (here shown for the quarks):
\ba
\label{yukawa}
{\cal L}_{Yukawa} = -\lambda(\bar{u}\,\bar{d})_L \Phi'(x) u_R
-\lambda^{*}\bar{u}_R {\Phi'}^{\dagger}(x)
\left( 
\begin{array}{l}
u 
\\
d
\end{array}
\right)_L.
\ea
Relation (\ref{yukawa}) couples the $SU_L(2)$ 
doublets to $SU_L(2)$ singlets with coupling $\lambda$,
also preserving hypercharge $Y$. If we choose the coupling 
$\lambda$ to be real we can have mass terms after spontaneous 
symmetry breaking, for example for the $u$ quark: 
\ba
\label{umass}
{\cal L}_{Yukawa} = -\frac{\lambda v}{\sqrt{2}}\bar{u}{u}.
\ea
The factor $(-\lambda v/\sqrt{2})$ ($\lambda < 0$) 
in (\ref{umass}) can be interpreted as mass term $m_u$.
This can be done in an equivalent manner for the $d$ quark 
and the leptonic case.

In reality this picture has to be slightly modified 
when adding the full particle content to the theory:
The number of particle families is increased to three,
each family providing one left-handed lepton,  
one doublet of an up and down-type quark, and the 
corresponding right-handed $SU_L(2)$ singlets. 
The six different quark flavors are $(u,d,s,c,b,t)$
accompanied by three leptons $(e,\mu,\tau)$
and the associated neutrinos $(\nu_e,\nu_\mu,\nu_\tau)$.
The interactions of the additional two particle 
generations with the gauge fields are simply copied
from the first family. The mass generation in
(\ref{doublet2}) to (\ref{umass}) can of course also 
be repeated analogously for the other particle generations. 

The larger particle content 
allows mixing between the massive quarks, which is
really observed in nature, while the leptons do
not mix as the neutrinos are considered massless
in a minimal {\tt SM} description.\footnote{
This fact has changed recently since the 
observation of neutrino oscillations at the 
Super Kamiokande experiment \cite{Fukuda:1998mi} 
has given evidence to neutrino masses and 
neutrino-lepton mixing.} 
This produces for the charged current interactions 
flavor changing transitions because the eigenstates
of the weak interaction Hamiltonian and the mass 
eigenstates do not coincide anymore. 
In the original Lagrangian the three-dimensional 
unitary transformations of the weak eigenstates of the 
up- and down-type quark fields into their mass
eigenstates are combined to one, 
in general complex but unitary $3\times 3$ 
mixing matrix $\left(V_{ij}\right)$, $i,j=1,2,3$.
This introduces to the quark sector the 
{\it Cabbibo-Kobayashi-Maskawa mixing matrix} (CKM)
$\left(V_{ij}\right)$ \cite{Cabibbo:1963,Kobayashi:1973}, 
in the charged current Lagrangian ${\cal L}_{CC}$ 
in (\ref{charcurr2}). Including all particle generations,
it reads (over index $j$ is summed):
\ba
\label{charcurr2}
{\cal L}_{CC} = 
\frac{g}{2\sqrt{2}}
\left\{
W^{\dagger}_\mu\, \left[
\sum\limits_{ij}\,
\bar{u}_i\,\gamma^\mu\, \frac{1-\gamma_5}{2}\, 
V_{ij} d_j
\,+\,
\sum\limits_{l}\,
\bar{\nu}_l\,\gamma^\mu\, \frac{1-\gamma_5}{2}\, l
\right]\, +\, \mbox{h.c.}
\right\}.
\nonumber\\
\ea
For massless neutrinos there is no meaning in a
distinction between interaction and mass eigenstates 
and a corresponding mixing matrix between leptons and 
neutrinos can always be chosen unitary.

Furthermore, the existence of at least three particle 
generations allows to introduce one complex phase to the 
{\it CKM} mixing matrix $\left(V_{ij}\right)$ in (\ref{charcurr2})
while all other matrix elements can be chosen real through
suitable field redefinitions. This introduces {\it CP violation} 
in the {\tt SM} at the tree level.

From (\ref{neutralL3}) one can also see that 
flavor changing neutral current transitions 
do not exist in the {\tt SM} at tree-level,
i.e.~when replacing the Cabibbo-rotated eigenstates 
in (\ref{neutralL3}) by the quark mass eigenstates,
the neutral current Lagrangian ${\cal L}_{NC}$ remains diagonal 
in the quark flavors. This {\it GIM mechanism} \cite{Glashow:1970st} 
predicted for example the existence of an additional 
fourth quark, the charm quark. 

Finally, {\it anomalies} from 
quantum effects which break the original 
symmetry of the Lagrangian
\cite{Adler:1969,Bell:1969,Bardeen:1969}
and therefore could destroy the renormalizability 
of the theory do not have an effect in the {\tt SM}:
%This is a direct consequence of the 
%$SU_c(3)\times SU_L(2)\times U_Y(1)$ gauge structure
%together with the lepton-quark family structure of the {\tt SM}:
Such anomalies can occur in chiral gauge symmetries which
contain both axial and vector currents, leading to
divergent loop contributions when one axial current
couples to two vector currents.
The number of particle flavors and quark colors, however,
is balanced in such a way that the sum of these single 
divergent contributions exactly cancel. 
So, the complete $SU_c(3)\times SU_L(2)\times U_Y(1)$ 
gauge symmetry and lepton-quark family structure 
of the theory is needed in order to have 
a consistent and renormalizable description of 
electroweak interactions.

The determination of weak neutral current 
parameters at $e^+e^-$ annihilation experiments like  
LEP or SLC now forms one key test to the {\tt SM} description
of electroweak interactions and shall be discussed in 
the next Chapter.
%

%======================================================================
\chapter{Precision Physics on the $Z$ Boson Resonance
%Precision Physics and Fermion Pair Production
%on the $Z$ Boson Resonance
\label{ch_lep1slc}
}
%======================================================================

In the past decade one of the great tasks of phenomenological
particle physics was to unravel experimentally the physics of 
electroweak interactions, so successfully described by 
the Standard Model ({\tt SM}). For this, especially 
$e^+e^-$ colliding experiments like LEP at CERN or SLC at SLAC 
have provided a detailed view into the nature 
of electroweak interactions of the {\tt SM}. During the 
starting phase of LEP and at SLC the main focus 
was on precision physics at the $Z$ boson resonance:
With center-of-mass energies on resonance, i.e.~for 
$\sqrt{s}\approx M_Z\pm 1.8\,\mbox{GeV}$,
the main task of both experiments was to measure the 
properties of the neutral, massive vector boson $Z^0$ 
with very high precision 
\cite{Grunewald:1998kw,Grunewald:1999wn,Barate:1999ce,Abbiendi:1999eh,Abreu:1998kh,Acciarri:2000ai,Abbaneo:2000aa,Abe:2000dq,Abe:2000ey}.
%\cite{Abbaneo:2000aa,Acciarri:2000ai,Brau:1999bg,Serbo:1999nw}.
These are basically the mass and the total 
 decay width of the $Z$ boson, $M_Z$ and $\Gamma_Z$,  
the partial decay widths $\Gamma_f$ into 
different leptonic and hadronic decay channels
($f=e,\mu,\tau,\nu_e,\nu_\mu,\nu_\tau,u,d,s,c,b$), 
and the fermionic vector and axial-vector 
couplings $v_f$ and $a_f$ to the $Z$ boson. 
The increase of precision by experiment over the 
last 10 years was for example summarized 
in \cite{Swartz:1999xv} and is shown here 
in Table \ref{Zprecision}:
%
%--------------------------------------------------------------------------------
\begin{table}[htbp]
\begin{center}
\begin{tabular}{|c||c|c|}
\hline
Quantity & LP~89 (233 events) & LP~99 
($17\times 10^6$ events)
\\
\hline
$M_Z$ (GeV)  &  $91.17\pm 0.18$  &  $91.1871\pm 0.0021$
\\
$\Gamma_Z$ (GeV)  &  $1.95^{+0.40}_{-0.30}$  &  $2.4944\pm 0.0024$
\\
$N_\nu$ (light) &  $3.0 \pm 0.9$  &  $2.9835 \pm 0.0083$
\\ 
\hline 
\end{tabular}
\caption[High precision measurements at the $Z$ peak]
{\sf
Examples for the development of high 
precision measurements at the $Z$ boson resonance 
\cite{Swartz:1999xv}.
}
\label{Zprecision}
\end{center}
\end{table}
%--------------------------------------------------------------------------------

%-----------------------------------------------------
\section{Electroweak precision observables 
\label{sec_lep1slc_precobs}
}
%-----------------------------------------------------
%
We see from Chapter (\ref{ch_sm}) that
the unified description of electromagnetic and weak 
interactions in the {\tt SM} neglecting masses and mixing 
in the fermionic sector is completely determined by 
$g$ and $g'$ as electroweak couplings and $v$ as vacuum 
expectation value of a {\tt SM} Higgs doublet.
{For} the three couplings generally the following 
three up-to-now experimentally best known electroweak 
parameters are used as input \cite{PDG:1998aa}:
\\

\begin{tabular}{cccc}
\label{sminput}
\hspace*{-0.5cm} $\alpha(0)^{-1}$ 
&=& $137.0359895\pm 0.0000061$ 
& {\it the Feinstructure constant};
\\
\\
\hspace*{-0.5cm} $G_\mu$ 
&=& $(1.16637\pm 0.00001) \times 10^{-5}\,\mbox{GeV}^{-2}$  
& {\it the Muon decay constant};
\\
\\
\hspace*{-0.5cm} $M_Z$ 
&=& $(91.1871\pm 0.0021)\,\mbox{GeV}$ 
& {\it the $Z$ boson mass}. 
\\
\end{tabular}
\vspace*{0.5cm}

\noindent
Further input values are the top quark mass $m_t$ and 
the (still unknown) Higgs mass $M_H$. 
The Higgs mass $M_H$ is used as free input 
value which leads to small logarithmic corrections 
to different observables, while for $m_t$ the experimental 
results from direct top quark production can be plugged in.

The Feinstructure constant in the Thomson limit $\alpha(0)^{-1}$ 
is most precisely extracted from comparisons of the experimental
and theoretical values for the electron anomalous 
magnetic moment $a_e^\gamma$
which has now been calculated up to 4-loop order 
\cite{Kallen:1955ks,Kallen:1968,Czarnecki:1998nd}.
The muon decay constant $G_\mu$ is determined from the muon
lifetime $\tau_\mu$, where the complete two-loop 
electromagnetic corrections have been calculated \cite{Berman:1958,Kinoshita:1959,vanRitbergen:1998hn,vanRitbergen:1998yd,Freitas:2000ll} 
%\cite{Berman:1958,Kinoshita:1959,Berman:1962,vanRitbergen:1998hn,vanRitbergen:1998yd}, 
while the remaining electroweak corrections are 
contained for historical reasons in the muon decay constant 
$G_\mu$ itself. The $Z$ boson mass with the other neutral 
current properties are determined with high precision from 
measurements of cross sections and asymmetries at the $Z$ peak.

As starting point for the latter point, 
the differential fermion pair production 
Born cross sections are 
given below in (\ref{difxsgen}). In case of unpolarized 
$e^+$ and $e^-$ beams with $h_f=\pm 1$ as the two 
helicities and $\cos\vartheta$ as the scattering angle 
of the produced fermion $f$ with respect to 
the $e^-$ beam, they are:
\ba
\frac{d{\sigma}}{d{\Omega}} &=& \frac{\alpha^2}{8 s} 
{N_c}_f\left\{
A (1+\cos^2\vartheta) + B \cos\vartheta
- h_f [ C (1+\cos^2\vartheta) + D \cos\vartheta ]
\right\},
\label{difxsgen}
\nonumber\\
\\
A &=& Q_e^2 Q_f^2 
+ 2 Q_e Q_f v_e v_f {\Re e}(\chi) + (v_e^2+a_e^2)(v_f^2+a_f^2)|\chi|^2,
\label{difA}
\\
B &=& 2 Q_e Q_f a_e a_f {\Re e}(\chi) + 4 v_e a_e v_f a_f |\chi|^2,
\label{difB}
\\
C &=& 2 Q_e Q_f v_e a_f {\Re e}(\chi) + 2 (v_e^2+a_e^2) v_f a_f |\chi|^2,
\label{difC}
\\
D &=& 2 Q_e Q_f a_e v_f {\Re e}(\chi) + 4 v_e a_e (v_f^2+a_f^2) |\chi|^2,
\label{difD}
\ea
with
\ba
\chi = \frac{G_{\mu} M_Z^2}{2\sqrt{2}\pi\alpha}
\frac{s}{s-M_Z^2+i s \Gamma_Z/M_Z}.
\label{propz}
%\\
%v_{e} &=& -\frac{1}{2} + 2 \siw,\qquad\qquad a_{e} = \frac{1}{2},
%\label{veae}
%\\ 
%v_{f} &=& I_3^f-2 Q_f \sin^2{\theta}_W,\quad a_f = I_3^f.
%\label{vfaf}
%\\ 
\ea
The values ${N_c}_f$ are the fermion colour factor (${N_c}_q = 3$, ${N_c}_l = 1$) 
and $Q_e$ and $Q_f$ the electric charges. The vector and axial-vector
couplings were already given in (\ref{veae}) and (\ref{vfaf}).

In a lowest order approximation, (\ref{difxsgen}) yields {\tt SM} cross sections
$\sigma^0(s)$, 
\ba
\sigma^0(s) &=& 
\sigma^0_Z(s) +  \sigma^0_{\gamma Z}(s) + \sigma^0_{\gamma}(s) 
= \frac{4\pi\alpha^2}{3 s} {N_c}_f A ,
\label{xs}
\ea
with
\ba
\sigma^0_Z(s) &=& 
\frac{12\pi}{M_Z^2}
\frac{\Gamma_e\Gamma_f}{\Gamma_Z^2}
\frac{s\Gamma_Z^2}{(s-M_Z^2)^2+s^2\Gamma_Z^2/M_Z^2},
\label{xsres}
\\
\sigma^0_{\gamma Z}(s) &=& 
\frac{4\pi\alpha^2}{3} J_f
\frac{s-M_Z^2}{(s-M_Z^2)^2+s^2\Gamma_Z^2/M_Z^2},
\label{xsint}
\\
\sigma^0_{\gamma}(s) &=& 
\frac{4\pi\alpha^2}{3 s} Q_e^2 Q_f^2 {N_c}_f,
\label{xsqed}
\ea
for s-channel processes $e^+e^-\to \bar{f}f$, $f\neq e,\nu_e$, 
together with the three contributions $\sigma^0_Z(s)$, 
$\sigma^0_{int}(s)$, and $\sigma^0_{\gamma}(s)$ from 
the $Z$ resonance term, the $\gamma Z$ interference, and 
pure $\gamma$ exchange. For a constant width description  
see also (\ref{simple_peak1}) to (\ref{simple_qed}) in 
the Introduction.
Different asymmetries ${\cal A}_{a}(s)$ 
calculated from (\ref{difxsgen}) are:
\ba
{\cal A}_{FB}(s) &=& \frac{N_F-N_B}{N_F+N_B} = \frac{3}{8}\frac{B}{A},
\label{afb}
\\
{\cal A}_{pol}(s) &=& \frac{\sigma^{(h_f=+1)}-\sigma^{(h_f=-1)}}
{\sigma^{(h_f=+1)}+\sigma^{(h_f=-1)}}
 = -\frac{C}{A},
\label{apol}
\\
{\cal A}_{FB,pol}(s) &=& \frac{
N_F^{(h_f=+1)}-N_F^{(h_f=-1)}
-N_B^{(h_f=+1)}+N_B^{(h_f=-1)}
}
{
N_F^{(h_f=+1)}+N_F^{(h_f=-1)}
+N_B^{(h_f=+1)}+N_B^{(h_f=-1)}
}
= -\frac{3}{8}\frac{D}{A}.
\label{afbpol}
\nonumber\\
\label{asym}
\ea
The numbers $N_F$ and $N_B$ are the number of particles scattered into 
the forward and backward hemisphere with respect to the $e^-$ beam. 
The coupling combinations $A$, $B$, $C$, and $D$ 
containing the $Z$ propagator $\chi(s)$ are 
defined in (\ref{difA}) to (\ref{difD}).

The $Z$ boson mass $M_Z$ and the total and partial decay widths
$\Gamma_Z$ and $\Gamma_f$ can now e.g. be extracted from 
{\tt SM} fits to measured leptonic and hadronic total cross 
sections on the $Z$ boson resonance 
(see (\ref{born_peak}) or apply (\ref{xs})):
\ba
\sigma^{0,f} = \sigma^0(M_Z^2) 
\approx 
\frac{12\pi}{M_Z^2}\frac{\Gamma_e\Gamma_f}{\Gamma_Z^2}.
\label{peakxs}
\ea
In (\ref{peakxs}) the fact was used that exactly on the $Z$ peak
the $\gamma Z$ interference term $\sigma^0_{int}$ (\ref{xsint})
does not contribute, while the pure QED term $\sigma^0_{\gamma}$ (\ref{xsqed})
only yields a small correction there.
In the special case of $b$ and $c$ quark final states, 
the branching ratios $R_b$ and $R_c$ are measured instead: 
\ba
R_b = \frac{\Gamma_b}{\Gamma_{had}}
\quad,\quad      
R_c = \frac{\Gamma_c}{\Gamma_{had}}. 
\label{rbrc}
\ea
The Born fermionic decay widths $\Gamma_f=\Gamma_f(s=M_Z^2)$ 
in (\ref{xsres}) and (\ref{peakxs}) and the $\gamma Z$ 
interference contribution $J_f$ in (\ref{xsint}) contain the vector 
and axial-vector coupling dependence of the cross sections:
\ba
\Gamma_f &=& 
\frac{G_{\mu} M_Z^3}{6\sqrt{2}\pi} {N_c}_f (v_f^2 + a_f^2),
\label{zwidthsj}
\\
J_f &=& 
\frac{G_{\mu} M_Z^3}{\sqrt{2}\pi\alpha} Q_e Q_f v_e v_f.
\label{jf}
\ea
While $\Gamma_l$ and $J_l$ can be determined separately for each
lepton flavor, tricky flavor tagging techniques are utilized 
in order to identify exclusive quark flavors. For the heavy 
$b$ and $c$ quarks this works fairly well, whereas 
for the light quark sector $q=u,d,s$ a sum is performed 
over the different quark flavors. Thus, $\Gamma_{had}$ is
defined as $\Gamma_{had}=\sum\limits_q \Gamma_{q}$ with 
$q=u,d,c,s,b$.

When determining parameters like $M_Z$ and $\Gamma_Z$  
in a model-independent approach, the strong correlation 
between different parameters as e.g. $M_Z$ and the $\gamma Z$ 
interference term ${\cal J}$ has to be dealt with 
\cite{Leike:1991pq,Riemann:1997kt,Riemann:1997tj}:
\ba
\sigma_T^0(s) \sim \frac{\alpha^2(M_Z)}{s} + 
\frac{ {\cal R}s + {\cal J}(s-M_Z^2) }
{ | s - M_Z^2 + i M_Z \Gamma_Z(s) |^2}.
\label{smeq}
\ea
In the {\tt SM} case we would have in (\ref{smeq}):  
${\cal R}= 12\pi\Gamma_e\Gamma_f/M_Z^2$ 
and ${\cal J}= J_f$.
This correlation has been studied at LEP energies.
The hadron production data allow to deduce from (\ref{smeq})
for example for the L3 experiment~\cite{Acciarri:2000ai}:
\ba
M_Z &=& 91\,188 \pm 3 \pm 2.7~~ \mathrm{MeV}.
\label{smmz}
\ea
When determined from $Z$ peak data alone, the error in
(\ref{smmz}) is $\pm 3 \pm 13$ MeV. The {\tt SM} fit yields 
$M_Z = 91187.1 \pm 2.1$ MeV \cite{Abbaneo:2000aa}, where the 
$Zf{\bar f}$ couplings and thus $\cal J$ in (\ref{smeq}) are 
fixed and lepton universality is assumed. The very good 
agreement of the two fit procedures is a valuable test
of the {\tt SM}.  

Furthermore, the invisible $Z$ decay width for 
neutrino pair production is defined as 
\ba
\frac{\Gamma_{inv}}{\Gamma_l} = 
\frac{N_\nu\,\Gamma(Z\to \bar{\nu}\nu)}{\Gamma_l} = 
\frac{N_\nu}{2(|v_l|^2+|a_l|^2)},
\label{invwidth}
\ea
from which the total number $N_\nu$ of light {\tt SM} neutrinos 
can be extracted, together with neutrino counting measurements
using the channel $e^+e^-\to\nu\bar{\nu}\gamma(\gamma)$ \cite{Berends:1988zz}.
%\cite{Caffo:1987kk,Berends:1988zz}.
The value $N_\nu$ is given at the $95\%$ C.L. in 
Table \ref{Zprecision}.
%
%The value $N_\nu$ is usually calculated from the 
%ratio 
%
%\ba
%N_\nu = \left(\frac{\Gamma_{\rm inv}}{\Gamma_{ll}}\right)_{\rm exp} 
%\left/ \left(\frac{\Gamma_{\nu\nu}}{\Gamma_{ll}}\right)_{\rm th} \right.
%\label{nunumber}
%\ea
%
%which is given at the $95\%$ C.L.
%in Table \ref{Zprecision} \cite{Quast:2000ll}.
%
%The L3 experiment at LEP, for example, obtains 
%at the $95\%$ C.L.~\cite{Acciarri:2000ai}:
%
%\ba
%N_\nu=2.978\pm 0.014.
%\label{nunumber}
%\ea
%
On the $Z$ boson resonance 
the asymmetries of (\ref{asym}) yield the fermionic 
asymmetries ${\cal A}_{a}^{0,f}$,

\ba
{\cal A}_{FB}^{0,f} &=& {\cal A}_{FB}(M_Z^2) = \frac{3}{4} A_e A_f,
\label{peakasymfb}
\\
{\cal A}_{pol}^{0,f} &=& {\cal A}_{pol}(M_Z^2) = A_f,
\label{peakasympol}
\\
{\cal A}_{FB,pol}^{0,f} &=& {\cal A}_{FB,pol}(M_Z^2) = \frac{3}{4}A_e,
\label{peakasymfbpol}
\\
\mbox{with} &&  A_f = -\frac{2 v_f a_f}{v_f^2+a_f^2},
\label{fermasym}
\ea
from which together
with the partial widths $\Gamma_f$ (see (\ref{zwidthsj}))
the couplings $v_f$ and $a_f$ can now be determined for the 
leptonic case $f=l=e,\mu,\tau$ and for the heavy quarks 
$f=b,c$.

Utilizing a polarized beam, which is e.g. done for the 
electron beam at the SLC experiment, we can also use the 
left-right asymmetry ${\cal A}_{LR}^0$ to determine $A_e$ 
from a simple ratio of total cross sections for left-
and right-handed polarized $e^-$.
\ba
{\cal A}_{LR}^{0} = {\cal A}_{LR}(M_Z^2) 
= \frac{\sigma_L(M_Z^2)-\sigma_R(M_Z^2)}
{\sigma_L(M_Z^2)+\sigma_R(M_Z^2)}\frac{1}{P_e} = - A_e.
\label{peakalr}
\ea
In (\ref{peakalr}) the dependence on the beam polarization $P_e$, 
which can be measured separately, has been divided out for this.

With the definitions of $v_f$ and $a_f$ in (\ref{vfaf})
one sees immediately that the asymmetries $A_f$
are therefore especially sensitive to the
weak mixing angle $\sin^2\theta_W$:
\ba
\sin^2\theta_W = 
\frac{1}{4}\left[1-\frac{v_f}{a_f}\right].
\label{sinthetawtree}
\ea

{From} the three precisely known values $\alpha(0)^{-1}$, $G_\mu$, 
and $M_Z$ one can also evaluate tree-level results for 
the $W$ boson mass $M_W$ and the sine squared of the weak 
mixing angle $\sin^2\theta_W$:
\ba
\label{mwsinthw}
M_W^2 &=& \frac{M_Z^2}{2}
\left\{
1+\sqrt{1-\frac{4 A}{M_Z^2}}
\right\} = 80.94 \,\mbox{GeV},
\\
\nonumber\\
\label{sinthetawtree2}
\sin^2\theta_W &=& \frac{1}{2}\left\{
1-\sqrt{1-\frac{4 A}{M_Z^2}} 
\right\} = 0.2121,
\\
\nonumber\\
\label{Arad}
\mbox{with}\quad A &=& \frac{\pi\alpha}
{\sqrt{2}G_\mu} = [(37.2802\pm 0.0003) \,\mbox{GeV}]^2.
\ea

Up to now only relations valid at Born level between 
couplings, masses, widths, and different cross section observables  
have been presented. The outcome of the measurements 
performed on the $Z$ boson resonance, however, was that 
tree-level relations as in (\ref{sinthetawtree}),
(\ref{mwsinthw}), and (\ref{sinthetawtree2}) 
do not correctly reproduce the measured values, but 
some large deviations from these are experimentally observed. 
The great success of the {\tt SM} lies in the amazingly accurate 
prediction of measured observables when taking 
into account all quantum corrections,
for example using perturbation theory.

One example for the importance of radiative corrections 
at the $Z$ resonance is the running of the QED coupling 
$\alpha=\alpha(s)$ which develops an $s$-dependence
due to vacuum polarization effects to the
photon propagator. In fact, in the relations given 
up to now the coupling $\alpha(0)$, valid only at low 
momentum transfers, has to be replaced by the renormalized 
coupling $\alpha(M_Z)$ at the $Z$ boson mass.
There exists a measured $6\%$ 
enhancement of the Feinstructure constant $\alpha(M_Z)$ 
with respect to the Thomson limit \cite{Kunszt:2000jz}:
\ba
\alpha(M_Z) = \frac{\alpha(0)}{1-\Delta\alpha}
\label{alphamz}
\ea

This correction $\Delta\alpha$ is a large effect, 
where the main contributions are proportional 
to logarithmic mass terms $\ln(s/m_f^2)$, arising from self-energy 
corrections to the photon propagator including leptons and light quarks.
In (\ref{alphamz}) the leading logarithmic terms have been resummed.
While the calculation of the corrections by the leptonic loops 
can be dealt with straightforwardly in 
perturbation theory \cite{Steinhauser:1998rq}, 
the inclusion of light quark effects is much more 
involved due to non-perturbative contributions at small loop
momenta. In the latter case one needs to partly rely on 
dispersion relations utilizing low-energy data on
hadronic cross sections $e^+e^-\to\mbox{hadrons}$ 
and on $\tau$ decay data \cite{Eidelman:1995nyx,Eidelman:1998vc}.
%\cite{Jegerlehner:1986gq,Eidelman:1995nyx,VAP-Davier:1997,VAP-Davier:1998,Eidelman:1998vc}. 

Another example, where additionally electroweak corrections 
beyond the running of $\alpha$ have to be included, 
is the correct determination of the electroweak parameters 
$\Delta r$ and $\Delta\hat r$. They are defined to summarize 
radiative corrections to the sine squared of the weak mixing angle 
in two different renormalization schemes \cite{Sirlin:1999zc}
($A$ defined in (\ref{Arad})):

\ba
s^2 c^2 = \frac{A^2}{M_Z^2 (1-\Delta r)}
\quad , \quad
{\hat s}^2 {\hat c}^2 = \frac{A^2}{M_Z^2 (1-\Delta\hat r)},
\label{sinthetawloop}
\ea
In (\ref{sinthetawloop}) $s^2 = \sin^2\theta_W\equiv 1-M_W^2/M_Z^2$
and ${\hat s}^2 = \sin^2\hat\theta_W(M_Z)$ introduce
the renormalized effective weak mixing angles in the 
on-shell and $\overline{\mbox{MS}}$
renormalization schemes, respectively ($c^2 = 1 - s^2$, 
$\hat c^2 = 1 - \hat s^2$). In both cases, the measured 
correction factors $\Delta r-\Delta\alpha$ and 
$\Delta\hat r-\Delta\alpha$, where the contributions 
from running $\alpha$ have been removed, 
deviate considerably from the tree-level value zero. 
It is an impressive $9.7\,\sigma$ effect in the first 
case, and $9.9\,\sigma$ in the latter \cite{Sirlin:1999zc}. 
These large deviations from the tree-level prediction 
can only be explained accurately if both 
the numerically leading fermionic contributions 
and the subleading corrections by 
vector and Higgs boson loops are taken into account. 

In cross section observables, defined at Born level 
e.g.~in (\ref{xs}), the one-loop electroweak radiative 
corrections consist of self-energy corrections
to the vector boson propagators, virtual corrections 
to the $\gamma f\bar{f}$ and $Z f\bar{f}$ vertices,
and weak box corrections with $WW$ or $ZZ$ exchange.
In general, due to a non-decoupling of heavy gauge 
fields in a spontaneously broken gauge symmetry like the 
electroweak sector of the {\tt SM} \cite{Appelquist:1975tgx} 
the main corrections to observables arise from heavy fermion 
doublets with large mass-splitting.
This stands in contrast to exact gauge symmetries like QED and QCD
where the heavy degrees of freedom decouple from the low
energy part. That is, the main corrections arise there from
vacuum polarization effects with leptons and light quarks.
In the electroweak sector, however, we obtain important, 
$m_t^2$-dependent terms from the large $m_t^2-m_b^2$ mass 
splitting in electroweak radiative corrections. 
Corrections from virtual Higgs bosons do not contribute 
quadratically, but only logarithmically due to an extra global, 
i.e.~{\it custodial SU(2) symmetry} of the electroweak {\tt SM} 
with its Higgs mechanism \cite{Veltman:1977rt}. 

The first complete one-loop calculation of electroweak 
and QED radiative corrections without a treatment of
the $Z$ boson resonance, of QCD and of higher order electroweak 
corrections, and of hard bremsstrahlung
had been derived in \cite{Passarino:1979jh},
and later including a correct $Z$ resonance treatment in 
\cite{Wetzel:1983mh,Brown:1984jv,Bohm:1984rn,Lynn:1985rk,Bardin:1989di}. 
Now also the leading and sub-leading two-loop corrections 
proportional to $G_\mu^2 m_t^4$ and $G_\mu^2 m_t^2 M_Z^2$
to relations between electroweak observables are known
\cite{Barbieri:1992dq,Fleischer:1993ub,Degrassi:1996mg,Degrassi:1997ps,Degrassi:1997iy}. 

\hspace*{-0.5cm} QCD corrections, including mixed QED$\otimes$QCD and 
non-factorizable EW$\otimes$QCD corrections
to the vector boson self energies or $Z q \bar{q}$ 
vertex also play an important role and are included up 
to $O(G_\mu m_t^2\alpha_S^2)$ \cite{Kniehl:1990yc,Czarnecki:1996ei} 
%\cite{Kniehl:1990yc,Chetyrkin:1994js3,Czarnecki:1996ei} 
for the pure QCD terms, and at $O(\alpha\alpha_S)$ and $O(\alpha\alpha_S^2)$ 
\cite{Djouadi:1988di,Avdeev:1994db,Chetyrkin:1995js,Harlander:1998zb}
for the mixed QED$\otimes$QCD contributions.
The remaining QCD contributions from real gluon bremsstrahlung 
for hadronic final states have been treated up to 
$O(\alpha_S^3)$ and $O(\alpha\alpha_S)$
\cite{Kataev:1992dg,vanRitbergen:1997va}.

In \cite{Consoli:1989fg,Consoli:1989pc} it has been shown, 
if only considering leading terms from the virtual  
corrections that the relations (\ref{mwsinthw}) and (\ref{sinthetawtree2})
can be generalized for the corrected $W$ boson mass 
$M_W$ and $\sin^2\theta_W$ to ($A$ defined in (\ref{mwsinthw})):
 
\ba
\label{mwsinthwr}
M_W^2 &=& \frac{\rho M_Z^2}{2}
\left\{
1+\sqrt{1-\frac{4 A^2}{\rho M_Z^2}\frac{1}{1-\Delta\alpha}}
\right\},
\label{mwcorr}
\\
\nonumber\\
\sin^2\theta_W(M_Z^2) &=& 1-\frac{M_W^2}{\rho M_Z^2} 
= \frac{1}{2}\left\{
1-\sqrt{1-\frac{4 A^2}{\rho M_Z^2}\frac{1}{1-\Delta\alpha}} 
\right\}.
\label{swcorr}
\ea
The parameter $\rho$ relates the charged to neutral current
couplings, introduced in \cite{Veltman:1977rt} and given 
in \cite{Consoli:1989fg} for the main corrections from charge 
renormalization and from heavy fermion corrections to the weak 
gauge boson propagators.
In the $\rho$ parameter for example 
the dependence of electroweak radiative 
corrections on the top mass $m_t$ 
was first examined in \cite{Veltman:1977fy}. 
It can be used to indirectly determine the top mass $m_t$, 
which poses an alternative to its direct measurement 
performed at the Tevatron experiment \cite{Abe:1995hr,Abachi:1995iq}. 
A thorough discussion of radiative corrections and their importance 
for electroweak precision physics was presented in 
\cite{Hollik:1990ii,Jegerlehner:1991ed}.  

As illustrated in the Introduction, the dominant radiative 
corrections to cross sections and asymmetries arise now from QED 
bremstrahlung. On resonance, for example, the hadronic peak 
cross section is lowered by some $30\%$, mainly through multiple soft 
and virtual photon corrections. The large numerical effect develops 
from logarithmic mass terms $\ln(s/m_e^2)$ due to collinear photon 
emission from the initial state. So, if one wants to
successfully test the {\tt SM} by checking the predicted EW or QCD 
quantum effects, experiment needs precise theoretical tools
in order to accurately remove the even larger bulk of corrections 
introduced by QED bremsstrahlung. 

On the $Z$ resonance the interference between initial and 
final state radiation and the added interference from Born 
and $\gamma\gamma$- and $\gamma Z$-exchange box diagrams 
contribute few per mil changes for loose kinematical cuts, 
but may carry more weight with tighter 
cuts or at higher energies. The small final state 
corrections can usually be approximated by a global
correction factor to cross sections 
\cite{Bardin:1989qr,Bohm:1989pb,Montagna:1998sp}.
With per mil level measurements performed by experiments
on the $Z$ resonance, the initial state bremsstrahlung 
has to be known there at $O(\alpha^2)$ exactly
\cite{Berends:1988ab} and in leading $O(\alpha^3)$ 
approximation \cite{Skrzypek:1992vk,Montagna:1997jv}.
Also QED corrections from the creation of fermion pairs 
after photon emission from the initial state are known
up to leading fourth order in $\alpha$ now 
\cite{Berends:1988ab,Kniehl:1988id,Jadach:1992aa,Arbuzov:1999uq} 
and will be or are already included in the experimental
analysis \cite{Quast:2000ll}.

As the full results for cross section observables including
all electroweak and QCD corrections with QED bremssstrahlung 
are in general quite complicated and lengthy, a short and 
handy description of the main corrections seems
quite attractive, especially when having data-fitting routines 
with limited CPU time in mind. Fortunately, due to a 
factorization of electroweak and QCD corrections on the 
$Z$ boson resonance these can be treated in 
an {\it effective} or {\it improved Born approximation}: 
Vector and axial-vector couplings $v_f$ and $a_f$ may be 
replaced by effective, in general complex valued couplings 
$\bar{v}_f$ and $\bar{a}_f$ \cite{Bardin:1989di}, while for 
the $Z$ boson width \cite{Wetzel:1983mh,Akhundov:1986fc,Jegerlehner:1986vs,Beenakker:1988pv,Bernabeu:1988me,Lynn:1990hd}
an $s$-dependence can be introduced, $\Gamma_Z=\Gamma_Z(s)$ 
\cite{Akhundov:1986fc,Bardin:1988xt,Bardin:1989di,Riemann:1997kt,Riemann:1997tj}.
Experimentally, the real parts of $\bar{v}_f$ and $\bar{a}_f$ 
define the effective vector and axial-vector couplings 
$v^{eff}_f$ and $a^{eff}_f$ which may be measured,
while the imaginary parts are small at the $Z$ peak 
and usually calculated within the {\tt SM}.
The weak box contributions are non-resonant near the 
$Z$ peak and produce only small corrections there 
with their different angular dependence being
neglected \cite{Consoli:1989pc}. They are infrared-finite 
and added at $O(\alpha)$. QCD effects can be 
parameterized by effective color factors 
to $\bar{v}_f$ and $\bar{a}_f$, i.e.~factors of the type:
$N_q = N_c\left\{1+\frac{\alpha_s}{\pi}+\ldots\right\}$.
See also \cite{Bardin:1999yd-orig} and references therein.

The pure QED corrections which form a gauge-invariant 
subset of the radiative corrections can then be described 
by convoluting QED {\it flux functions} ({\it radiators}) 
with the improved Born observables containing the effective 
couplings and the added weak box corrections.
A common convolution integral of the initial and final state 
radiators can be performed with the leading logarithmic soft 
and virtual corrections resummed and the QED interference 
and box parts usually added at $O(\alpha)$. Infrared singularities 
from soft photon corrections cancel completely. 
The full QED description of such an effective Born 
approach to cross sections and asymmetries will be presented 
in detail in the next Section \ref{sec_lep1slc_radcuts}.
%
%--------------------------------------------------------------------------------
\begin{figure}[htbp] 
\begin{flushleft}
\vspace*{-0.5cm}
\hspace*{0.5cm}
\mbox{
 \epsfig{file=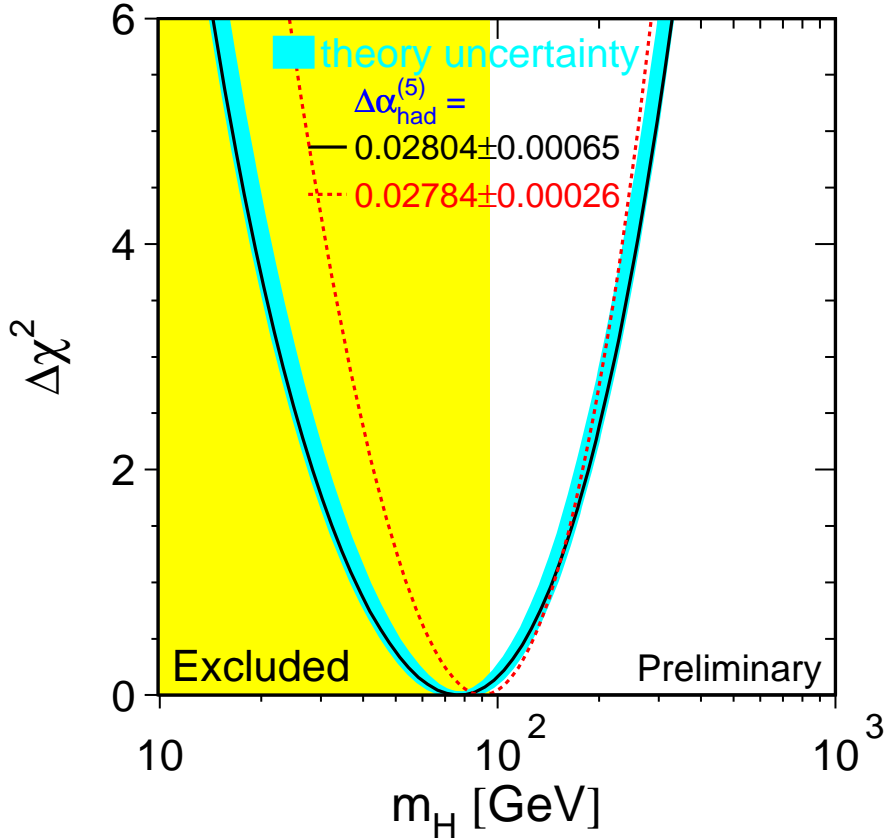,width=12cm}
}
\end{flushleft}
\vspace*{-1.5cm}
\caption[$\Delta\chi^2$ distribution of LEP data fit]
{\sf
The $\Delta\chi^2$ distribution of the total LEP data fit
depending on the Higgs boson mass $m_H$ compared with 
program {\tt ZFITTER} 
\cite{Grunewald:1999wn,Abbaneo:2000aa}.
\label{higgs_lep_data} 
} 
\end{figure} 
%--------------------------------------------------------------------------------

An important application of these calculations
is of course the investigation of indirect effects from 
a {\tt SM} Higgs boson through virtual corrections:
Bounds on the mass $M_H$ of a {\tt SM} Higgs boson
can be derived from the logarithmic Higgs mass dependence
of different electroweak precision observables, using the 
directly measured value for $m_t$ from the Tevatron. 
These corrections were calculated and analyzed in 
\cite{vanderBij:1984aj,vanderBij:1984bw,vanderBij:1987hy,Barbieri:1993ra,Fleischer:1995cb}.

Such an indirect determination of $M_H$ is for example 
illustrated in Fig.~\ref{higgs_lep_data}, 
depicting the $\Delta\chi^2$ distribution of a {\tt SM} fit
to LEP data on different electroweak precision observables
with $M_H$ as a free parameter. 
The minimum of the curve clearly indicates a low-mass {\tt SM} Higgs 
boson. The solid and hatch-marked lines together with 
the shaded error band were calculated  using 
the programs {\tt ZFITTER} \cite{Bardin:1987hva,Bardin:1989cw,Bardin:1989di,Bardin:1991de,Bardin:1991fu,Bardin:1992jc2,Christova:1999cc,Bardin:1999yd-orig} 
and {\tt TOPAZ0} \cite{Montagna:1995b,Montagna:1998kp}, which calculate 
{\tt SM} observables at the $Z$ peak within an effective Born 
approximation decribed above.
The depiceted error band is obtained 
changing the values of the {\tt SM} input like 
e.g.~the hadronic contributions $\Delta\alpha^{(5)}_{had}$ 
as main uncertainties 
to the running electromagnetic coupling.
At the $95\%$ C.L.~one has from a recent analysis 
\cite{Grunewald:1999wn},
\ba
\label{higgsbound}
M_H < 262\,\mbox{GeV},
\ea
while a lower bound of at least 
$M_H > 90\,\mbox{GeV}$ ($95\%$ C.L.) 
is determined from direct searches at LEP 
which could increase up to 
$M_H > \sqrt{s}-M_Z\approx 109\,\mbox{GeV}$
for $\sqrt{s}\approx 200\,\mbox{GeV}$ at the end of LEP
\cite{Grunewald:1999wn}.
%
%total hadronic cross sections  
%\cite{Dine:1979qh,Chetyrkin:1979bj,Celmaster:1980xr,Gorishnii:1991hw,Kataev:1992dg}, 
%the $Z$ boson decay width 
%\cite{Kniehl:1990qu,Chetyrkin:1996ia,Harlander:1998zb}, 
%and for heavy quark asymmetries
%\cite{Jersak:1979uv,Djouadi:1990uk,Djouadi:1995wt,Ravindran:1998jw,Ravindran:1998qz},
%together with the dominant corrections by heavy quark masses 
%\cite{Fleischer:1992fq,Chetyrkin:1993jp,Degrassi:1999jd}
%\cite{Veltman:1977kh,Veltman:1980fk,Djouadi:1988di,Kniehl:1990yc,Chetyrkin:1996ii}
%\cite{Degrassi:1996mg,Degrassi:1997iy,Degrassi:1997ps}.
%\cite{Veltman:1977fy,vanderBij:1987hy,Djouadi:1987gn,Avdeev:1994db,Fleischer:1995cb,Chetyrkin:1995ix,Chetyrkin:1995js}
%\cite{vanderBij:1984aj,vanderBij:1984bw,Barbieri:1993ra}.

%=========================================================================
\section{Realistic observables and the {\tt ZFITTER} approach
%Realistic observables to $e^{+}e^{-}\to \bar{f}f$
%and the {\tt ZFITTER} approach:
%\\ 
%Radiative corrections and kinematical cuts 
\label{sec_lep1slc_radcuts}
}
%-------------------------------------------------------------------------
%
In the Introduction it was illustrated that  
the main corrections to fermion pair production 
cross sections and asymmetries arise from QED bremsstrahlung.
While multiple soft and virtual corrections 
can be treated universally with a suitable
resummation procedure, especially important on the $Z$ peak,
hard photon emission is strongly suppressed there.
This is due to a strong decrease of the effective
Born cross section or asymmetry as soon as the radiation 
of a hard photon from the initial state 
shifts the effective center-of-mass
energy $s'$ after photon emission away from the $Z$ boson 
resonance. Although these hard photon effects, which 
are strongly cut-dependent, are small on the peak they 
are still important for the per mil and sub per mil level 
analysis of experimental data performed at LEP and SLC.

At higher energies the photonic corrections 
lead to the observed pronounced radiative tail 
for total cross sections $\sigma_T$ and forward-backward
asymmetries $A_{FB}$, with $s'$ being shifted to $M_Z^2$ 
due to initial state hard photon emission.
While $\sigma_T$ is strongly enhanced, 
$A_{FB}$ is decreased by a correction factor.
This effect, the {\it radiative return to the $Z$},
can be removed completely or approximately by sufficiently
strong kinematical cuts to the 
hard photon phase space. This is very often preferred
by experiment in order to probe the interesting electroweak
or `New Physics' sector.

Experiments at LEP~1, SLC, LEP~2, and those planned at 
a future $e^+e^-$ Linear Collider aim at
precisions well below a per cent and need theoretical predictions with an
accuracy of the order of 0.1~\% or better.  
A basic ingredient of the predictions is the complete $O(\alpha)$ 
photonic correction including initial and final state radiations and 
their interference:
\ba
\label{sigma}
\sigma(s) &=& \sigma^0(s) + \sigma^{ini}(s) + \sigma^{int}(s) +
\sigma^{fin}(s) .
\ea
These corrections have to be determined for two basic quantities: 
The total cross section $\sigma_T(s)$ and the forward-backward asymmetry
$A_{FB} = \sigma_{FB}/\sigma_T$; 
other asymmetries may then easily be derived from them. 

Basically, there are two experimental set-ups to be treated:
\begin{itemize}
\item[(i)] a lower cut on the final state fermions' invariant mass squared,
$s'$, $s'_{\min}\geq 4m_f^2$,
\item[(ii)] or combined cuts on the maximal {\it acollinearity angle} 
  ${\theta_{\rm acol}^{\max}}\leq 180^{\circ}$,
  and on the {\it minimal energy} of the fermions $E_{min}\geq m_f$.
\end{itemize}
The {\it acollinearity angle} defines the small angular deviation 
from the back-to-back scattering situation in case of hard photon emission.  
It is discussed in more detail in Appendix \ref{crossphase}
and shown in Fig.~\ref{phfig3} there.

Both cut settings (i) and (ii) may be combined with an 
acceptance cut $c$, $c\leq1$, on the cosine of the fermionic 
production angle $\vartheta$ ({\it acceptance cut}).  
While case (i) is suitable to describe hadronic events, where
due to jets from the hadronization of the final state quarks 
a definition of angular cuts is not easily possible, case (ii)  
provides an alternative to remove hard photon effects from 
leptonic final states instead of the kinematically simpler $s'$-cut.

In order to illustrate the effects of QED corrections
at LEP~1 and LEP~2 energies, the semi-analytical program 
{\tt ZFITTER} \cite{Bardin:1987hva,Bardin:1989cw,Bardin:1989di,Bardin:1991de,Bardin:1991fu,Bardin:1992jc2,Christova:1999cc,Bardin:1999yd-orig} is used.
The semi-analytical approach of the {\tt ZFITTER} code 
consists of a fast, one-dimensional numerical integration
of analytical formulae for different observables like cross sections, 
asymmetries, and angular distributions with the inclusion 
of different experimentally relevant cut options. 
The $O(\alpha)$ soft and virtual QED terms are resummed
and dominant higher order effects included. 
Corresponding numerical programs for fermion pair production 
like {\tt TOPAZ0} \cite{Montagna:1995b,Montagna:1998kp}, 
{\tt ALIBABA} \cite{Beenakker:1991mb}, 
or {\tt KORALZ/KK2f} \cite{Jadach:1994yv,Jadach:1998jb,Jadach:1999tr,Jadach:1999kkkz},
are in this respect complementary to our approach 
as they can in principle treat multi-differential observables 
with more complex cuts to the final state phase space, 
but this at the expense of a clear increase in computing time.

With {\tt ZFITTER} three different cut options are available 
\cite{Bardin:1992jc2,Bardin:1999yd-orig}:
(i) no cuts \cite{Bardin:1987hva,Bardin:1989cw},
(ii) cuts on $s'$ and on the scattering angle $\vartheta$ 
of one fermion \cite{Bardin:1991de,Bardin:1991fu}, or  
(iii) cuts on the fermions' acollinearity 
angle $\theta_{\rm acol}$ on their energies $E^{f}=E^{\bar f}$ 
and on $\cos \vartheta$ \cite{Christova:1999cc}.
The effective Born cross sections $\sigma^0(s')$ may 
also be chosen according to following approaches:
(A) {\tt SM}, (B) Model Independent,
(C) Others \cite{Bardin:1989di,Bardin:1992jc2,Bardin:1999yd-orig}.

Without cuts, i.e.~in case of complete acceptance and including all 
hard photon effects, total cross section formulae with the
exact $O(\alpha)$ initial state radiation have been calculated in
\cite{Bonneau:1971mk,Bardin:1989cw}: In analogy to (\ref{crossini}),
the initial state corrected total cross section $\sigma_T^{ini}(s)$
may be written as a convolution integral of the (effective) Born 
cross section $\sigma^0(s')$ with a {\it flux function (radiator)} 
describing the photon-emission over the (normalized) invariant mass 
squared $R\equiv s'/s$ of the final state fermion pair:

\ba
  \label{eq:2}
 \sigma_T^{ini}(s) = \int dR~ \sigma^0(s')~ \rho_T^{ini}(R).  
\ea
Including a resummation of soft and virtual 
photonic higher order corrections \cite{Greco:1975rm,Greco:1980}, 
the initial state radiator function $\rho_T^{ini}(R)$
can be written as:
\ba
  \label{eq:3}
  \rho_T^{ini}(R) &=& 
\left(1+{\bar S}^{ini}\right)\beta_e (1-R)^{\beta_e-1} 
   + {\bar H}_{T}^{ini}(R),
\ea
with
\ba
  \label{eq:4}
         {\bar S}^{ini} &=& \frac{3}{4}\beta_e +\frac{\alpha}{\pi}Q_e^2
\left(\frac{\pi^2}{3} - \frac{1}{2}\right) ,~~~~~ 
  \beta_e = \frac{2\alpha}{\pi} Q_e^2 \left( \ln \frac{s}{m_e^2}-1\right) ,
\ea
and
\ba
  \label{eq:3a}
 {\bar H}_{T}^{ini}(R)&=&\left[ H_{BM}(R)-\frac{\beta_e}{1-R}\right],
\ea
where 
\ba
  \label{eq:bm}
  H_{BM}(R) = \frac{1}{2}~\frac{1+R^2}{1-R}~\beta_e
\ea
is the Bonneau-Martin term for the one-loop hard photon 
correction \cite{Bonneau:1971mk}, while $(1+\bar{S})$ 
are the regularized, infrared-finite soft and virtual 
contributions. 

Historically, effects from soft and virtual photonic 
corrections to total cross sections 
are already known since \cite{Bloch:1937}.
%
%The first to discuss effects of QED radiative corrections, especially
%of multiple soft photon emission, on resonance shapes were 
%\cite{Etim:1966}
%%\cite{Etim:1966,Greco:1967,Etim:1967,Pancheri:1969}, 
%with a more up-to-date summary given in 
%\cite{Bardin:1989qr,Bohm:1989pb} and a discussion of the 
%electroweak radiative corrections in 
%\cite{Consoli:1989pc,Hollik:1990ii,Jegerlehner:1991ed}.  
%%\cite{Altarelli:1986kq,Barroso:1987ae,Bardin:1989qr,Bohm:1989pb}.  
The function $\rho_T^{ini}(R)$ in (\ref{eq:2}) and (\ref{eq:3}) 
is a regularized and infrared-finite flux function 
as according to \cite{Bloch:1937} all infrared 
divergences from soft and virtual photon contributions have to
cancel exactly to arbitrary perturbative order. The same holds
for unphysical singularities arising for vanishing fermion
masses which have to disappear   
in the complete scattering amplitude of a process
\cite{Kinoshita:1962ur,Lee:1964is}.
Only logarithmic mass terms of the type $\ln(s/m^2)$ ($m=m_e,m_f$) 
from collinear photon-emission from the initial or final state  
fermions are allowed to survive.
Also any unphysical distinction between soft and 
hard photon phase space, typically denoted with an arbitrary 
soft photon cut-off parameter $\varepsilon$, has to 
disappear in the final results at any given order
\cite{Bloch:1937,Kinoshita:1962ur,Lee:1964is}.

That the leading logarithmic soft and virtual terms
exponentiate, in (\ref{eq:3}) given by the 
term $(1+\bar{S})\beta_e (1-R)^{\beta_e-1}$,
%with a corresponding
%term from the unregularized hard photon radiator $H(v)$,   
was first shown in \cite{Yennie:1961ad}. The divergent 
contributions proportional to $(\ln\varepsilon)^n$ from 
the soft and hard photon radiators of course have to cancel 
to all orders. 
%This non-perturbative inclusion 
%of all soft and virtual higher order effects can be shown to be 
%absolutely crucial in order to correctly reproduce the measured 
%peak cross section observable \cite{Bardin:1989qr,Bohm:1989pb}. 
Technically, also other resummation procedures   
than a {\it soft photon exponentiation} 
\cite{Greco:1975rm,Greco:1980}   
are possible 
\cite{Yennie:1961ad,Kuraev:1985hb,Nicrosini:1987sm}.
Analytically, they have to exactly reproduce 
the calculated two-loop results \cite{Berends:1988ab} 
after expansion in the coupling, and numerically the 
observed $Z$ boson line shape \cite{Bardin:1989qr,Bohm:1989pb}.

Having in mind kinematical cuts to the hard photon phase space, 
a generalization of the Bonneau-Martin formula in 
(\ref{eq:3a}) with (\ref{eq:bm}) with soft photon
exponentiation, may be found in \cite{Bardin:1989cw,Bardin:1991de}. 
There, all three corrections -- the initial state, the final state, 
and the initial-final state interference -- are treated for a cut 
on $s'$, without \cite{Bardin:1989cw} or with an additional 
acceptance cut $c$ \cite{Bardin:1991de}. 
The extremely compact expressions 
with $c=1$ get quite involved when the acceptance cut is applied. 
The corresponding formulae are contained in the program {\tt ZFITTER}
\cite{Bardin:1992jc2,Bardin:1999yd-orig}. 

In the initial and final state soft and vertex 
corrections infrared singularities cancel each other 
completely; for this the necessary counter term diagrams from 
self-energy corrections to external fermion legs have to be
included. This cancellation occurs separately for the initial 
and the final state radiator functions. The corresponding Feynman 
diagrams are shown in Fig.~\ref{phfig1}, \ref{Fig.virtph}, 
and \ref{fig.fsesm} in the Appendix.
The QED box contributions consist in first order 
approximation of the interference terms between the 
Born diagram and the $\gamma\gamma$ and $\gamma Z$ exchange, 
direct and crossed box diagrams (see Fig.~\ref{fig.virtbx} in 
the Appendix). 
They have to be combined with the initial-final 
state interference in order to obtain infrared-finite results.
%While the contributions from initial and final state
%radiation to total or forward-backward cross sections 
%are proportional to an (improved) Born cross section,
%or Born asymmetry respectively, the vice versa is true for
%the interference and box corrections due to their 
%different angular dependence. 

The slightly more involved 
treatment of higher order soft and virtual photonic effects 
is given for initial or final state radiation in {\tt ZFITTER}.
Alternatively, also a common treatment of the initial and final state 
soft photon emission is possible \cite{Bardin:1991de}. 
The interference and QED box corrections are added exactly at $O(\alpha)$. 
In principle, following \cite{Greco:1975rm,Greco:1980},
an analogous resumming procedure for soft interference and 
box corrections 
is possible \cite{Bardin:1991fu,Aversa:1989xh,Greco:1989gk}. 
\vfill\eject
%--------------------------------------------------------------------

A realistic description 
of total or forward-backward cross sections 
$\sigma_A(c)$, $A=T,FB$ ($\bar{A}=FB,T$),  
which is calculated by {\tt ZFITTER} with 
acceptance cut, is:
\ba 
&&\sigma_A(c) =  \left(\int\limits_{0}^{c}\pm\int\limits_{-c}^{0}\right)
 \, d{\cos\vartheta} \,  \frac{d{\sigma}}{d{\cos\vartheta}}
\nl
&=&  \int\limits_0^{1-4m_f^2/s} dv \,  
\Biggl\{ 
\left[ \sigma_A^{0}(s',c)
\left(1+{\bar{S}}^{ini}\right)
\beta_e v^{\beta_e-1} + \sigma_A^0(s') {\bar{H}}_A^{ini}
(v,c) \right]  
{\bar R}_A^{fin}(v) 
\nl \nl 
&&+~ \sigma_{\bar{A}}^0(s,s') \left[ H_A^{int}(v,c) - 
\frac{\sigma_{\bar{A}}^0(s)}{\sigma_{\bar{A}}^0(s,s')} 
H_A^{int,sing}(v,c)\right] 
\Biggr\} 
\nl \nl && 
+~ \sigma_{A}^0(s,c) {\bar{S}}_A^{int}  +
\sum_{m,n=\gamma,Z}  
\sigma_{\bar{A}}^0(s,s,m,n) B_A(c,m,n) . 
\label{generic_zf} 
\ea 
%--------------------- 
%
The convolution integral in (\ref{generic_zf}) has  
$v\equiv 1-R = 1-s'/s$ as integration variable which 
corresponds to the normalized energy of the emitted
or virtual photon. 
The (effective) Born expressions $\sigma_{A}^0(s,s')$ 
in (\ref{generic_zf}) are introduced in (\ref{effbornt}) and
(\ref{effbornfb}):
\ba
\label{effbornt}
\sigma_T^0(s,s') &=& \sum_{V_i,V_j=\gamma, Z} 
\sigma_{T}^0(s,s',i,j)
=\frac{4\pi\alpha^2}{3s'}~ {\cal V},
\\
\label{effbornfb}
\sigma_{FB}^0(s,s') &=& \sum_{V_i,V_j=\gamma, Z} 
\sigma_{FB}^0(s,s',i,j) =
\frac{\pi\alpha^2}{s'}~ {\cal A},
\ea
with the combinations ${\cal V}$ and ${\cal A}$ for the
neutral current couplings $v_e$, $a_e$, $v_f$, and $a_f$ 
from (\ref{veae}) and (\ref{vfaf}) 
and the $Z$ boson propagator, defined in (\ref{propz}):

\ba
{\cal V} &=&  
Q_e^2 Q_f^2 +   Q_e Q_f v_e v_f \Re e [\chi(s)+\chi(s')] 
+ (v_e^2+a_e^2)(v_f^2+a_f^2)
\Re e[\chi(s)\chi^*(s')], 
\nonumber\\
\\
{\cal A} 
&=&
 Q_e Q_f a_e a_f \Re e [\chi(s)+\chi(s')] +
4 v_e a_e v_f a_f  \Re e[\chi(s)\chi^*(s')].
\ea
The general formulae (\ref{effbornt}) 
and (\ref{effbornfb}) hold exactly as effective 
Born total and forward-backward cross sections for the 
initial-final state interference. The corresponding 
initial state results $\sigma_{T,FB}^0(s')$ and 
final state results $\sigma_{T,FB}^0(s)$ can be obtained
easily from (\ref{effbornt}) and (\ref{effbornfb})
setting $s=s'$, or $s'=s$ respectively:
\ba
\label{effbornini}
\sigma_A^{0,ini} &=& \sigma_A^0(s') = \sigma_A^0(s',s'),
\\
\label{effbornfin}
\sigma_A^{0,fin} &=& \sigma_A^0(s) = \sigma_A^0(s,s). 
\ea
In (\ref{generic_zf}) all (regularized) 
hard, soft, and virtual photonic, together with the 
QED box corrections are contained in form of the radiator 
functions:
\ba
\label{radiators}
H({\bar{H}})_A^{a}, 
\quad
S({\bar{S}})^{a}, 
\quad 
\mbox{and} 
\quad
B_{\bar{A}}(c,m,n),
\quad a = ini, int,
\quad m,n = \gamma, Z.
\ea
The (un)barred radiators in (\ref{radiators}) 
are (un)regularized functions, respectively.
The corresponding final state radiator ${\bar R}_A^{fin}(v)$ 
is completely analogous to the initial state term in front
with the necessary substitutions $s/m_e^2,Q_e\to s'/m_f^2,Q_f$. 
It can be convoluted together with the initial state radiator
over $v=1-s'/s$ in a common soft photon exponentiation formula. 
This is done in (\ref{generic_zf}) introducing the factor 
$\beta_e v^{\beta_e-1}$ for the resummed leading higher 
order effects to the initial state from soft and virtual photons. 
The $O(\alpha)$ QED interference and box corrections are
regularized and added.
Also note the antisymmetric angular dependence of the
QED interference and box terms with respect to the 
initial and final state radiators which are multiplied with 
the effective Borns $\sigma_{FB}^0(s,c)$ in $\sigma_{T}(c)$ 
and $\sigma_{T}^0(s,c)$ in $\sigma_{FB}(c)$.
Formula (\ref{generic_zf}) is thus exact 
at first order in {\tt ZFITTER}, but may include
the exact two- \cite{Berends:1988ab} and leading 
three-loop \cite{Skrzypek:1992vk,Montagna:1997jv} 
contributions for initial state bremsstrahlung.

\begin{figure}[htp]
\begin{flushleft}
\vspace*{-0.5cm}
\begin{tabular}{ll}
%\vspace*{-1.0cm}
%\hspace*{-0.5cm}
%\mbox{
% \epsfig{file=sigspbbpeak.ps,width=7.5cm}}
%&
\hspace*{2.0cm}
\mbox{
 \epsfig{file=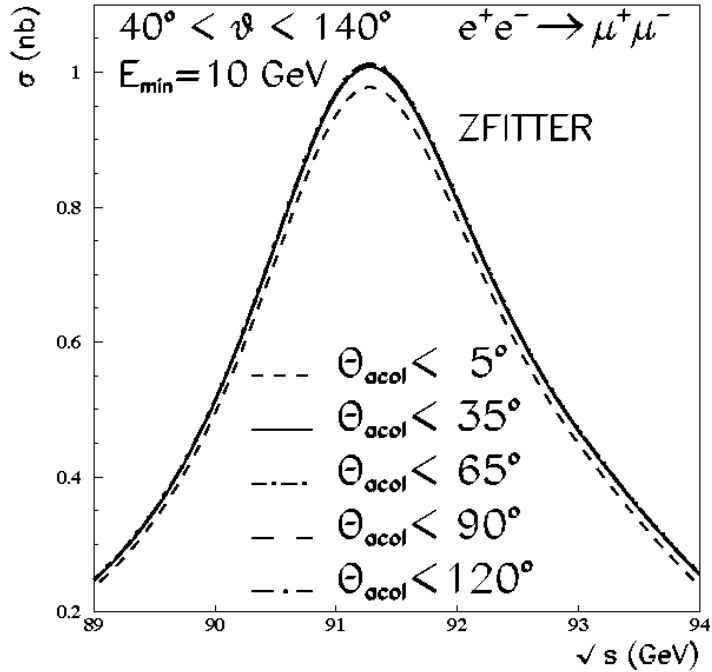,width=10cm}}
\end{tabular}
\end{flushleft}
\vspace*{-0.5cm}
\caption[Muon pair production cross sections at the $Z$ peak]
{\sf
Muon pair production cross sections from 
{\tt ZFITTER} \cite{Bardin:1992jc2,Bardin:1999yd-orig} 
with different cuts on maximal acollinearity $\theta_{\rm acol}$ 
at the $Z$ boson resonance \cite{Christova:1999gh}.
\label{sig_peak_zf}
}
\end{figure}

\begin{figure}[htp] 
\begin{flushleft}
\vspace*{-0.25cm}
\begin{tabular}{ll}
%\vspace*{-0.25cm}
\hspace*{-0.25cm}
\mbox{
\epsfig{file=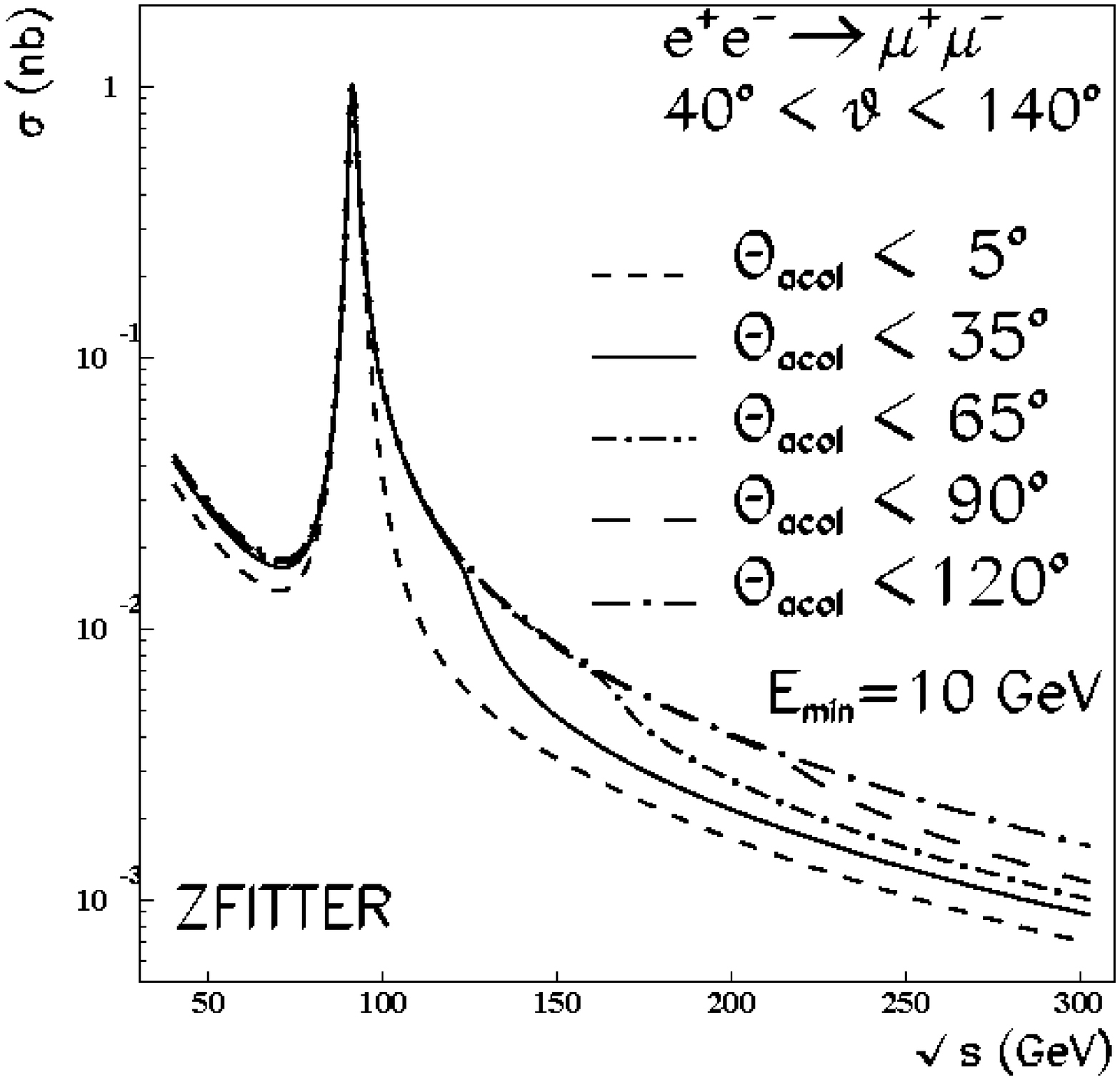,width=7.5cm}}
    &
%\vspace*{-0.25cm}
\hspace*{-1.cm}
\mbox{
 \epsfig{file=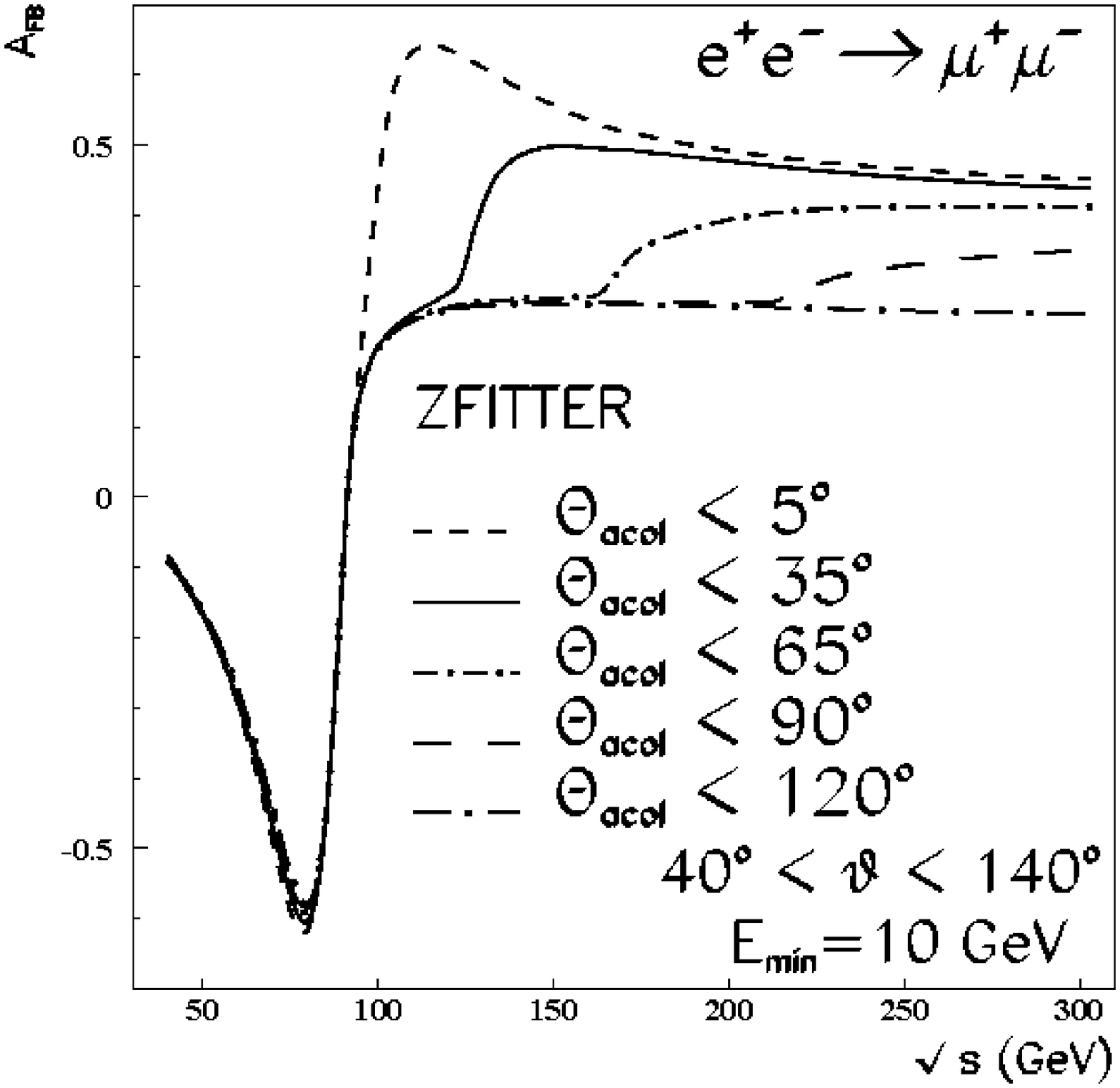,width=7.5cm}}
\\
%\vspace*{-0.25cm}
\hspace*{-0.25cm}
\mbox{
\epsfig{file=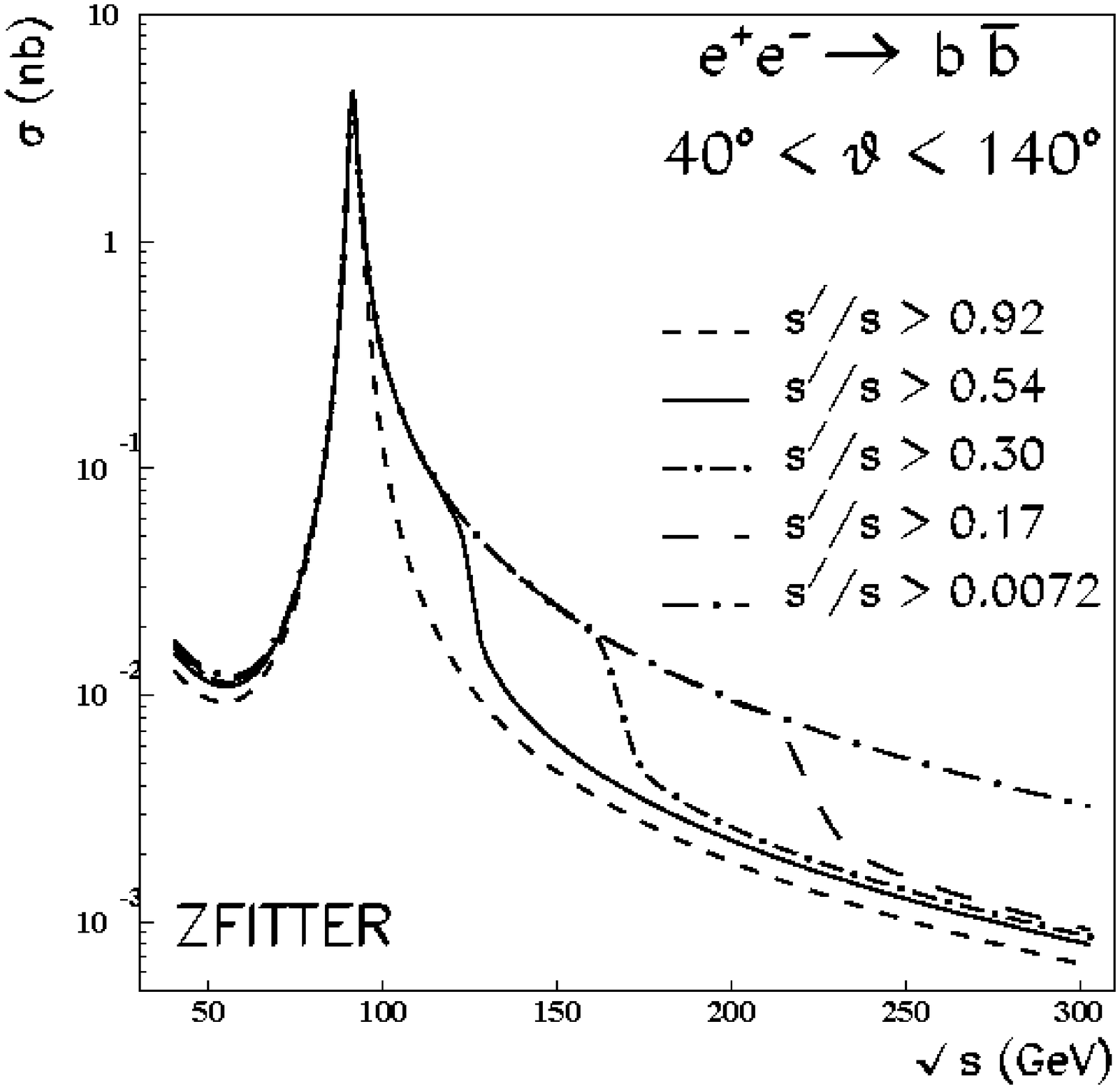,width=7.5cm}}
    &
%\vspace*{-0.25cm}
\hspace*{-1.cm}
\mbox{
\epsfig{file=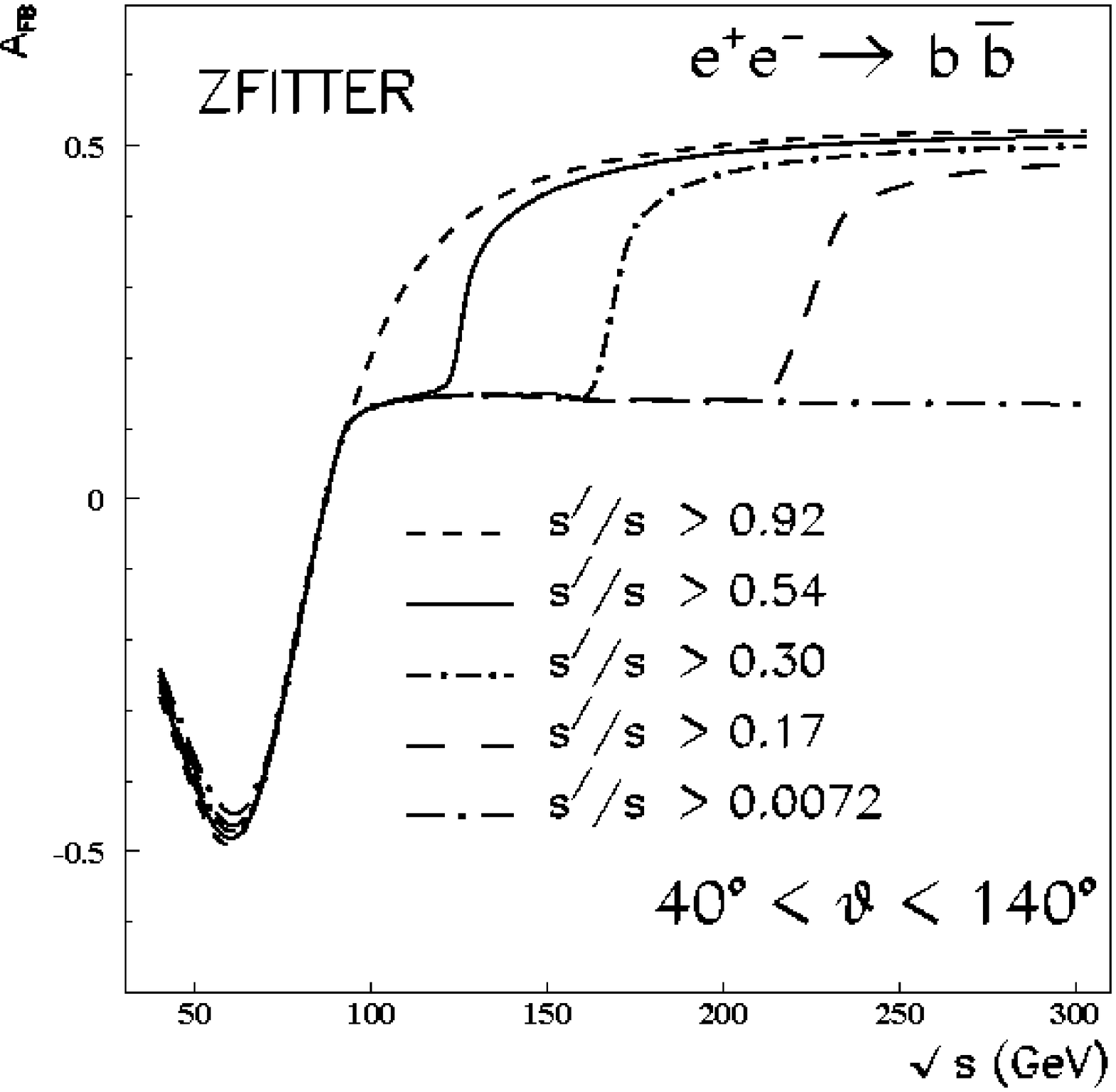,width=7.5cm}}
\\
\hspace*{-0.65cm}
\end{tabular}
\end{flushleft}
\vspace*{-1cm}
\caption[Muon pair and $b\bar{b}$ production cross sections and 
asymmetries with cuts above the $Z$ peak]
{\sf
Cross sections and forward-backward asymmetries
from {\tt ZFITTER} \cite{Bardin:1992jc2,Bardin:1999yd-orig} 
for muon pair and $b\bar{b}$ production with different cuts 
on a.~the maximal acollinearity angle $\theta_{\rm acol}$ 
or b.~minimal invariant mass squared $s'$ \cite{Christova:1999gh}.
\label{sig_afb_zf} 
} 
\end{figure}

If the radiative return is prevented, the influence of hard photonic
corrections will be much larger at higher energies than it is near the $Z$
resonance where hard bremsstrahlung is strongly suppressed.      
Fig.~\ref{sig_peak_zf} demonstrates that different portions of hard photon
emission lead to nearly identical cross sections unless the region is
reached where even soft photon emission is touched (lowest lying curve).
%
%--------------------------------------------------------------------------------
%--------------------------------------------------------------------------------
%
The excellent precision of {\tt ZFITTER} at the $Z$ peak, however,
does not automatically guarantee a sufficient accuracy at higher
energies, especially since the hard photonic contributions 
including higher order corrections are no longer suppressed. 
{For} this compare Fig.~\ref{sig_afb_zf} with Fig.~\ref{sig_peak_zf}.
It is well-known that deviations up to several per cent may result from
different treatments of radiative corrections.  
%
%--------------------------------------------------------------------------------
%--------------------------------------------------------------------------------

Fig.~\ref{sig_afb_zf} shows muon pair cross section predictions
for different acollinearity cuts at LEP~1 and LEP~2 energies.
The radiative return is prevented if $\sqrt{s'} > M_Z$. 
Depending on the value of the hard photon cut, this implies 
a certain center-of-mass energy $\sqrt{s^{\min}}$ 
above which the radiative return is suppressed 
(see Section (\ref{sub_lep1slc_phasesp} and Table \ref{rxivalues} 
for this).
For comparison, the corresponding curves for $b\bar{b}$ 
production with $s'$-cut are shown as well.
The $s'$-cut values given in the figure 
corresponds approximately to the acollinearity cut values
in the left-hand plots.

Traditionally, 
an accuracy of {\tt ZFITTER} at LEP~1 energies of the order of
$0.5 \%$ was aimed at. 
The successful running of LEP~1 together with the precise knowledge of 
the beam energy, however, made an even higher precision necessary
\cite{Christova:1998tc}: For the final measurements relative errors 
for total cross sections and absolute errors for asymmetries are expected 
up to $0.15\%$ at the $Z$ peak and of up to $0.5\%$ at 
$\sqrt{s} = M_Z~ \pm$ several GeV.
Aiming from the theoretical side ideally at a tenth of these values
for the errors of single corrections, limits of $0.015\%$, 
and $0.05\%$ respectively, can be estimated.
%~\footnote
%{This might not even be sufficient if the Giga-Z option of the TESLA  
%project will be realised (see e.g.~\cite{Moenig:1999aa}), 
%with a factor of 10 or 100 more $Z$ bosons produced than at LEP~1.
%} 
%Similar claims may be found in \cite{Bardin:1999gt}.
First applications of {\tt ZFITTER} at energies above the $Z$ resonance have 
become relevant since data from LEP~1.5 and LEP~2 are being analyzed.

Recent studies for a cut on $s'$ \cite{Placzek:1999xc}
claim for the Bhabha scattering process
$e^+e^-\to e^+e^-(n\gamma)$ that an accuracy 
of $0.3\%$ for $O(\alpha)$ corrections and of $1\%$ 
for the complete corrections has been reached at LEP~2 energies as long
as the radiative return to the $Z$ peak is prevented by cuts.
Similarly, a comparison of codes {\tt ALIBABA} and {\tt TOPAZ0}
had delivered for LEP~1 energies maximal theoretical 
uncertainties of $0.6$ per mil \cite{Beenakker:1997fi}. 
These conclusions were also drawn for $s$-channel 
fermion pair production processes in 
\cite{Bardin:1995aa,Montagna:1997jt,Bardin:1998nm,Bardin:1999gt,Jadach:1999pp,Jadach:1999gz,Passarino:1999kv}.

%So, two things have to be realized from theory 
%when calculating the important QED radiative corrections 
%to {\tt SM} observables: a.~the exact description of 
%all photonic radiative corrections and their  
%important higher order corrections and b.~the correct
%%inclusion of experimentally relevant kinematical cuts
%to the final state.
%
%=======================================================
\section{QED bremsstrahlung  
with {\it acollinearity cut}
%QED bremsstrahlung to 
%$e^{+}e^{-}\to \ell^{+}\ell^{-}$ 
%with {\it acollinearity cut}
\label{sec_lep1slc_corracol}   
}
%=======================================================
%
In the previous Section we briefly introduced the two 
different cut options which are semi-analytically
treatable: (i) kinematical cuts on the final state invariant 
mass squared $s'$, and on the minimal scattering angle $\vartheta$;
or (ii) cuts on the final state maximal aollinearity 
angle $\theta_{\rm acol}$, on their minimal energies 
$E_{{\bar f},f}$, and on the minimal scattering angle $\vartheta$.

{For} the kinematically simpler $s'$-cut the correct $O(\alpha)$ 
photonic corrections with or without a cut on one scattering 
angle are given in
\cite{Bardin:1989cw,Bardin:1991de,Bardin:1991fu}.
The {\tt ZFITTER} program relies on these duplicated analytical 
calculations, with its $O(\alpha)$ corrections basically remaining 
untouched since about 1989. Numerical comparisons with other 
two-fermion codes showed the reliability of the predictions 
at LEP energies; see e.g.~\cite{Bardin:1995aa,Bardin:1998nm,Bardin:1999gt} 
for the LEP~1 and \cite{Boudjema:1996qg} for the LEP~2 case. 

Concerning the acollinearity cut, which is experimentally 
interesting for leptonic final states as alternative to 
an $s'$-cut, the situation was not so clear:
There has been no independent check until recently
for the acollinearity cut branch and only little
literature available on the exact $O(\alpha)$ final state 
corrections to the total cross section
and forward-backward asymmetry \cite{Montagna:1993mf}. 

The corresponding part of {\tt ZFITTER} \cite{MBilenky:1989ab} was 
never checked independently and is not documented, except for a
collection of some formulae related to the initial state
corrections and its combined exponentiation with final state 
radiation for the angular distribution in \cite{Bilenkii:1989zg}.
For total cross sections only the final state
corrections are analytically known \cite{Montagna:1993mf}. 

Furthermore, when comparing cross section results from {\tt ZFITTER} 
with those from program {\tt ALIBABA} \cite{Beenakker:1991mb}, first 
deviations were observed at LEP~1 energies, but especially at 
intermediate energies above the $Z$ resonance where effects from a 
radiative return to the $Z$ could not be prevented completely by the 
applied cut. These deviations were of the order of several per cent.
Slightly later it was observed in \cite{Bardin:1999gt} that the 
perfect agreement of many predictions of {\tt ZFITTER} v.5.20 
and {\tt TOPAZ0} v.4.3 at LEP~1 energies of about typically $0.01\%$ 
could not be reproduced when an acollinearity cut was applied and the 
initial-final state interference was taken into account.

So, a recalculation and documentation of the 
acollinearity cut situation was absolutely mandatory with 
the main focus first at energies around the $Z$ boson resonance.\footnote{
The slightly more involved Bhabha scattering case with extra 
$t$ channel contributions is kept for a later analysis.
} 
First compact formulae and numerical comparisons were published in 
\cite{Christova:1998tc} for initial state bremsstrahlung without
acceptance cut, and for all first order corrections 
in \cite{Christova:1999cc}.
Before analytical results are presented, 
the hard photon phase space and a suitable parameterization 
for an acollinearity cut shall be briefly described. 

%=======================================================
\subsection{The phase space for hard photon emission
%The phase space for the
%$O(\alpha)$ hard photon emission
\label{sub_lep1slc_phasesp}
}
%=======================================================
%
It was already mentioned earlier that a three- , or respectively
two-fold analytical integration of the squared matrix 
elements has to be performed in order to calculate total
cross section observables like $\sigma_{T,FB}(s)$, or angular
cross section distributions respectively. The final 
integration over $R=s'/s$ is then performed numerically. 
We follow the phase space parameterization presented in \cite{Passarino:1982}.

\begin{figure}[htp]
  \begin{center}
\vspace*{-0.5cm}
%--- 
  \mbox{%
  \epsfig{file=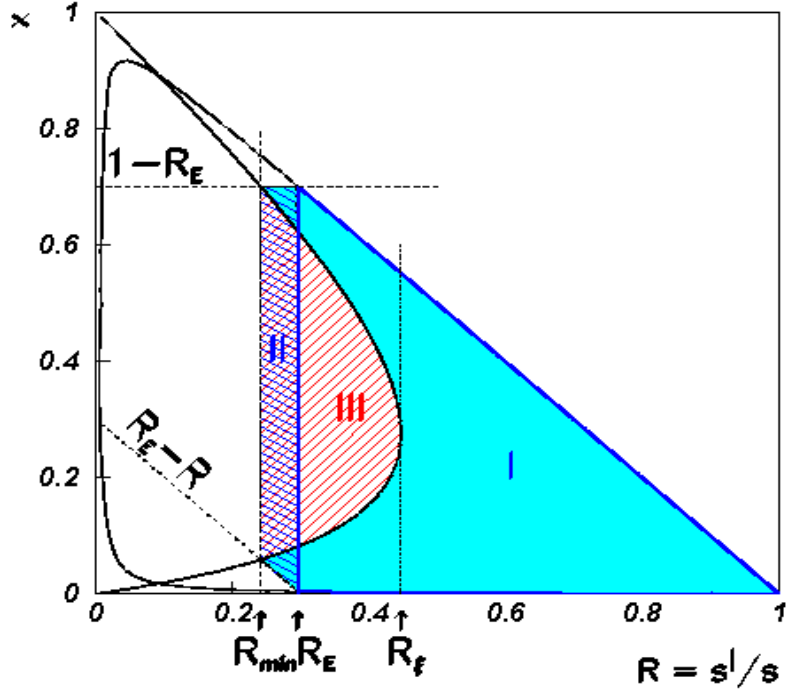,height=11cm,width=12cm}}
\vspace*{-0.5cm}
\caption[Phase space with acollinearity cut]
{\sf
Phase space with cuts on maximal acollinearity and minimal 
energy of the fermions with $x\equiv 2 p_\gamma p_{\bar{f}} / s$
\cite{Christova:1998tc,Christova:1999cc,Christova:1999gh}.
\label{dalitz} 
}
  \end{center}
\end{figure}

\vfill\eject
%--------------------
The Dalitz plot given in Fig.~\ref{dalitz} may help to 
understand the relation between a kinematically simple $s'$-cut 
and a more involved acollinearity cut.
%
%--------------------------------------------------------------------------------
%--------------------------------------------------------------------------------
%
The variable shown in Fig.~\ref{dalitz} besides $R$ is $x$, 
the normalized invariant mass of the (${\bar f}+\gamma$) rest system. 
The phase space is naturally split 
into three separate parts due to the cuts applied, 
each region described by a different and only one cut: 
\begin{itemize}
\label{regions_sum}
\item The main triangular region I is constrained for the variable $x$
      by the kinematical boundary approximately given by $(1-R)$. 
      It is equivalent to a phase space 
      with a cut on $s'$; the minimal $R$-value $R_E$ is defined
      by the minimal fermion energy, while maximal $R\to 1-\varepsilon$ 
      touches the soft photon corner of phase space.

\item Then a trapezoidal region II is added. The bounds for 
      $x$ are defined by the minimal energies cut $R_E = 2 E_{\min}/s$;
      the minimal $R$-value $R_{\min} < R_E$ is given below in (\ref{rmin})
      with $R_E$ as upper bound of $R$.       

\item Region III with a curved boundary from 
      the cut on maximal acollinearity $\theta^{\max}_{acol}$,
      is subtracted. The variable $R$ ranges between   
      $R_{\min}$ and a maximal value $R_{\theta_{\rm acol}}$
      defined in (\ref{eq:rx}).

\end{itemize}
Depending on the cuts applied, special cases can arise, if one 
uses sufficiently strong cuts on acollinearity or energies, where 
only regions I and II or regions I and III are kinematically allowed.

As Fig.~\ref{dalitz} shows, we have to determine cross sections 
in three different regions of phase space with different boundary 
values of $x$ with $R$ fixed:

\ba
\frac{d\sigma^{hard}}{d\cos\vartheta} 
= \left[ \int\limits_{\mathrm{I}} +
\int\limits_{\mathrm{II}}   
- \int\limits_{\mathrm{III}} \right]~ dR ~ dx ~
\frac{d\sigma^{hard}}{dR dx d\cos\vartheta} .
\label{sig}
\ea
Region I corresponds to a simple $s'$-cut.
The integration over $R$ extends from $R_{\min}$ to 1 with
\ba
\label{rmin}
R_{min} &=& R_E \left(1 - \frac{\sin^2(\theta_{\rm acol}^{\max}/2)}
{1-R_E\cos^2(\theta_{\rm acol}^{\max}/2)} \right) . 
\ea
The $R$ value $R_E$ is defined in (\ref{eq:re}).
The soft photon corner of the phase space resides at $R=1$.
Thus, the additional contributions related to the acollinearity cut are
exclusively due to hard photons. The boundaries for the integration over 
$x$ are, for a given value of $R$:
\ba
\label{ibound}
x_{max,min}(R) &=& \frac{1}{2} (1-R)\, \left[ 1 \pm A(R) \right],
\ea
where the value $A=A(R)$ depends in every region on only one of
the cuts applied:  
\ba
A_{\mathrm{I}}(R) &=& \sqrt{1-\frac{R_{m}}{R}} \approx 1,
\label{A1}
\\
A_{\mathrm{II}}(R) &=& \frac{1+R-2R_E}{1-R}, 
\label{A2}
\\
A_{\mathrm{III}}(R) &=& 
\sqrt{1 - \frac{R(1-R_{\theta_{\rm acol}})^2}
{R_{\theta_{\rm acol}}(1-R)^2}},  
\label{A3}
\ea
with
\ba
  \label{eq:rs}
R_{m}  &=& \frac{4m_f^2}{s},
\\
  \label{eq:re}
R_E &=& \frac{2E_{min}}{\sqrt{s}}, 
\\
  \label{eq:rx}
R_{\theta_{\rm acol}} &=&
\frac{1-\sin(\theta_{\rm acol}^{\max}/2)}
{1+\sin(\theta_{\rm acol}^{\max}/2)}.     
\ea
Here, $m_f$ and $E_{min}$ are the final state fermions' mass and a cut on
their individual energies in the cms. For simplicity, equal energy cuts
are used for both fermions.

An acollinearity cut may act as a simple cut on invariant masses and
thus it may prevent the radiative return 
of $\sqrt{s'}$ to the $Z$ peak (and the development of the radiative tail) 
for measurements at higher $\sqrt{s}$.
In Fig.~\ref{dalitz} it may be seen that a reasonable analogue of a cut
value $\sqrt{s^{min}}$ is the upper value of $s'$ of region III,
$R_{\theta_{\rm acol}}$, defined in (\ref{eq:rx}):
\ba
\nonumber
{\DD\sqrt{s^{\min}}}
&=&
 \frac{M_Z}{\sqrt{R_{\theta_{\rm acol}}}}.
\ea
The relations are also visualized in Table (\ref{rxivalues})
and the effects of some cut values have been shown in Fig.~\ref{sig_afb_zf}.
%
%--------------------------------------------------------------------------------
\begin{table}[htb]
\begin{center}
\begin{displaymath}
\begin{array}{|r@{.}l|r@{.}l|r@{.}l|} 
\hline
\multicolumn{2}{|c|}{}&\multicolumn{2}{c|}{}&\multicolumn{2}{c|}{}
\\
 \multicolumn{2}{|c|}{\raisebox{1.5ex}{${\DD {\theta_{\rm acol}}}$}} 
& \multicolumn{2}{c|}{\raisebox{1.5ex}[-1.5ex]
{
$
{\DD R_{\theta_{\rm acol}} 
%= \frac{1-\sin{\frac{{\DD {\theta_{\rm acol}}}}{{\DD 2}}}}
%{1+\sin{\frac{{\DD {\theta_{\rm acol}}}}{{\DD 2}}}}
}
$
}} 
& \multicolumn{2}{c|}{\raisebox{1.5ex}
{
${{\DD\sqrt{s^{\min}}
%= \frac{M_Z}{\sqrt{R_{\theta_{\rm acol}}}}
}}
$
}}
\\\hline\hline
     ~~~0&0^\circ~   & ~1&0000~   & ~~~~91&2~\GeV~~
\\\hline
     2&0^\circ   & 0&9657   & 92&8~\GeV
\\\hline
     5&0^\circ   & 0&9164   & 95&3~\GeV
\\\hline
    10&0^\circ   & 0&8397   & 99&5~\GeV
\\\hline
    15&0^\circ   & 0&7691  & 104&0~\GeV
\\\hline
    20&0^\circ   & 0&7041  & 108&7~\GeV
\\\hline
    25&0^\circ   & 0&6441  & 113&6~\GeV
\\\hline
    30&0^\circ   & 0&5888  & 118&8~\GeV
\\\hline
    45&0^\circ   & 0&4465  & 136&5~\GeV
\\\hline
    60&0^\circ   & 0&3333  & 157&9~\GeV
\\\hline
    75&0^\circ   & 0&2432  & 184&9~\GeV
\\\hline
    90&0^\circ   & 0&1716  & 220&1~\GeV
\\\hline
   120&0^\circ   & 0&0718  & 340&3~\GeV
\\\hline
   150&0^\circ   & 0&0173  & 692&6~\GeV
\\\hline
   180&0^\circ   & 0&     &\multicolumn{2}{c|}{\infty}
\\\hline
\end{array}
\end{displaymath}
\caption[Minimal center-of-mass energies without radiative return
to the $Z$ peak]
{\sf
The 
minimal center-of-mass energy $\sqrt{s^{\min}}$
at which the radiative return
to the $Z$ peak is prevented by an acollinearity cut given as a
function of this cut \cite{Christova:1999gh}.
\label{rxivalues}
}
\end{center}
\end{table}
%--------------------------------------------------------------------------------
%
%=======================================================
\subsection{Initial state radiation 
and mass singularities
\label{sub_lep1slc_mass_sing}   
}   
%=======================================================
%
In the last section we saw that the Dalitz plot 
in Fig.~\ref{dalitz} describing the hard photon phase space 
is independent of the scattering angle $\cos\vartheta$.
Here it will be shown that the integrations in regions II and III are
nevertheless crucially influenced by $\cos\vartheta$. 
We will have to deal with artificially arising mass singularities
when neglecting masses for analytical integration and will have to
adjust our phase space treatment accordingly. The final results,
however, can be shown to be finite when going to the continuous 
phase space limit.
We want to illustrate this for the integration of the hard 
initial state radiation. The treatment of the QED initial-final 
state interference is then completely analogous, the final state
is safe of these mass singularities. 

First, we want to denote with $k_1$, $k_2$, $p_1$, $p_2$, and $p$ 
the 4-momenta of $e^-$, $e^+$, $f^-$, $\bar{f}$ and of the hard photon,
respectively. 
We have two photon angles $\varphi_\gamma$ and $\theta_\gamma$
and one fermionic scattering angle $\vartheta$.
The variables depicted in Fig.~\ref{dalitz} were $R=s'/s$ as
normalized final state invariant mass squared ($s'=m_{f\bar{f}}^2$) 
and $x = 2 p_2 p/s$ as normalized invariant mass squared of 
the $(\bar{f}+\gamma)$ subsystem. A detailed description
of the phase space and complete kinematics is given in 
Appendix \ref{crossphase}. The analytical integration will be performed
over the three variables $\varphi_\gamma$, $x$, which is 
affine linear in $\cos\vartheta_\gamma$, and $\cos\vartheta$, if 
interested in flux functions for total cross sections and asymmetries. 
The final results for cross sections and their angular distributions 
over $\cos\vartheta$ are obtained from a numerical integration over 
$v\equiv 1-R = 1-s'/s$ (or equivalently over $R$) after regularization 
with the first order soft and virtual corrections.  

Starting with the bremsstrahlung contribution from 
{\it initial state radiation}, the corresponding squared 
matrix element contains the electron (positron) propagator, 
and these terms are proportional to first and second powers of 
\ba
\frac{1}{Z_{1(2)}} &=& - \frac{1}{(k_{1(2)}-p)^2-m_e^2} 
    = \frac{1}{2k_{1(2)}p}
\label{z12a}
\\
&=& \frac{1}{A_{1(2)} \pm b \cos\varphi_\gamma},
\label{z12b}
\ea
with
\ba
A_{1(2)} &=& \frac{s}{2}  
(1-R)(1 \pm \beta_0 
\cos\vartheta 
\cos\theta_{\gamma} ),
\\
B &=& \frac{s}{2}  
(1-R)\beta_0 \sin\vartheta \sin\theta_{\gamma},
\label{A12B}
\ea
and
\ba
\beta_0 &=& \sqrt{1- 4m_e^2/s},
\\
\cos\theta_{\gamma}&=& \frac{\lambda_1 - \lambda_2 -\lambda_p }
{2\sqrt{\lambda_2\lambda_p}},
\ea
and
with $\sqrt{\lambda_1} = (1-x)s$, $\sqrt{\lambda_2} = (x+R)s$, 
and $\sqrt{\lambda_p} = (1-R)s$.
Final state mass effects have to be neglected
for a complete analytical integration over all angles of 
phase space.

The first analytical integration is now performed over 
$\varphi_\gamma$, the photon production angle in the 
($f+\gamma$) rest system \cite{Passarino:1982}:
\ba
\label{firstint}
\frac{d\sigma^{hard}}{{d{R}}\,{d{x}}\,{d{\cos\vartheta}}} &=& 
  \int\limits_{0}^{2\pi} d{\varphi_\gamma}\, 
\frac{d\sigma^{hard}}{{d{R}}\,{d{x}}\,{d{\cos\vartheta}}
\,{d{\varphi_\gamma}}}.
\ea
It is important to take care of the electron mass $m_e$
in order to regularize mass singularities from collinear 
initial state photon emission.
This has to be done e.g.~in the following contribution:

\ba
\label{integral}
\int\limits^{2\pi}_0 d{\varphi_\gamma}\,\frac{1}
{Z_i(R,\cos\vartheta,x,\varphi_\gamma)} 
&=&
\frac{2\pi\sqrt{\lambda_2}}{\sqrt{C_i}},
\ea
\ba
\label{integral2}
C_i &=& s^2 a_i x^2 - 2 s b_i x  + c_i,
\\
a_i &=& s^2 ( z_i^2-R\,\eta_0^2),
\\
b_i &=& s^3[ R\, z_i\,(1- z_i)-\frac{1}{2}\,R(1-R)\,\eta_0^2],
\\
c_i &=& s^4\,R^2\,(1- z_i)^2,
\\
z_{1(2)} &=& \frac{1\mp\beta_0\cos\vartheta}{2}
+ R\frac{1\pm\beta_0\cos\vartheta}{2},
\\
\eta_0^2 &=& 1-\beta_0^2;\quad \beta_0^2 = 1-\frac{4 m_e^2}{s}.
\ea
In the second step we integrate over $x$ with limits given in 
(\ref{ibound}). One of the basic integrals arising is for example:
\ba 
\label{integral3}
I_i^0(R,\cos\vartheta)
&=&
s\int\limits_{x_{min}}^{x_{max}}
\frac{d x}{\sqrt{C_i}} 
\nonumber\\
&=&
\frac{1}{\sqrt{a_i}}
\left.
\ln\left[\sqrt{a_i} C_i^{\frac{1}{2}}+ (s a_i x - b_i) \right]
\right|^{x_{max}}_{x_{min}},
\ea
with
\ba
\label{integral4}
C_i^{\frac{1}{2}} |_{x_{max,min}}
&=& 
\frac{1}{2}\,s^2(1-R)
\sqrt{(y_i\pm A z_i)^2+R\,(1-A^2)\,\eta_0^2},
\\
\nonumber\\
\label{integral5}
(s a_i x - b_i)\Biggr|^{x_{max}}_{x_{min}}
&=& (1-R) (y_i \pm A z_i) + O(\eta_0^2),
\ea
and
\ba
\label{yi}
y_{1(2)} 
&=& \frac{1\mp\beta_0\cos\vartheta}{2}
- R\, \frac{1\pm\beta_0\cos\vartheta}{2}.
\ea
Again, in order to be able to integrate analytically
over $\cos\vartheta$, we are now interested in the limit of 
vanishing electron mass $m_e$ for the subsequent integrations.
In this limit, there will occur zeros in arguments of logarithms 
like the one in (\ref{integral3}) at four different locations 
in the remaining $\cos\vartheta$-$R$ phase space.
These locations are defined by the conditions:
\ba
\label{yaz}
y_i \pm A z_i = 
y_i(R,\cos\vartheta) \pm A(R)\, z_i(R,\cos\vartheta) = 0, ~~~~~i=1,2.
\ea
These zeros appear as functions of $\cos\vartheta$ with parameters
$R$ and $A=A(R)$ at certain values $\cos\vartheta = c_i^{\pm}$ ($i=1,2$):
\ba
\label{c1+}
c_1^{+}(R) &=& -\frac{1-R+A(R) (1+R)}{1+R+A(R)(1-R)} \le 0 \quad 
\forall\, R\,\epsilon\,[0;1],
\\
\label{c1-}
c_1^{-}(R) &=& -\frac{1-R-A(R) (1+R)}{1+R-A(R) (1-R)},
\\
\vspace*{0.4cm}
\label{c2+}
c_2^{+}(R) &=& - c_1^{+}(R)\ge 0 \quad
\forall\, R\,\epsilon\,[0;1], 
\\
\vspace*{0.4cm}
\label{c2-}
c_2^{-}(R)  &=& - c_1^{-}(R).
\ea
The relations $c_1^{+} \le c_1^{-}$ and 
$c_2^{+}  \ge c_2^{-}$ are also fulfilled.

In the course of integration, different 
analytical expressions have to be used in different
kinematical regions when neglecting $m_e$ wherever
possible, except for logarithmic terms proportional to  
$L_e = \ln({s}/{m_e^2})$ and 
$L_{\beta}=\ln(1\pm\beta_0\cos\vartheta)$
from collinear photon emission.
This results in cutting the remaining phase space for 
the $\cos\vartheta$-integration, at fixed $R$ and for 
given $i$, into three different regions.
This splitting of phase space of course translates into
a phase space splitting for each one of the three regions 
I, II, and III parameterized by $A(R)$ in (\ref{A1}) 
to (\ref{A3}), depending on the specific value of $R$ 
and depicted in Fig.~\ref{dalitz}.

The differential cross section  
${d{\sigma}}/({d{R}\,d{\cos\vartheta}})$ is a double-sum over $i=1,2$, 
but for the $s'$-cut-like region I the conditions
(\ref{c1+}) to (\ref{c2-}) become trivial,
\ba 
\label{cI}
c_2^{+}(I) = - c_1^{+}(I) = - c_1^{-}(I) = - c_2^{-}(I)= 1,
\ea
only leaving one case to be studied ($-1\leq\cos\vartheta \leq 1$).
In regions II and III, however,  
${d{\sigma}}/({d{R}d{\cos\vartheta}})$ will consist of
different analytical expressions for each combination of the 
kinematical ranges defined by (\ref{c1+}) to (\ref{c2-}).
The final result for e.g.~$I^0_i$ after integration over $x$,
setting $m_e=0$, becomes ($i=1,2$):

\vfill\eject
%----------------
\begin{itemize}
\item[(i)] For $|\cos\vartheta| < |c^{-}_{i}|$ 
with $ y_i\pm A z_i>0 $ (case $(i)^{+\,+}$):  
\ba
\label{I0i}
I_i^0 = \frac{1}{s z_i}\ln\left(\frac{y_i+A z_i}
{y_i-A z_i}\right),
\ea
\item[(ii)] for $|c^{-}_{i}|<|\cos\vartheta|<|c^{+}_{i}|$ 
with $y_i+A z_i> 0$ and
$y_i-A z_i< 0 $ (case $(i)^{+\,-}$):
\ba
\label{I0ii}
I_i^0 = \frac{1}{s z_i}\left\{\ln\left[\frac{ z_i^2(y_i+A z_i)
(A z_i-y_i)}{R^2(1-\beta_0^2\cos^2\vartheta)}\right]
+\ln\left(\frac{s}{m_e^2}\right)\right\},
\ea
\item[(iii)] for $|\cos\vartheta|>|c^{+}_{i}|$ 
with $y_i\pm A z_i< 0$ (case $(i)^{-\,-}$): 
\ba
\label{I0iii}
I_i^0 =  -\frac{1}{s z_i}\ln\left(\frac{y_i+A z_i}
{y_i-A z_i}\right).
\ea
\end{itemize}

It can be shown that the resulting number of cases for the 
angular distribution, depending on the value of $\cos\vartheta$ 
with respect to $c^{\pm}_{i}$ and on $R$, is at most four in 
regions II and III, as only certain combination of signs 
of $c^{\pm}_{i}$ are possible (see also Appendix \ref{hardini}).
These are for $\cos\vartheta\ge 0$ with the abbreviations given 
in (\ref{I0i}) to (\ref{I0iii}) \cite{Christova:1999gh}:
\ba
\label{case1}
\mbox{a.} && (1)^{+\,+}\quad \mbox{combined with}\quad (2)^{+\,+},
\label{case1a}
\\
\nonumber\\
\mbox{b.} && (1)^{+\,-}\quad \mbox{combined with}\quad (2)^{+\,-},
\label{case1b}
\\
\nonumber\\
\mbox{c.} && (1)^{+\,+}\quad \mbox{combined with}\quad (2)^{+\,-},
\label{case1c}
\\
\nonumber\\
\mbox{d.} && (1)^{-\,-}\quad \mbox{combined with}\quad (2)^{+\,+}.
\label{case1d}
\\
\nonumber
\ea
{For} $\cos\vartheta<0$, cases c.~and d.~are exchanged 
by {c.'} and {d.'} (formally by interchanging indices $(1)$ and $(2)$):\footnote{
{For} region I in (\ref{A1}), only case b.~is possible.
}
\ba
\label{case2}
\mbox{a.\,} && (1)^{+\,+}\quad \mbox{combined with}\quad (2)^{+\,+},
\label{case2a}
\\
\nonumber\\
\mbox{b.\,} && (1)^{+\,-}\quad \mbox{combined with}\quad (2)^{+\,-},
\label{case2b}
\\
\nonumber\\
\mbox{c.'} && (1)^{+\,-}\quad \mbox{combined with}\quad (2)^{+\,+},
\label{case2c}
\\
\nonumber\\
\mbox{d.'} && (1)^{+\,+}\quad \mbox{combined with}\quad (2)^{-\,-}.
\label{case2d}
\ea

{For} the hard photon parts of the total cross section $\sigma_T(c)$ or 
forward-backward asymmetry $A_{FB}(c)$ 
with acceptance cut $c$, we finally have to integrate 
over $\cos\vartheta$ within cut boundaries $\pm c$:

\ba
\label{xsc}
\sigma_T^{hard}(c) &=& \int\limits^c_{-c}{d\cos\vartheta}\, 
\frac{d\sigma^{hard}}{d\cos\vartheta};
\\
\label{afbc}
\sigma_{FB}^{hard}(c) &=& 
\left(
\int\limits^c_{0}-\int\limits^{0}_{-c}
\,\right) 
{d\cos\vartheta}\,\frac{d\sigma^{hard}}{d\cos\vartheta}
\\
&=& 
\left(
\int\limits^c_{0}+\int\limits^{-c}_{0}
\,\right) 
{d\cos\vartheta}\,\frac{d\sigma^{hard}}{d\cos\vartheta}.
\ea

Looking at the possible logarithmic expressions 
which have to be integrated, 
like the ones in (\ref{I0i}) to (\ref{I0iii}),
one immediately sees that these are,
except for some rational factors in $R$, merely
reproduced with the variable $\cos\vartheta$
now replaced by $c$ (see Appendix \ref{hardini}). 
The treatment of mass singularities for $m_e\to 0$ 
and the distinction of different regions in phase space 
therefore has to be repeated for the totally integrated case 
where the cut-off $c$ now plays the role of $\cos\vartheta$.
Depending on the relative position of $c$ with respect to
the values $c_i^{\pm}$, we have to integrate over different 
expressions of the angular distribution 
in the $\cos\vartheta$-$R$ phase space
(see cases a.~to d.~or a.~to {d.'} above).

For the corresponding hard radiator functions
$H_T(R,A,c)$ and $H_{FB}(R,A,c)$, 
defined for example by (\ref{generic_zf}),
one finally gets at most four or respectively six 
different analytical expressions from different 
regions of phase space. This is because of symmetric 
cancellations when integrating over $\cos\vartheta$
for $\sigma_T^{hard}$, while in the definition of
$\sigma_{FB}^{hard}$ there is the additional occurance 
of $c=0$ which leads to more cases; see (\ref{xsc}) and
(\ref{afbc}).

If the acceptance cut is omitted, i.e.~setting $c=1$, 
only case d. (or respectively d.') from above remains for  
$\sigma_T^{hard} = \sigma(1) - \sigma(-1)$ 
in regions II and III because then $-1< c^{\pm}_{i} < 1$.
For $\sigma_{FB}^{hard}$ with 
$\sigma_{FB}^{hard} = \sigma(1)-2 \sigma(0)+ \sigma(-1)$,
two cases are left because the additional integrated 
contributions from $\cos\vartheta=0$ depend on whether 
$c^{-}_{2}> 0$ from (\ref{c2-}) or not.
The conditions (\ref{c1+}) to (\ref{c2-}) are fulfilled for
$\cos\vartheta = 0$ with 
\ba
\label{a0}
A_0(R) &=& \frac{1-R}{1+R},
\ea
so that, depending on the sign of $(A(R)-A_0(R))$, 
one or the other analytical expression has to be used.
This will lead for $c=1$ to quite compact results for 
$\sigma_{T}$ and $\sigma_{FB}$,
presented in the next Subsection \ref{sub_lep1slc_formacol}   
and published in \cite{Christova:1999cc}.

One can check that 
the integrated results $\sigma_{T,FB}^{hard}$ are continuous when 
$c\rightarrow c_i^{\pm}$, while 
$d\sigma^{hard}/d\cos\vartheta$ can be regularized at
$\cos\vartheta=c_i^{\pm}$, 
taking the exact logarithmic results in $m_e$ for the integrals.  
So, the artificially introduced mass `singularities' when 
neglecting $m_e$ have to cancel.
Also, as a further check, the contributions proportional to
the Born cross section $\sigma^0$ and Born asymmetry $A_{FB}$ are 
(anti)symmetric respectively, as it should be for the one loop
corrected initial state results. 

The phase space splitting discussed above also has an influence 
on the initial-final state interference corrections 
since there the initial state propagators with $Z_{1,2}^{-1}$ 
appear linearly. The discussion of the different steps of integration 
can be done completely analogously as for the initial state case.
These propgators, however, do not contribute in the 
{\it final state expressions} so that the phase space 
splitting is not necessary there.

Summarizing, we observe that neglecting the initial and final state masses 
at the mentioned high energies necessitates a separation of the 
phase space formed by the cosines of the remaining two angles of integration 
$\cos\vartheta$ and $\cos\theta_{\rm acol}$ into several different regions. 
Only where necessary, the masses are kept in order to regularize 
the mass singularities from collinear radiation of bremsstrahlung
photons. This splitting of phase space 
delivers for each region different analytical
expressions for the calculated observables \cite{Christova:1999gh}.
{For} the special cases of either full angular 
acceptance, i.e.~no cut on $\cos\vartheta$, or no cut 
on the acollinearity angle $\theta_{\rm acol}$, the number of different
expressions can be substantially reduced and very compact formulae 
can be obtained \cite{Christova:1999cc}. 

%=======================================================
\subsection{Cross section formulae
%Formulae for the total cross section $\sigma_T$ 
%and forward-backward asymmetry $A_{FB}$ with acollinearity cut
\label{sub_lep1slc_formacol}   
}   
%======================================================

Beginning with the soft and virtual photon corrections,
these are of course independent of the applied cuts and 
we can use the results for the initial, final, interference,
and box corrections from \cite{Bardin:1989cw,Bardin:1991de,Bardin:1991fu}. 
Concerning the hard corrections
with acollinearity cut, see the phase space discussion
in the previous Subsection (\ref{sub_lep1slc_phasesp}).
%
%Before presenting and discussing the three different radiators to
%initial state and final state bremsstrahlung and QED interference,
%we show generically the principal structure of the flux functions
%using the initial state radiator as an example. It is completely
%analogous for the other hard contributions.
%
We have seen that the hard photon part of the total 
cross section including initial state 
radiation can be written as the sum of different 
contributions from three regions in phase space:

%---------------------
\ba
\sigma_T^{hard}(s) = \left[ \int\limits_{\mathrm{I}} + \int\limits_{\mathrm{II}}  
- \int\limits_{\mathrm{III}} \right] ~ ds' ~ dx ~ 
d{\cos\vartheta}\, \frac{d\sigma(A)}{ds' dx d\cos\vartheta},
\label{sigt}
\ea
%---------------------
with the parameter $A=A(s'/s)$ and its different meanings 
in these regions given in (\ref{A1}), (\ref{A2}), and (\ref{A3}). 
%---------------------
%\ba
%A_{\mathrm{I}} &=& 1,
%\\
%A_{\mathrm{II}} &=& (1+R-2R_E)/(1-R), 
%\\
%A_{\mathrm{III}} &=& [1 - R(1-R_{{\theta_{\rm acol}}})^2/(R_{{\theta_{\rm acol}}}(1-R)%^2)]^{1/2}, 
%\label{A}
%\ea
%---------------------
%with $R_E = 2E_{min}/\sqrt{s}$, 
%$R_{{\theta_{\rm acol}}} = (1-\sin({\theta_{\rm acol}}/2))/(1+\sin({\theta_{{\rm acol}%}}/2))$, and $R=s'/s$. 
%
%We would like to remind that 
%(\ref{sig} and (\ref{sigtA} are only meant to be generic formulae 
%to demonstrate the description of the hard photon effects, 
%with a suitable cut-off to remove the soft photonic phase space 
%region. A derivation of the complete regularized radiator 
%functions with the $O(\alpha)$ real hard and soft and virtual 
%photon contributions will be given. The infrared poles will be
%cancelled and the arbitrarily chosen, but unphysical cut-off between 
%hard  and soft photon phase space will also cancel. 

%====================================================================
\subsubsection*{Initial state radiation }
\label{ssub_lep1slc_ini}
%====================================================================

For the total cross section, the analytical formula with cuts on
acollinearity and minimal fermion energy is remarkably compact
for the full angular acceptance ($c=1$). 
For the initial state hard radiator function
$H_T^{ini}(R,A)$ we have, replacing the Bonneau-Martin function  
$H_{BM}(R)$ from (\ref{eq:bm}) in (\ref{eq:3}) and (\ref{eq:3a}):

%------------------
\ba
\label{init1}
H_T^{ini}(R,A) &=& 
\frac{3\alpha}{4\pi} Q_e^2 
\left[
\left(A+\frac{A^3}{3}\right) \frac{1+R^2}{1-R}
\left( \ln\frac{s}{m_e^2}-1\right) + (A-A^3) \frac{{\cal B} R}{1-R}
\right],
\nonumber\\
\ea
%------------------
with ${\cal B}=2$.
In $\sigma_{FB}^{ini}(s)$, the corresponding hard radiator part is:
%------------------------------------------------------------------ 
\ba
\label{inifb1}
H^{ini}_{FB}(R,A\ge A_0)&=&
\frac{\alpha}{\pi} Q_e^2
\Biggl\{
\frac{1+R^2}{1-R}\Biggl[
\frac{4 R}{(1+R)^2} \left( \ln\frac{s(1+R)^2}{4m_e^2R} - 1 \right)
\nonumber\\
&&
  -~\frac{1}{(1+R)^2}
\left[
y_+ y_- \ln\left|y_+ y_-\right| 
+
{4 R}\ln(4 R)
\right]
\nonumber\\
&&-~(1-A^2) \left( \ln\frac{s}{4m_e^2(1+A)^2R} - 1\right)
\Biggr]
+\frac{4A(1-A)R}{1-R}
\Biggr\}, 
\nonumber\\
\\
%==============================================
\label{inifb2} 
H^{ini}_{FB}(R,A<A_0)&=&
\frac{\alpha}{\pi} Q_e^2
\Biggl\{
\frac{1+R^2}{1-R}
\left[
-\frac{y_+ y_-}{(1+R)^2}
\ln\left|\frac{y_+}{y_-}\right|
+(1-A^2)\ln\frac{1+A}{1-A}\right]
\nonumber\\
&&
+~\frac{8 A R^2}{(1+R)(1-R)}
\Biggr\}.
%------------------------------------------------------------------ 
\ea
%------------------------------------------------------------------ 
%
The following definition is used:
\ba
%  \label{eq:a0}
%A_0(R) &=& \frac{1-R}{1+R},
%\\
  \label{eq:yy}
y_{\pm} &=& (1-R) \pm A(1+R).
\ea
In region I, ($A \to 1$), the above expressions (\ref{init1}) and (\ref{inifb1})
reduce to those known from \cite{Bonneau:1971mk} and \cite{Bardin:1989cw}.
The phase space regions II and III do not contribute there.
In this region the radiators diverge for $R \to 1$, and soft photon
exponentiation and the subtraction $\beta/(1-R)$ is applied there 
in order to get ${\bar H}_B^{ini}(R)$, $B=T,FB$; see (\ref{eq:3a}).
For phase space regions II and III, i.e.~$A\neq 1$, safe of infrared
divergent contributions we immediately 
have ${\bar H}_{B}^{ini}(R,A) = H_{B}^{ini}(R,A)$.
 
The additional contributions for $A=1$ from final state radiation 
and the initial-final state interference to $\sigma_T$ 
(and also those to $\sigma_{FB}$) may be found in \cite{Bardin:1989cw}. 
Important to note is that, differing from (\ref{init1}), 
the coding in the program {\tt ZFITTER} corresponds to  
${\cal B}= 4/3$ if one looks there into the limit $c=1$.
The resulting numerical deviations are typically of the 
order of $0.5\%$ to $2\%$. They will not lead to drastical 
improvements in the comparisons shown later in 
Section \ref{sec_lep1slc_codes}.

%====================================================================
\subsubsection*{Initial-final state interferences}
\label{ssub_lep1slc_int}
%====================================================================
%
In the initial-final state interferences,
the effective Born cross sections depend on both $s$ and $s'$
as well as on the type of exchanged vector particles $V_i$ (e.g.~photon and or
$Z$):
\ba
\label{intro_box}
\sigma_B^{int}(s) &=& \int dR \sum_{V_i,V_j=\gamma, Z} 
\sigma_{\bar B}^0(s,s',i,j)~ \rho_B^{int}(R,A,i,j).
\ea
%
%For $B$=$T$ it is ${\bar B}$=$FB$ and vice versa.
%
The radiator functions are:
\ba
\label{eq:rhoint}
\rho_B^{int}(R,A;i,j) &=& \delta(1-R) \left[S_B + b_B(i,j) \right] 
+ \theta(1-R-\epsilon) H_{B}^{int}(R,A).
\nonumber\\
\ea
The soft corrections, already known from 
\cite{Bardin:1989cw,Bardin:1991fu},
we give explicitly:
\ba
   \label{eq:intsoftt}
S_T^{int}  &=& 8 \frac{\alpha}{\pi} Q_eQ_f
\left( 1-\ln\frac{2\epsilon}{\lambda}\right),
\\
     \label{eq:intsoftfb} 
S_{FB}^{int} &=& \frac{\alpha}{\pi} Q_eQ_f
\left[
-\left(1+8\ln 2\right) \ln\frac{2\epsilon}{\lambda} +4\ln^2 2 + \ln 2 +
\frac{1}{2} + \frac{1}{3}\pi^2    \right] .
 \ea
The box contributions  $b_T(i,j)$ may be taken from equations (116)
and (118) (to be multiplied by 4/3) of
\cite{Bardin:1991fu} and the $b_{FB}(i,j)$ from equations (123) and
(126). Finally, the hard radiator parts are:

\ba
  \label{eq:intht}
  H^{int}_{T}(R,A) &=&  
-\frac{\alpha}{\pi} Q_eQ_f
\frac{4 A R (1+R)}{1-R},
\ea
and 
\ba
  \label{eq:inthfb1}
 H^{int}_{FB}(R,A\geq A_0)&=&
\frac{\alpha}{\pi} Q_eQ_f 
\Biggl\{
\frac{3R}{2}\left[\ln\frac{z_+}{z_-}
+\frac{2-R+\frac{5}{3}R^2}{1-R}\ln R 
\right]
\nonumber\\
&&-~\frac{1+R}{2(1-R)}
(5-2 R+5 R^2)
\ln\frac{(1+R)(1+A)}{2} 
\nonumber\\
&&+~\frac{1}{4(1-R)}\Biggl[
\frac{(1-4R+R^2)[A(1+R)^2-(1-R)^2]}{1+R}
\nonumber\\ &&+~ 2 A (1-A) (1+R^3)\Biggr]
\Biggr\},
\\
\nonumber\\
  \label{eq:inthfb2}
H^{int}_{FB}(R,A<A_0)&=&
\frac{3\alpha}{2\pi} Q_eQ_f ~ R
\Biggl\{
\ln\frac{z_+}{z_-}
-
\frac{2-R+\frac{5}{3}R^2}{1-R}
\ln\frac{1+A}{1-A} 
+ A (1-R)  
 \Biggr\},
\nonumber\\
\ea
with
\ba
z_{\pm} = (1+R) \pm A(1-R).  
\label{eq:zz}
\ea
Again, for $A\to 1$ the radiators 
$H^{int}_{T}(R,A)$ and $H^{int}_{FB}(R,A\geq A_0)$
approach the known expressions of the $s'$-cut given in
\cite{Bardin:1989cw}. A misprint could be found 
in eq. (22) of \cite{Bardin:1989cw}: 
The non-logarithmic terms there have to be multiplied by $1/(1+R)$. 

%====================================================================
\subsubsection*{Final state radiation}
\label{ssub_lep1slc_fin}
%====================================================================
%
The final state corrections to order $O(\alpha)$ are:

\ba
\label{fins}
\sigma_B^{fin}(s) &=& \sigma_B^0(s) \int dR~ \rho_B^{fin}(R,A),
\ea
with
\ba
  \label{eq:f1}
\rho_B^{fin}(R,A) &=& \delta(1-R) S_f + \theta(1-R-\epsilon)
H_{B}^{fin}(R,s,A), 
\\
S_f &=& {\bar S}_f + \beta_f \ln \epsilon,
\ea
where ${\bar S}_f$ and $\beta_f$ can trivially be obtained 
from the initial state terms ${\bar S}$ and $\beta$,
replacing $s/m_e^2$ by $s'/m_f^2$ and $Q_e$ by $Q_f$.
The hard radiators are:

\ba
  \label{eq:fint}
H^{fin}_{T}(R,s,A)&=& 
\frac{\alpha}{\pi} Q_f^2
\left[
\frac{1+R^2}{1-R} \ln\frac{1+A}{1-A}
-\frac{8 A m_f^2/s}{(1-A^2)(1-R)}-A(1-R)
\right],
\nonumber\\
\\
%------------------------------------------------------------------ 
  \label{eq:finfb}
H^{fin}_{FB}(R,s,A)
&=& 
H^{fin}_{T}(R,A)
+
\frac{\alpha}{\pi} Q_f^2
\left[
A(1-R)
-(1+R)\ln\frac{z_+}{z_-}
\right].
%------------------------------------------------------------------  
\ea
Some analytical formulae for the final state corrections are also 
given in \cite{Montagna:1993mf}. This will be treated in more
detail in Appendix \ref{fin}.

If one is interested in considering the leading higher order effects
of multiple soft photon emission and virtual corrections, a 
{\it common initial and final state soft photon exponentiation} may be
performed, following \cite{Nicrosini:1988sw,Bardin:1991fu}:

\ba
  \label{eq:comex}
 \sigma_B^{ini+fin}(s) &=& 
\int dR ~\sigma^0(s') ~\rho_B^{ini}(R,A) ~{\bar \rho}_B^{fin}(R,s',A), 
\ea
with 
\ba
  \label{eq:rbar}
  {\bar \rho}_B^{fin}(R,s',A)  &=& 
%\int\limits_{R_{min}/R}^{1} du ~\rho_B^{fin}(u,s',A').
(1-R_E)^{\beta'_f}(1+S'_f) 
\\
&&+ \int\limits_{R_{min}/R}^{1} du 
\left[H^{fin}_{B}(u,s',A')-\frac{\beta'_f}{1-u}
\theta(R-R_E)\right].
\ea
The soft part of $\rho_B^{fin}(u,s',A')$, $A'=A(u)$, and $\beta'_f$
are derived from (\ref{eq:3}) by replacing there $Q_e$ by $Q_f$ and 
$s/m_e^2$ by $s'/m_f^2$. Such a procedure for resumming the leading 
photonic higher order corrections is straightforward and for 
the theoretical confirmation of the experimental precision results 
on the $Z$ peak absolutely mandatory there.

It shall be mentioned here that in the hard radiators 
the integration over $u$ may also be performed analytically. 
In region III one has to interchange for this the order of
integration over $u$ and $x$ \cite{Montagna:1993mf}. 
There the over $R=s'/s$ analytically fully integrated result 
for $\sigma_B^{fin}(s)$ was calculated. The recalculation
could also correct for some smaller misprints there.
The general results with acceptance cut in 
\cite{Montagna:1993mf} can be easily compared 
with the results here setting $c=1$. 
A complete derivation of the angular distribution
$d{\sigma^{fin}}/{d{\cos\vartheta}}$ is illustrated
in Apendix \ref{fin}.

%====================================================================
\subsubsection*{General example: Hard photon initial state 
{\it radiator} with general cuts}
\label{ssub_lep1slc_gen}
%====================================================================
%
We want to present here as an example of one of the main results
obtained during this dissertation. These are radiator functions  
\ba
\label{hardradini}
{H}^{ini}_{T,FB} = {H}^{ini}_{T,FB}(R,{\theta_{\rm acol}^{\max}},E_{\min},c)
\ea
for the hard initial state bremsstrahlung 
to total and forward-backward cross sections 
$\sigma_{T,FB}({\theta_{\rm acol}^{\max}},E_{\min},c)$.
We cut on the maximal acollinearity angle
${\theta_{\rm acol}}$ of the final state fermions, 
on the minimal energy of the fermions $E_{\min}$, and
on the scattering angle $\vartheta$ of one fermion.
As general convolution integrals with all cuts we have:
\ba
\sigma_{T}^{ini}({\theta_{\rm acol}^{\max}},E_{\min},c) 
&=& 
\int dR~ \sigma^0(s')\,
\rho_{T,FB}^{ini}(R,A,c),
\label{sigtini_gen}
\\
\nonumber\\
\sigma_{FB}^{ini}({\theta_{\rm acol}^{\max}},E_{\min},c) 
&=& 
\int dR~ \sigma^0(s')\,
\rho_{T,FB}^{ini}(R,A,c),
\label{sigfbini_gen}
\\
\nonumber\\
\rho_{T,FB}^{ini}(R,A,c) &=& 
\left(1+{\bar S}\right)\beta_e (1-R)^{\beta_e-1} 
   + {\bar H}_{T,FB}^{ini}(R,A,c).
\label{rhoini_gen}
\ea
The soft photon part $\bar{S}$ can be looked up with 
the factor $\beta_e$ in (\ref{eq:4}), the function $A$
depending on  
\ba
\label{Afunction}
A = A(R,\theta_{\rm acol}^{\max},E_{\min}).
\ea
For each region of phase space depending on $R$ for fixed
cut-off value $c$ we then have to insert the appropiate 
hard photon flux function ${\bar H}_{T,FB}^{ini}(R,A,c)$.
For region I, $A\approx 1$, the regularized result 
for the radiator is given by
\ba
\bar{H}_{T,FB}^{ini}(R,1,c) &=& H_{T,FB}(R,1,c)-\frac{\beta_e}{1-R},
\label{hard_reg}
\ea
while for regions II and III, safely away from the 
soft photon region, we 
have ${\bar H}_{T,FB}^{ini}(R,A,c)= H_{T,FB}^{ini}(R,A,c)$.

Following the discussion in Section 
\ref{sub_lep1slc_mass_sing} for the treatment of the 
phase space with respect to mass singularities, 
we have to distinguish 4 regions 
for ${H}^{ini}_{T,FB}$
%, each for positive or negative 
%sign of $c$, 
with a further separation into two cases 
for the antisymmetric radiator ${H}^{ini}_{FB}$.
This is due to additional contributions from $\sigma(0)$ from 
the lower bound of the integration over $\cos\vartheta$
which cancel in ${H}^{ini}_{T,FB}$. 
The generic structure of ${H}^{ini}_{T,FB}$ 
with cut-off $c\ge 0$ can be written as (i = 0,\,1): 
\ba
\label{hard_gen}
\hspace*{-0.1cm} {H}^{ini}_{T}(R,A,c) 
\hspace*{-0.2cm}&=&\hspace*{-0.2cm} 
\frac{3 \alpha}{4 \pi}\, Q_e^2\,
\Biggl\{
{\cal F}_{ii}(R,A,c)\mp {\cal F}_{ii}(R,A,c)
+ {\cal C}_0(R,A,c) 
\biggl[\pm {\cal F}_{10}(R) \biggr]
\Biggr\},
\label{hard_gent}
\nonumber\\
\\
\hspace*{-0.1cm} {H}^{ini}_{FB}(R,A,c)
\hspace*{-0.2cm} 
&=&\hspace*{-0.2cm}
\frac{\alpha}{\pi}\, Q_e^2\,
\Biggl\{
{\cal G}_{ii}(R,A,c)\pm {\cal G}_{ii}(R,A,c) 
+ {\cal G}_{0,1}(R,A)
\biggl[\pm (2) {\cal G}_{10}(R,A) \biggr]
\Biggr\},
\label{hard_genfb}
\nonumber\\
\ea
with functions ${\cal F}_{ii}(R,A,c)$ 
and ${\cal G}_{ii}(R,A,c)$ which
depend on the acceptance cut $c$
and further functions independent of $c$, 
${\cal F}_{10}(R)$, ${\cal G}_{0}(R)$, ${\cal G}_{1}(R)$, 
and ${\cal G}_{10}(R)$. 
The functions ${\cal F}_{ij}$ and ${\cal G}_{ij}$, 
$i,j = 0,1$ contain different logarithmic expressions
\ba
\label{logs}
{\cal L}^{\pm}_{Ac}(R,A,c), 
\quad
L_z(R,c),  
\quad
L_{m_e}(R),  
\quad
L^{\pm}(c),  
\quad
\mbox{and} 
\quad
L^{\pm}(A).
\ea
These depend on $R=s'/s$ as last variable for the numerical 
integration and on the cuts $c$, ${\theta_{\rm acol}^{\max}}$, 
and $E_{\min}$, the latter two contained in the function $A$. 
%
%\ba
%\label{Afunction}
%A = A(R,\theta_{\rm acol}^{\max},E_{\min}).
%\ea
%
The detailed structure of the hard radiators ${H}^{ini}_{T,FB}$
we give below following the distinction of cases 
given in (\ref{case1}) ($c >0$):
%
%\vfill\eject
%----------------
\ba
\label{hard_fg_1}
I.\qquad \mbox{Case}&& (1)^{+\,+}\leftrightarrow~ (2)^{+\,+}  
\qquad (\Rightarrow A\ge A_0(R)) :
%1.\qquad I=0\quad &,& J=0\quad (\Rightarrow K =0) : 
\nonumber\\
\nonumber\\
{H}^{ini}_{T} (R,A,c)& = & 
\frac{3\alpha}{4\pi} Q_e^2
\Biggl\{
{\cal F}_{00}(R,A,c) - {\cal F}_{00}(R,A,-c) + {\cal C}_0(R,A,c)
\Biggr\},
\label{hard_fg_12}
\\
\nonumber\\
{H}^{ini}_{FB}(R,A,c)& = & 
\frac{\alpha}{\pi} Q_e^2
\Biggl\{
{\cal G}_{00}(R,A,c) + {\cal G}_{00}(R,A,-c) + {\cal G}_0(R,A)
\Biggr\},
\label{hard_fg_13}
\\
\nonumber\\
II.\qquad \mbox{Case}&&  (1)^{+\,-}\leftrightarrow~ (2)^{+\,-}  
\qquad (\Rightarrow A< A_0(R)) :
%2.\qquad I=1\quad &,& J=1\quad (\Rightarrow K = 1) : 
\nonumber\\
\nonumber\\
{H}^{ini}_{T} (R,A,c)& = & 
\frac{3\alpha}{4\pi} Q_e^2
\Biggl\{
{\cal F}_{11}(R,A,c) - {\cal F}_{11}(R,A,-c) + {\cal C}_0(R,A,c)
\Biggr\},
\label{hard_fg_14}
\\
\nonumber\\
{H}^{ini}_{FB}(R,A,c)& = & 
\frac{\alpha}{\pi} Q_e^2
\Biggl\{
{\cal G}_{11}(R,A,c) + {\cal G}_{11}(R,A,-c) + {\cal G}_1(R,A)
\Biggr\},
\label{hard_fg_15}
\\
\nonumber\\
III.\qquad  \mbox{Case}&& (1)^{+\,+}\leftrightarrow~ (2)^{+\,-} :  
%3.\qquad I=1\quad &,& J=0\quad (c\ge 0) : 
\nonumber\\
\nonumber\\
{H}^{ini}_{T} (R,A,c)& = &  
\frac{3\alpha}{4\pi} Q_e^2
\Biggl\{
{\cal F}_{11}(R,A,c) - {\cal F}_{00}(R,A,-c) 
+ {\cal F}_{10}(R) + {\cal C}_0(R,A,c)
\Biggr\},
\nonumber\\
\label{hard_fg_16}
\\
a.~  A< A_0(R) &:&
\nonumber\\
{H}^{ini}_{FB}(R,A,c)& = & 
\frac{\alpha}{\pi} Q_e^2
\Biggl\{
{\cal G}_{11}(R,A,c) + {\cal G}_{00}(R,A,-c) 
+ {\cal G}_1(R,A) + {\cal G}_{10}(R,A)
\Biggr\},
\nonumber\\
\label{hard_fg_17}
\\
b.~ A\ge A_0(R) &:&
\nonumber\\
{H}^{ini}_{FB}(R,A,c)& = & 
\frac{\alpha}{\pi} Q_e^2
\Biggl\{
{\cal G}_{11}(R,A,c) + {\cal G}_{00}(R,A,-c) 
+ {\cal G}_0(R,A) - {\cal G}_{10}(R,A)
\Biggr\},
\nonumber\\
\label{hard_fg_18}
\\
IV.\qquad \mbox{Case}&& (1)^{-\,-}\leftrightarrow~ (2)^{+\,+} : 
%4.\qquad I=1\quad &,& J=-1\quad (c\ge 0) : 
\nonumber\\
\nonumber\\
{H}^{ini}_{T} (R,A,c)& = & 
\frac{3 \alpha}{4\pi} Q_e^2
\Biggl\{
{\cal F}_{11}(R,A,c) + {\cal F}_{11}(R,A,-c) 
+ {\cal C}_0(R,A,c)
\Biggr\},
\label{hard_fg_19}
\\
a.~  A< A_0(R) &:&
\nonumber\\
{H}^{ini}_{FB}(R,A,c)& = & 
\frac{\alpha}{\pi} Q_e^2
\Biggl\{
{\cal G}_{11}(R,A,c) - {\cal G}_{11}(R,A,-c) 
+ {\cal G}_1(R,A)
\Biggr\},
\\
\label{hard_fg_110}
b.~  A\ge A_0(R) &:&
\nonumber\\
{H}^{ini}_{FB}(R,A,c)& = & 
\frac{\alpha}{\pi} Q_e^2
\Biggl\{
{\cal G}_{11}(R,A,c) - {\cal G}_{11}(R,A,-c) 
+ {\cal G}_0(R,A) - 2\,{\cal G}_{10}(R,A)
\Biggr\}.
\label{hard_fg_111}
\nonumber\\
\ea
${\cal F}_{00}$, ${\cal G}_{00}$, ${\cal F}_{11}$, and ${\cal G}_{11}$ 
are now illustrated below:

\ba
\label{hard_fg_2}
{\cal F}_{00}(R,A,c) &=& \frac{1}{v}\,\left\{
(c+\frac{1}{3}c^3)\,\left[L_z(R,c)+L_{m_e}(R)\right]
+\frac{1}{3}c(1-c^2)\,\ln\left(1-c^2\right)\right.
\nonumber\\
&&-\left.\frac{4}{3}\left[(1+c)\, L^{+}(c)-(1-c)\, L^{-}(c)\right]
\right\}
\nonumber\\
\nonumber\\
&&\,+\, \frac{1}{6v}\,\left[ (3+c^2+A^2)\, {\cal L}^{+}_{Ac}(R,A,c)
-A c\, {\cal L}^{-}_{Ac}(R,A,c) \right]
\nonumber\\
\nonumber\\
&&\,+\, f_1( z,R,A,c)\, {\cal L}^{+}_{Ac}(R,A,c)
+A\, f_2( z,R,c)\, {\cal L}^{-}_{Ac}(R,A,c)
\nonumber\\
\nonumber\\
&&\,+\,\frac{2}{3} (1+R)\,\left[L^{+}(c)-L^{-}(c)\right]
\nonumber\\
\nonumber\\
&&\,+\, f_{01}(z,R,c)\,\left[L_z(R,c)+L_{m_e}(R)-\ln(1-c^2)\right]
\nonumber\\
\nonumber\\
&&\,+\, f_{02}(z,R,c) + f_{03}(R,A,c),
\ea
\ba
\label{hard_fg_3}
{\cal F}_{11}(R,A,c) &=& 
 \frac{1}{2}\frac{1+R^2}{v}
\,\left\{(A+\frac{1}{3}A^3)\, L_{m_e}(R)
+\frac{1}{3} A(1-A^2)\,\ln(1-A^2)\right.
\nonumber\\
&&-\left. \frac{4}{3}\left[(1+A)\, L^{+}(A)
-(1-A)\, L^{-}(A)\right]\right\}
\nonumber\\
\nonumber\\
&&\,+\,\frac{1}{6v}\,\left\{
(3+c^2+A^2)\, {\cal L}^{-}_{Ac}(R,A,c)
-A c\, {\cal L}^{+}_{Ac}(R,A,c)\right\}
\nonumber\\
\nonumber\\
&&\,+\,f_1( z,R,A,c)\, {\cal L}^{-}_{Ac}(R,A,c)
+ A\, f_2( z,R,c)\, {\cal L}^{+}_{Ac}(R,A,c)
\nonumber\\
\nonumber\\
&&\,+\,f_{11}(z,R,A,c) + f_{12}(R,A,c),
\\
\nonumber\\
\label{hard_fg_4}
{\cal G}_{00}(R,A,c) &=& \frac{1}{2v}\,\left\{
c\, {\cal L}^{+}_{Ac}(R,A,c)-A\, {\cal L}^{-}_{Ac}(R,A,c)\right.
\nonumber\\
&&-\left. 2 (1-c^2)\,\left[L_z(R,c)+L_{m_e}(R)-\ln(1-c^2)\right]\right\}
\nonumber\\
\nonumber\\
&&\,+\, g_1( z,R,c)\, {\cal L}^{+}_{Ac}(R,A,c)
+A\, g_2( z,R,c)\, {\cal L}^{-}_{Ac}(R,A,c)
\nonumber\\
\nonumber\\
&&\,+\, g_{01}( z,R,c)\,\left[L_z(R,c)+L_{m_e}(R)-\ln(1-c^2)\right]
\nonumber\\
\nonumber\\
&&\,+\, g_{02}(R,A,c),
\\
\nonumber\\
\label{hard_fg_5}
{\cal G}_{11}(R,A,c) &=& \frac{1}{2v}\,
\left[ c\, {\cal L}^{-}_{Ac}(R,A,c)
-A\, {\cal L}^{+}_{Ac}(R,A,c)\right]
\nonumber\\
\nonumber\\
&&\,+\,g_1( z,R,c)\, {\cal L}^{-}_{Ac}(R,A,c)+
A\, g_2( z,R,c)\, {\cal L}^{+}_{Ac}(R,A,c)
\nonumber\\
\nonumber\\
&&\,+\, g_{11}(R,A,c),
\\
\nonumber\\
\label{hard_fg_5a}
{\cal C}_0(R,A,c) &=& \frac{c}{R}\, c_0(R,A) = - 2 A c\, (1-R), 
\ea
with the coefficient functions
\ba
\label{coef_fun1}
f_1( z,R,A,c) 
&=&
\frac{1}{6 z}\,
\biggl\{
\frac{2(1-c^2)}{z}
\left[
c + \frac{y + 2 R (1+R)}{z}
\right]
\nonumber\\
&& 
-\left[(1-c)^2(1+c)+4(c+R)\right]-A^2\,(c+R)
\biggr\},
\label{coef_funfy1}
\\
g_1( z,R,c) &=& 
\frac{1}{2 z}\,\left[ 1-c^2-(c+R)\,\frac{y}{ z}\right],
\label{coef_fungy1}
\\
f_2( z,R,c) &=&
-\frac{1}{3}\, g_1( z,R,c),
\label{coef_funy2} 
\\
g_2( z,R,c) &=& 
\frac{1}{2 z}\, (c+R),
\label{coef_fungy2}
\quad \mbox{with}\quad y\equiv y(R,c)\,,\quad  z\equiv  z(R,c),
\ea
and other simple coefficient functions 
$f_{01}$, $f_{02}$, $f_{03}$, $f_{11}$, $f_{12}$
and $g_{01}$, $g_{02}$, $g_{11}$, 
as rational functions in terms of $z(R,c)$, $R$, $A$, and $c$.
They are listed in Appendix (\ref{hardini}), 
(\ref{coef_fun2a}) to (\ref{coef_fun4}).
In (\ref{hard_gent}) and (\ref{hard_genfb})  
we have also introduced the additional expressions:  
\ba
\label{hardradtot4}
{\cal F}_{10}(R) &=& -\frac{2}{3}\frac{1+R^2}{v}\,\ln{R},
\\
\nonumber\\
{\cal G}_0(R,A) &=&
\frac{1+R^2}{1+R}\,
\left[
\frac{1}{v}\, A\, {{\cal L}}^{-}_{A}
\,-\,\frac{1}{1+R}\, {{\cal L}}^{+}_{A}
\right]
\nonumber\\
&&+\frac{4R}{1+R}\,\frac{1+R^2}{1+R}\,\frac{1}{v}
\,\left[L_0(R)+L_{m_e}(R)\right],
\\
\nonumber\\
{\cal G}_1(R,A) &=&\frac{1+R^2}{1+R}\, 
\left[
\frac{1}{v}\, A\, {{\cal L}}^{+}_{A}
\,-\,\frac{1}{1+R}\, {{\cal L}}^{-}_{A}
\right]
\,-\,\frac{4 R}{1+R}\, A,
\\
\nonumber\\
{\cal G}_{10}(R,A) &=& \frac{1}{2}\frac{1+R^2}{v}\, (1-A^2)\,
\left[L_{m_e}(R)-\ln(1-A^2)\right]
+\frac{2 A^2 R}{v},
\nonumber\\
\ea
with following basic logarithms,
\ba
\label{hard_fg_6}  
{{\cal L}}^{\pm}_{A}(R,A) &=& y(R,A)\,\ln\biggl|y(R,A)\biggr|
\pm y(R,-A)\,\ln\biggl|y(R,-A)\biggr|,
\\
\nonumber\\  
{{\cal L}}^{\pm}_{Ac}(R,A,c) 
&=&\quad [y(R,c)\,+\,A\, z(R,c)]
\,\ln\biggl|y(R,c)\,+\,A\, z(R,c)\biggr|
\nonumber\\ 
&&\pm\, [y(R,c)\,-\,A\, z(R,c)]
\,\ln\biggl|y(R,c)\,-\,A\, z(R,c)\biggr|,
\\
\nonumber\\
&=&\quad [y(R,A)\,+\,c\, z(R,A)]
\,\ln\biggl|y(R,A)\,+\,c\, z(R,A)\biggr|
\\ 
&&\pm\, [y(R,-A)\,+\,c\, z(R,-A)]
\,\ln\biggl|y(R,-A)\,+\,c\, z(R,-A)\biggr|,
\nonumber\\
\nonumber\\
L^{\pm}(c) &=& \ln(1\pm c),\qquad L^{\pm}(A) = \ln(1\pm A),
\\
\nonumber\\ 
L_{ z}(R,c) &=& \ln\left[\frac{ z^2(R,c)}{4 R}\right],
\quad
L_0(R) = \ln\left[\frac{(1+R)^2}{4 R}\right]= L_{ z}(R,0),
\\
\nonumber\\
L_{m_e}(R) &=& \ln\left(\frac{s}{m_e^2}\right)-1-\ln(4R),
\\
\nonumber\\
y(R,c) &=& (1-R)\, +\, c\,(1+R)\,,
\quad z(R,c) = (1+R)\,+\,c\,(1-R).
\nonumber\\
\ea

If we only consider an acollinearity and energy cut, $A$, 
without an acceptance, i.e.~$c=1$, only the case 
$(1)^{-\,-}\leftrightarrow~ (2)^{+\,+}$ 
in (\ref{hard_fg_19}) to (\ref{hard_fg_111}) is possible.
{For} ${H}^{ini}_{FB}$ there is still the further distinction 
into cases $A(R)\ge A_0(R)$ and $A(R) < A_0(R)$.
We then obtain the very compact results already 
shown in (\ref{init1}), (\ref{inifb1}), and (\ref{inifb2})
and published in \cite{Christova:1999cc}. Equivalently, one can 
show for omitted acollinearity cut, i.e.~for $A=1$, corresponding 
to a simple $s'$-cut that the results with acceptance cut $c$ and 
$s'$-cut given in \cite{Bardin:1991fu} 
can be obtained.

A nice consistency check can also be done for the forward-backward
radiators when setting 
\ba
\label{y1acheck1}
y(R,-A) = 0\qquad\leftrightarrow
\qquad A=A_0(R)=\frac{1-R}{1+R}.
\ea
Relation (\ref{y1acheck1}) immediately gives 
\ba
\label{y1acheck1b}
1 / v \cdot A_0(R)\, {{\cal L}}^{\pm}_{A}
 - 1 / (1+R)\cdot {{\cal L}}^{\mp}_{A} = 0,
\ea
and therefore:
\ba
\label{y1acheck2}
 H^{ini,A< A_0}_{FB}(R,1) &=& H^{ini,A\ge A_0}_{FB}(R,1) 
\nonumber\\
&=&
\frac{\alpha}{\pi} Q_e^2
\left\{
-\frac{4 R (1+R^2)}{(1+R)^2}\,\frac{1}{v}\,\ln(R)
+\frac{8 R^2}{(1+R)^2}
\right\}.
\ea
That is, the results $H^{ini,A< A_0}_{FB}(R,1)$ and 
$H^{ini,A\ge A_0}_{FB}(R,1)$ are continuous at the phase space 
boundaries for $A\rightarrow A_0(R)$.
Without any angular or energy cuts at all, i.e.~$c=A=1$,
we can at last reproduce the classical results
for a simple cut on $s'$ with the Bonneau-Martin 
term from (\ref{eq:bm}) \cite{Bonneau:1971mk,Bardin:1989cw}.\footnote{
In this limit, only region $A\ge A_0(R)$ is possible,
i.e.~only one function $H^{ini,A\ge A_0}_{FB}(R,1)$ 
remains for the forward-backward radiator.
}
\ba
\label{hard_class}
{H}^{ini}_{T} (R,1)&=& 
\int\limits^1_0 d{\cos\vartheta}
\,\,\,
h^{ini}_{T}(R,\cos\vartheta)
=
\frac{\alpha}{\pi} Q_e^2
\frac{1+R^2}{v}
\,\left[\ln\left(\frac{s}{m_e^2}\right)-1\right],
\nonumber\\ 
\\
{H}^{ini}_{FB}(R,1)&=& 
\int\limits^1_0 d{\cos\vartheta}
\,\,\,
h^{ini}_{FB}(R,\cos\vartheta)
\nonumber\\ 
&=&
\frac{\alpha}{\pi} Q_e^2
\frac{4 R (1+R^2)}{(1+R)^2}\frac{1}{v}
\,\left\{\ln\left(\frac{s}{m_e^2}\right)-1
+\ln\left[\frac{(1+R)^2}{4 R}\right]\right\}. 
\nonumber\\ 
\ea
For a description of the calculation of the radiators 
${H}^{ini}_{T,FB}$ please refer to Appendix \ref{hardini}.
The evaluation of the interference results 
${H}^{int}_{T,FB}$ is absolutely analogous to the
initial state case with the final expressions 
summarized in \ref{int}.

All analytical formulae for the hard radiators 
presented in this Chapter and contained 
in the Appendix were numerically checked by 
comparing them with the squared matrix element, 
numerically integrated over the angular phase space.\footnote{
For simplicity, the straightforward integration over the 
azimuthal photon angle $\phi_\gamma$ was done analytically 
in order to get shorter expressions which were easier 
to numerically integrate.
}
This was done for each hard flux function separately.
Different acollinearity, energy, and acceptance 
cuts were applied for values of 
$s'/s = m_f^2/s\ldots(1-\varepsilon)$ at different center-of-mass 
energies $\sqrt{s} = 30\ldots 10^3 \,\mbox{GeV}$.
Also the single steps of the analytical integration
were numerically checked for each integrand listed 
in Appendix \ref{hardini}.

While in the flux functions all mass terms except for the logarithmic 
terms are omitted, the squared matrix element contains the mass 
terms in the propagators in order to numerically regularize
the mass singularities described above. 
The analytically and numerically integrated results
agree at the $10^{-8}$ to $10^{-4}$ level, depending on the 
functions compared, the phase space region examined, and the cuts 
applied. The agreement is naturally restricted due to the partly 
omitted mass terms and deteriorates from the final state to the 
interference term, and is worst for the initial state results. 
This is due to the critical, numerically instable propagators 
of order $1/Z_{1,2} = -1/(2 p k_{1,2})$    
for the interference terms and even of order 
$1/Z_{1,2}^2$ for the initial state terms which are 
regularized by the initial state mass terms $m_e^2/s$.
These propagators are not contained in the hard photon final 
state radiators.  

%==========================================================================
\section{Numerical results of {\tt ZFITTER} and comparisons
\label{sec_lep1slc_codes}
}
%-------------------------------------------------------------------------
%
%==========================================================================
\subsection{Hard bremsstrahlung corrections in {\tt ZFITTER}  
\label{sub_lep1slc_zf}
}
%-------------------------------------------------------------------------
%
The corrections of the new 
formulae on the $O(\alpha)$ bremsstrahlung to $e^+e^-\to
\bar{f}f$ implemented in the semi-analytical program 
{\tt ZFITTER} versions v.6 were compared with the results 
of versions v.5.20 and earlier. 

The main modifications in the new coding are corrected terms 
in the QED initial state and interference radiator parts. 
{For} the case of initial state radiation, it was possible to trace back the
origin of the numerical inaccuracies related to the acollinearity cut
of {\tt ZFITTER} below version 6. It is the result of leaving out a certain 
class of non-logarithmic, simple terms of order $O(\alpha)$.
{For} $\sigma_T$, polynomials proportional to $\cos\vartheta$ (and their
integrals) are concerned, and for $\sigma_{FB}$ polynomials of the type
$(a+b\cos^2\vartheta)$ (and their integrals). At first glance the 
corresponding contributions seem to vanish for symmetric acceptance cuts. 
But this is not the case !
As was explained in Section (\ref{sub_lep1slc_mass_sing}), the cross section 
formulae lose the usual simple symmetry/anti-symmetry behaviour under the
transformation $\cos\vartheta \leftrightarrow (-\cos\vartheta)$ in
regions II and III of the phase space since different analytical
expressions may be needed depending on the location of the parameters
$c_i^{\pm}$  in (\ref{c1+}) to (\ref{c2-}).
Then, the symmetry behaviour as a function of $\cos\vartheta$ is 
`hidden' since different regions contribute differently to the net 
result. Omitting these terms in earlier versions is justified with the then 
anticipated experimental precision of only $5\times 10^{-3}$ at LEP~1,
but not with the higher accuracy now at the $Z$ resonance peak.
{For} the hard final state radiators implemented in {\tt ZFITTER}
some misprints could be corrected for. These modifications are completely 
negligible for total cross sections (only at the level $10^{-5}$ 
or less), and only of minor importance for the forward-backward asymmetries,
i.e.~always stay below per mil level. The final state radiators 
$H_{T,FB}^{fin}(R,A,c)$  
and the over $R$ analytically integrated results we could also 
compare -- for the general case with acceptance cut -- with the 
results of \cite{Montagna:1993mf}, obtaining complete agreement 
except for some smaller misprints there.

The corresponding Fortran package is {\tt acol.f}. 
We merged package {\tt acol.f} with photonic corrections for the integrated
total cross section and the integrated forward-backward asymmetry (with and
without acceptance cut) into {\tt ZFITTER} v.5.21, thus creating 
{\tt ZFITTER} v.6.04/06 \cite{zfitter:v6.0406,Bardin:1999yd-orig} 
onwards. The angular distribution is available 
in v.6.2 \cite{Bardin:1999yd-orig} onwards. 
The remarkably compact expressions for the case 
that no angular acceptance cut is applied are 
published \cite{Christova:1999cc} and also implemented 
in an extra branch of {\tt ZFITTER} \cite{Bardin:1999yd-orig} 
for quick cross section or asymmetry evaluations in the case of 
less cuts. A complete collection of the most 
general analytical expressions with cuts on maximal acollinearity, 
minimal energies, and minimal acceptance is given in the Appendix.

Numerical predictions of {\tt ZFITTER} v.5.20 
\cite{Bardin:1992jc2,zfitter:v5.20} and 
{\tt ZFITTER} v.6.11 \cite{zfitter:v6.11} 
were systematically compared with default flag settings.
Version 5.20 was used as released, while version 6.11 was prepared
such that the changes due to the recalculation of initial state
corrections, final state corrections, their interferences, and the 
net effect could be isolated. To look at the single cross section
and asymmetry contributions is also interesting from the point of view
that experimentalists very often use two different approaches to data: 
Sometimes the initial-final state interference contributions are 
subtracted from measured data, and sometimes the interference effects 
remain in the data sample. We begin with a study of the changes related 
to {\it initial state radiation}. 

%==========================================================================
\subsubsection*{Initial state corrections 
\label{ssub_lep1slc_ini_num}
}
%-------------------------------------------------------------------------
%
For $\sigma_T$, the changes are at most one unit in the fifth 
digit at LEP~1 energies and thus considered to be completely 
negligible. In Tables \ref{tab10sigifi} and \ref{tab10afbifi} 
the predictions for $\sigma_T$ and 
for $A_{FB}$ are shown for two acollinearity cuts
$\theta_{\rm acol}<10^{\circ},25^{\circ}$ and three different 
acceptance cuts $\theta_{\rm acc}=0^{\circ},20^{\circ},40^{\circ}$.
%
%--------------------------------------------------------------------------------
\begin{sidewaystable}
\renewcommand{\arraystretch}{1.1}
\begin{tabular}{|c||c|c|c|c|c||c||c|c|c|c|c|}
\hline
\multicolumn{6}{|c|}
{$\sigma_{\flm}\,$[nb] with $\theta_{\rm acol}<10^{\circ}$} &
\multicolumn{6}{|c|}
{$\sigma_{\flm}\,$[nb] with $\theta_{\rm acol}<25^{\circ}$} \\
\hline
$\theta_{\rm acc}$
& $\mz - 3$ & $\mz - 1.8$ & $\mz$ & $\mz + 1.8$ & $\mz + 3$  &
$\theta_{\rm acc}$
& $\mz - 3$ & $\mz - 1.8$ & $\mz$ & $\mz + 1.8$ & $\mz + 3$  \\
\hline\hline
{\tt Z6} \hfill $0^{\circ}$
  & 0.21928  & 0.46285  & 1.44780  & 0.67721  & 0.39360 &
{\tt Z6} \hfill $0^{\circ}$ 
  & 0.22328  & 0.46968  & 1.46598  & 0.68688  & 0.40031 \\
  & 0.21772  & 0.46082  & 1.44776  & 0.67898  & 0.39489 &
  & 0.22228  & 0.46836  & 1.46602  & 0.68816  & 0.40128 \\
  &-7.16     &-4.41     &-0.03     &+2.60     &+3.27    &
  &-4.51     &-2.82    & +0.03    & +1.86    & +2.41   \\
\hline
{\tt Z5} \hfill $0^{\circ}$  
  & 0.21928  & 0.46285  & 1.44781  & 0.67722  & 0.39361 &
{\tt Z5} \hfill $0^{\circ}$ 
  & 0.22328  & 0.46968  & 1.46598  & 0.68688  & 0.40031 \\
  & 0.21852  & 0.46186  & 1.44782  & 0.67814  & 0.39429 &
  & 0.22281  & 0.46905  & 1.46603  & 0.68754  & 0.40081 \\
  &-3.48     &-2.14     &+0.01     &+1.36     &+1.72    &
  &-2.11     &-1.34     &+0.03     &+0.96     &+1.25    \\
\hline\hline
{\tt Z6} \hfill $20^{\circ}$
  & 0.19987  & 0.42205  & 1.32053  & 0.61756  & 0.35881 &
{\tt Z6} \hfill $20^{\circ}$ 
  & 0.20357  & 0.42834  & 1.33718  & 0.62647  & 0.36505 \\  
  & 0.19869  & 0.42046  & 1.32018  & 0.61877  & 0.35972 &
  & 0.20281  & 0.42729  & 1.33689  & 0.62731  & 0.36572 \\
  & -5.96    &-3.79     &-0.27     &+1.95     &+2.53    &
  & -3.74    & -2.46    & -0.21    & +1.35    & +1.83   \\
\hline
{\tt Z5} \hfill $20^{\circ}$ 
  & 0.19987  & 0.42205  & 1.32053  & 0.61756  & 0.35881 &
{\tt Z5} \hfill $20^{\circ}$ 
  & 0.20357  & 0.42833  & 1.33718  & 0.62647  & 0.36505 \\
  & 0.19892  & 0.42075  & 1.32021  & 0.61857  & 0.35959 &
  & 0.20321  & 0.42781  & 1.33689  & 0.62684  & 0.36536 \\
  &-4.78     &-3.09     &-0.24     &+1.63     &+2.17    &
  &-1.77     &-1.22     &-0.22     &+0.59     &+0.85    \\
\hline\hline
{\tt Z6} \hfill $40^{\circ}$ 
  & 0.15032  & 0.31760  & 0.99416  & 0.46475  & 0.26983 &
{\tt Z6} \hfill $40^{\circ}$ 
  & 0.15318  & 0.32243  & 1.00682  & 0.47164  & 0.27477 \\ 
  & 0.14974  & 0.31675  & 0.99349  & 0.46515  & 0.27019 &
  & 0.15280  & 0.32183  & 1.00619  & 0.47188  & 0.27502 \\
  &-3.88     &-2.72     &-0.67     &+0.87     &+1.32    &
  & -2.48    & -1.88    & -0.62    & +0.51    & +0.91   \\ 
\hline
{\tt Z5} \hfill $40^{\circ}$ 
  & 0.15032  & 0.31760  & 0.99415  & 0.46474  & 0.26983 &
{\tt Z5} \hfill $40^{\circ}$ 
  & 0.15318  & 0.32243  & 1.00682  & 0.47164  & 0.27477 \\
  & 0.14978  & 0.31680  & 0.99350  & 0.46511  & 0.27016 &
  & 0.15287  & 0.32192  & 1.00619  & 0.47180  & 0.27496 \\
  &-3.61     &-2.53     &-0.65     &+0.80     &+1.22    &
  &-2.03     &-1.58     &-0.63     &+0.34     &+0.69    \\
\hline 
\end{tabular}
\caption[Corrections to initial state radiation and 
initial-final state interference in $\sigma_T$
for {\tt ZFITTER} at the $Z$ peak]
{\sf
%THIS IS similar to TABLE 37 OF PCPR ---
Comparison of {\tt ZFITTER} v.6.11 \cite{Bardin:1999yd-orig} (first row)  
with {\tt ZFITTER} v.5.20 \cite{Bardin:1992jc2,zfitter:v5.20} (second row)
for muon pair production cross sections
% of complete RO (CA3-mode) 
with angular acceptance cuts 
($\theta_{\rm acc}=0^{\circ},20^{\circ},40^{\circ}$) 
and acollinearity cut ($\theta_{\rm acol}<10^{\circ},25^{\circ}$).
First row is without initial-final state interference, second row with,
third row the relative effect of that interference in per mil.
Final state radiation is treated as in v.5.20 \cite{Christova:1999gh}.
\label{tab10sigifi}
}
\end{sidewaystable}
%--------------------------------------------------------------------------------
%
%-------------------------------------------------------------------------------- 
\begin{sidewaystable}
\renewcommand{\arraystretch}{1.1}
\hspace*{-0.5cm}
\begin{tabular}{|c||c|c|c|c|c||c||c|c|c|c|c|}
\hline
\multicolumn{6}{|c||}{$\afba{\flm}$ with $\theta_{\rm acol}<10^{\circ}$} &
\multicolumn{6}{|c|}{$\afba{\flm}$ with $\theta_{\rm acol}<25^{\circ}$} \\
\hline
$\theta_{\rm acc}$
& $\mz - 3$ & $\mz - 1.8$ & $\mz$ & $\mz + 1.8$ & $\mz + 3$  &
$\theta_{\rm acc}$
& $\mz - 3$ & $\mz - 1.8$ & $\mz$ & $\mz + 1.8$ & $\mz + 3$  \\
\hline\hline
{\tt Z6} \hfill $0^{\circ}$  
  & -0.28462 & -0.16916  & 0.00024  & 0.11482  & 0.16063 &
{\tt Z6} \hfill $0^{\circ}$ 
  & -0.28651 & -0.17051 & -0.00043  & 0.11292  & 0.15680 \\ 
  & -0.28187 & -0.16689  & 0.00083  & 0.11379  & 0.15907 &
  & -0.28554 & -0.16960 & -0.00000  & 0.11285  & 0.15669 \\
  & +2.75    & +2.27     & +0.60    &-1.03     &-1.56    &
  & +0.97    & +0.91    & +0.43     & -0.06    & -0.11   \\
\hline
{\tt Z5} \hfill $0^{\circ}$  
  & -0.28453 & -0.16911  & 0.00025  & 0.11486  & 0.16071 &
{\tt Z5} \hfill $0^{\circ}$ 
  & -0.28647 & -0.17049 & -0.00043  & 0.11293  & 0.15682 \\
  & -0.28282 & -0.16783  & 0.00070  & 0.11475  & 0.16059 &
  & -0.28555 & -0.16975 & -0.00005  & 0.11307  & 0.15701 \\
  & +1.71    & +1.28     &+0.45     &-0.11     &-0.12    &
  & +0.92    & +0.74    & +0.48     &+0.14     &+0.19    \\
\hline \hline
{\tt Z6} \hfill $20^{\circ}$
  & -0.27521 & -0.16355  & 0.00032  & 0.11141  & 0.15602 &
{\tt Z6} \hfill $20^{\circ}$ 
  & -0.27727 & -0.16499 & -0.00038  & 0.10942  & 0.15201 \\ 
  & -0.27285 & -0.16167  & 0.00080  & 0.11053  & 0.15467 &
  & -0.27659 & -0.16436 & -0.00006  & 0.10943  & 0.15199 \\
  & +2.35    & +1.88     &+0.47     & -0.89    &-1.35    &
  & +0.68    & +0.63    & +0.32     & +0.00    & -0.02   \\
\hline
{\tt Z5} \hfill $20^{\circ}$  
  & -0.27506 & -0.16347  & 0.00035  & 0.11148  & 0.15616 &
{\tt Z5} \hfill $20^{\circ}$ 
  & -0.27722 & -0.16497 & -0.00037  & 0.10944  & 0.15204 \\
  & -0.27408 & -0.16261  & 0.00070  & 0.11133  & 0.15594 &
  & -0.27657 & -0.16447 & -0.00009  & 0.10963  & 0.15229 \\
  & +0.98    & +0.86     &+0.35     &-0.15     &-0.22    &
  & +0.65    & +0.50    & +0.28     &+0.19     &+0.25    \\
\hline\hline
{\tt Z6} \hfill $40^{\circ}$
  & -0.24230 & -0.14398  & 0.00045  & 0.09881  & 0.13868 &
{\tt Z6} \hfill $40^{\circ}$ 
  & -0.24452 & -0.14549 & -0.00027  & 0.09675  & 0.13449 \\
  & -0.24063 & -0.14277  & 0.00073  & 0.09825  & 0.13780 &
  & -0.24423 & -0.14527 & -0.00010  & 0.09687  & 0.13464 \\ 
  & +1.67    & +1.22     & +0.28    & -0.56    & -0.88   &
  & +0.29    & +0.22    & +0.17     & +0.12    & +0.15   \\
\hline
{\tt Z5} \hfill $40^{\circ}$ 
  & -0.24207 & -0.14386  & 0.00050  & 0.09893  & 0.13891 &
{\tt Z5} \hfill $40^{\circ}$ 
  & -0.24445 & -0.14545 & -0.00026  & 0.09678  & 0.13454 \\
  & -0.24151 & -0.14343  & 0.00069  & 0.09890  & 0.13888 &
  & -0.24444 & -0.14542 & -0.00011  & 0.09700  & 0.13483 \\
  & +0.56    & +0.43     &+0.19     &-0.03     &-0.03    &
  & +0.01    & +0.03    & +0.15     &+0.22     &+0.29    \\
\hline 
\end{tabular}
\caption[Corrections to initial state radiation and 
initial-final state interference in $A_{FB}$
for {\tt ZFITTER} at the $Z$ peak]
{\sf
%THIS IS TABLE 38 OF PCPR ---
Comparison of {\tt ZFITTER} v.6.11 (first row) 
with {\tt ZFITTER} v.5.20 (second row)
for the muonic forward-backward asymmetry 
% of complete RO (CA3-mode) 
with angular acceptance cuts 
($\theta_{\rm acc}=0^{\circ},20^{\circ},40^{\circ}$) and acollinearity cut 
($\theta_{\rm acol}<10^{\circ},25^{\circ}$). First row is without
initial-final state interference, second row with,
third row the relative effect of that interference in per mil.
Final state radiation is treated as in v.5.20 \cite{Christova:1999gh}.
\label{tab10afbifi}
}
\end{sidewaystable}
%--------------------------------------------------------------------------------
%
The numbers are given in the first row of each box
for a certain acceptance angle $\theta_{\rm acc}$, for 
both versions v.6.11 and v.5.20. 
The changes are less than the theoretical accuracies demanded.
It was checked that the numbers for flag value 
{\tt ICUT} = 0 (see \cite{Bardin:1999yd-orig}) agree with 
the {\tt ZFITTER} predictions shown in Tables 26 and 27 of \cite{Bardin:1999gt}.

%==========================================================================
\subsubsection*{Initial-final state interference corrections 
\label{ssub_lep1slc_int_num}
}
%-------------------------------------------------------------------------
%
The {\tt ZFITTER} v.5 predictions of photonic corrections from the 
{\it initial-final state interference} also receive modifications  
after the recalculation for the versions v.6. The explanation given above  
for the case of initial state radiation is also applicable for a part 
of the deviations here. The codings for the initial-final state interference 
also show additional deviations in the hard photonic corrections and the
resulting numerical differences are much larger.   

The absolute values of $\sigma_T$ and $A_{FB}$ are also listed
in Tables \ref{tab10sigifi} and \ref{tab10afbifi}
in the second row of each box, while the third rows show 
the shifts when switching on the initial-final state interference.
For the cross sections these are relative shifts in per mil,
for the asymmetries they are absolue values also in per mil.
The two tables are the analogues to 
Tables 37--40 of \cite{Bardin:1999gt} where 
{\tt TOPAZ0} v.4.3 \cite{Montagna:1995b,Montagna:1998kp} 
and {\tt ZFITTER} v.5.20 were compared. 
At the $Z$ peak, the predictions for the influence of the initial-final state 
interference from {\tt ZFITTER} v.5.20 and {\tt ZFITTER} v.6.11 deviate 
from each other only
negligibly, with maximal deviations of up to 0.015\%.
At the wings, the situation is quite different, we observe
deviations of up to several per mil for cross sections and 
up to a per mil for asymmetries.
The deviations between the two codings decrease if the
acollinearity cut is weakened.  

%==========================================================================
\subsubsection*{Final state corrections 
\label{ssub_lep1slc_fin_num}
}
%-------------------------------------------------------------------------
%
{For} LEP~1, the numerical outcome of the minor 
improvements to the code is shown in Tables \ref{tab1025fin}
for $A_{FB}$ with $\theta_{\rm acol}<10^{\circ}, 25^{\circ}$ 
and several different acceptance cuts:
$\theta_{\rm acc}=0^{\circ},20^{\circ},40^{\circ}$.
Again, the numbers for {\tt ICUT} = 0 agree with those 
shown in Tables 26 and 27 of \cite{Bardin:1999gt}.
All the changes are though visible, but negligible.
{For} the cross sections, the differences are completely negligible and
not tabulated here.
%
%--------------------------------------------------------------------------------
\begin{table}[htb]
\begin{center}
\begin{tabular}{|c||c|c|c|c|c|}
\hline
\multicolumn{6}{|c|}{$\afba{\flm}$ with $\theta_{\rm acol}<10^{\circ}$} \\
\hline
$\theta_{\rm acc}$
& $\mz - 3$ & $\mz - 1.8$ & $\mz$ & $\mz + 1.8$ & $\mz + 3$  \\
\hline\hline
$0^{\circ}$ 
      &-0.28487 &-0.16932 & 0.00025 & 0.11500 & 0.16091 \\
      &-0.28453 &-0.16911 & 0.00025 & 0.11486 & 0.16071 \\
\hline
$20^{\circ}$
      &-0.27539 &-0.16367 & 0.00035 & 0.11162 & 0.15635 \\
      &-0.27506 &-0.16347 & 0.00035 & 0.11148 & 0.15616 \\
\hline
$40^{\circ}$
      &-0.24236 &-0.14404 & 0.00050 & 0.09905 & 0.13908 \\
      &-0.24207 &-0.14386 & 0.00050 & 0.09893 & 0.13891 \\
\hline 
\hline
\multicolumn{6}{|c|}{$\afba{\flm}$ with $\theta_{\rm acol}<25^{\circ}$} 
\\
\hline
$\theta_{\rm acc}$
 & $\mz - 3$ & $\mz - 1.8$ & $\mz$ & $\mz + 1.8$ & $\mz + 3$  \\
\hline\hline
$0^{\circ}$ 
      &-0.286732 &-0.170647 &-0.000428 & 0.113029 & 0.156963 \\
      &-0.286474 &-0.170493 &-0.000427 & 0.112927 & 0.156821 \\
\hline
$20^{\circ}$
      &-0.277471 &-0.165114 &-0.000370 & 0.109537 & 0.152173 \\
      &-0.277221 &-0.164965 &-0.000370 & 0.109438 & 0.152036 \\
\hline
$40^{\circ}$
      &-0.244669 &-0.145582 &-0.000255 & 0.096867 & 0.134658 \\
      &-0.244449 &-0.145451 &-0.000255 & 0.096780 & 0.134537 \\
\hline 
\end{tabular}
\caption[Corrections to final state radiation 
for {\tt ZFITTER} at the $Z$ peak]
{\sf
%THIS IS similar to TABLE 26 OF PCPR ---
Comparison of {\tt ZFITTER} v.6.11 (first row) with {\tt ZFITTER} v.5.20
\\ 
(second row)
for the muonic forward-backward asymmetry with
  angular acceptance cut ($\theta_{\rm acc}=0^{\circ},20^{\circ},40^{\circ}$)
  and acollinearity cuts ($\theta_{\rm acol}<10^{\circ}$) and
  ($\theta_{\rm acol}<25^{\circ}$).
The initial-final state interference is switched off and only
final state radiation is corrected \cite{Christova:1999gh}.
\label{tab1025fin}
}
\end{center}
\end{table}%26
%--------------------------------------------------------------------------------

{For} the case of {\it final state radiation}, common soft photon
exponentiation together with initial state radiation 
is foreseen in {\tt ZFITTER}. 
{For} an $s'$-cut, {\tt ZFITTER} follows \cite{Montagna:1993mf}.
As may be seen from \cite{Bilenkii:1989zg} (for the angular
distributions) or from \cite{Christova:1999cc} 
(for integrated observables), the predictions for common
soft photon exponentiation include one additional integration, namely 
that over the invariant mass of the final state fermion pair at a
given reduction of $s$ into $s'$ after initial state
radiation.\footnote{With acollinearity cut, there remains some
arbitrariness in the choice of the region with exponentiation.
In {\tt ZFITTER}, the acollinearity cut is simulated by an
effective $s'$-cut for this purpose. 
For details see also \cite{Bardin:1999yd-orig}.
}
%
%~\footnote{
%We did not change what was realized in {\tt ZFITTER} v.5:
%initial state radiation is exponentiated for $R > \max(R_E,R_{{\theta_{\rm acol}}})$
%and the final state radiation, at given $R$, for 
%$R' > \max(R_{min}/R,R_E)$. 
%A preferred condition might be $R' > \max(R_{min}/R,R_E,R_{{\theta_{\rm acol}}})$. 
%In this case, for $R_{min}/R < R_{{\theta_{\rm acol}}}$, the
%non-exponentiated hard photonic corrections from region III would have to be 
%left out in order to avoid double counting.
%} 
This additional integration is done partly analytically,
for not too involved integrands, and partly numerically 
using a Lagrange interpolating formula for the integrand.
Common exponentiation was always included in the comparisons 
shown here.

%==========================================================================
\subsubsection*{Net corrections 
\label{ssub_lep1slc_net_num}
}
%-------------------------------------------------------------------------
%
In case of the net corrections, first the corrections for the 
initial and final state calculations in v.6.11 were combined 
and compared with the old coding in v.5.20. Then the corrected
initial-final state interference was added and compared 
with the earlier formulae to see the overall effect of the 
modifications.
%
%--------------------------------------------------------------------------------
\begin{table}[th]
\begin{center}
\begin{tabular}{|c||c||c|c|c|c|c|}
\hline
\multicolumn{6}{|c|}{$\sigma_{\flm}\,$[nb] 
with $\theta_{\rm acol}<10^{\circ}$} 
\\
\hline
\hline
$\theta_{\rm acc}$
 & $\mz - 3$ & $\mz - 1.8$ & $\mz$ & $\mz + 1.8$ & $\mz + 3$  \\
\hline\hline
$0^{\circ}$ 
      & 0.21772 & 0.46081 & 1.44776 & 0.67898 & 0.39489 \\
      & 0.21852 & 0.46186 & 1.44782 & 0.67814 & 0.39429 \\
\hline
$20^{\circ}$
      & 0.19869 & 0.42046 & 1.32018 & 0.61877 & 0.35972 \\
      & 0.19892 & 0.42075 & 1.32021 & 0.61857 & 0.35959 \\
\hline
$40^{\circ}$
      & 0.14974 & 0.31675 & 0.99349 & 0.46515 & 0.27019 \\
      & 0.14978 & 0.31680 & 0.99350 & 0.46511 & 0.27016 \\
\hline\hline
\multicolumn{6}{|c|}{$\afba{\flm}$ with $\theta_{\rm acol}<10^{\circ}$} 
\\
\hline
$0^{\circ}$ 
      &-0.28222 &-0.16710 & 0.00083 & 0.11392 & 0.15926 \\ 
      &-0.28282 &-0.16783 & 0.00070 & 0.11475 & 0.16059 \\
\hline
$20^{\circ}$
      &-0.27319 &-0.16187 & 0.00080 & 0.11066 & 0.15486 \\
      &-0.27408 &-0.16261 & 0.00070 & 0.11133 & 0.15594 \\
\hline
$40^{\circ}$
      &-0.24093 &-0.14294 & 0.00074 & 0.09837 & 0.13797 \\
      &-0.24151 &-0.14343 & 0.00069 & 0.09890 & 0.13888 \\
\hline 
\end{tabular}
\caption[Net corrections for {\tt ZFITTER} at the $Z$ peak]
{\sf
%THIS IS similar to TABLE 26 OF PCPR ---
Comparison of net corrections from 
{\tt ZFITTER} v.6.11 (first row) with {\tt ZFITTER} v.5.20 (second row)
for muon pair production with
  angular acceptance cut ($\theta_{\rm acc}=0^{\circ},20^{\circ},40^{\circ}$)
  and acollinearity cut ($\theta_{\rm acol}<10^{\circ}$).
The initial-final state interference is switched on \cite{Christova:1999gh}.
\label{tab10acolc}
}
\end{center}
\end{table}%26
One can state that the net corrections without initial-final state 
interference are negligible for the cross section. 
The corrections to the numerical output from {\tt ZFITTER} with
acollinearity cut increased when the corrected initial-final state
interference is taken into account.
The resulting net corrections for the total cross section
and the forward-backward asymmetry at LEP~1 are shown in 
Table \ref{tab10acolc} for $\theta_{\rm acol}<10^{\circ}$
and different acceptance cuts.

%==========================================================================
\subsection{Comparisons of {\tt ZFITTER} with different numerical programs 
\label{sub_lep1slc_num}
}
%-------------------------------------------------------------------------
%
The $s'$-cut was studied for LEP~1 in \cite{Bardin:1995aa} and more recently by
\cite{Beenakker:1997fi,Montagna:1997jt,Placzek:1999xc,Bardin:1999gt,Jadach:1999pp,Jadach:1999gz,Passarino:1999kv}. 
The agreement of two-fermion codes for $s$-channel observables is now 
better than $0.1$ per mil on resonance and quite sufficient 
for experimental applications at LEP~1 \cite{Consoli:1989pc,Bardin:1995aa}.

Numerical results with acollinearity cut are given in
Table 3 of \cite{Beenakker:1997fi} for the $s$-channel part of Bhabha
scattering for programs {\tt TOPAZ0} and {\tt ALIBABA}.
In \cite{Christova:1998tc} this was compared 
with {\tt ZFITTER} v.5.14 \cite{zfitter:v5.14} 
and one gets in both cases roughly a 3 per mil agreement. 
%
%--------------------------------------------------------------------------------
%\begin{figure}[htp] 
%\begin{center} 
%---
%\hspace*{-0.5cm} 
%\mbox{ 
%       \epsfig{file=sigz0acol.ps,
%                    height=8.cm,% 
%                   clip=% 
%}} 
%\caption[Comparison of $s$-channel Bhabha scattering cross sections at LEP~1]
%{\sf
%Ratios of $s$-channel contributions to Bhabha scattering at LEP~1 
%\cite{Christova:1998tc}: 
%{\tt ZFITTER} v.5.14 \cite{Bardin:1992jc2,zfitter:v5.14}
%versus {\tt TOPAZ0} \cite{Montagna:1995b}
%and {\tt ALIBABA} \cite{Beenakker:1991mb}.
%\label{sigma9} 
%} 
%\end{center} 
%\end{figure}
%-------------------------------------------------------------------------------- 
%
Out to the wings of the $Z$ resonance region, 
i.e.~$\sqrt{s}=M_Z\pm 3\,\mbox{GeV}$, 
this numerical agreement remains for the comparison with {\tt TOPAZ0},
but deteriorates to several per mil for the {\tt ALIBABA} case.
{For} the accuracies demanded from theory by experiment at the starting 
phase of LEP~1 in 1989, i.e.~roughly $0.5\%$ \cite{Consoli:1989pc}, this was 
quite sufficient, but not anymore now with the better than per mil precision 
at the $Z$ peak \cite{Grunewald:1998kw,Grunewald:1999wn,Abreu:1998kh,Barate:1999ce,Abbiendi:1999eh,Acciarri:2000ai,Abbaneo:2000aa}.

After the update of the {\tt ZFITTER} code for 
combined cuts on energies, acollinearity, and acceptance angle 
\cite{Christova:1999cc,Bardin:1999yd-orig} and 
after comparisons with other numerical programs 
\cite{Beenakker:1991mb,Montagna:1995b,Montagna:1998kp}, 
the situation for LEP~1 energies shown in Fig.~\ref{top-zf-peak}
and Table \ref{tab-acol10-th40} can be stated as quite 
satisfactory \cite{Christova:1998tc,Bardin:1999gt,Christova:1999gh,Jack:1999xc,Jack:1999af,Christova:2000zu}: 
%
%--------------------------------------------------------------------------------
\begin{figure}[htp] 
\begin{flushleft}
%---
\vspace*{0cm} 
\begin{tabular}{ll}
\hspace*{-0.25cm}
  \mbox{%
\epsfig{file=%
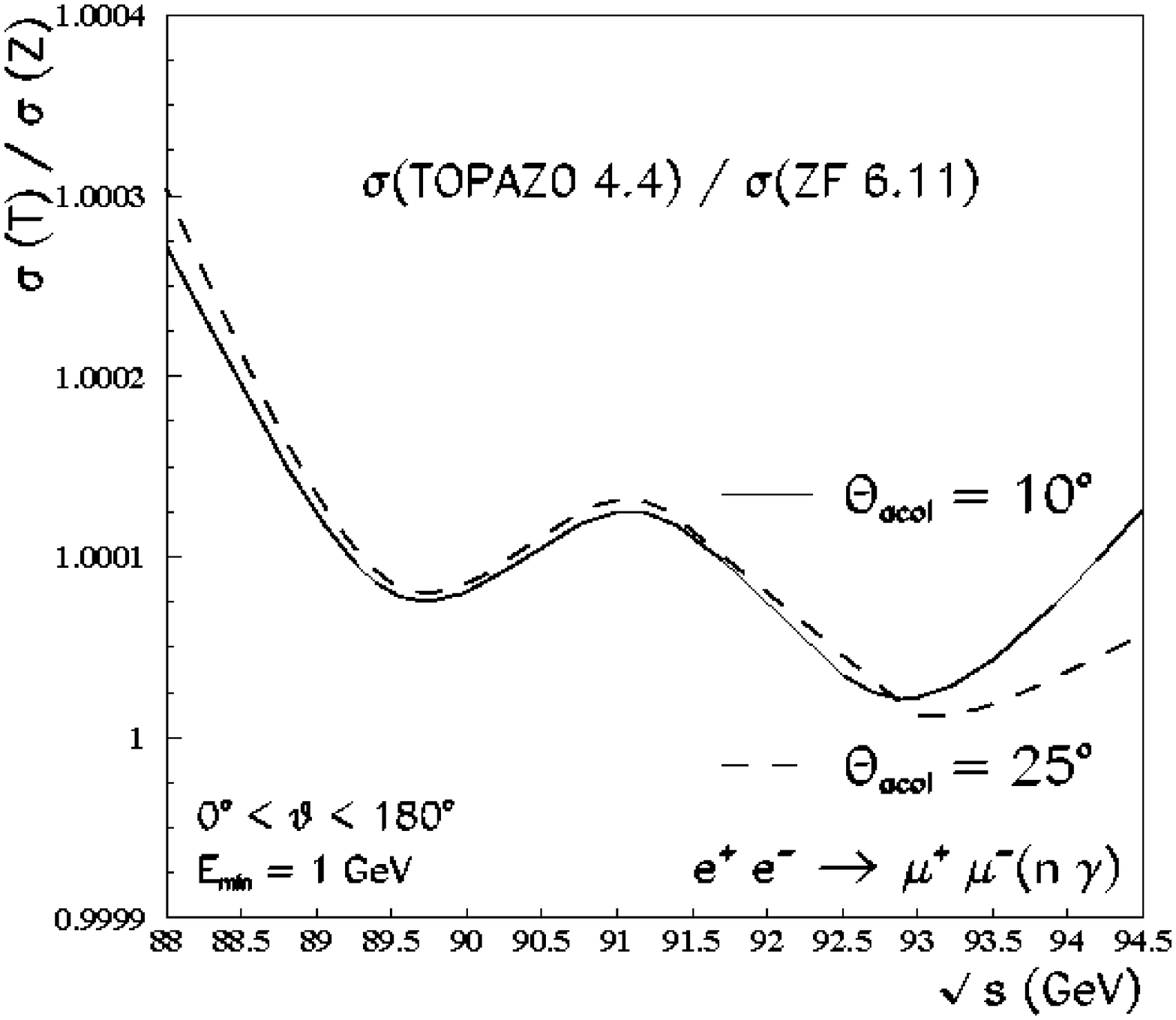
           ,width=7.5cm   % this is the width of the figure (optional)
         }}%
&
\hspace*{-0.8cm}
  \mbox{%
\epsfig{file=%
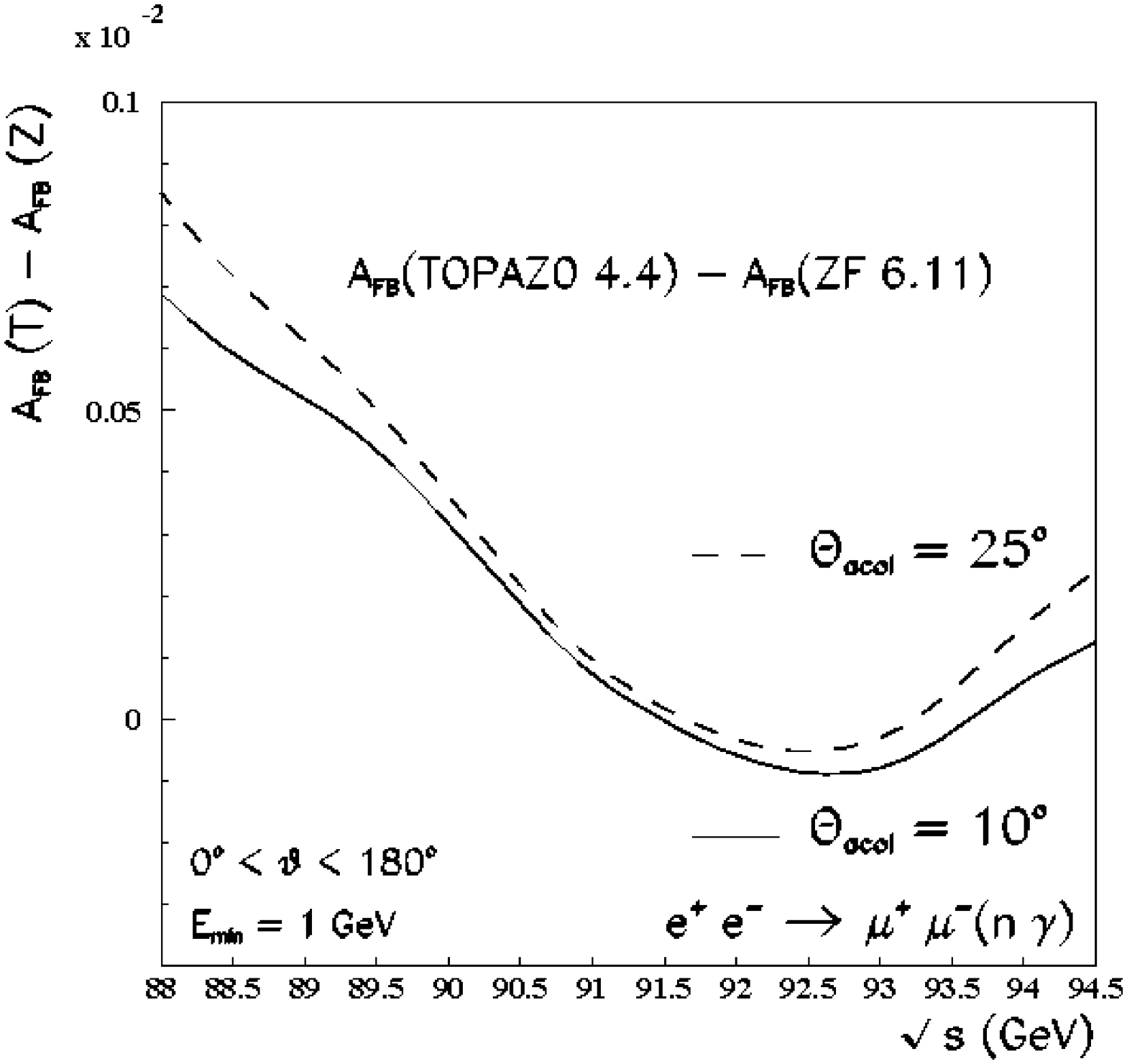
           ,width=7.5cm   % this is the width of the figure (optional)
         }}%
\\
\end{tabular}
\vspace{-1cm}
\caption[Ratios of muon pair cross sections from 
{\tt ZFITTER} and {\tt TOPAZ0}]
{\sf
Ratios of muon pair production cross sections and differences of 
forward-backward asymmetries at the $Z$ resonance: 
{\tt ZFITTER} v.6.11 \cite{zfitter:v6.11,Bardin:1999yd-orig} 
vs.~{\tt TOPAZ0} v.4.4 \cite{Montagna:1998kp} for different 
acollinearity cuts (full acceptance) \cite{Jack:1999af}.
\label{top-zf-peak}
}
\end{flushleft}
\end{figure}
%--------------------------------------------------------------------------------
%
%Total cross sections and forward-backward asymmetries 
%are now treated better than at the per mil level 
%for both cut options -- the invariant mass and the acollinearity cut --
%around the $Z$ boson resonance, and even better than $10^{-4}$
%at the $Z$ peak itself.

The numerical comparison of the newly updated {\tt ZFITTER} version 
v.6.11 \cite{Bardin:1999yd-orig} with {\tt TOPAZ0}'s latest release version 
v.4.4 \cite{Montagna:1998kp} now delivers for LEP~1 energies the same 
high level of agreement as for the $s'$-cut (Fig.~\ref{top-zf-peak}): 
At the peak itself we have a deviation of the codes of 
$O(10^{-4})$ or less for $\sigma_T$ and $A_{FB}$, with 
an acceptable increase to maximally $3\times 10^{-4}$ 
for $\sigma_T$ and a slightly worse value for $A_{FB}$ 
of $7\times 10^{-4}$ at $\sqrt{s} = M_Z\pm 3$ GeV
for a maximal acollinearity angle of $10^\circ$. 
This is also listed in more detail in Table \ref{tab-acol10-th40}.
The agreement at the $Z$ peak is now as good as for the 
$s'$-cut which was checked recently in \cite{Bardin:1995aa,Bardin:1999gt}.

{For} LEP~2, the  $s'$-cut is estimated to be `under control' in
\cite{Boudjema:1996qg}, while a warning was given there
that `the agreement between {\tt TOPAZ0} and {\tt ZFITTER} somehow
degrades when implementing an acollinearity cut'.
However, one should mention here that both programs were originally 
designed for applications around the $Z$ boson resonance and using them 
at higher energies deserves dedicated checks and, if necessary, 
further improvements. A detailed comparison of the situation of the codes 
at LEP~2 energies, before and after our corrections to the {\tt ZFITTER}
code will follow in the next Chapter. 
%
%\vfill\eject
%--------------------------------------------------------------------------------
\begin{table}[htp]
\begin{center}
\begin{tabular}{|c||c|c|c|c|c|}
\hline
\multicolumn{6}{|c|}{
{$\sigma_{\mu}\,$[nb] with $\theta_{\rm acol}<10^\circ$}}
\\ 
\hline
$\theta_{\rm acc} = 0^\circ$& $\mz - 3$ & $\mz - 1.8$ & $\mz$ & $\mz + 1.8$ &
$\mz + 3$  \\ 
\hline\hline
  & 0.21932  & 0.46287  & 1.44795  & 0.67725  & 0.39366 \\
{{\tt TOPAZ0}}  & 0.21776  & 0.46083  & 1.44785  & 0.67894  & 0.39491 \\
  & 
{\bf --7.16}    & 
{\bf --4.43}     &
{\bf --0.07}     &
{\bf +2.49}     &
{\bf +3.17}    
\\ 
\hline
  & 0.21928  & 0.46284  & 1.44780  & 0.67721  & 0.39360 \\
{{\tt ZFITTER}}  & 0.21772  & 0.46082  & 1.44776  & 0.67898  &
0.39489 \\ 
  &
{\bf --7.16}     &{\bf --4.40}     &{\bf --0.03 }    &{\bf +2.60}
&{\bf +3.27}   
\\ 
\hline 
\hline
\multicolumn{6}{|c|}{
{${A_{FB}}_{\mu}$ with $\theta_{\rm acol}<10^{\circ}$}} 
\\
\hline
$\theta_{\rm acc} = 0^{\circ} $& $\mz - 3$ & $\mz - 1.8$ &
$\mz$ & $\mz + 1.8$ & $\mz + 3$  \\ 
\hline\hline
  & --0.28450 & --0.16914  & 0.00033  & 0.11512  & 0.16107 \\
{{\tt TOPAZ0}}  & --0.28158 & --0.16665  & 0.00088  & 0.11385  &
0.15936 \\ 
  & 
{\bf +2.92}    & {\bf +2.49}     &
{{\bf +0.55}}     &{\bf --1.27}
&
{{\bf --1.71}}    \\ 
\hline
  & --0.28497 & --0.16936  & 0.00024  & 0.11496  & 0.16083 \\
{{\tt ZFITTER}}  & --0.28222 & --0.16710  & 0.00083  & 0.11392  & 0.15926 \\
  & {\bf +2.75}    & {\bf +2.27}     & {\bf +0.60}    &{\bf --1.03}
&{\bf --1.56}
%----------------------------
\\
\hline 
\end{tabular}
\caption[Muon pair cross sections from 
{\tt ZFITTER} and {\tt TOPAZ0} at the $Z$ peak]
{\sf
A comparison of predictions from {\tt ZFITTER} v.6.11 
\cite{zfitter:v6.11,Bardin:1999yd-orig} 
and {\tt TOPAZ0} v.4.4 \cite{Montagna:1998kp} 
for muonic cross sections and forward-backward
asymmetries around the $Z$ peak.
First row is without initial-final state interference, second row with,
third row the relative effect of that interference in per mil
\cite{Christova:1999gh,Jack:1999af,Christova:2000zu}.
\label{tab-acol10-th40}
}
\end{center}
\end{table}
%--------------------------------------------------------------------------------

%==========================================================================
\subsection{Effects on the experimental analysis at LEP
\label{sub_comp_exp}
}
%-------------------------------------------------------------------------
%
The LEP~1 data taking phase happened between 1989 and 1995 
with about 17 million $Z$ decays reported, i.e.
$15.5\times 10^6$ hadronic and $1.7\times 10^6$
leptonic decays. Several energy scans were undertaken
around the $Z$ peak during this period 
with a high statistics run exactly on the peak 
in 1992 and 1994 \cite{Grunewald:1999wn,Quast:2000ll}. 
The theoretical description of the experimental results
by available two-fermion programs taking into account 
all radiative corrections was reported to be in good 
shape then in \cite{Bardin:1995aa}. Since then, there
has been theoretical progress in several branches 
of the programs {\tt ZFITTER} \cite{Bardin:1999yd-orig},
{\tt TOPAZ0} \cite{Montagna:1998kp}, and others,
especially for the description of hard bremsstrahlung 
and the inclusion of higher order corrections. 
As an example, the effects of the latest changes can be 
discussed for the typical peak cross section parameters 
$M_Z$, $\Gamma_Z$, and 
$\sigma^{0}_{\rm had}$ as total hadronic peak cross section.
The LEP average value for the latter 
is $\sigma^{0}_{\rm had}=41.491\pm 0.058$ nb \cite{Grunewald:1999wn,Quast:2000ll},
while $M_Z$, $\Gamma_Z$ were given in Table \ref{Zprecision}. 
%
%\ba
%M_Z\quad \mbox{and} \quad \Gamma_Z, 
%\label{mzgz}
%\\
%&&
%\sigma^{0}_{\rm had} = 
%\frac{12\pi}{M_Z^2}
%\frac{\Gamma_{e e} \Gamma_{\rm had}}
%{\Gamma_Z^2},
%\label{sighad}
%\\
%R_e = \frac{\Gamma_{\rm had}}{\Gamma_{e e}}, 
%\quad
%R_\mu = \frac{\Gamma_{\rm had}}{\Gamma_{\mu \mu}}, 
%\quad
%R_\tau = \frac{\Gamma_{\rm had}}{\Gamma_{\tau \tau}}.
%\label{Rlep}
%\ea
%%
%with $\sigma^{0}_{\rm had}$ as total hadronic peak cross section
%and $R_e$, $R_\mu$, and $R_\tau$ providing the leptonic contributions 
%to the total decay width $\Gamma_Z$.

To start with, in 1998 there were some 
changes to the extracted values of $M_Z$ and $\Gamma_Z$ using 
programs {\tt ZFITTER} \cite{Bardin:1999yd-orig} 
and {\tt TOPAZ0} \cite{Montagna:1998kp}
due to newly implemented leading logarithmic 
$O(\alpha^3)$ QED corrections to the initial state 
\cite{Skrzypek:1992vk,Montagna:1997jt}.  
There are in principle two different schemes for the estimate 
of the higher order QED corrections which are included in the 
programs and used by the experiments. Either the exact two-loop
results for initial state bremsstrahlung \cite{Berends:1988ab} 
are applied together with a soft and virtual photon exponentiation
including the third order results by \cite{Skrzypek:1992vk,Montagna:1997jt}.
Or secondly, there is the possibility of an inclusive exponentiation
scheme via Yennie-Frautschi-Suura \cite{Yennie:1961ad} treated
by \cite{jadach:1991,Skrzypek:1992vk}. The differences of both schemes, 
however, are documented to be small with $0.1$ MeV uncertainties 
to $M_Z$ and $\Gamma_Z$ and a $0.01\%$ correction to $\sigma^{0}_{\rm had}$.  
Much larger theoretical errors are estimated to arise from the inclusion
of corrections due to QED initial state pair creation.
These uncertainties are $\Delta M_Z\approx \pm 0.3$ MeV,   
$\Delta \Gamma_Z\approx \pm 0.2$ MeV, and
$\Delta\sigma^{0}_{\rm had}\approx \pm 0.02\%$ \cite{Quast:2000ll}. 

Since 1999, the version {\tt ZFITTER} v.6.10/6.11 \cite{zfitter:v6.10,zfitter:v6.11}
are used together with {\tt TOPAZ0} \cite{Montagna:1998kp} 
by all four LEP collaborations. 
The addition of new leading logarithmic $O(\alpha^3)$ and $O(\alpha^4)$ 
QED radiators calculated recently in \cite{Arbuzov:1999uq} will lead to 
the main uncertainties from QED corrections. 
For example, for $M_Z$ and $\Gamma_Z$ 
these new terms will lead to changes of $+0.5$ MeV with a final theoretical 
error of roughly $\pm 0.3$ MeV \cite{Quast:2000ll}.   

In \cite{Bardin:1998nm,Bardin:1999gt} there was a detailed comparison 
of both programs at the precision level $10^{-4}$ to estimate the effect 
on cross section and {\tt SM} observables by the different approaches 
used for {\tt SM} or model independent calculations.
The main focus there was on the $s'$-cut branch for total cross sections 
and asymmetries. Adding the comparisons with the 
new results on the acollinearity cut in the previous sections
and partly presented in 
\cite{Christova:1998tc,Christova:1999gh,Jack:1999xc,Jack:1999af,Christova:2000zu}, 
one can summarize as general outcome of all analysis 
that the errors from these different approaches are negligible 
with respect to the uncertainties from the higher order QED 
effects stated above. They are only at the order $\pm 0.1$ MeV
for $M_Z$ and $\Gamma_Z$ and not more than $\pm 1$ pb for 
$\sigma^{0}_{\rm had}$. The new hard photon calculation therefore
guarantees that both codes perfectly describe with their 
predictions for cross sections and other observables the 
experimental results, for both cut options using different 
approaches.

%==========================================================================
\section{Conclusions
\label{sec_conc_lep1slc}
}
%-------------------------------------------------------------------------
%
A derivation of analytical formulae for the $O(\alpha)$
hard QED bremsstrahlung corrections to $e^+e^-\to \bar{f} f$
was presented with cuts 
to the fermions' acollinearity angle and energies ($f\neq e$).
Very compact formulae can be obtained and pose an alternative 
for lepton pair final states to the usually applied kinematically 
simpler cut on the final state fermions' invariant mass squared $s'$. 
The hard radiators for the integrated total and forward-backward 
cross sections with an additional cut on angular acceptance 
also have been derived within this dissertation 
(see Section \ref{sub_lep1slc_formacol} 
and Appendix \ref{hardini}, \ref{int}, and \ref{fin}).

This was done in the context of the semi-analytical program {\tt ZFITTER} 
\cite{Bardin:1999yd-orig}
which calculates radiatively corrected observables with realistic 
experimental cuts, e.g.~for LEP/SLC applications.
Several numerical applications of the above formulae
were applied. {For} this purpose, the package {\tt acol.f} was added.
As a result, it was concluded that older versions of {\tt ZFITTER}, 
i.e.~versions v.5.20 \cite{zfitter:v5.20} and earlier,
derive the $O(\alpha)$ QED corrections to the total cross section 
$\sigma_T$ with acollinearity cut with a numerical accuracy of about 
$0.4 \%$ in the $Z$ resonance region ($M_Z\pm 3\,\mbox{GeV}$),
and similarly for the forward-backward asymmetry $A_{FB}$ with about 
$0.13 \%$. 

The determined modifications to hard photon radiators for the
initial and final state radiation and its 
interference are certain non-logarithmic terms. 
When the code was created in 1989 \cite{MBilenky:1989ab}, 
an accuracy of $0.5 \%$ at LEP~1 was assumed to be needed,
so these corrected terms could be neglected then.
This accuracy is not sufficient anymore now with the high
level of experimental precision (e.g.~$\delta{M_Z}/M_Z 
= 2.2\times 10^{-5}$ \cite{Swartz:1999xv})
and demanded the recalculation
of the hard photonic corrections.
The new and improved coding starting from 
{\tt ZFITTER} v.6.11 \cite{Bardin:1999yd-orig} with {\tt acol.f} 
gave a numerical agreement of $\sigma_T$ for leptons, with 
${\theta_{\rm acol}}\leq 10^{\circ}$ and $E_{min}=1$ GeV, with predictions 
from {\tt TOPAZ0} v.4.4 \cite{Montagna:1998kp} at LEP~1 of $0.03 \%$ 
(at the wings) or better (at resonance). {For} $A_{FB}$, the accuracy
at LEP~1 energies is now estimated to be better than $0.1\%$ for the 
same cuts.

The numerical limitations at LEP~1 before had 
been mainly due to the initial-final state interference which is now
corrected after the recalculation. At the $Z$ peak itself, however, the 
accuracy had already been quite satisfactory before, i.e.~better than 
$10^{-4}$, due to suppressed hard photon radiation. The new coding in 
{\tt ZFITTER} now reproduces the very nice agreement with program 
{\tt TOPAZ0}, already obtained for the $s'$-cut \cite{Bardin:1999gt}. 
These findings were partly published 
in~\cite{Christova:1998tc,Christova:1999gh,Jack:1999xc,Jack:1999af,Christova:2000zu}.
The influence of higher order corrections in these and other 
two-fermion codes has been recently  
treated in~\cite{Jadach:1999pp,Jadach:1999gz,Arbuzov:1999uq,Passarino:1999kv}.

%#############################################################################
\chapter{Fermion Pair Production at {LEP~2} Energies	
\label{ch_lep2}
}
%#############################################################################
%
While data taking at energies around the $Z$ boson resonance ended
at LEP~1 in 1995, a new phase started after a brief run around 
an intermediate energy scale of $\sqrt{s}\approx 135\,\mbox{GeV}$:
The LEP~2 went into operation with center-of-mass energies at the $W$ pair 
production threshold $\sqrt{s}\approx 161\,\mbox{GeV}$ 
\cite{Boudjema:1996qg,Grunewald:1999wn}
and is expected to reach up to $208\,\mbox{GeV}$ at the final end 
of LEP \cite{Grunewald:2000}. The above said 
already marks the main physics goal at LEP~2: High precision 
physics to the charged weak gauge bosons $W^\pm$, 
is the main objective, i.e.~determining 
the mass $M_W$, width $\Gamma_W$, and properties of the $W$ boson.  
Especial focus is here also put on details of triple gauge 
boson couplings (TGC) in the electroweak sector of the {\tt SM}.
Especially interesting are contributions
from the $(\gamma,Z) W W$ vertex which is contained in 
the $s$-channel part of the processes $e^+e^-\to W^+W^-$.
Also additional $Z$ pair production sets in above 
$\sqrt{s} = 2 M_Z$ for which additional searches 
of anomalous TGC, not described in the {\tt SM}, 
are conducted \cite{Boudjema:1996qg}. 
As the massive gauge bosons are instable particles and 
cannot be observed directly, one has to reconstruct  
the $W$ pair decays from the corresponding 4-fermion
final states arising from decays of the $W$ and $Z$ pairs. 
But this demands the inclusion of 4-fermion final states 
which did not originate from the $W$ and $Z$ pair decays
as background processes \cite{Boudjema:1996qg}. 
%
%Another important issue in 
%4-fermion physics is the investigation 
%of 4-jet events for QCD studies
%at higher energies, e.g.~in order to test the running 
%of the strong coupling constant $\alpha_S(Q^2)$ from 
%$Q^2=M_Z^2$ to $Q^2\approx (200\,\mbox{GeV})^2$.

But though the main interest at LEP~2 naturally 
lies on the above sketched physics program, two-fermion physics 
still is an interesting branch which is quite actively pursued
\cite{Barate:1999qx,Acciarri:1999rw}.
Fermion pair production is still one of the most copious 
processes at these energies having to be treated as possible 
background to 4-fermion final states. Moreover, precision physics 
with fermion pairs at LEP~2 extends the indirect searches 
for `New Physics' already undertaken on the $Z$ boson resonance
\cite{Abbiendi:1999wk,Abbiendi:1999wm,Abreu:2000kg}.
%More on this issue can be seen in Chapter \ref{ch_linac}.
%\footnote{With the $\nu\bar{\nu}\gamma$ final state it could be 
%confirmed that not more than the three particle generations 
%with light {\tt SM} neutrinos exist in nature. Any sequential
%extension of the {\tt SM} can only contain heavy neutral fermions.}

The main introductory facts to fermion pair production
at LEP~2 energies sufficiently above the $Z$ boson resonance can be
summarized as follows \cite{Montagna:1998sp}:
\begin{itemize}
\item The effective Born cross section $\sigma^0(s)$ for 
      $e^+e^-\to \bar{f}f$ (including electroweak
      and QCD corrections) drop by roughly three orders of magnitude 
      from the peak cross section $\sigma^0_{peak}(M_Z^2)$ down 
      to $O(10\,\mbox{pb})$. 

\item The estimated statistical error by experiment will be 
      approximately $\Delta\sigma\approx 1\%$, while at LEP~1 we
      typically had $\Delta\sigma\approx 10^{-3}$. 

\item For loose cuts, one observes in cross section distribuitions 
      for different c.m.~energies a second peak arising from
      events with hard photons emitted with energies 
      $E_\gamma\approx \sqrt{s}-M_Z$ (radiative return 
      events). This can be explained, 
      by a shift of the gauge boson 
      propagator onto the $Z$ boson resonance after 
      initial state hard photon emission. This gives 
      a strong increase of $\sigma_T(s)$ by the resonant 
      effective Born term. 

\item The weak box corrections from box diagrams with 
      $WW$ and $ZZ$ exchange, which are usually only 
      approximately treated at energies around the 
      $Z$ resonance, may grow, depending on the calculational 
      gauge chosen 
      and the cuts applied, roughly up to 1 or $2\%$ effects.
      Also other weak corrections from massive top quark 
      or weak gauge boson loops play an important role.
       
\item Comparing LEP~2 fermion pair production data 
      with theory, can also be used for {\tt SM} 
      Higgs boson or New Physics searches.
      In this context, also looking at processes 
      $e^+e^-\to \gamma\gamma$ 
      and $e^+e^-\to \nu\bar{\nu}\gamma(\gamma)$ 
      allows further search options.
      More on this general issue can be seen in Chapter \ref{ch_linac}.
%      This is equivalent to a peaking 
%      at $M_{f\bar{f}}^2\approx M_Z^2$ neglecting final state 
%      photon effects.
\end{itemize}
%
%-----------------------------------------------------------
\section{Physics effects in virtual radiative corrections 
%Physics effects in virtual radiative 
%corrections to $e^+e^-\to \bar{f}f$
\label{sec_lep2_virt}
}
%------------------------------------------------------------
%
First let us have a look at two examples on virtual 
radiative corrections with special focus on the 
interplay between pure QED and electroweak corrections
and their relative importance. We will especially 
have to realize that QED and electroweak corrections 
may start to be of similar importance 
at center-of-mass energies well above 
the $Z$ boson resonance and it will give us some feeling
on the importance of precisely knowing the different QED
contributions better than one per cent also at LEP~2
energies. This shall be discussed before we go to the 
details of the recalculated QED radiation with cuts 
at higher energies.
%
%-----------------------------------------------------------
\subsection*{The $Z b\bar b$ Vertex at LEP~2
\label{sub_lep2_zbb}
}
\begin{figure}[th] 
\begin{center} 
%--- 
  \mbox{%
  \epsfig{file=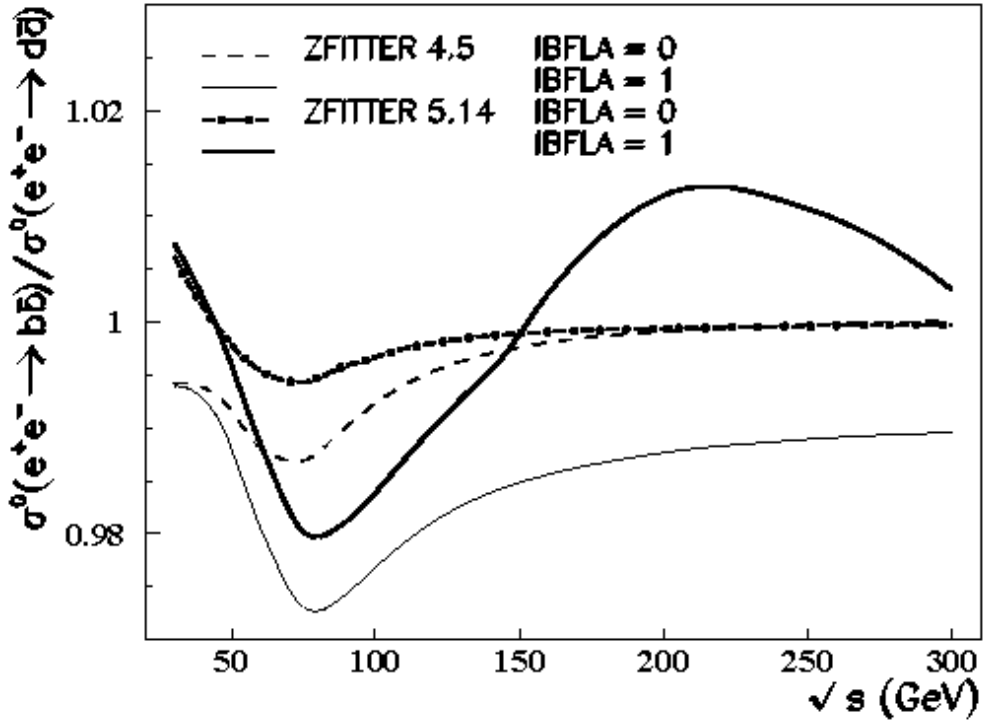
          ,height=10.cm  % this is the height of the figure (optional)
         }%
  }
\vspace*{-0.5cm}
\caption[Ratios of improved Born cross sections for $b\bar b$ and $d \bar d$
production] 
{
Ratios of improved Born cross sections for $b\bar b$ and $d \bar d$
production from {\tt ZFITTER} v.4.5 (1992) \cite{Bardin:1992jc2} and v.5.14 
(1998) \cite{zfitter:v5.14} ; the latter has the correct $t$ mass dependence 
at LEP~2 \cite{Christova:1998tc}.    
\label{fig-zbb} 
}
\end{center}
\end{figure}

%------------------------------------------------------------
%
An instructive example for different behaviour of weak corrections on and
off the $Z$ peak is $b\bar b$ production.
The corrections differ from those to $d\bar d$ production due to the huge
$t$-quark mass, which is depicted in Fig.~\ref{fig-zbb}. 
%
%xxxxxxxxxxxxxxxxxxxxxxxxxxxxxxxxxxxxxxxxxxxxxxxxxxxxxxxxxxxxxxxxx 
%xxxxxxxxxxxxxxxxxxxxxxxxxxxxxxxxxxxxxxxxxxxxxxxxxxxxxxxxxxxxxxxxx 
%
Only in the $b\bar{b}$ case, we have large per cent level 
corrections from the large mixing to virtual top quarks 
in the loop, with $|V_{tb}|^2\approx 1$ for the top-bottom mixing 
matrix element squared. The one-loop 
results to the $Z\bar{b}b$ vertex including the effects
by a heavy quark exchange were first treated in
\cite{Akhundov:1986fc,Jegerlehner:1986vs,Beenakker:1988pv,Bernabeu:1988me,Lynn:1990hd}. 

While at LEP~1, the off-resonant $WW$ box corrections 
and the $\gamma b\bar{b}$ vertex corrections are negligible   
compared to the $Z b\bar{b}$ vertex, they become more
and more important with increasing c.m.~energy.
Further, one has to correctly include the $s$-dependence of 
the vertices and for the box also its different angular dependence.
The net effect is taken into account in {\tt ZFITTER} since v.5.12
\cite{zfitter:v5.12},
which contains the complete one-loop virtual EW corrections to the 
$(\gamma,Z) f\bar{f}$ vertex,
and is shown in Fig.~\ref{fig-zbb}. 
It may be switched off with flag {\tt IBFLA}=0 
\cite{Bardin:1992jc2,Bardin:1999yd-orig}.
It amounts up to about 2--4 \%
and is thus of the order of the statistical error \cite{Christova:1998tc}.
Deviations from {\tt SM} predictions for corrections to the 
$Z b\bar b$ vertex would of course immediately imply effects 
from {\it New Physics} through virtual corrections. A brief  
report on such searches for non-{\tt SM} physics   
in the next Chapter will be given in connection with 
the possibilities at a future $e^+e^-$ linear colliding machine.
%
%i.e.~at an $e^+e^-$ {\it Linear Collider} like the {\tt TESLA} project 
%\cite{Accomando:1997wt}. 
%
%==========================================================================
\subsection*{The QED interference and electroweak box 
corrections -- a comparison
\label{sub_lep2_ewboxes}
}
%--------------------------------------------------------------------------
%
%As already mentioned above,
%the semi-analytical program {\tt ZFITTER} was originally developed for 
%{\tt SM} predictions of cross sections and asymmetries at LEP~1 energies.
%Observables like total cross sections and asymmetries 
%can be calculated in an {\it effective Born description},
%as is done in the {\tt ZFITTER} approach: 
%EW and QCD corrections are described as 
%effective couplings in {\it effective Born observables} which are 
%convoluted with the photonic corrections as flux functions. Higher 
%order QED effects can then partly be described by resumming finite 
%soft and virtual corrections ({\it soft-photon exponentiation}).   
%This was illustrated e.g.~in 
%\cite{Bardin:1989di,Bardin:1989cw,Bardin:1991de,Bardin:1991fu,Christova:1999cc,Bardin:19%99yd-orig,Christova:1998tc,Christova:1999gh,Jack:1999af,Christova:2000zu}.
%
While at LEP~1 the EW and QCD corrections can in general be 
considered as small in comparison to the QED bremsstrahlung,
this observation is not necessarily valid anymore at higher energies, 
where EW and QED corrections can grow to comparable magnitudes.
%
%xxxxxxxxxxxxxxxxxxxxxxxxxxxxxxxxxxxxxxxxxxxxxxxxxxxxxxxxxxxxxxxxx
\begin{figure}[htb]
\vspace*{-0.5cm} 
\hspace*{-2cm} 
\begin{flushleft}
%--- 
\begin{tabular}{ll}
\hspace*{-0.25cm} 
  \mbox{%
  \epsfig{file=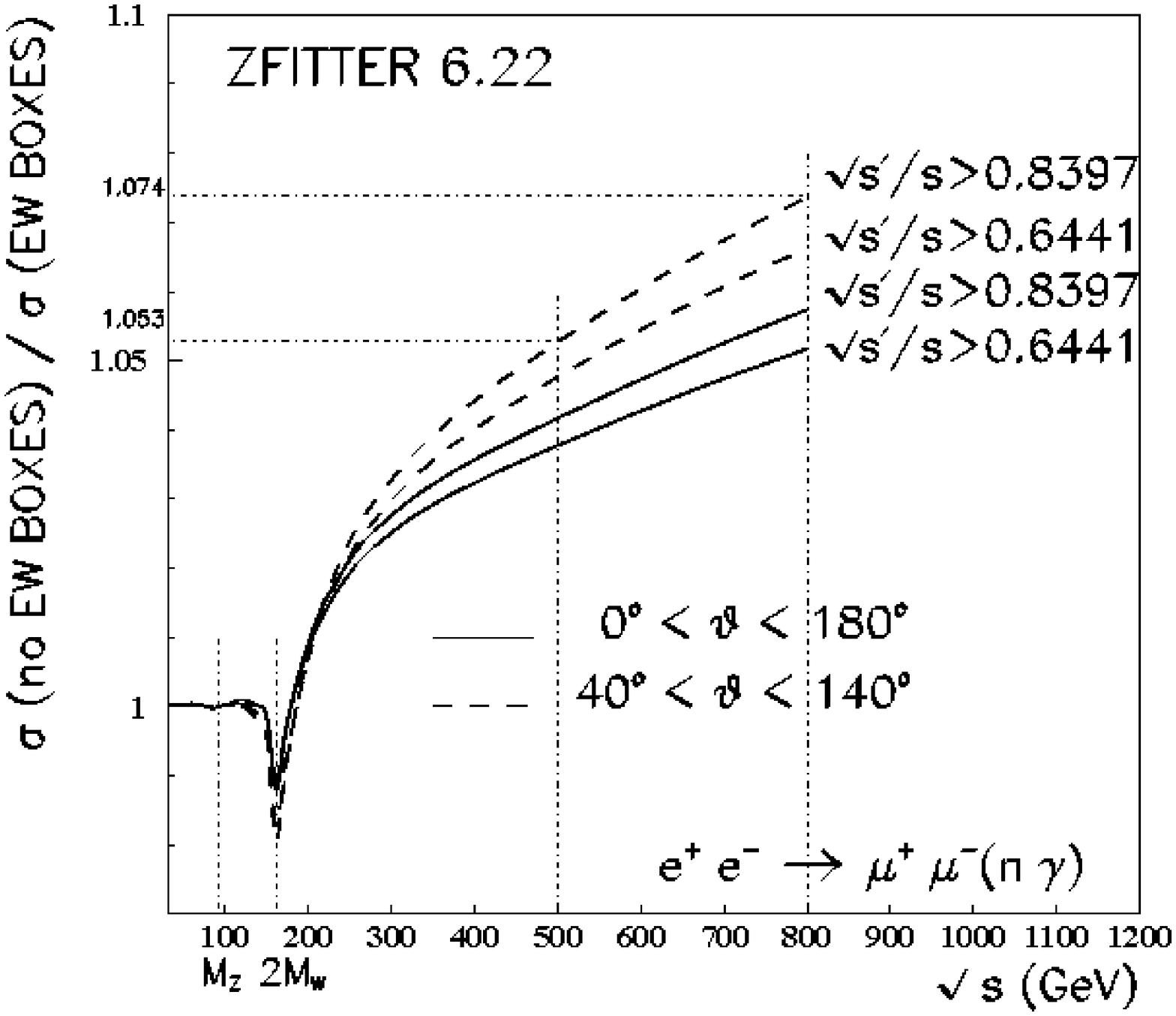,width=7.5cm}}
&
\hspace*{-0.75cm} 
  \mbox{%
  \epsfig{file=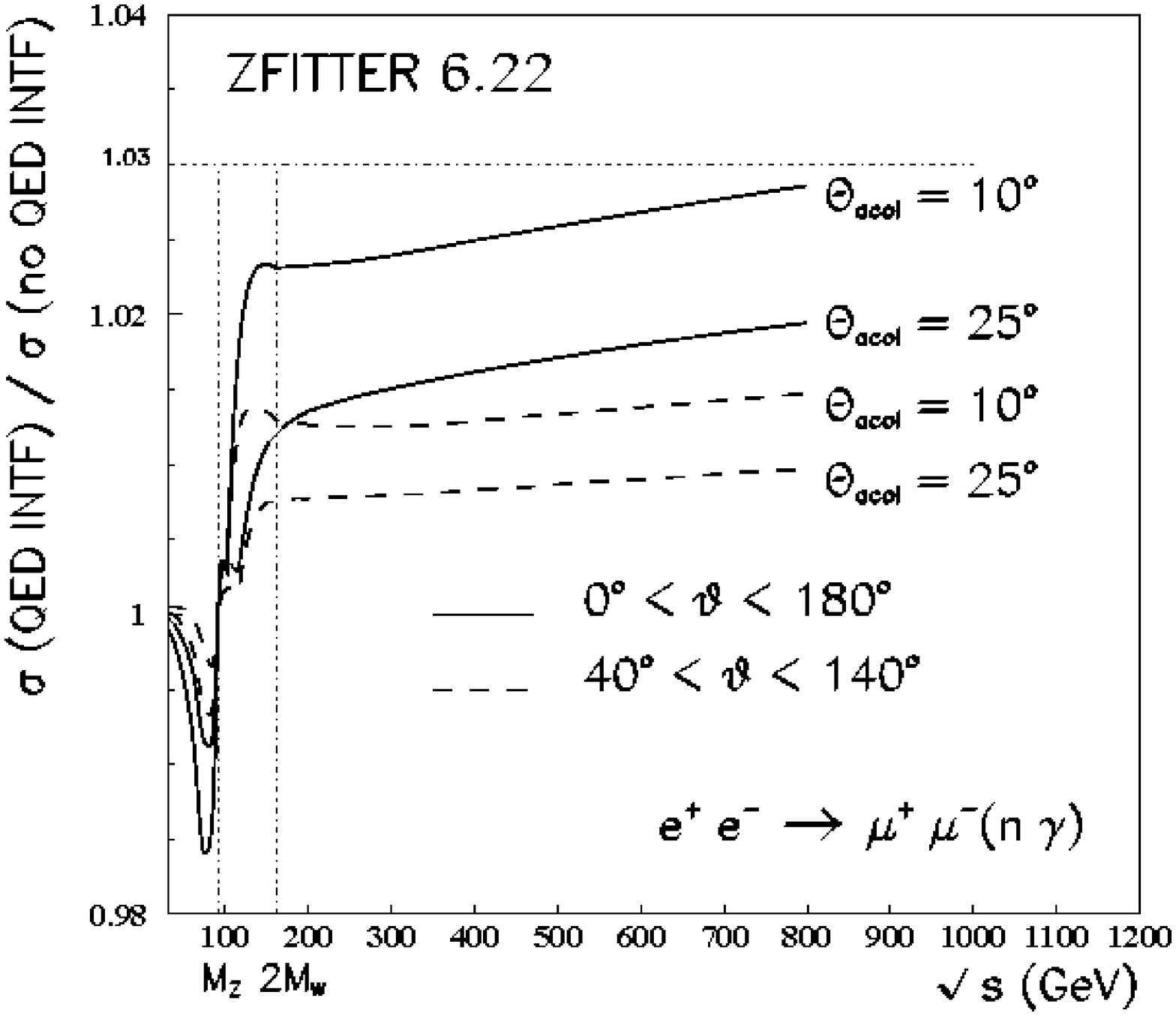,width=7.5cm}}
\\
\end{tabular}
\vspace*{-0.25cm} 
\caption[Comparison of electroweak box and QED corrections]
{\sf
a.~Effect of electroweak box corrections at LEP and LC energies,
b.~QED radiative corrections from initial-final state 
interference, calculated by {\tt ZFITTER} 
v.6.22 \cite{Bardin:1999yd-orig},
with cuts on $s' = M^2_{\mu^+\mu^-}$,
on maximal acollinearity $\theta_{\rm acol}$, and
on total acceptance $\vartheta$ \cite{Jack:1999af}.
\label{sig_EW_QED}}
\end{flushleft}
\end{figure}
%xxxxxxxxxxxxxxxxxxxxxxxxxxxxxxxxxxxxxxxxxxxxxxxxxxxxxxxxxxxxxxxxx 
%
In order to underline this, in Fig.~\ref{sig_EW_QED}
the effect of virtual $ZZ$ and $WW$ box corrections as 
important EW corrections was compared with corrections from 
the QED initial-final state interference 
for muon pair production cross sections $\sigma_T$.
In Fig.~\ref{sig_EW_QED}a, the $ZZ$ and $WW$ 
box corrections are switched off in order to visualize a positive effect.
Correspondingly, the QED interference was switched on and off 
in the right-hand plot (Fig.~\ref{sig_EW_QED}b). 
The net effect of these EW box corrections 
grows with increasing c.m.~energy roughly up to per cent 
level at LEP~2 energies, with the QED interference corrections
being slightly larger depending on the cut applied.
At LC energies, however, the EW contributions can even surpass
the QED interference contribution by roughly a factor of 2, while 
the effect from the QED interference approaches a more or less 
constant value of 2 to 3\%. 
For this comparison, the {\tt ZFITTER} code version v.6.22 
\cite{Bardin:1999yd-orig} was run `blindly' as it stands, i.e.~without 
considering possible extra effects above the $t\bar{t}$ threshold 
due to the top quark mass \cite{Christova:1998tc}. 

At higher energies, all virtual corrections will start to become
equally relevant introducing large gauge cancellations.
Logarithmic and double logarithmic `Sudakhov-type' 
contributions could lead to measurable $1\%$ or larger 
effects at a LC for $\sigma_T(b\bar{b})$ and $R_b$ 
at $500\,\mbox{GeV}$ or higher, but only to per mil level
modifications for different ${b\bar{b}}$ asymmetries 
\cite{Beccaria:1999xd} (see (\ref{rbrc}) for $R_b$).
For an estimate of the EW situation at energies up to 1 TeV 
also consult for example \cite{Ciafaloni:1999it}.
%
%======================================================================
\section{Photonic corrections above the $Z$ resonance 
%with {\it acollinearity cut} 
\label{sec_lep2_radacol}
}
%======================================================================
%
The cross section ratios and the absolute 
differences of the asymmetries of {\tt ZFITTER} v.6.11
\cite{zfitter:v6.11,Bardin:1999yd-orig}, containing the 
new results, are compared with v.5.20 \cite{zfitter:v5.20,Bardin:1992jc2}, 
still with the old coding, for different acollinearity cut values.
First the single corrected contributions  
in the new code were compared with the old code, i.e.~when the initial state
corrections were compared, the final state radiation and the
QED interference were switched off and so on.
%
%======================================================================
\subsection*{Initial state corrections 
\label{sub_lep2_ini}
}
\begin{figure}[htp] 
%---
%\begin{center}
\begin{flushleft}
\vspace*{-0.5cm}
  \begin{tabular}{ll}
\hspace*{-0.25cm}
\vspace*{-0.25cm}
  \mbox{%
  \epsfig{file=%
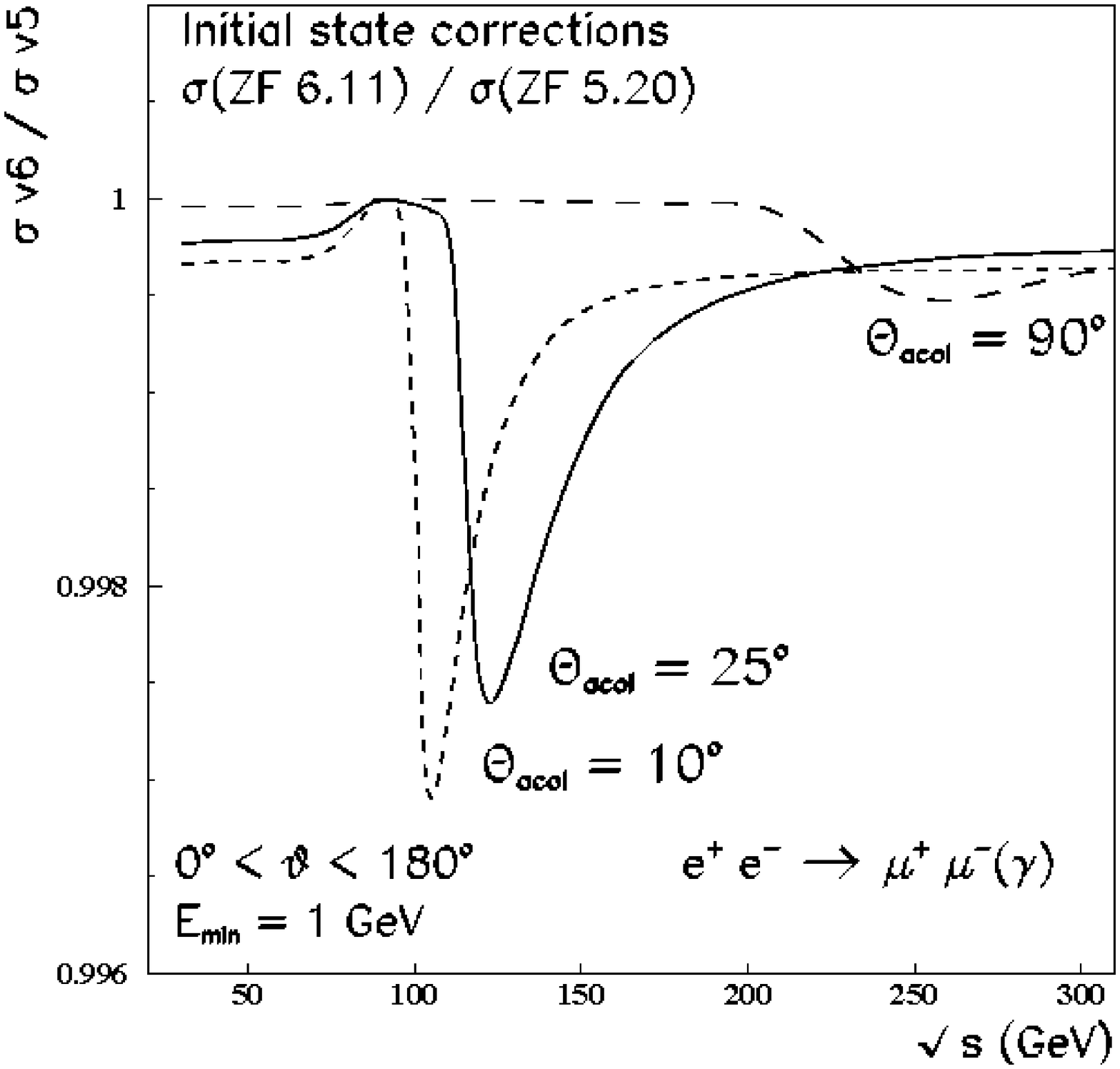
           ,width=7.5cm   % this is the width of the figure (optional)
         }}%
&
\hspace*{-0.75cm}
  \mbox{%
  \epsfig{file=%
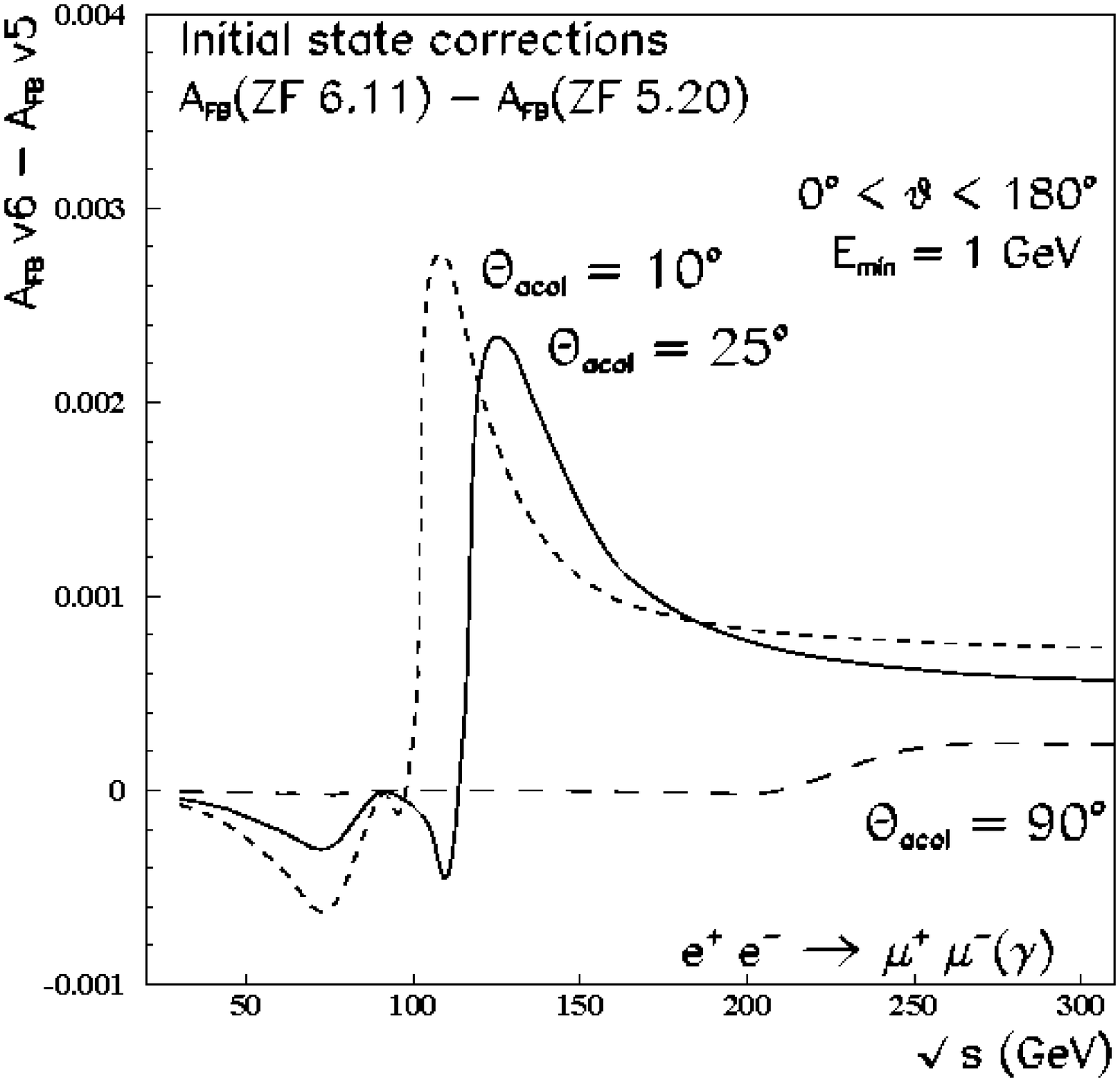
           ,width=7.5cm   % this is the width of the figure (optional)
         }}%
\end{tabular}
\vspace*{-0.25cm}
\caption[Initial state and  
interference corrections for {\tt ZFITTER} at LEP~2]
{\sf 
Ratios of muon pair production cross sections
and differences of forward-backward asymmetries
predicted by {\tt ZFITTER} v.6.11 \cite{Bardin:1999yd-orig}
and v.5.20 \cite{Bardin:1992jc2,zfitter:v5.20} 
without and with acceptance cut and with three different
acollinearity cuts: $\theta_{\rm acol} < 10^{\circ}, 25^{\circ}, 90^{\circ}$;
$E_{min}=1$ GeV; programs differ by initial state radiation 
\cite{Christova:1999gh}. 
\label{xsafbc_ini} 
}
\end{flushleft}
%\end{center}
\end{figure}

%======================================================================
%
In Fig.~\ref{xsafbc_ini}, 
%
%3333333333xxxxxxxxxxxx4xxxxxxxxxxxxxxxxxxxxxxxxxxxxxxxxxxxxxxxxxxxxxxxxxxxxx 
%xxxxxxxxxxxxxxxxxxxxxxxxxxxxxxxxxxxxxxxxxxxxxxxxxxxxxxxxxxxxxxxxx 
%
the ratios of total cross sections $\sigma_T$ and 
the differences of forward-backward asymmetries $A_{FB}$ 
for muon pair production are plotted from the two codes. 
The figure shows a wide range of c.m.~energies from
$30$ to $300\,\mbox{GeV}$. We only discuss the energy regime 
above the $Z^0$ resonance, i.e.~at c.m.~energies of roughly
$\sqrt{s}>100\,\mbox{GeV}$, 
as the resonance region was already discussed in 
the preceding Chapter \ref{ch_lep1slc}, 
while the low energy region below $\sqrt{s}=M_Z$ is not
interesting here and just given for completeness.

While at energies slightly above the $Z$ peak
the differences of the predictions show local peaks, at
LEP~2 energies and beyond they are negligible for 
$\sigma_T$ and amount to only $0.1\%$ -- $0.2\%$ for $A_{FB}$.
The peaking structures in Fig.~\ref{xsafbc_ini} 
disappear at energies for which the radiative
return is prohibited by the cuts.
Depending on the acollinearity cut, this happens for energies 
$\sqrt{s} > \sqrt{s^{min}}$ with $\sqrt{s^{min}}$ being an 
effective $s'$-cut defined by the acollinearity cut. 
The values of $s^{min}$ were given in Table \ref{rxivalues}.
%
%======================================================================
\subsection*{Initial-final state interference corrections
\label{sub_lep2_int}
}
\begin{figure}[htp] 
%---
%\begin{center}
\begin{flushleft}
\vspace*{-0.5cm}
  \begin{tabular}{ll}
\hspace*{-0.25cm}
\vspace*{-0.25cm}
  \mbox{%
\epsfig{file=%
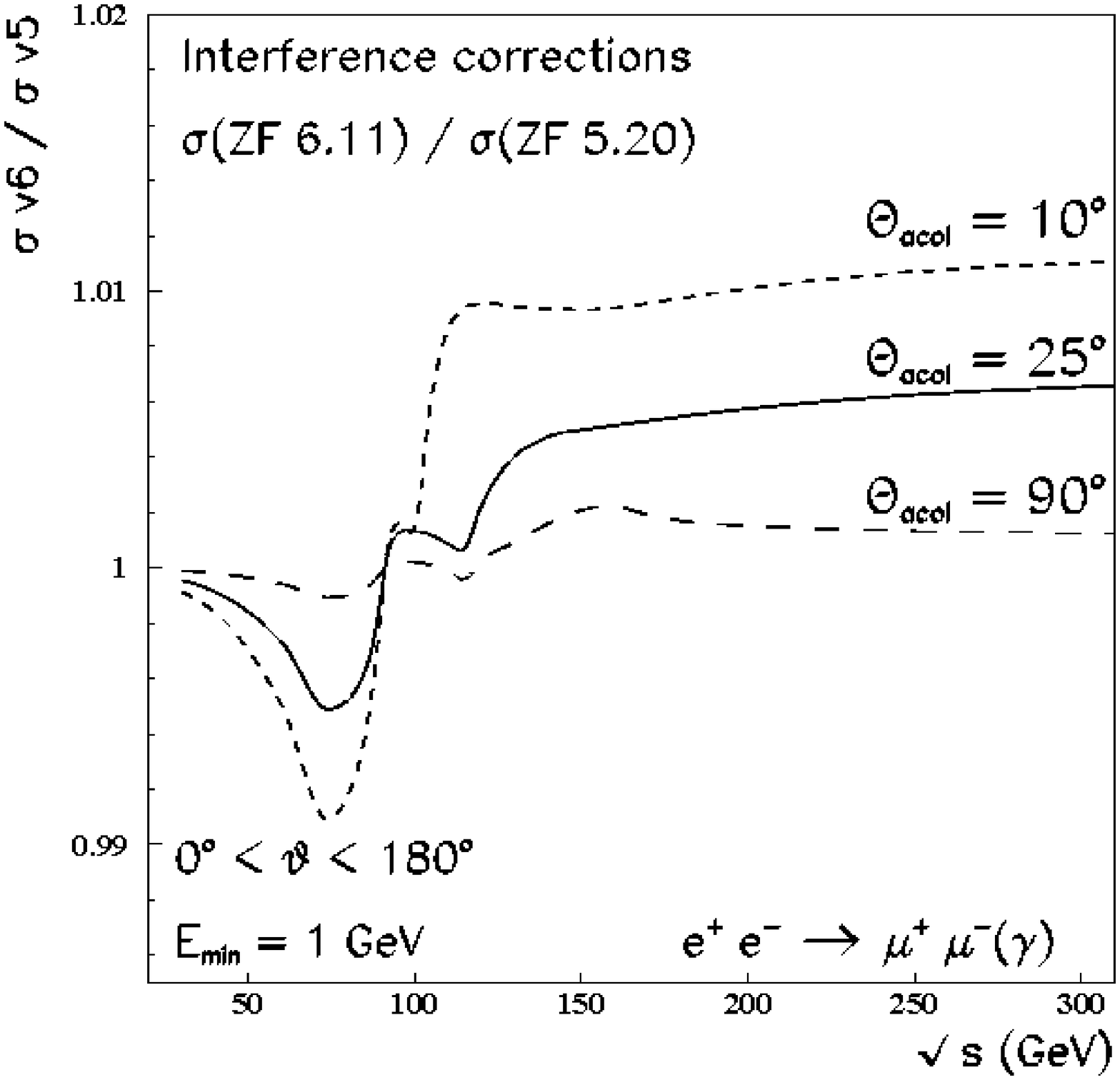
           ,width=7.5cm   % this is the width of the figure (optional)
         }}%
&
\hspace*{-0.75cm}
  \mbox{%
\epsfig{file=%
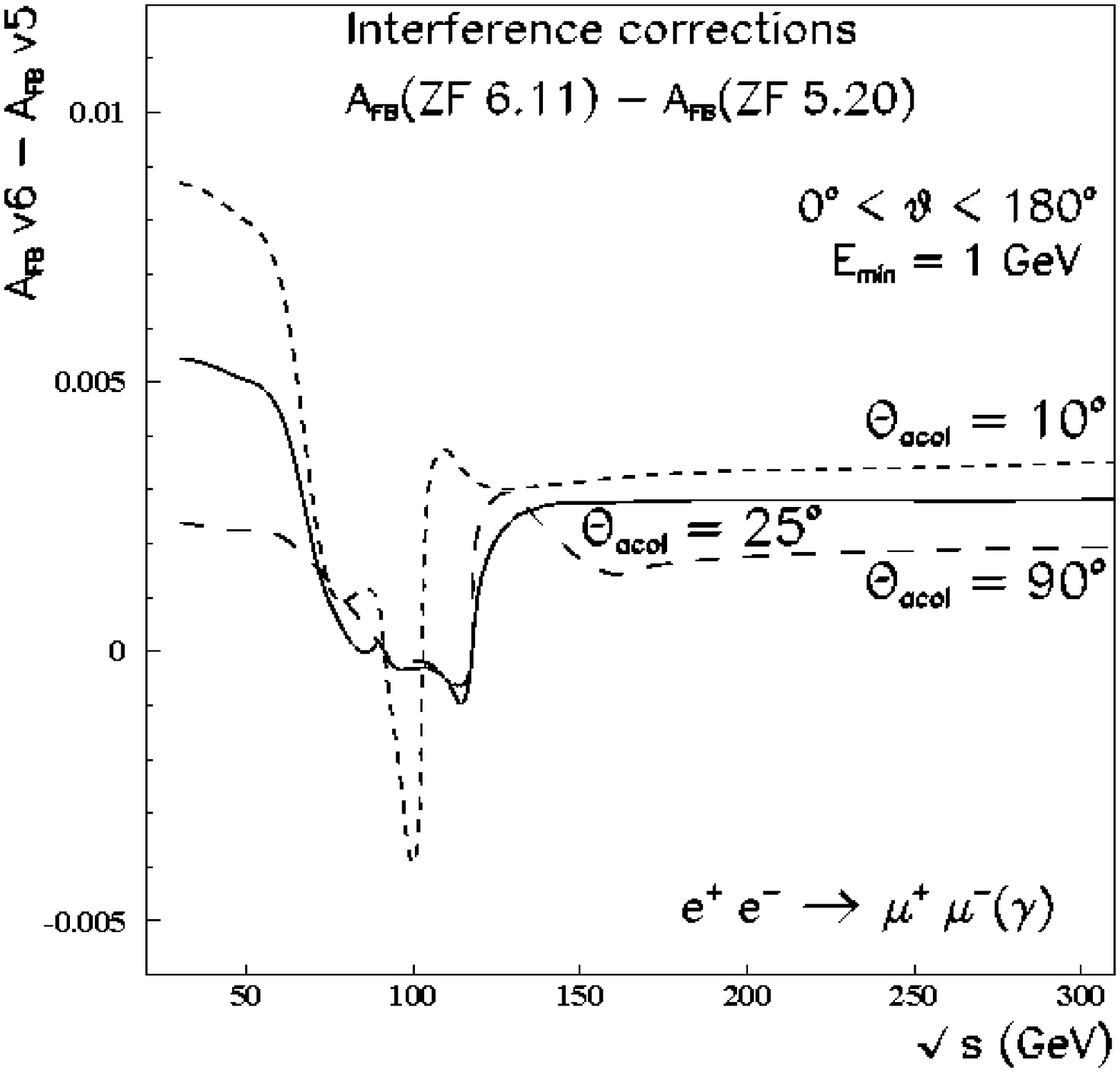
           ,width=7.5cm   % this is the width of the figure (optional)
         }}%
\end{tabular}
\vspace*{-0.25cm}
\caption[QED interference corrections for {\tt ZFITTER} at LEP~2]
{\sf 
Ratios of muon pair production cross sections
and differences of forward-backward asymmetries
predicted by {\tt ZFITTER} v.6.11 \cite{Bardin:1999yd-orig}
and v.5.20 \cite{Bardin:1992jc2,zfitter:v5.20},
cuts as in Fig.~\ref{xsafbc_ini};
%without and with acceptance cut and with three different
%acollinearity cuts: $\theta_{\rm acol} < 10^{\circ}, 25^{\circ}, 90^{\circ}$;
%$E_{min}=1$ GeV; 
programs differ 
by the initial-final state interference \cite{Christova:1999gh}. 
\label{xsafbc_int} 
}
\end{flushleft}
%\end{center}
\end{figure}

%======================================================================
%
In Fig.~\ref{xsafbc_int} the corresponding changes to the code 
are shown for the QED interference corrections. 
%
%3333333333xxxxxxxxxxxx4xxxxxxxxxxxxxxxxxxxxxxxxxxxxxxxxxxxxxxxxxxxxxxxxxxxxx 
%xxxxxxxxxxxxxxxxxxxxxxxxxxxxxxxxxxxxxxxxxxxxxxxxxxxxxxxxxxxxxxxxx 
%
{For} total cross sections, the deviations may reach at most up to $1\%$
at LEP~2 energies, while for asymmetries they stay below $0.5\%$ 
there. Both shifts are more than the precision we aim at for the 
theoretical predictions. 

%======================================================================
\subsection*{Final state corrections 
\label{sub_lep2_final}
}
\begin{figure}[htbp] 
\begin{flushleft}
%--- 
\begin{tabular}{ll}
\hspace*{-0.25cm}
  \mbox{%
  \epsfig{file=%
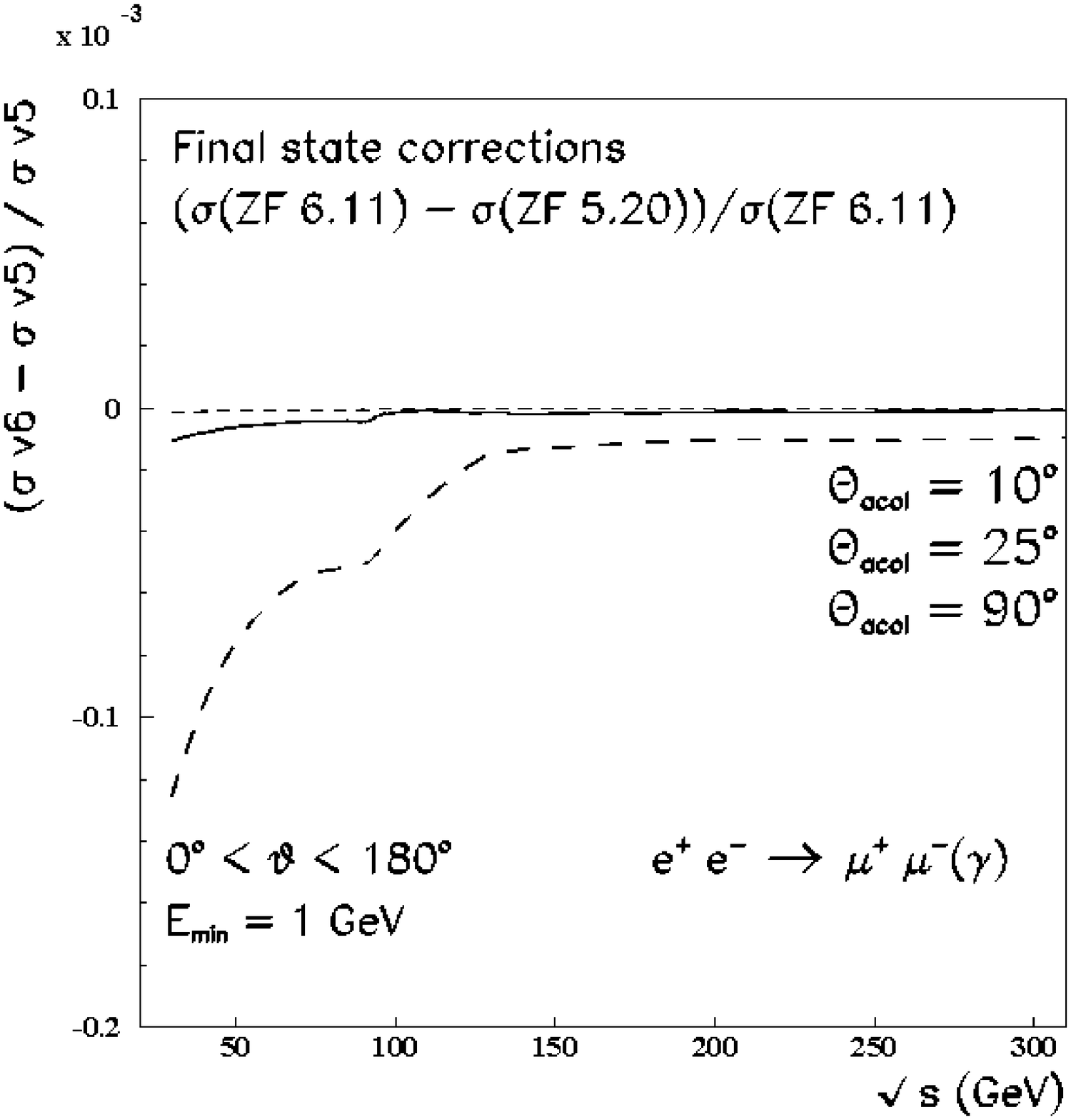
           ,width=7.5cm   % this is the width of the figure (optional)
         }}%
&
\hspace*{-0.75cm} 
  \mbox{%
  \epsfig{file=%
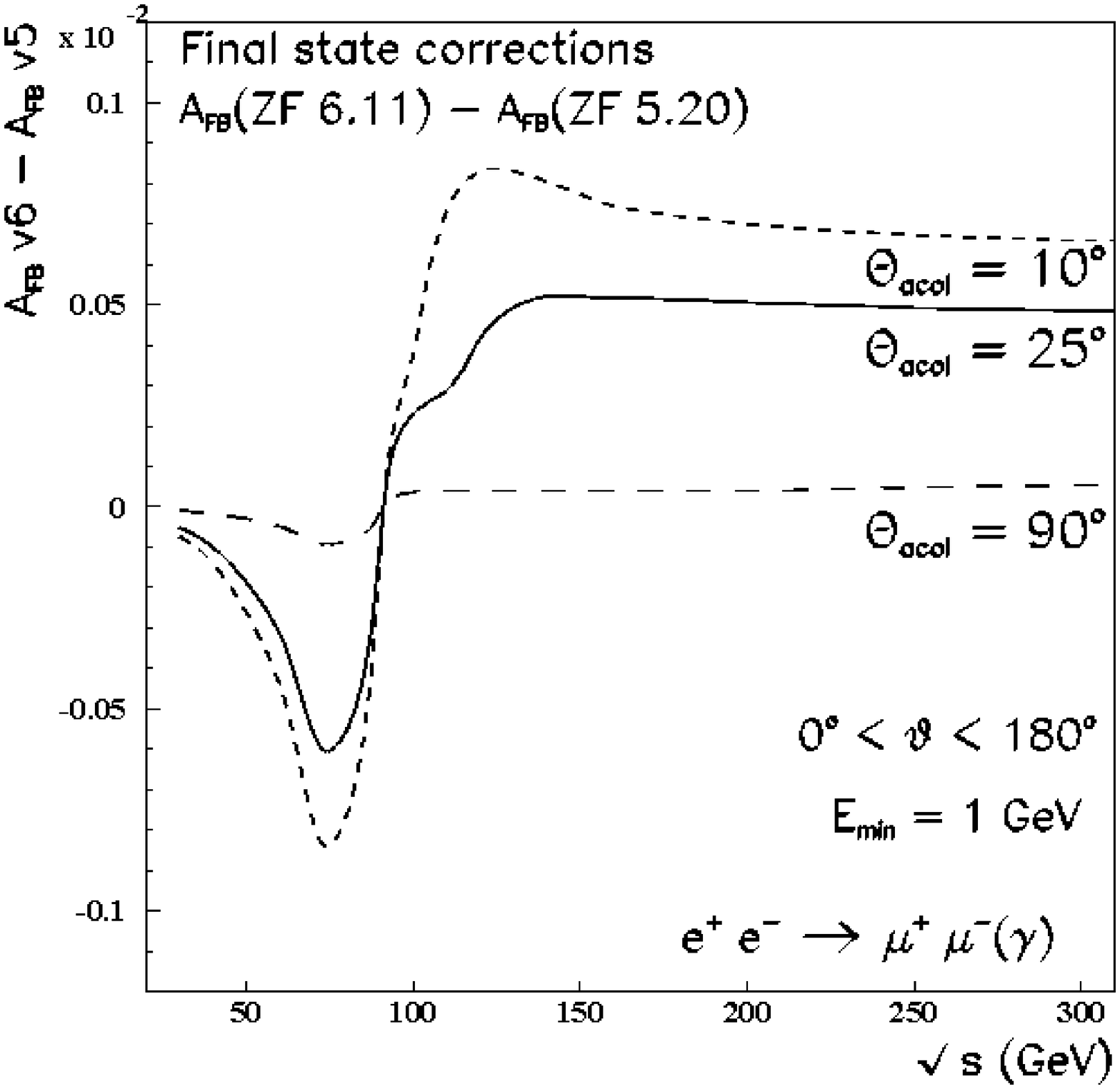
           ,width=7.5cm   % this is the width of the figure (optional)
         }}%
\end{tabular}
\caption[Final state corrections for {\tt ZFITTER} at LEP~2]
{\sf
Ratios of muon pair production cross sections and 
differences of forward-backward asymmetries 
predicted by {\tt ZFITTER} v.6.11 \cite{Bardin:1999yd-orig}
and v.5.20 \cite{Bardin:1992jc2,zfitter:v5.20},
cuts as in Fig.~\ref{xsafbc_ini};
%without acceptance cut and with three different
%acollinearity cuts: $\theta_{\rm acol} < 10^{\circ}, 25^{\circ}, 90^{\circ}$,
%$E_{min}=1$ GeV; 
programs differ by final state radiation 
\cite{Christova:1999gh}.
\label{xsafbc_fin} 
}
\end{flushleft}
\end{figure}

%======================================================================
%
In Fig.~\ref{xsafbc_fin}, omitting an acceptance cut,
we see that the deviations from the corrected final state radiators
are negligible in the wide energy range, never exceeding $0.01\%$ 
for the cross section and $0.1\%$ for the asymmetry. 
If an acceptance cut is applied, the changes are yet smaller.
%
%xxxxxxxxxxxxxxxxxxxxxxxxxxxxxxxxxxxxxxxxxxxxxxxxxxxxxxxxxxxxxxxxx 
%xxxxxxxxxxxxxxxxxxxxxxxxxxxxxxxxxxxxxxxxxxxxxxxxxxxxxxxxxxxxxxxxx 
% 
%======================================================================
\subsection*{Net corrections
\label{sub_lep2_net}
}
\begin{figure}[htp] 
%---
\begin{center}
\vspace*{-0.5cm}
\begin{flushleft} 
\begin{tabular}{ll}
\hspace*{-0.25cm}
\vspace*{-0.25cm}
  \mbox{%
  \epsfig{file=%
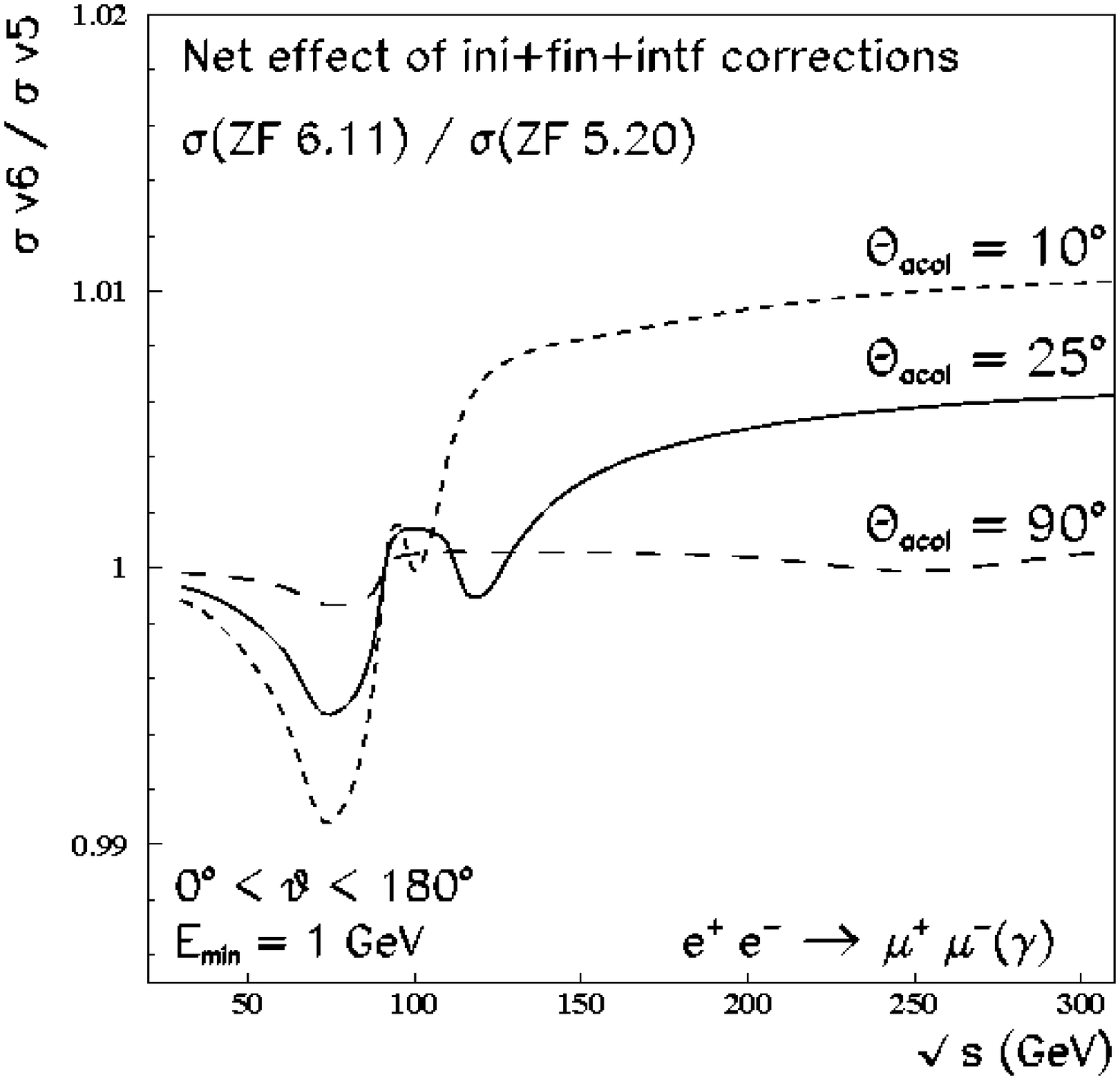
           ,width=7.5cm   % this is the width of the figure (optional)
         }}%
&
\hspace*{-0.75cm}
  \mbox{%
  \epsfig{file=%
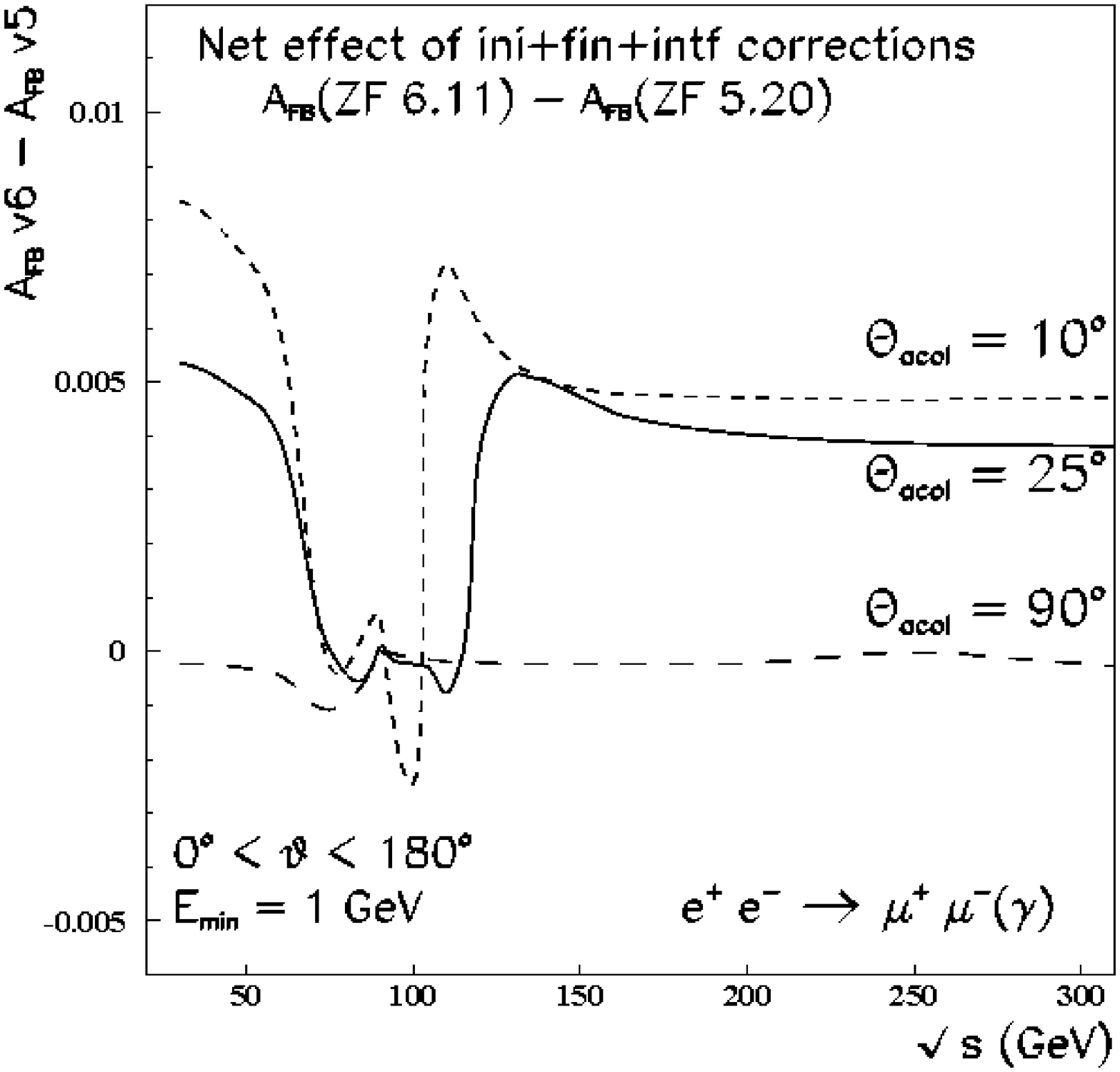
           ,width=7.5cm   % this is the width of the figure (optional)
         }}%
\\
\end{tabular}
\vspace*{-0.25cm}
\caption[Net corrections for {\tt ZFITTER} at LEP~2]
{\sf
Net ratios of muon pair production cross sections and 
differences of forward-backward asymmetries
predicted by {\tt ZFITTER} v.6.11 \cite{Bardin:1999yd-orig}
and v.5.20 \cite{Bardin:1992jc2,zfitter:v5.20},
cuts as in Fig.~\ref{xsafbc_ini};
%without acceptance cut and with three different
%acollinearity cuts: $\theta_{\rm acol} < 10^{\circ}, 25^{\circ}, 90^{\circ}$;
%$E_{min}=1$ GeV; 
initial-final state interference included
\cite{Christova:1999gh,Jack:1999xc,Jack:1999af}.
\label{xsafbc_net_int}
}
\end{flushleft}
\end{center}
\end{figure}

%======================================================================
%
We have to distinguish two different approaches to data. 
Sometimes experimentalists subtract the initial-final state 
interference contributions from measured data, and sometimes the 
interference effects remain in the data sample. 
The resulting effects of all photonic corrections discussed 
in the foregoing sections are shown in Fig.~\ref{xsafbc_net_int}.
The net corrections for the muon pair production cross section
and the forward-backward asymmetry at LEP~1 we had shown in 
Table \ref{tab10acolc}.

%xxxxxxxxxxxxxxxxxxxxxxxxxxxxxxxxxxxxxxxxxxxxxxxxxxxxxxxxxxxxxxxxx 
%xxxxxxxxxxxxxxxxxxxxxxxxxxxxxxxxxxxxxxxxxxxxxxxxxxxxxxxxxxxxxxxxx 
%
For the wider energy range, 
the changes without initial-final interferences 
are below what is expected to be relevant at LEP~2 energies.  
The corrections to the numerical output from {\tt ZFITTER} with
acollinearity cut, however, increased when the corrected initial-final state
interference is taken into account (see Fig.~\ref{xsafbc_net_int}).

The numerical effects are dominated by the initial-final state
interference and never exceed $1\%$ 
at LEP~2 energies \cite{Christova:1999gh,Jack:1999xc,Jack:1999af}.
The corrected initial state and final state terms
only have minor effects on $\sigma_T$ and $A_{FB}$ 
and amount at most to corrections at the order of 
$0.1\%$ -- $0.2\%$ for $A_{FB}$ at LEP~2 energies for
different cuts. The net corrections are largest where the radiative 
return to the $Z$ starts to be prevented by the acollinearity 
cut. {For} $\theta_{acol} < 10^\circ$ or $25^\circ$ this sets in 
at roughly $\sqrt{s} > 100\,\mbox{GeV}$, or $115\,\mbox{GeV}$
respectively. These corrections to the code are at most 
roughly $0.5\%$ for $\sigma_T$ and $1\%$ for $A_{FB}$ and 
shrink below $1\%$ at higher energies.
A detailed analysis of all new modifications to the code
can be found in \cite{Christova:1999gh}. 

%===================================================================
\section{Comparisons with different programs
\label{sec_lep2_codes}
}
\begin{figure}[bhtp]
\begin{center} 
%---
\vspace*{-0.75cm} 
\mbox{ 
       \epsfig{file=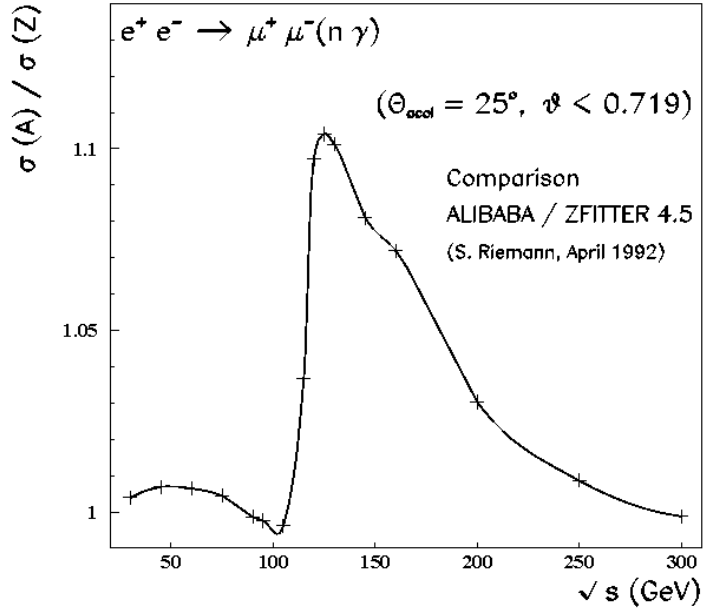,%
height=10.cm,width=10.cm% 
}}
\vspace*{-1.25cm}
\caption[Early comparison of {\tt ALIBABA} and 
{\tt ZFITTER} above the $Z$ peak]
{\sf
Muon pair production cross section ratios 
{\tt ALIBABA} v.1 (1990) \cite{Beenakker:1991mb} 
versus {\tt ZFITTER} v.4.5 (1992) \cite{Bardin:1992jc2}.
An acollinearity cut was applied with $\theta_{\rm acol}=25^{\circ}$ 
\cite{Christova:1999gh,Jack:1999af}.
\label{compar1992} 
} 
\end{center} 
\end{figure}

%======================================================================
%
In 1992, a comparison of {\tt ALIBABA} v.1 (1991)
\cite{Beenakker:1991mb}
and {\tt ZFITTER} v.4.5 (1992) \cite{Bardin:1992jc2} showed
deviations between the predictions of the two programs of up to $10\%$ cent
\cite{Riemann:1992up}; one of the plots of that study
is shown in Fig.~\ref{compar1992}. 
%
%xxxxxxxxxxxxxxxxxxxxxxxxxxxxxxxxxxxxxxxxxxxxxxxxxxxxxxxxxxxxxxxxx 
%xxxxxxxxxxxxxxxxxxxxxxxxxxxxxxxxxxxxxxxxxxxxxxxxxxxxxxxxxxxxxxxxx 

These deviations were observed only above the $Z$ peak and only when
an acollinearity cut on the fermions was applied;
the agreement was much better without this cut.
The comparison was repeated in 1998 with {\tt ALIBABA} v.2 (1991) 
\cite{Beenakker:1991mb}, 
{\tt TOPAZ0} v.4.3 (1998) 
\cite{Montagna:1995b,Montagna:1998kp,Passarino:199800},
and {\tt ZFITTER} v.5.14 (1998) \cite{Bardin:1992jc2,zfitter:v5.14}.
{\tt ALIBABA} v.2 was used with the default settings and in
{\tt ZFITTER} v.5.14 one flag was modified ({\tt PHOT2}=2). 
{\tt TOPAZ0} v.4.3 was run in accordance with {\tt ZFITTER} v.5.14.
The outcome was basically unchanged compared to 1992 as may be seen in
Fig.~3 of reference \cite{Christova:1998tc} which 
shows cross section ratios as functions of $s$
with a cut on the maximal acollinearity angle $\theta_{\rm acol}$
between the fermions.
In addition, selected predictions had been shown 
in \cite{Christova:1998tc} at 120 GeV arising from a
variation of flags ({\tt{IORDER,NONLOG,IFINAL}}) in {\tt ALIBABA}:
upper ones at (4,n,m), lower ones at (3,n,m), with n=0,1, m=1,2 (best
choice: (4,1,2)), with the deviations strongly depending on the 
calculated higher order corrections.   
All numbers were produced with the default settings of the 
programs.

\begin{figure}[thb] 
\begin{flushleft}
\begin{tabular}{ll}
\hspace*{-0.25cm}
  \mbox{%
  \epsfig{file=%
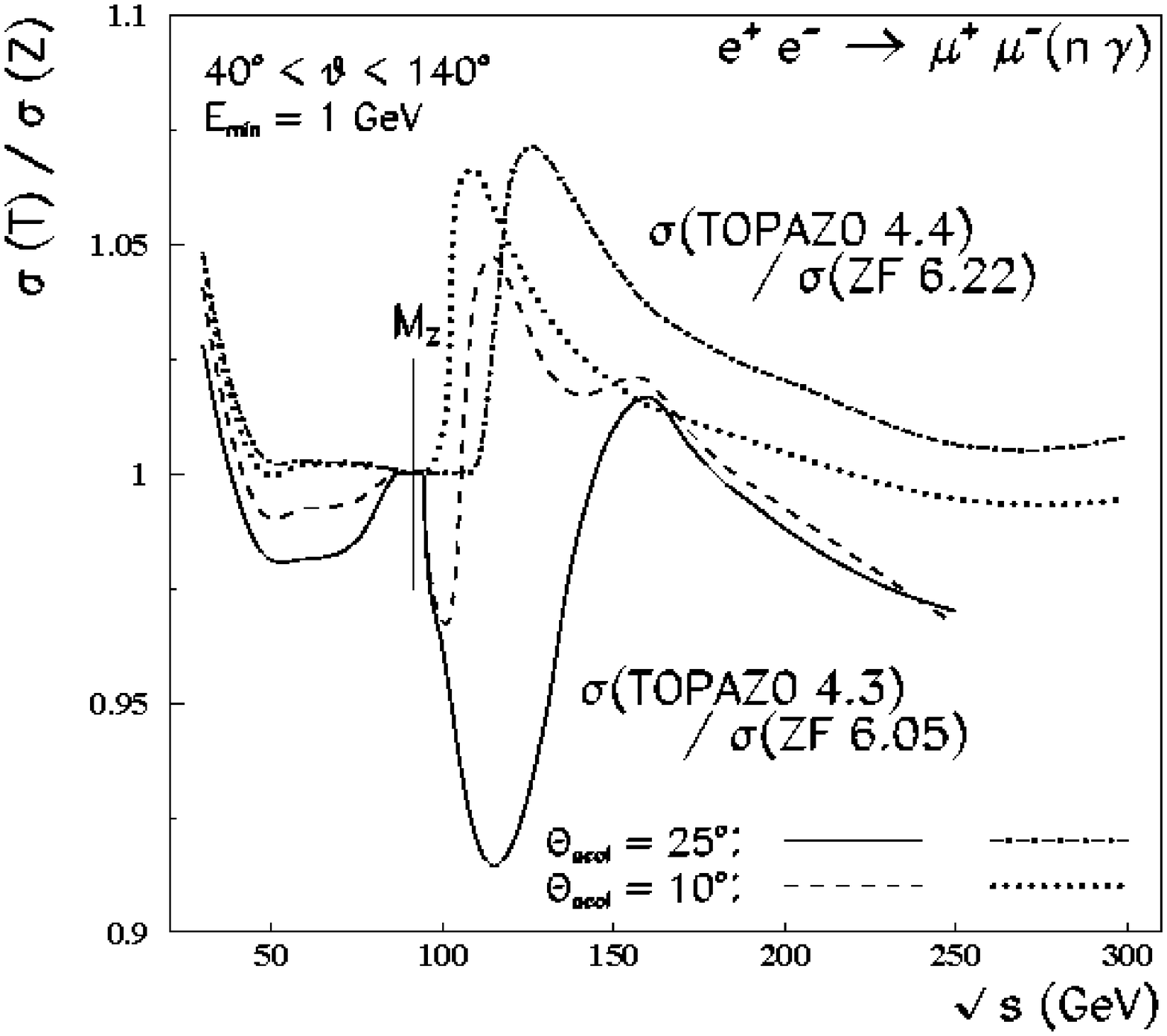
           ,width=7.5cm   % this is the width of the figure (optional)
         }}%
&
\hspace*{-0.75cm}
  \mbox{%
  \epsfig{file=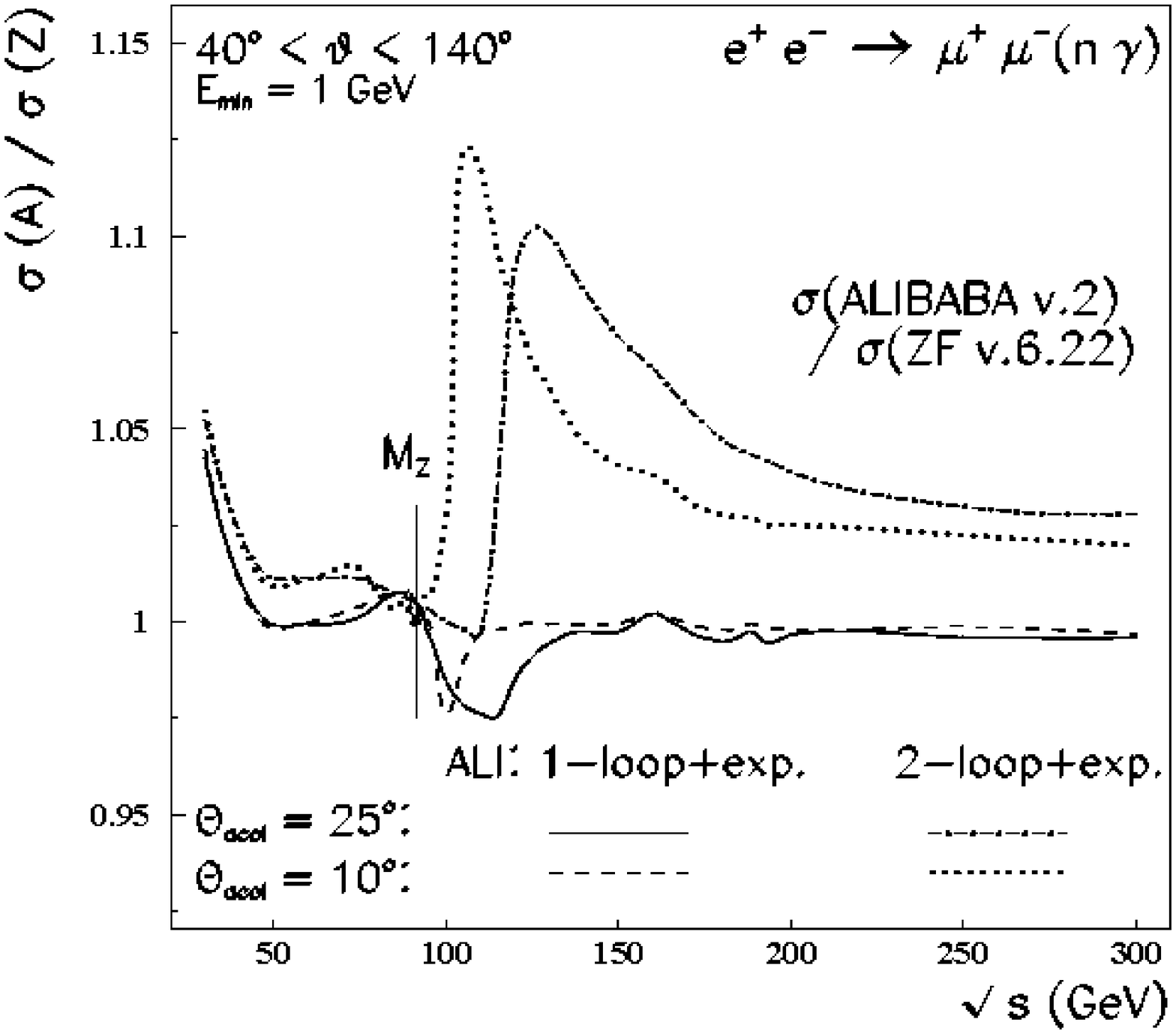,width=7.5cm}}
\\
\end{tabular}
\caption[Comparison of {\tt TOPAZ0}, {\tt ALIBABA}, and {\tt ZFITTER} 
above the $Z$ peak]
{\sf
Muon pair production cross section ratios with 
$\theta_{\rm acol}=10^{\circ}, 25^{\circ}$ and 
$\theta_{\rm acc}=40^{\circ}$; a.~{\tt TOPAZ0} v.4.3 and v.4.4 
\cite{Montagna:1995b,Montagna:1998kp} versus 
{\tt ZFITTER} v.6.04/06 \cite{zfitter:v6.0406} 
and v.6.22 \cite{Bardin:1999yd-orig} (1999),  
b.~{\tt ALIBABA} v.2 (1990) \cite{Beenakker:1991mb}
versus {\tt ZFITTER} v.6.22 \cite{Christova:1999gh,Jack:1999xc,Jack:1999af}.
Flag setting: {\tt ISPP}=0 \cite{Bardin:1999yd-orig}.
\label{compar-tzaz}
}
\end{flushleft}
\end{figure}

Finally, the same version of {\tt ALIBABA} was compared with 
{\tt ZFITTER} v.6.22 (1999) \cite{zfitter:v6.11,Bardin:1999yd-orig}
and the same was done for {\tt TOPAZ0} v.4.3 and v.4.4 
\cite{Montagna:1995b,Montagna:1998kp} and {\tt ZFITTER} v.6.04/06 
\cite{zfitter:v6.0406} and v.6.22 (1999) \cite{Bardin:1999yd-orig}, 
all depicted in Fig.~\ref{compar-tzaz}.
%
%xxxxxxxxxxxxxxxxxxxxxxxxxxxxxxxxxxxxxxxxxxxxxxxxxxxxxxxxxxxxxxxxx 
%xxxxxxxxxxxxxxxxxxxxxxxxxxxxxxxxxxxxxxxxxxxxxxxxxxxxxxxxxxxxxxxxx 

Concerning the {\tt TOPAZO} v.4.4 ratios, we register a different 
behaviour (compared to v.4.3) for $\theta_{\rm acc}=40^{\circ}$ 
which is now much closer to the {\tt ALIBABA} ratios
at energies above roughly $100\,\mbox{GeV}$.
Between about 100 GeV and 200 GeV, the deviations in the predictions 
from different programs are huge and  heavily depending on 
the maximally allowed acollinearity angle $\theta_{\rm acol}$, 
here shown for $\theta_{\rm acol}=10^{\circ},25^{\circ}$.
The ratios stabilize at higher (or smaller) energies.
The {\tt ZFITTER} numbers are produced with the default settings (if not
otherwise stated).

At LEP~2 energies the deviation of 
{\tt ZFITTER} v.6.22 and {\tt TOPAZ0} v.4.4 is at the order of $1\%$
or less for different acollinearity cuts and an 
acceptance cut of $40^{\circ} < \vartheta < 140^{\circ}$. 
In both cases, however, there is a clear peak of the cross section ratios 
at energies where the $Z$ radiative return is not prevented by the 
cuts. While for the $s'$-cut this discrepancy stays moderate at the 
per cent level, it grows up to several per cent for the acollinearity 
cut.\footnote{The flip of sign of these effects compared to the 
older versions, {\tt TOPAZO} v.4.3 and {\tt ZFITTER} v.6.04/06, is mainly 
due to a corrected interference contribution in the {\tt TOPAZ0} code.
Changes to code {\tt ZFITTER} v.6.04/06 were negligible here.}  
Corrections by initial state pair production or different 
exponentiation of initial and final state higher orders, however, 
do not have a large effect here \cite{Jack:1999xc,Jack:1999af}.\footnote{
In {\tt ZFITTER} , the treatment of some higher order corrections was
varied via flags {\tt FOT2} and {\tt PAIRS}.
}
%
%xxxxxxxxxxxxxxxxxxxxxxxxxxxxxxxxxxxxxxxxxxxxxxxxxxxxxxxxxxxxxxxxx 
\begin{figure}[th]
\begin{flushleft}
\begin{tabular}{ll}
\hspace*{-0.25cm}  
  \mbox{%
  \epsfig{file=%
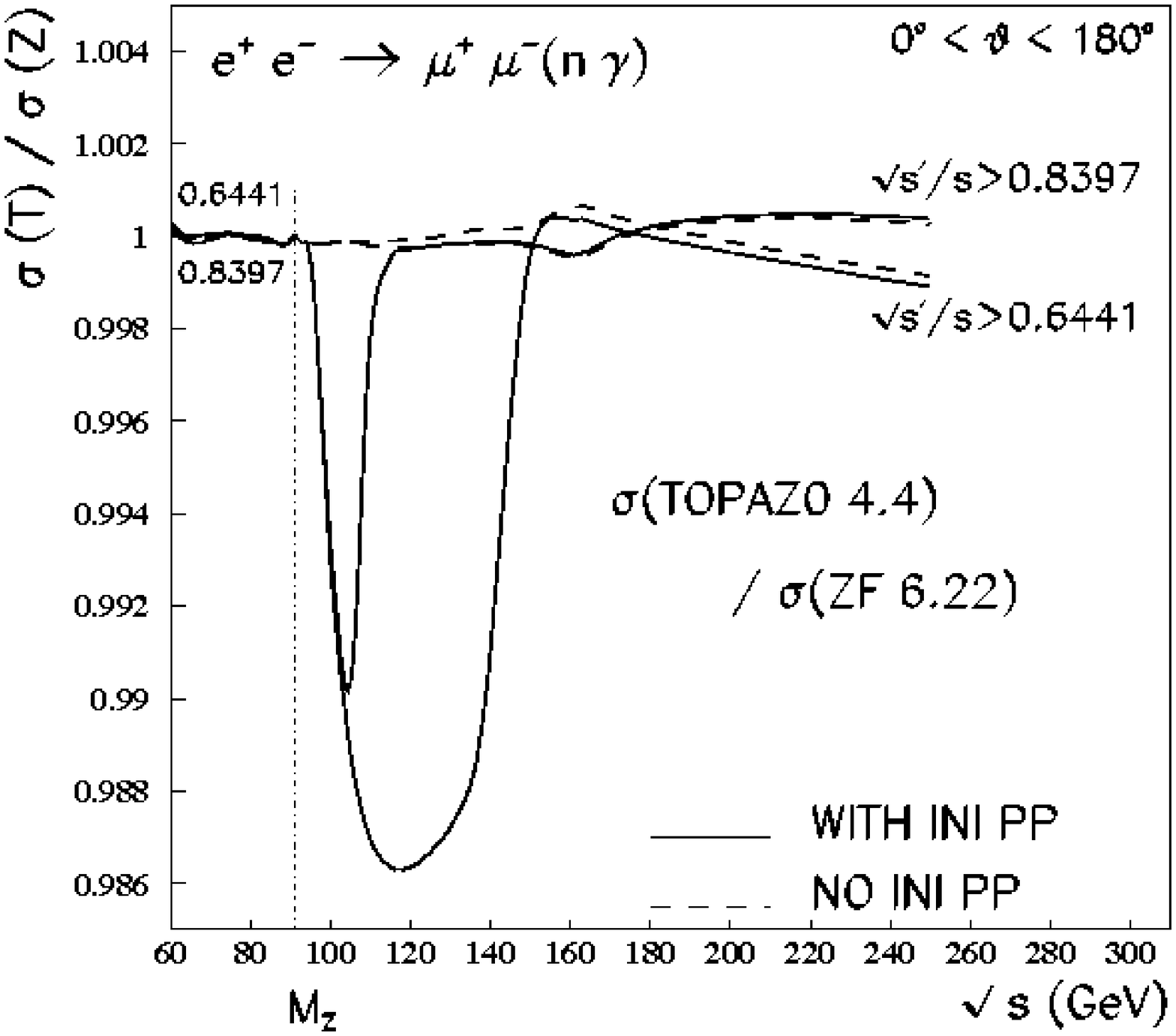
,width=7.5cm   % this is the width of the figure (optional)
         }}%
&
\hspace*{-0.75cm}
  \mbox{%
  \epsfig{file=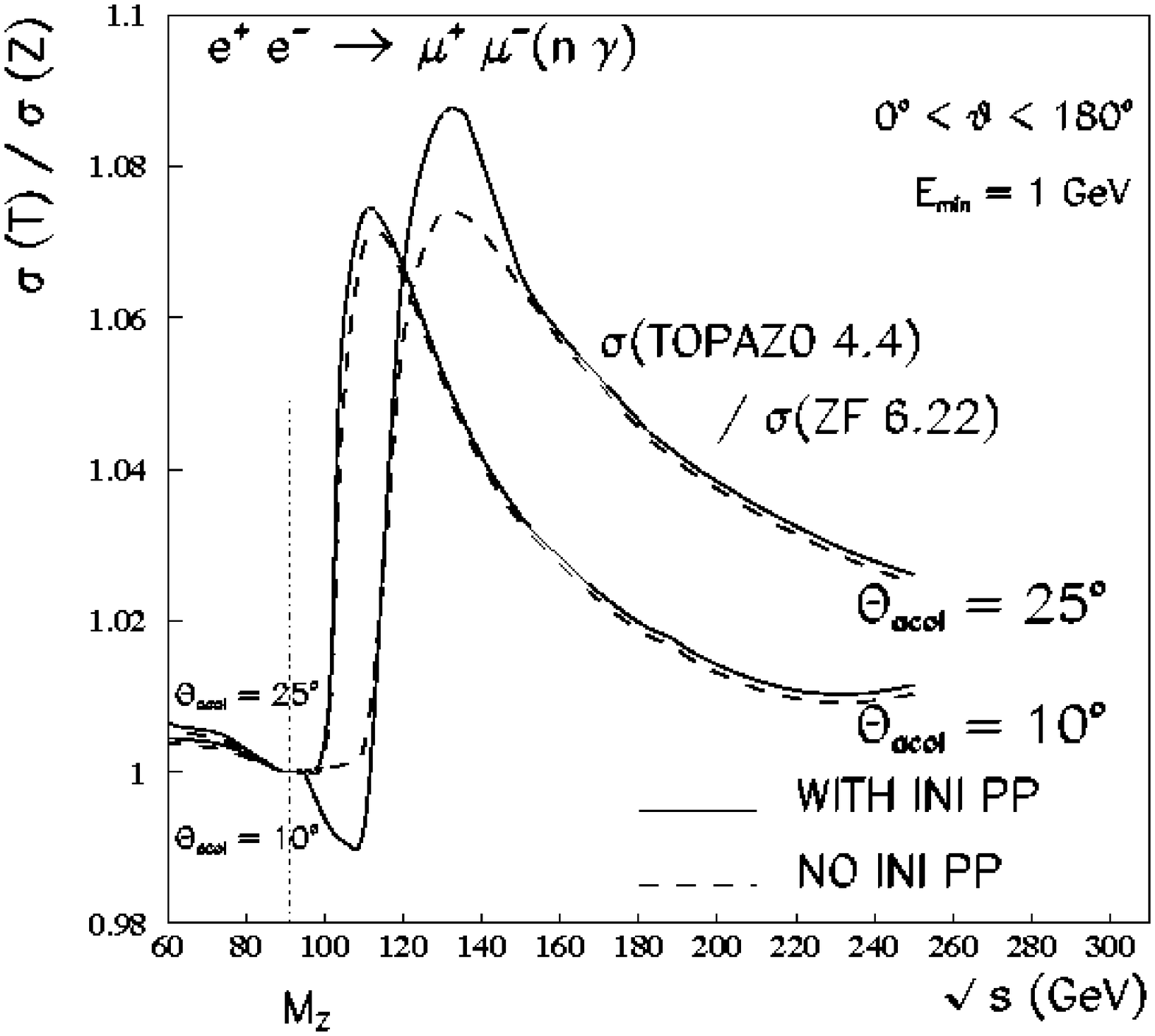,width=7.5cm}}
\\
\end{tabular}
\caption[Comparison of $\sigma_T$ for {\tt TOPAZ0} and {\tt ZFITTER} 
above the $Z$ peak]
{\sf
Comparison of predictions from {\tt ZFITTER} v.6.22 
\cite{Bardin:1999yd-orig}
and 
{\tt TOPAZ0} v.4.4 \cite{Montagna:1998kp} 
for muon pair production cross section ratios
with a.~an $s'$-cut or b.~an acollinearity cut \cite{Jack:1999af}
(Flag setting: {\tt ISPP}=0,1, {\tt FINR}=0; further: 
{\tt SIPP}={\tt S\_PR} \cite{Bardin:1999yd-orig}). 
\label{xs-top-zf}
}
\end{flushleft}
\end{figure}
%xxxxxxxxxxxxxxxxxxxxxxxxxxxxxxxxxxxxxxxxxxxxxxxxxxxxxxxxxxxxxxxxx 

Furthermore, when the two-loop contributions in {\tt ALIBABA} 
(with setting\\ {\tt IORDER=3}) were switched off, the agreement 
improved considerably. This visualizes the strong dependence of 
predictions on the details of the theoretical input chosen, 
e.g.~the treatment of higher order contributions or the correct 
inclusion of non-logarithmic $O(\alpha)$ corrections.
Evidently, the largest deviations arise from the radiative return 
of $\sqrt{s'}$ to the $Z$ boson resonance due to hard initial state
radiation. Interest in the high energy part of the data anyhow means 
to cut this away and so there should not be a serious problem.
If instead one is interested in the radiative return, one has to be
concerned about accuracies.
These observations confirm similar statements from other
studies \cite{Montagna:1997jt}.
%
%xxxxxxxxxxxxxxxxxxxxxxxxxxxxxxxxxxxxxxxxxxxxxxxxxxxxxxxxxxxxxxxxx 
\begin{figure}[th]
\begin{flushleft}
\begin{tabular}{ll}
\hspace*{-0.25cm}  
  \mbox{%
  \epsfig{file=%
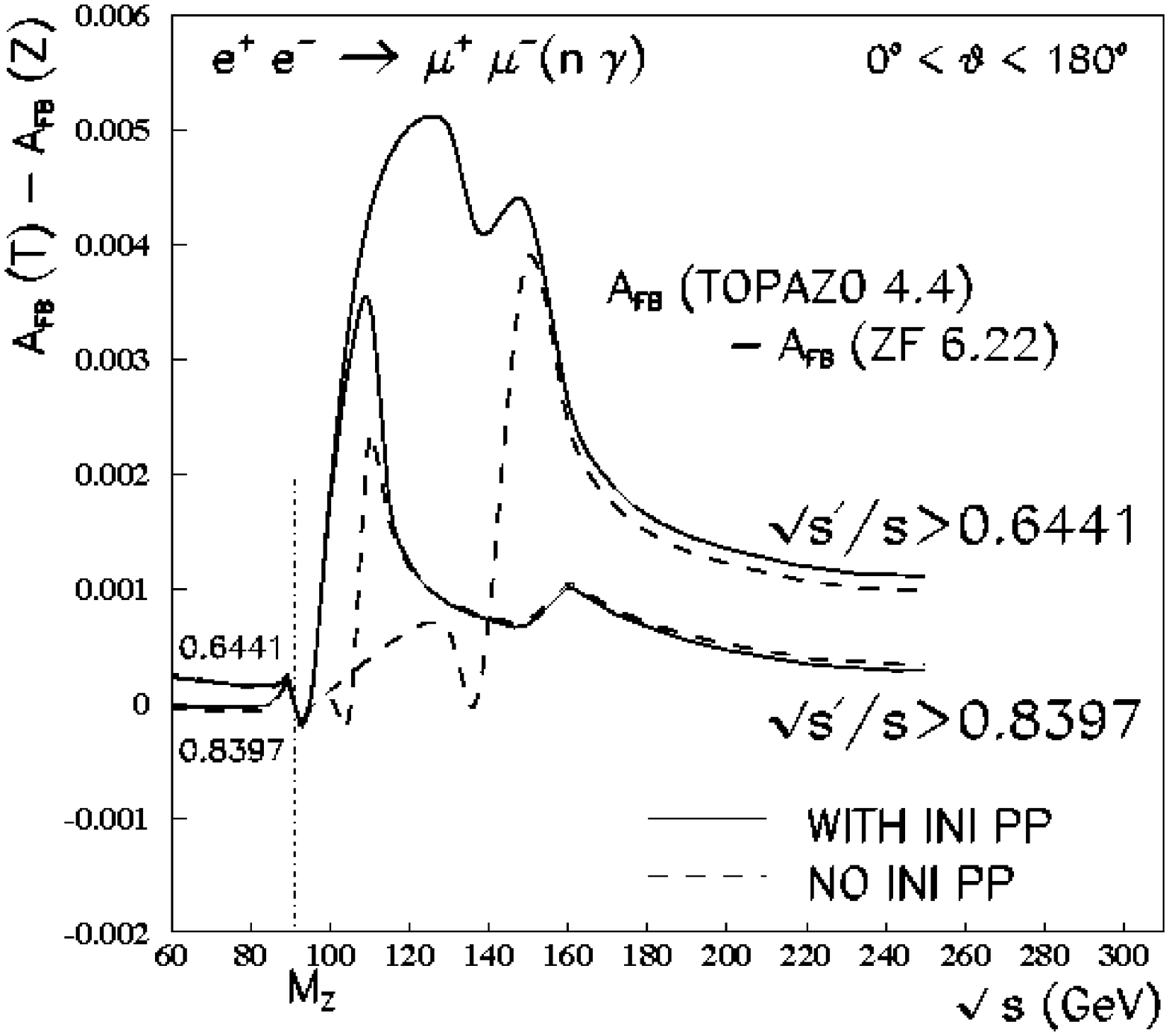,width=7.5cm   % this is the width of the figure (optional)
         }}%
&
\hspace*{-0.75cm}
  \mbox{%
  \epsfig{file=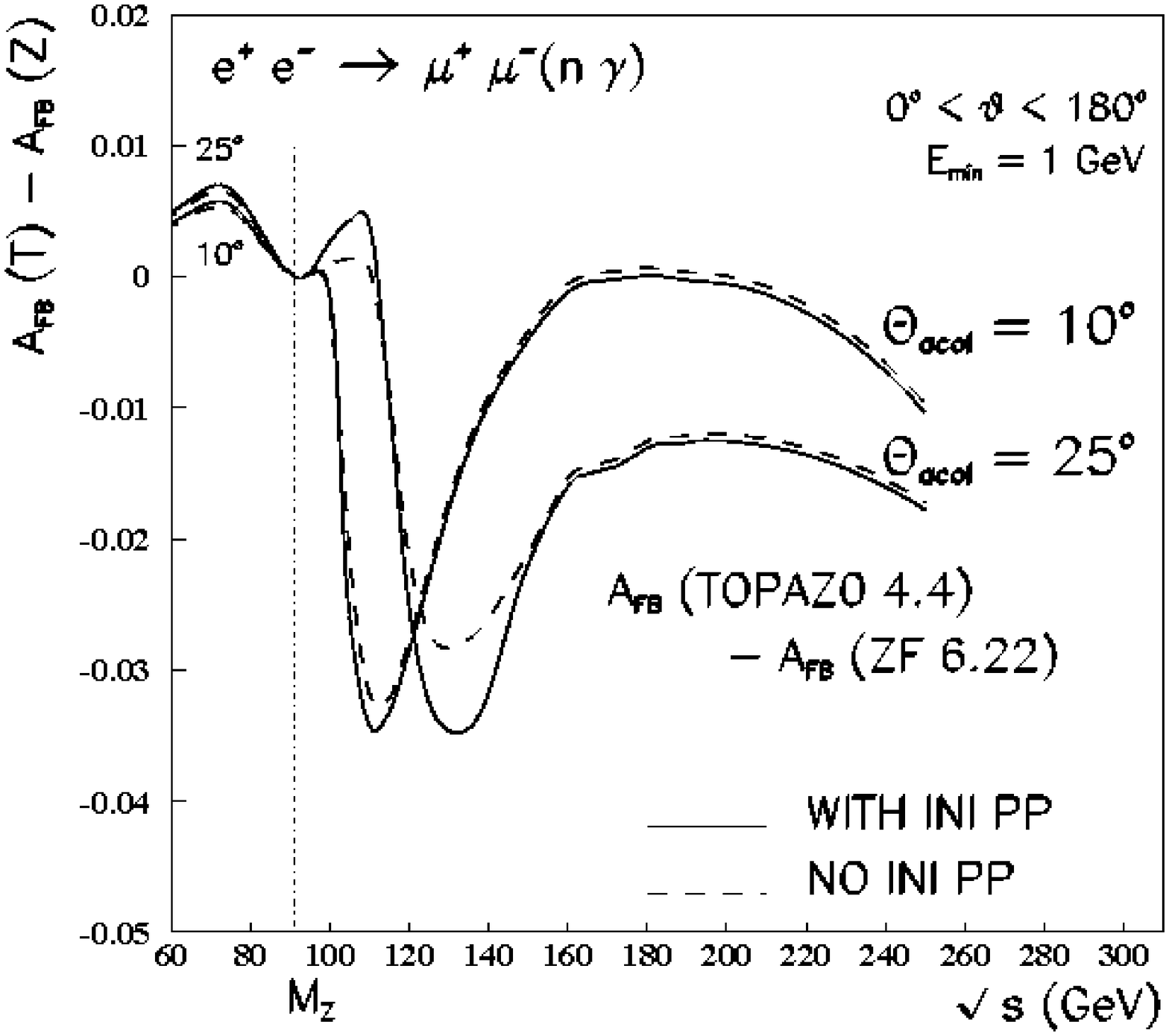,width=7.5cm}}
\\
\end{tabular}
\caption[Comparison of $A_{FB}$ for {\tt TOPAZ0} and {\tt ZFITTER} 
above the $Z$ peak]
{\sf
Comparison of predictions from  {\tt ZFITTER} v.6.22 \cite{Bardin:1999yd-orig} 
and {\tt TOPAZ0} v.4.4 \cite{Montagna:1998kp} 
for differences of muon pair forward-backward
asymmetries with a. an $s'$-cut or b. an acollinearity cut \cite{Jack:1999af}
(Flag setting: {\tt ISPP}=0,1, {\tt FINR}=0; further: 
{\tt SIPP}={\tt S\_PR} \cite{Bardin:1999yd-orig}). 
\label{afb-top-zf}
}
\end{flushleft}
\end{figure}
%xxxxxxxxxxxxxxxxxxxxxxxxxxxxxxxxxxxxxxxxxxxxxxxxxxxxxxxxxxxxxxxxx 

On the other hand, a cross check of the {\tt ZFITTER} and {\tt TOPAZ0} 
programs applying $s'$-cuts comparable to the acollinearity cuts
show a very high level of agreement between the two, at 
LEP~1 ($< O(3\cdot 10^{-4})$), but also at LEP~2 energies 
at the order of less than a per mil~\footnote{For this, a sufficiently large invariant mass cut 
preventing the radiative return to the $Z$ boson resonance 
is applied. For LEP~2 energies flag {\tt FINR} was set to 0 
for the final state corrections which is the recommended choice.}
and is under control with respect to the experimentally 
demanded accuracy \cite{Bardin:1999gt}. 
Initial state pair production and exponentiation of higher orders 
do not spoil this high level of agreement for the $s'$-cut.
For the acollinearity cut branch, 
the deviation of the codes increases to few per cent, 
even with stringent hard photon cuts \cite{Jack:1999xc,Jack:1999af,Christova:2000zu}.
This may be seen in Fig.~\ref{xs-top-zf} and \ref{afb-top-zf}. 

Our $s'$-cut dependent ratios deviate from unity mostly in the regions 
where the radiative return is not prevented.
The same is true for the ratios with acollinearity cut; since this cut is
not as effective in preventing the radiative return as the $s'$-cut,
the deviations survive at higher energies to some extent.
This fact and the higher order corrections, which remained untouched
by our study, seem to be the main sources of the remaining deviations
between the different programs.

Preliminary studies show that a correct description of hard  
two-loop QED corrections, especially for the acollinearity cut 
option in the {\tt ZFITTER} code, together with a correct
resummation of the soft and virtual initial-final state 
interference contribution, not contained in the {\tt ZFITTER} 
code so far, seem to play a key role here.
Since the acollinearity cut is not as effective in preventing the 
radiative return to the $Z$ boson as the $s'$-cut, these deviations also  
survive more profoundly for the acollinearity cut than for the $s'$-cut.
%
%======================================================================= 
\section{Conclusions
\label{sec_lep2_conc}
}
%======================================================================
%
Analytical formulae were derived for the photonic corrections with
acollinearity cut and substantial deviations were obtained 
from the coding in {\tt ZFITTER} until version 5. 
The essentials of the changes have been described 
and numerical comparisons were performed in great detail. 

At LEP~1 energies and for the $s'$-cut branch we had shown
in Chapter \ref{ch_lep1slc} that the situation of the {\tt ZFITTER} 
code up to versions v.5.x (1998) in comparison with the code 
{\tt TOPAZ0} can be stated as quite satisfactory. 
At $M_Z\pm 3\,\mbox{GeV}$ the agreement is better than $10^{-4}$
\cite{Bardin:1995aa,Bardin:1999gt}. 
And also at LEP~2 energies and higher we can meet the demands 
by experiment with a deviation of the codes of not more than 1 
or 2 per mil for different cut values and with a substantial 
decrease of this difference below 1 per mil in case of
a sufficiently large $s'$-cut.
This situation does not change when an extra cut on the maximal 
scattering angle $\cos\vartheta$ is applied 
\cite{Christova:1999gh,Jack:1999xc,Jack:1999af}. 

If one introduces an acollinearity cut instead of the 
$s'$-cut, a comparable agreement was obtained around the $Z$ resonance 
as long as only initial state bremsstrahlung was considered. 
But as soon as we include the initial-final state corrections 
this agreement deteriorates to $O(3\times 10^{-3})$ which grows to a 
large discrepancy between the two codes of several 
per cent at larger energies ($\sqrt{s}\approx 100\ldots
200\,\mbox{GeV}$) \cite{Christova:1998tc}.
An earlier comparison of the {\tt ZFITTER} code with the {\tt ALIBABA} code 
for the s-channel part of the Bhabha scattering 
branch had already shown similar deviations \cite{Riemann:1992up,Christova:1999gh}. 
At higher energies, $\sqrt{s}> 200\,\mbox{GeV}$, the agreement 
with $s'$-cut is better than per mil, but still only 1 to $2\%$
for the acollinearity cut. Especially in the intermediate energy range 
where the $Z$ radiative return events are not prevented this discrepancy  
peaks and amounts up to several per cent. A similar effect is also
visible for the $s'$-cut, although to a much lesser extent (below $1\%$).

We can therefore conclude that in the case of a cut on the 
acollinearity angle the correct treatment of the higher
order hard photonic corrections 
is crucial in order to obtain better than per 
cent predictions. This is especially important at 
energies above the $Z$ peak were the $Z^0$ radiative 
return effect is just approximately prevented by the 
applied cut. Due to the specific phase space for the 
acollinearity cut (see Fig.~\ref{dalitz}), a complete  
removal of the $Z^0$ radiative return via hard photon 
emission is not possible, no matter how strong the applied 
cut. These effects therefore also survive stronger 
at the 1 to 2 per cent level at LEP~2 energies 
$\sqrt{s}\approx 200\,\mbox{GeV}$ than for an $s'$-cut
where the deviations are only few per mil.

In the {\tt ZFITTER} code an approximation is implemented 
for the inclusion of hard two-loop corrections from 
initial state bremsstrahlung \cite{Berends:1988ab} with 
acollinearity cut, including an exponentiation of soft and 
virtual photonic corrections. For this, the acollinearity cut 
is simulated by an effective $s'$-cut which seems to be too 
crude at these energies. A delicate cancellation of the 
infrared divergences is at play here demanding a precise 
matching of the soft, virtual, and hard higher order corrections.
The approximation is especially problematic at energies 
where the hard photonic corrections are strongly reduced. 
This has to be done as an analytical calculation for the
hard two-loop radiator functions with acollinearity cut 
is not available. Cuts on the minimal and maximal scattering 
(acceptance) angle of 
$\vartheta_{min}<\vartheta<180^{\circ}-\vartheta_{min}$ with 
$\vartheta_{min} = 20^{\circ}$ and $40^{\circ}$ had no effect
on the outcome of these comparisons.

Fortunately, we may conclude that the numerical changes 
at the interesting LEP~2 energy range above roughly $160\,\mbox{GeV}$ 
are not as big as one could expect. 
Applying an extra $s'$-cut on top of the acollinearity 
cut for the intermediate energy region, could possibly 
reduce the large discrepancies between the codes. This 
would perhaps be a possibility for experimental fitting 
procedures in order to assure the necessary level of agreement 
between the different codes. If one is interested in performing 
investigations in this kinematical region, further studies with 
the codes are necessary.

%##########################################################################
\chapter{The $e^{+}e^{-}$ Linear Collider and Fermion Pairs
\label{ch_linac}
}
%##########################################################################

With the experience of LEP in mind, there appear two 
equally fascinating opportunities for studying fermion pair 
production processes at a future $e^+e^-$ Linear Collider (LC). 
One option, the Giga-Z option, would be to run 
with high luminosity on the $Z$ boson resonance.
This may be performed in a quick and feasible few months run 
in order to pin down the symmetry breaking 
mechanism of the electroweak sector by indirectly 
determining the masses of a light {\tt SM} or {\tt MSSM} 
Higgs boson or looking for effects of supersymmetric particles 
from virtual corrections \cite{Moenig:1999aa,Heinemeyer:1999aa}. 
The main motivation to build such a machine of course is to look
for such particles or other `New Physics' in direct 
particle production at energies typically reaching the TeV scale
like the {\it Tesla project} \cite{Brinkmann:1997nb}.

These two scenarios for the LC shall be sketched at the beginning
of this Chapter. We then want to put particular emphasis 
on what this means for the QED description to fermion pair
production and on what is provided in this respect by 
the semi-analytical program {\tt ZFITTER} \cite{Bardin:1992jc2,Bardin:1999yd-orig}  
in comparison with the numerical programs {\tt TOPAZ0} 
\cite{Montagna:1995b,Montagna:1998kp} and {\tt KK2f} 
\cite{Jadach:1994yv,Jadach:1999tr,Jadach:1999kkkz}.
%
%==========================================================================
\section{Searches for {\it Physics beyond the Standard Model}
\label{sec_lc_searches}
}
%--------------------------------------------------------------------------
%
%==========================================================================
\subsection*{High precision measurements to the {\tt SM} and {\tt MSSM}
\label{sub_lc_sm_mssm}
}
%--------------------------------------------------------------------------
%
Starting with the Giga-Z option,
it was demonstrated in \cite{Moenig:1999aa} that with a factor of 100 or so 
higher statistics than at LEP running on the $Z$ boson resonance,\footnote{
In comparison to SLD at SLAC, it would be even a factor
of roughly 2000.
}
which corresponds to a luminosity of 
${\cal L}\sim 5\cdot 10^{33}cm^{-2}s^{-1}$ or roughly $10^9$ 
hadronic $Z$ boson decays after just a few months of running,
especially fermion pair production asymmetries like $A_{LR}$ 
or the polarized $b\bar{b}$ forward-backward asymmetry 
$A^b_{FB}$ could be measured with very high precision
when using one or both beams polarized. This latter condition 
together with good $b$-tagging techniques and 
the collected experiences at LEP and SLD 
should help to keep the systematic errors 
under control.
The implication of this from the theoretical side 
on extracted {\tt SM} parameters, like e.g.~on the $W$ boson
mass $M_W$ or the effective weak mixing angle 
$\sin^2\theta_{eff}$ was illustrated in \cite{Heinemeyer:1999aa}:
With expected experimental accuracies at the Giga-Z 
one obtains $\Delta M_W = 6\,\mbox{MeV}$ 
or $\Delta\sin^2\theta_{eff} = 4\times 10^{-5}$. 
This is to be compared with the presently achievable 
total experimental errors by the end of LEP 
of $\Delta M_W = 40\,\mbox{MeV}$ or 
$\Delta\sin^2\theta_{eff} = 1.8\times 10^{-4}$ \cite{Moenig:1999aa}.
Due to loop corrections $M_W$ and $\sin^2\theta_{eff}$  
are sensitive to the mass of a light Higgs boson $M_H$, the 
top quark mass $m_t$, and in the supersymmetric case, 
also on the mass scale $M_{susy}$. These much 
improved experimental values 
for $M_W$ and $\sin^2\theta_{eff}$ thus allow at the Giga-Z, 
together with the precise knowledge of $m_t$, an indirect determination
of the mass of a light Higgs boson in the {\tt SM} at the 10\% 
level. Moreover, strong consistency checks can be performed 
on the {\tt SM/MSSM} values of $M_W$ and $\sin^2\theta_{eff}$ 
with $m_t$ and supersymmetric masses as input parameters 
\cite{Heinemeyer:1999aa}.
%\cite{Heinemeyer:1999aa,Djouadi:2000gu}.

%==========================================================================
\subsection*{Virtual corrections and New Physics Phenomena
\label{sub_lc_newphysics}
}
%--------------------------------------------------------------------------
%
Probably one of the most fascinating applications of fermion pair production 
processes at higher energies is then the search for 
`New Physics Phenomena' (NPP), i.e.~new effects which would be 
observed, but not described by the {\tt SM} \cite{Settles:1997wj,Brinkmann:1997nb}.
This is of course quite actively pursued already at existing  
$e^+e^-$ high energy facilities, giving quite stringent bounds 
on masses and couplings of 
exchanged `exotic' particles or minimal interaction scales of NPP.
With a future LC, however, reaching much higher energies close to the
$\mbox{TeV}$ scale and using high luminosities,  
there is the justified hope of really uncovering this `beyond the {\tt SM}' 
domain of particle physics. Examples of such investigations  
are e.g.~setting lower limits on four-fermion contact interaction scales 
or on masses and couplings of extra heavy neutral or 
charged gauge bosons, $Z'$ and $W'$, 
\cite{Leike:1992uf,Djouadi:1992sx,Erler:1999nx,Czakon:1999ha},
of {\tt susy} particles in ${\cal R}$ parity violating 
supersymmetric models, or for interaction-unifying models ({\tt GUTs}). 
Also searches for excited leptons, leptoquarks, preons, or heavy
fermions from Technicolor models could be conducted or
\cite{Heyssler:1999my}, 
one could look for effects in angular cross section distributions
from spin-2 boson exchanges predicted in string-inspired,
low-scale quantum gravity models
\cite{Hewett:1998sn}. 

\vfill\eject
%-------------------------------------------------------------------
LEP~1 and LEP~2, for example, already have some potential for the 
observation of new virtual effects in the $2f$ final state
\cite{Acciarri:1999aa,Abbiendi:1999wk,Abbiendi:1999wm}: 
\begin{itemize}
\item
Heavy neutral $Z'$ bosons may be searched for in two
respects:
At the $Z$ peak, 
limits on a $ZZ'$ mixing angle may be derived, typically
$|\theta_{M}| <  0.003$ \cite{Leike:1992uf,Djouadi:1992sx}. 
While, at LEP~2 limits on the mass 
of a $Z'$ boson are obtained in the range
$M_{Z'}> 250 - 725$ GeV depending on the models studied 
\cite{Acciarri:1999aa}.
\item
A limit on the energy scale $\Lambda$ at which contact interactions
could appear is $\Lambda > 4 - 10$ TeV.
Typical limits from atomic parity violation searches are $\Lambda
> 15$ TeV. They are not sensitive to the $\cal P$ conserving 
$VV, AA, LL+RR, LR+RL$ type models \cite{Abbiendi:1999wm}.
\item
Leptoquarks and also sneutrinos and squarks from
supersymmetric theories with $\cal R$-parity breaking 
may be exchanged in addition to $\gamma$
and $Z$. The leptoquark mass limits, $m_{LQ} > 120 - 430$ GeV,
are for some models competitive with direct searches 
\cite{Abbiendi:1999wk}. 
\item
Extra dimensions naturally arising in quantum gravity 
models could be probed through the relation
$M_{Pl} = {\cal R}^n M_S^{n+2}$ which relates
the Planck mass scale $M_{Pl}\approx 10^{19}\,\mbox{GeV}$
with an effective gravity scale $M_S$ in usual 
4-dimensional space-time assuming a maximal spatial 
extension $R$ of the extra dimensions \cite{Hewett:1998sn,Abbiendi:1999wm}.
This delivers lower bounds on the energy scale 
at which gravity effects could appear of
$M_S > 0.7 - 1$ TeV. 
\end{itemize}
With a LC, the so far
checked energy region for NPP from LEP or SLC can be
extended from typically $O(\mbox{few TeV})$ up to several  
tenths of $\mbox{TeV}$ at a LC. A complete presentation of these 
activities is given in \cite{Settles:1997wj}.

Recently collected evidence of neutrino-oscillations 
at the Super Kamiokande experiment
\cite{Fukuda:1998mi} makes another application in the 
context of the Giga-Z option quite interesting:
looking for lepton flavor
number violating $Z$ decays like $Z\to\mu\tau$, $e\tau$, or $e\mu$ when
heavy neutrinos are exchanged in virtual corrections (Dirac or Majorana type).
The estimated branching ratios in the case of $Z\to e\tau$ or $\mu\tau$
could be large enough in some models to be observable at the Giga-Z. 
A first calculation had been done in \cite{Mann:1984dv} and
studies for the LC were presented in \cite{Illana:1999aa}.

%==========================================================================
\section{Higher order QED corrections
%Higher order QED corrections with cuts and different codes 
\label{sec_lc_qed}   
}
%--------------------------------------------------------------------------
%
In order to successfully look at the Giga-Z for such small corrections by NPP,
the large QED corrections have to be subtracted reliably from data
with an exact treatment of the other radiative corrections. 
This means that the {\tt SM} corrections have to be known 
precisely at least at the level of the New Physics effects. The per mil 
precision obtained at LEP~1 and SLC naturally sets a lower benchmark for 
the expected accuracies at the Giga-Z. For the {\tt ZFITTER} program and 
other two-fermion codes it was shown in Chapter \ref{ch_lep1slc} that  
a relative precision for cross section observables of the order $10^{-4}$
can already be guaranteed on the $Z$ peak. This could especially be proven
for the new QED bremsstrahlung calculation with general cuts on 
final state angles and energies, shown there.

At energies approaching the TeV scale, the QED interference between initial 
and final state radiation starts to become as equally important as the
initial state contribution and also higher order corrections grow
in importance. Furthermore, the removal of $Z$ radiative return
events from the experimental data through kinematical cuts to the 
hard photon phase space is important in order to remove most
of the, for particle searches uninteresting {\tt SM} background.

\begin{figure}[htb] 
\begin{flushleft}
%--- 
\begin{tabular}{ll}
\hspace*{0.5cm} 
  \mbox{%
  \epsfig{file=%
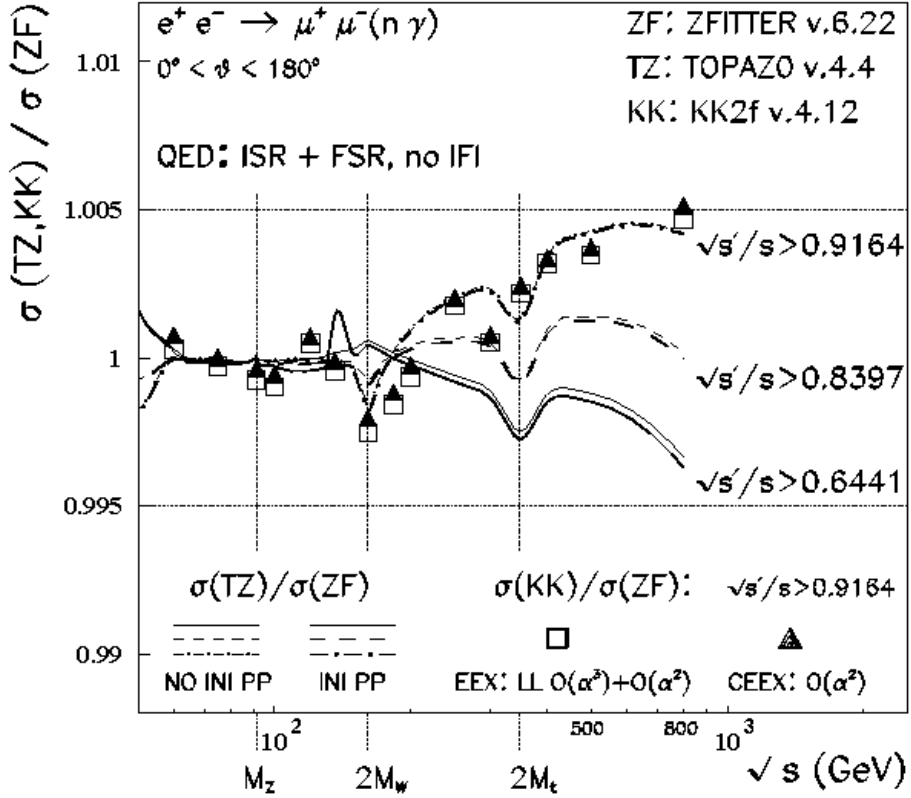,width=13.cm   % this is the width of the figure (optional)
         }}%
\end{tabular}
\vspace*{-1cm}
\caption[Comparison of {\tt TOPAZ0}, {\tt ZFITTER}, and {\tt KK2f}
at the Linear Collider without initial-final state interference]
{\sf
Cross section ratios for muon pair production with 
different $s'$-cuts for codes 
{\tt ZFITTER} v.6.22 \cite{Bardin:1999yd-orig}, 
{\tt TOPAZ0} v.4.4 \cite{Montagna:1998kp}, 
{\tt KK2f} v.4.12 \cite{Jadach:1999kkkz} (1999) 
from 60 to 800 GeV c.m.~energy;
without initial-final state interference \cite{Christova:2000zu}
({\tt INI PP}: initial state pair production; {\tt LL}: leading logarithmic terms). 
\label{top_zf_kk_codes_a}}
\end{flushleft}
\end{figure}

\begin{figure}[htb] 
\begin{flushleft}
%--- 
\begin{tabular}{ll}
\hspace*{0.5cm} 
  \mbox{%
  \epsfig{file=%
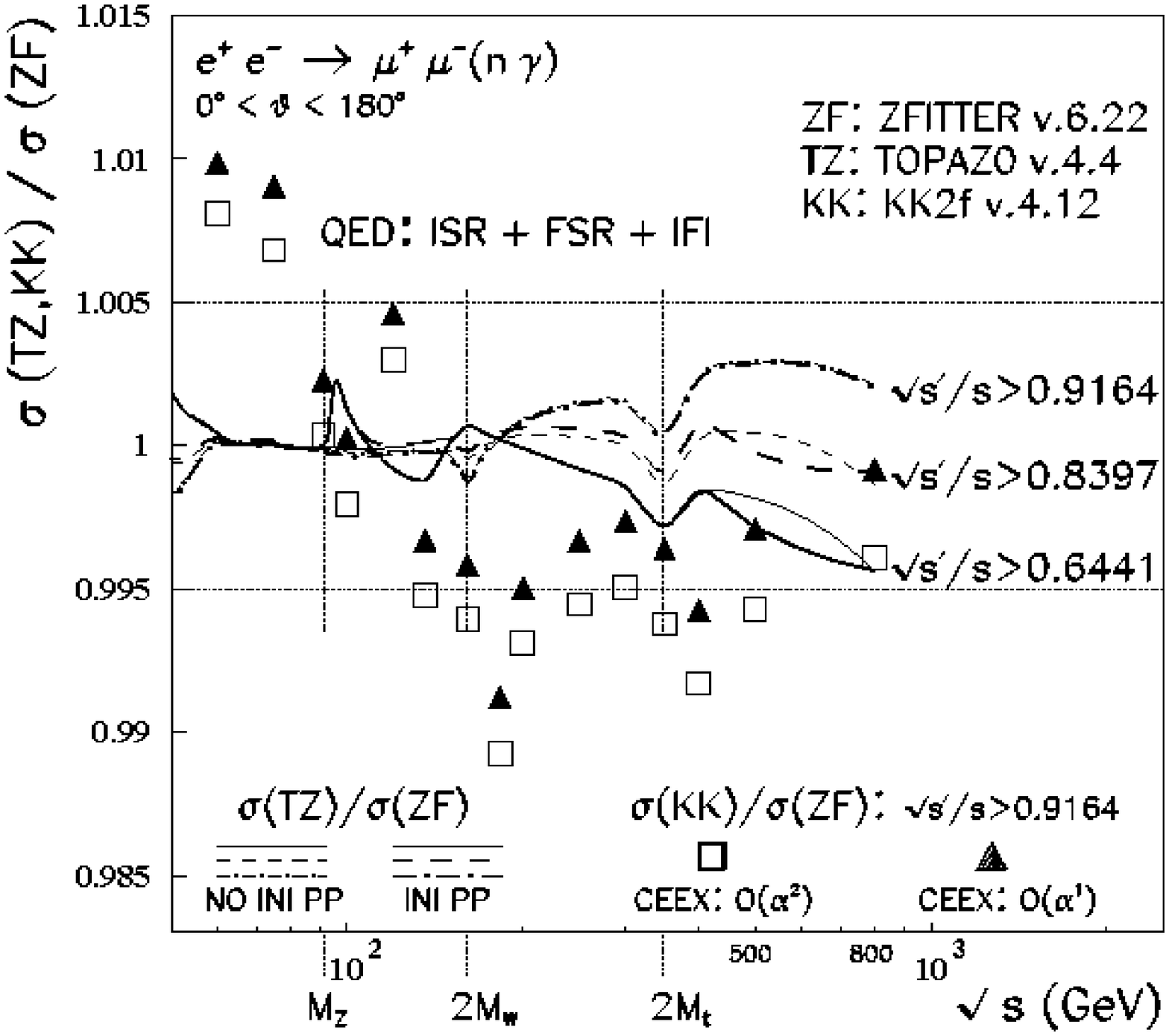,width=13.cm   % this is the width of the figure (optional)
         }}%
\end{tabular}
\vspace*{-1cm}
\caption[Comparison of {\tt TOPAZ0}, {\tt ZFITTER}, and {\tt KK2f}
at the Linear Collider with initial-final state interference]
{\sf
Cross section ratios for muon pair production with 
different $s'$-cuts for codes 
{\tt ZFITTER} v.6.22 \cite{Bardin:1999yd-orig}, 
{\tt TOPAZ0} v.4.4 \cite{Montagna:1998kp}, 
{\tt KK2f} v.4.12 \cite{Jadach:1999kkkz} (1999) 
from 60 to 800 GeV c.m.~energy;
with initial-final state interference \cite{Christova:2000zu}
({\tt INI PP}: initial state pair production; 
{\tt LL}: leading logarithmic terms). 
\label{top_zf_kk_codes_b}}
\end{flushleft}
\end{figure}

Therefore, the comparison of total cross section predictions by 
codes {\tt ZFITTER} v.6.22, {\tt TOPAZ0} v.4.4, and {\tt KK2f} v.4.12 
\cite{Jadach:1999kkkz} was now extended up to typical LC energies 
of 500 to 800 GeV for the invariant mass cut option. This is shown in 
Fig.~\ref{top_zf_kk_codes_a} and Fig.~\ref{top_zf_kk_codes_b}.
%
%xxxxxxxxxxxxxxxxxxxxxxxxxxxxxxxxxxxxxxxxxxxxxxxxxxxxxxxxxxxxxxxxx 
%xxxxxxxxxxxxxxxxxxxxxxxxxxxxxxxxxxxxxxxxxxxxxxxxxxxxxxxxxxxxxxxxx 
%
%xxxxxxxxxxxxxxxxxxxxxxxxxxxxxxxxxxxxxxxxxxxxxxxxxxxxxxxxxxxxxxxxx 
%xxxxxxxxxxxxxxxxxxxxxxxxxxxxxxxxxxxxxxxxxxxxxxxxxxxxxxxxxxxxxxxxx 
%
The general result of this analysis is that the deviation of the cross 
section predictions by the three codes is not more than 5 per mil 
for the complete energy range for the {\tt TOPAZ0}--{\tt ZFITTER} 
comparison. This observation also holds for the case of 
initial state QED bremsstrahlung (ISR) alone 
when comparing code {\tt KK2f} with {\tt ZFITTER}, 
applying sufficiently strong invariant mass cuts and taking 
into account different higher order corrections 
(Fig.~\ref{top_zf_kk_codes_a}). Including QED initial-final 
state interference (IFI), the comparisons with {\tt KK2f} 
delivered a maximal deviation of roughly 1 \%
(Fig.~\ref{top_zf_kk_codes_b}).

In detail, this meant:
The numerical precision of {\tt TOPAZ0} and {\tt ZFITTER}
was better than $10^{-5}$ everywhere, while the accuracy  
of the Monte Carlo (MC) event generator {\tt KK2f} was necessarily
restricted due to limited CPU time: Calculating ISR with an accuracy of 
at least $10^{-3}$ required samples of 100000 events for each energy point. 
When including the resummed IFI, smaller samples of 30000   
events had to be used, resulting in a lower precision of roughly $2\times 10^{-3}$. 
For ISR only, the typical CPU time per MC data point 
e.g.~on an HP-UX 9000 workstation was about 25 minutes, increasing to 
roughly 100 minutes if IFI is added for the event samples stated 
above. In comparison, {\tt TOPAZ0} calculated one cross section value 
in a few minutes, while {\tt ZFITTER} with its semi-analytical approach 
calculated all 32 cross section values for one cut in a few seconds.
On the other hand, when interested in more complex setups, 
i.e.~calculating multi-differential observables, using a wider variety of
cuts, or including extra higher order effects to the 
initial-final state interference, which {\tt ZFITTER} cannot or only 
partly provide, the numerical programs {\tt TOPAZ0}, or respectively 
{\tt KK2f}, clearly have their advantages.

The effect of ISR was compared alone (Fig.~\ref{top_zf_kk_codes_a}) or of 
ISR together with IFI (Fig.~\ref{top_zf_kk_codes_b}) for three different 
cut values: $\sqrt{s'/s} > 0.6441$, $0.8397$, and $0.9164$,
in the case of the {\tt TOPAZ0}--{\tt ZFITTER} comparison,  
and $\sqrt{s'/s} > 0.9164$ when comparing with {\tt KK2f}.\footnote{The cut values correspond approximately to a relatively strong 
cut on the maximal final state leptons' acollinearity angle of  
$25^{\circ}$, $10^{\circ}$, and $5^{\circ}$ respectively. 
}
The value $s'$ is defined here as the invariant mass squared of the $\gamma$ 
or $Z$ propagator after ISR, which is equal to the final state invariant 
mass squared including the emitted final state photons. For this, 
final state radiation (FSR) was treated in form of a global correction 
factor. Alternatively, cutting on the minimal final state invariant mass 
squared $M^2_{f\bar{f}}$ after FSR, though it slightly worsened 
the good agreement at LEP~1 energies \cite{Bardin:1999gt},
it did not change the overall agreement in 
Fig.~\ref{top_zf_kk_codes_a} and Fig.~\ref{top_zf_kk_codes_b} 
substantially. A recent discussion on this issue, defining 
kinematical cuts with radiative corrections  
for the experimental and computational situation
-- e.g.~when including mixed QED and QCD contributions from
photonic and gluonic emission in the case of hadronic final states --
was given in \cite{Bardin:1999gt}. 
In particular, at LEP~2 energies the predictions of the codes 
lie well inside the estimated experimental accuracies of  
e.g.~$\Delta\sigma_{\mu\mu}\approx 1.2\%$, 
$\Delta\sigma_{had}\approx 0.5\%$ for sufficiently 
strong cuts \cite{Bardin:1999gt}. 

Except where otherwise stated, the default settings of the 
programs were used, thus taking into account the $O(\alpha^2)$ 
photonic initial state corrections \cite{Berends:1988ab} with the 
leading logarithmic $O(\alpha^3)$ \cite{Montagna:1997jv}
corrections together with the exactly added $O(\alpha)$ 
IFI and QED box contribution 
\cite{Montagna:1998kp,Jadach:1999kkkz,Bardin:1999yd-orig}.
In {\tt ZFITTER} and {\tt TOPAZ0}, the two-loop corrections 
are complete. All three programs have installed higher order 
corrections to ISR where the finite soft and virtual 
photonic corrections are resummed.\footnote{In {\tt KK2f}, 
the Yennie-Frautschi-Suura (EEX) prescription was used
\cite{Jadach:1999kkkz}.
}
In contrast to codes {\tt ZFITTER} and {\tt TOPAZ0},
{\tt KK2f} also possesses a procedure to exponentiate 
IFI corrections with its newly implemented
{\it coherent exclusive exponentiation (CEEX)} 
\cite{Jadach:1998jb,Jadach:1999gz,Jadach:1999kkkz}.\footnote{
The initial-final state interference contribution in {\tt KK2f} 
is only available with the {\it CEEX} option; for {\it EEX}
it is neglected.
}
To be more precise, {\it CEEX}  
does not include $O(\alpha^3)$ contributions to initial state 
bremsstrahlung up to now, 
while the {\it EEX} option in {\tt KK2f} does not contain 
the second order, subleading $O(\alpha^2 L)$ corrections, 
so both options are complementary to each other when interested in 
estimating these higher order effects for the initial state
bremsstrahlung with {\tt KK2f} ($L=\ln(s/m_e^2)$). 
In both cases {\tt KK2f} lacks the $O(\alpha^2 L^0)$ terms which, 
however, are estimated to be of the order $10^{-5}$ and so do not 
play a visible role in this comparison \cite{Jadach:1999kkkz}.

In Fig.~\ref{top_zf_kk_codes_a}, the predictions by  
{\tt KK2f} were compared with those by {\tt ZFITTER} for ISR alone, first for 
the {\it EEX} option with the leading logarithmic (LL)  
$O(\alpha^3 L^3)$ and $O(\alpha^2 L^2)$
corrections ({\tt ZFITTER} flag values: {\tt FOT2} = 3, 5) 
\cite{Bardin:1999yd-orig}. Then we used the {\it CEEX} option 
for {\tt KK2f} and compared the $O(\alpha^2)$ results 
({\tt ZFITTER}: {\tt FOT2} = 2). The cross section ratios and the 
maximally 5 per mil deviation of the codes did not change considerably.
The values were calculated with a numerical uncertainty 
of $0.4\times 10^{-3}$ at the $Z$ peak, and $1\times 10^{-3}$ overall.

In Fig.~\ref{top_zf_kk_codes_b} the cross section 
ratios, now with the IFI contribution, are compared for {\it CEEX}
$O(\alpha^1)$ and {\it CEEX} $O(\alpha^2)$. There is roughly 
a 2 per mil shift of the central values -- always having in mind 
calculational uncertainties of $2\times 10^{-3}$ -- 
when going to the $O(\alpha^2)$ calculation, but they stay inside 
the overall, $\pm 1 \%$ margin.

Another, in {\tt ZFITTER} recently updated contribution 
are initial state pair corrections \cite{Bardin:1999yd-orig,Arbuzov:1999uq}. 
These originate from bremsstrahlung photons dissociating into light fermion pairs: 
{\tt TOPAZ0} and {\tt ZFITTER} versions contain the $O(\alpha^2)$ 
leptonic and hadronic initial state pairs and a realization for simultaneous 
exponentiation of the photonic and pair radiators 
\cite{Berends:1988ab,Kniehl:1988id,Jadach:1992aa}. 
According to \cite{Jadach:1999kkkz}, initial state pair corrections 
are not included in the {\tt KK2f} code. Since {\tt ZFITTER} v.6.20 
\cite{Bardin:1999yd-orig}, also the leading and subleading terms 
of $O(\alpha^3)$ and the LL $O(alpha^4)$ corrections 
from initial state pair emission
may be included through convolution of the 
photonic and pair flux functions \cite{Arbuzov:1999uq}. The effect of the
pair corrections is e.g.~with strong cuts roughly 2.5 per mil at the $Z$
peak, slightly decreases to approximately 2 per mil at LEP~2 energies,
and is not more than roughly 1 per mil at 500 to 800 GeV c.m.~energy. 
In Fig.~\ref{top_zf_kk_codes_a} and Fig.~\ref{top_zf_kk_codes_b}, switching on 
the pair corrections for different cuts, does not change the level of 
agreement between {\tt ZFITTER} v.6.22 and {\tt TOPAZ0} v.4.4
substantially. One interesting feature 
at lower energies between roughly 100 and 150 GeV -- just where the $Z$
radiative return is not prevented anymore by the applied cuts -- is 
the fact that the several per mil deviation of the two codes there
disappears when the pair corrections are switched off. {From} 
Fig.~\ref{top_zf_kk_codes_a} and Fig.~\ref{top_zf_kk_codes_b} it can 
also be seen that such deviations can also be 
prevented by a sufficiently large $s'$-cut of e.g.~$s'/s > 0.9$ 
if initial state pair corrections shall be included.

The inclusion of 4-fermion final states in this context, 
e.g.~from final state pair creation, with their rather large, per cent 
level corrections at LEP~2 and higher energies \cite{Passarino:1999kv}
is another task which has to be pursued 
for an update of the codes for the LC.
Especially, the definition of background and signal diagrams in the 
hadronic case together with kinematical cuts will be one of the major 
obstacles to overcome for experiment and theory \cite{Passarino:1999kv}.

%==========================================================================
\section{Conclusions
\label{sec_lc_conc}
}
%--------------------------------------------------------------------------

A possible quick usage of a future $e^+e^-$ Linear Collider (LC) 
as a high luminosity $Z$ factory in a Giga-Z mode leads to strong demands 
on theoretical cross section predictions by two-fermion codes 
like {\tt ZFITTER}. The analysis done in Chapter \ref{ch_lep1slc}
for LEP~1 and SLC precisions can of course also be applied here.
Cross section results are now calculated by {\tt ZFITTER} with 
$10^{-4}$ precision on the $Z$ resonance itself which is a good
start for the estimated high precisions at a Giga-Z with a factor
of $10^2$ or $10^3$ more $Z$ decays than at LEP. 

For typical LC energies, a first comparison of programs 
{\tt ZFITTER}, {\tt TOPAZ0}, and {\tt KK2f} for the standard
$s'$-cut option shows very nice agreement for the whole 
center-of-mass energy range from 
$\sqrt{s}\approx 60\,\mbox{GeV}$ to $800\,\mbox{GeV}$. 
The maximal deviation is $5$ per mil
for initial state radiation alone, which also holds for codes 
{\tt ZFITTER} and {\tt TOPAZ0} when switching on the QED interference.
The maximal deviation grows up to $1\%$ when comparing with {\tt KK2f},
but can be explained with a resumming of soft interference terms,
installed in {\tt KK2f} \cite{Jadach:1998jb,Jadach:1999kkkz}, 
but not yet for the other codes \cite{Christova:2000zu}.
This analysis included different initial state photonic corrections 
and corrections from initial state QED pair creation with different cuts. 
It can be concluded that the codes already fulfil the minimal precision 
requirements at the $O(1\%)$ or better for the higher energies at a $LC$.
Still a lot has to be done having in mind for example an update 
of codes for top pair production with final state masses and cuts
or for an experimentally realistic description of beamstrahlung 
effects with polarized beams, higher luminosities, and  
higher energies.

%
%-----------------------------------------------------------
\def\theequation{5.\arabic{equation}}
\def\thetable{5.\arabic{table}}
\def\thefigure{5.\arabic{figure}}
\setcounter{equation}{0}
\setcounter{table}{0}
\setcounter{figure}{0}
%-----------------------------------------------------------
%

%==========================================================================
\chapter*{Summary 
\label{ch_sum}
}
%==========================================================================
\addcontentsline{toc}{chapter}{Summary}
\pagestyle{myheadings}
\markright{\it SUMMARY}
%\pagenumbering{arabic}

%--------------------------------------------------------------------------
\section*{Results 
\label{sec_results}
}
%--------------------------------------------------------------------------
\addcontentsline{toc}{section}{Results}
\subsection*{QED flux functions with cuts on maximal 
acollinearity, minimal energies, and minimal and maximal acceptance
\label{sub_res_theo}
}
%--------------------------------------------------------------------------
%
Summarizing, new and more general formulae for flux functions for 
hard brems\-strah\-lung were derived for total and 
forward-backward cross sections $\sigma_T$, and
$\sigma_{FB}$ and the angular cross section 
distribution $d{\sigma}/d{\cos\vartheta}$, 
for $s$-channel fermion pair production, 
$e^+e^-\to \bar{f}f$, $f\neq e, \nu_e$. 
The semi-analytical results contain different 
kinematical cuts to the hard photon phase space
which had not or only incompletely been treated in the 
literature so far \cite{Bilenkii:1989zg}: 
\begin{itemize}
\item Cuts on the final state fermions' 
      maximal acollinearity angle $\theta_{\rm acol}^{\max}$
      and minimal energies $E_{\min}$:
      \ba
      \sigma_T &=&  \sigma_T(\theta_{\rm acol}^{\max},E_{\min}),
      \label{sum_xs}
      \\
      \sigma_{FB} &=& \sigma_{FB}(\theta_{\rm acol}^{\max},E_{\min}),
      \label{sum_xfb}
%      \\
%      \frac{d{\sigma}}{d{\cos\vartheta}}
%      &=& \frac{d{\sigma}}{d{\cos\vartheta}}
%      (\cos\vartheta;\theta_{\rm acol},E_{\min});
%      \label{sum_dsig}
%      \label{acol_sum}
      \ea

\item with an additional symmetrical cut $c$ on the cosine of the 
      scattering angle $\vartheta$ of one of the final state fermions:
      \ba
      \sigma_T(c,\theta_{\rm acol},E_{\min}) 
      &=& \int^{c}_{-c} \,{d{\cos\vartheta}}\,
      \frac{d{\sigma}}{d{\cos\vartheta}}(\cos\vartheta;
      \theta_{\rm acol}^{\max},E_{\min}),
      \label{sum_xsc}
      \\
      \sigma_{FB}(c,\theta_{\rm acol}^{\max},E_{\min}) 
      &=& \left(\int^{c}_{0}-\int^{0}_{-c}\right){d{\cos\vartheta}}\,
      \frac{d{\sigma}}{d{\cos\vartheta}}(\cos\vartheta;
      \theta_{\rm acol}^{\max},E_{\min}),
      \label{sum_xfbc}
      \nonumber\\
      \\
      \rightarrow A_{FB}(c,\theta_{\rm acol}^{\max},E_{\min}) &=&
      \frac{\sigma_{FB}}{\sigma_{T}}(c,\theta_{\rm acol}^{\max},E_{\min}). 
      \label{sum_afbc}
      \label{acc_sum}
      \ea
\end{itemize}

A set of simplified and handy
expressions for flux functions ({\it radiators}) 
was determined for $O(\alpha)$ hard QED corrections
to $s$-channel processes $e^+e^-\to \bar{f}f$
with the above listed general cuts. 
These are radiators $H^{a}_{A}$, $A = T,FB$, $a=ini,int,fin$, 
for the initial state, final state, 
and initial-final state interference hard photonic corrections.
They are regularized by cancelling all infrared 
singularities after adding the corresponding $O(\alpha)$ radiator 
functions from \cite{Bardin:1989cw,Bardin:1991de,Bardin:1991fu} 
for soft and virtual photon corrections
$S^{a}_{A}$, including corrections $B_{A}$ from 
$\gamma\gamma$ and $\gamma Z$ exchange box diagrams. 

%The corresponding electromagnetic one-loop soft 
%and virtual photonic corrections given in 
%\cite{Bardin:1989cw,Bardin:1991de,Bardin:1991fu}
%could be rederived. They consist of flux functions 
%$S^{a}_{A}$ and $B_{A}$ for the soft and virtual photon 
%contributions and the interference of the Born with 
%$\gamma\gamma$ and $\gamma Z$ exchange box diagrams.
%Adding them to the hard photon results $H^{a}_{A}$,
%any dependency on an arbitrarily chosen parameter 
%$\varepsilon$ separating soft and hard photon
%phase space can be removed while all infrared 
%divergences exactly cancel.
%
The complete, regularized radiator functions 
$R^{a}_{A}$, containing all soft, virtual, 
and hard photonic terms, can be integrated in an effective Born 
approximation together with improved Born observables over $s'$.
%
%This yields the cross sections $\sigma_{T,FB}$, 
%or differential cross section $d{\sigma}/d{\cos\vartheta}$
%in (\ref{sum_xs}) to (\ref{sum_dsig}) and
%in (\ref{sum_xsc}) to (\ref{sum_afbc}) respectively. 
%An additional cut on the minimal $s'$ value can be easily
%accounted for by adjusting the lower limit of the convolution
%integral. 
%
See Sections \ref{sec_lep1slc_radcuts} and 
\ref{sub_lep1slc_formacol} in Chapter \ref{ch_lep1slc} 
on these issues. For brevity, the radiators for 
$d{\sigma}/d{\cos\vartheta}$ were not presented here, but
are available and will be published later. 

The results were implemented in the semi-analytical program {\tt ZFITTER} 
for fermion pair production in $e^+e^-$ annihilation
from version 6.04/6.06 onwards \cite{zfitter:v6.0406,Bardin:1999yd-orig}.
The program with the new results is used by the experimental 
communities at the LEP and SLC experiments e.g. for an analysis
of the $Z$ line shape. The considered cuts are an 
experimentally requested option for leptonic final states 
as alternative to a kinematically simpler cut on $s'$
and on $\cos\vartheta$.

The semi-analytical calculation of the hard photon flux functions 
consisted in a two- or respectively three-fold analytical integration 
over three angles of phase space, while the convolution integrals 
over $s'$ are done numerically in the program {\tt ZFITTER}. 
%
%the azimuthal photon angle 
%$\phi_\gamma$, the Lorentz invariant $V_2 = 2 p p_2$,
%and the cosine of the scattering angle $\cos\vartheta$ 
%of the anti-fermion $\bar{f}$. 
%The momenta $p$ and $p_2$ denote the photon 
%and $\bar{f}$ momenta respectively. The invariant $V_2$
%is proportional to the polar photon angle $\cos\theta_\gamma$
%defined with $\phi_\gamma$ in the $(f\gamma)$ rest frame
%while $\cos\vartheta$ is given in the center-of-mass system.
%
The convolution integrals over $s'$ are done numerically. 
The phase space parameterization was done 
in accordance with \cite{Passarino:1982}.
There, the hard-photon corrections with acollinearity
and energies cut had been calculated completely numerically.
The phase space was shown 
for two variables of integration in Fig.~\ref{dalitz}. 
Details of the parameterization can be seen 
in Section \ref{sub_lep1slc_phasesp} and Appendix \ref{crossphase}.

It is is naturally split 
into three separate parts due to the cuts applied.
A nice observation is that all three regions of phase space
can be treated equivalently, using one general parameter 
$A=A_{I,II,III}(s'/s)$ as function of $R\equiv s'/s$, and for each 
region only depending on $m_f$ or on one of the two cuts applied.
This was first illustrated in \cite{Passarino:1982,Bilenkii:1989zg}.
The insertion of the specific value of $A(s'/s)$ for each region 
yields the corresponding expression of a radiator 
in each of the three phase space regions  
\cite{Christova:1999cc,Christova:1999gh,Jack:1999xc}.

In order to analytically integrate over the final 
two angles of phase space $\theta_\gamma$ and $\theta$, the 
fermion masses $m_e$ and $m_f$ had to be neglected,
which is safe, having applications at LEP or a LC in mind
and not considering top pair production.
The calculation demanded a separation of phase space 
into different regions providing different analytical 
expressions for each region. The necessity for this phase 
space splitting is due to artificial poles which arise
in the squared matrix element for collinear photon
emission from the initial state, i.e.~ in the 
case $\cos\theta_\gamma=\pm \cos\vartheta$ and when neglecting masses.
While this separation of phase space into different regions 
is necessary for the initial state and initial-final state 
interference terms, the critical electron and positron propagator
terms do not occur in the final state matrix element
and therefore no phase space splitting is necessary there. 

It can be shown that these unphysical poles cancel
when combining all terms, i.e.~the expressions from different 
phase space regions transform continuously 
into each other for $\cos\vartheta\to\pm\cos\theta_\gamma$.
Only logarithmic mass terms of the form $\ln(s/m_e^2)$
and $\ln[(1+c)/(1-c)]$ with $c$ as symmetric acceptance cut
remain, or $\ln(s'/m_f^2)$ respectively for the final state,
from collinear photon emission remain. 
This is discussed for the initial
state case in Section \ref{sub_lep1slc_mass_sing} 
and Appendix \ref{hardini}. The QED interference can be treated 
completely analogously.

Each flux function with general cuts consists of not more than
approximately 50 terms and can be straightforwardly implemented
into a semi-analytical Fortran program like {\tt ZFITTER} 
\cite{Bardin:1999yd-orig}.
Allowing full acceptance, i.e.~omitting the cut on the 
scattering angle $\cos\vartheta$ (c=1), strongly reduces the number 
of different expressions for the totally integrated radiators
-- only one expression remains for the $c$-even terms $H^{a}_{T}$, 
and only two are possible for the $c$-odd ones, $H^{a}_{FB}$.
This yields very compact formulae of less than 10 terms in each case.  
The short formulae were published in \cite{Christova:1999cc}, 
while a collection of all general radiator formulae are given  
in the Appendix \ref{int} and \ref{fin} or were presented for the 
initial state case in Chapter \ref{ch_lep1slc}.
The cross section distribution over $s'$, $d{\sigma^{fin}}/d{s'}$, 
can be analytically integrated further over $s'$ in order to
reproduce the results given in \cite{Montagna:1993mf},
where some misprints could be corrected there.

%--------------------------------------------------------------------------
\subsection*{Programming in {\it ZFITTER}
\label{sub_res_zf}
}
%--------------------------------------------------------------------------
%
The update to the {\tt ZFITTER} code from version 6.04/06 
\cite{zfitter:v6.0406,Bardin:1999yd-orig} onwards
consisted in a correction of the up-to-now mainly undocumented 
flux functions for hard brems\-strah\-lung to $e^+e^-\to \bar{f}f$
for the calculational branch with acollinearity cut. Results for the 
integrated initial state and initial-final state interference radiators 
had been given in \cite{Bilenkii:1989zg}, but with errors there. 

The new formulae implemented in the code now only contain
terms proportional to $P=1/(1-s'/s)$ which formally correspond to the 
infrared poles of the unregularized expressions for $s'\to s$. 
All unphysical powers $P^{n}$ for $n\geq 2$ could be removed
in these first order results. 
The formulae with all non-logarithmic fermion mass terms being 
neglected are still sufficiently compact for a convenient 
description in the Fortran code. The initial state and initial-final 
state interference contributions can be calculated in parallel 
in the different subroutines with a small distinction for
the soft and virtual photonic terms. For initial state
radiation the finite part is resummed in a soft-photon exponentiation,
while the QED interference radiators are added 
in first order approximation. The expressions in the code for
final state radiation could also be corrected.

The very short formulae on cross section and asymmetry flux 
functions without acceptance cut \cite{Christova:1999cc}
allow a numerically fast, additional branch of the code, 
when one is only interested in cuts on the maximal acollinearity 
angle and the minimal energies. 
All analytical formulae were numerically checked by 
comparing them with the squared matrix element which was 
numerically integrated over the angular phase space. 
Only the trivial integration over the photon 
angle $\varphi_\gamma$ was performed analytically.
We obtained complete agreement, solely restricted by the
numerical precision of the applied Simpson integration routine
and neglected mass terms in the analytical formulae. 
The Fortran package {\tt acol.f} was written containing
the new results and linked to {\tt ZFITTER}. This is
described in \cite{Bardin:1999yd-orig}.

%--------------------------------------------------------------------------
\subsection*{Numerical analysis of codes
\label{sub_res_num}
}
%--------------------------------------------------------------------------
%
The other major part of this thesis consisted of numerical
comparisons of the earlier {\tt ZFITTER} codes \cite{Bardin:1992jc2}
with the new versions from version 6.04/06 
\cite{zfitter:v6.0406,Bardin:1999yd-orig} onwards which were
updated with the new results.
This internal comparison was extended to other 
2-fermion programs with special focus on QED radiative corrections 
for center-of-mass energies of $\sqrt{s} = 30\ldots 800\,\mbox{GeV}$:
\begin{itemize}
\label{codes_sum}
\item[(1)] {\tt ALIBABA} versions 1 and 2 \cite{Beenakker:1997fi} 
      were compared with {\tt ZFITTER} v.4.5 to v.6.11 
      \cite{Bardin:1992jc2,Bardin:1999yd-orig}
      for the $s$-channel contribution to Bhabha scattering 
      $e^+e^-\to e^+e^-$ with cuts on 
      acollinearity, acceptance, and energies
      for $\sqrt{s} = 30\ldots 300\,\mbox{GeV}$
      \cite{Christova:1998tc,Christova:1999gh,Jack:1999xc,Jack:1999af,Christova:2000zu}.

\item[(2)] A comparison for {\tt TOPAZ0} versions 4.3 and 4.4 
      \cite{Montagna:1995b,Montagna:1998kp}  
      and {\tt ZFITTER} v.4.5 to v.6.11 was performed
      \cite{Bardin:1992jc2,Bardin:1999yd-orig}
      for $e^+e^-\to \mu^+\mu^-$ with cuts on 
      acollinearity, acceptance, and energies,
      or alternatively on $s'$ and acceptance 
      for $\sqrt{s} = 30\ldots 300\,\mbox{GeV}$
      \cite{Christova:1998tc,Christova:1999gh,Jack:1999xc,Jack:1999af,Christova:2000zu}.

\item[(3)] {\tt KK2f} version 4.12 \cite{Jadach:1999kkkz}
      was treated together with {\tt ZFITTER} v.6.22 \cite{Bardin:1999yd-orig}
      and {\tt TOPAZ0} v.4.4 \cite{Montagna:1998kp}  
      for muon pairs with cuts  
      on $s'$ and acceptance
      for $\sqrt{s} = 60\ldots 800\,\mbox{GeV}$
      \cite{Christova:2000zu}.
\end{itemize}

The main result of the numerical comparisons is 
that now a numerical agreement for codes {\tt ZFITTER} 
v.6.04/06 onwards and {\tt TOPAZ0} v.4.4 can be guaranteed 
of better than 1 per mil for the $Z$-boson resonance region: 
For $\sigma_T$, results agree now better than $0.3\times 10^{-3}$ 
for $\sqrt{s} = M_Z \pm 3\,\mbox{GeV}$, and better than $10^{-4}$ 
at the $Z$ peak itself. For $A_{FB}$ we have:
$\delta A_{FB}< 10^{-4}$ for $\sqrt{s} = M_Z$,
and $\delta A_{FB}< 10^{-3}$ for $\sqrt{s} = M_Z\pm 3\,\mbox{GeV}$.
The numerical accuracy of the codes 
now meets the precision already obtained earlier for the 
kinematically simpler $s'$-cut \cite{Bardin:1999gt}.
%
%This can be achieved for total cross sections 
%$\sigma_T$ and forward-backward asymmetries $A_{FB}$ 
%with strong cuts on the maximal acollinearity angle 
%$\theta_{acol}^{\max}$ and a cut on the minimal energy 
%$E_{\min}$ of the fermions. The muon pair production 
%channel was used as an example and typical experimental 
%cut values of $\theta_{acol}^{\max}=10^\circ$ and $25^\circ$
%and $E_{\min}=1\,\mbox{GeV}$ were applied. 
%
The results were presented in Fig.~\ref{top-zf-peak}
and Table \ref{tab-acol10-th40} in Chapter \ref{ch_lep1slc}. 
For the internal {\tt ZFITTER} comparisons 
please refer to Tables \ref{tab10sigifi}, 
\ref{tab10afbifi}, \ref{tab1025fin}, and \ref{tab10acolc}.
%The {\tt TOPAZ0} code \cite{Montagna:1998kp}
%determines observables with a numerical integration
%of the complete hard photon phase space
%while in {\tt ZFITTER} \cite{Bardin:1999yd-orig}
%a semi-analytical approach is used.
%
%The numerical accuracy of the codes for the acollinearity cut 
%option, illustrated in Chapter \ref{ch_lep1slc}, 
%now meets the precision already obtained earlier for the 
%kinematically simpler $s'$-cut \cite{Bardin:1999gt}:
%For $\sigma_T$, results agree now better than $0.3\times 10^{-3}$ 
%for $\sqrt{s} = M_Z \pm 3\,\mbox{GeV}$, and better than $10^{-4}$ 
%at the $Z$ peak itself. For $A_{FB}$ we have:
%$\delta A_{FB}< 10^{-4}$ for $\sqrt{s} = M_Z$,
%and $\delta A_{FB}< 10^{-3}$ for $\sqrt{s} = M_Z\pm 3\,\mbox{GeV}$.
%These results agree with other comparisons done recently
%for the $s'$-cut option, including different higher order 
%hard photonic corrections and QED initial state pair creation 
%\cite{Bardin:1999gt} and resummed soft and virtual photonic corrections
%\cite{Jadach:1999pp}.

At energies slightly above the $Z$ peak,
around $\sqrt{s}\approx 100 \ldots 130\,\mbox{GeV}$,
deviations were observed of several per cent between 
codes {\tt ZFITTER}, {\tt ALIBABA}, and {\tt TOPAZ0}. 
The energy interval describes the region where the $Z$ radiative 
return effect is still active for the applied cuts of 
$\theta_{acol}^{\max}=10^\circ, 25^\circ$, i.e.~where the invariant 
mass squared $s'$ is shifted back onto the $Z$ resonance
through initial state hard photon emission 
\cite{Christova:1999gh,Jack:1999xc,Jack:1999af}.  
%
%In detail this means: An earlier comparison of 
%{\tt ZFITTER} v.4.5 with {\tt ALIBABA} v.1 \cite{Riemann:1992up}
%was reproduced with the new versions {\tt ZFITTER} v.6.11 
%and {\tt ALIBABA} v.2 \cite{Christova:1999gh,Jack:1999xc,Jack:1999af}.
%We saw that the deviation between the codes peaks
%at energies around $100$ and $115\,\mbox{GeV}$ for the above 
%cut values with a maximal discrepancy of roughly $10\%$. 
%This was also observed for the same energy range, although to a 
%lesser extent, i.e.~with $\pm 5\ldots 6\%$ deviations, when 
%comparing codes {\tt TOPAZ0} v.4.3 and 4.4 with {\tt ZFITTER} 
%v.6.04/06 onwards, using the same cuts 
%\cite{Christova:1999gh,Jack:1999af}.
Switching off the resummed QED two-loop corrections
in {\tt ALIBABA} and just applying the exponentiated one-loop 
corrections, however, substantially reduces this discrepancy 
down to roughly 1 to 2 per cent which decreases to few per mil
above typical LEP~2 energies of roughly 160 GeV
\cite{Jack:1999xc,Jack:1999af}. 
Repeating these comparisons for similarly 
strong $s'$-cuts instead, shows an agreement of better than one 
per cent which does not depend on higher order corrections
\cite{Christova:1999gh,Jack:1999xc,Jack:1999af} and also
observed by \cite{Montagna:1997jt,Arbuzov:1999uq,Passarino:1999kv}.
These findings are summarized in Fig.~\ref{sig_EW_QED} 
%\ref{xsafbc_ini}, \ref{xsafbc_int}, 
%\ref{xsafbc_fin}, \ref{xsafbc_net_ini}, \ref{xsafbc_net_int}, 
%\ref{compar1992}, \ref{compar-tzaz}, \ref{xs-top-zf}, and 
to \ref{afb-top-zf} in Chapter \ref{ch_lep2}.

The studies conducted here show that higher order photonic corrections,
in particular the $O(\alpha^2)$ hard photon initial state contribution
together with a correct resumming of soft and virtual photons 
for an acollinearity cut seem to be the major underlying cause 
for the observed discrepancies \cite{Jadach:1998jb,Jadach:1999gz}.
Analytically, an exact soft  and virtual photon exponentiation 
procedure is only known for an $s'$-cut 
\cite{Greco:1975rm,Greco:1980,Berends:1988ab,Bardin:1991fu}, 
so an approximation has to be used for the semi-analytical approach 
in {\tt ZFITTER} \cite{Bardin:1992jc2,Bardin:1999yd-orig} 
for an acollinearity cut. The resummation is done here with an effective 
$s'$-cut which removes the bulk of the hard photon effects but
does not take into account some higher order hard photonic contributions,
still allowed by the phase space with acollinearity cut.
So, while this procedure exactly reproduces the first order results,
higher order photonic corrections are only approximately described
for the acollinearity cut.

This explains the large deviations of codes at energies, where the
$Z$ radiative return is approximately prevented because here a very 
delicate removal of hard photons has to occur by the applied cut 
which cannot be done exactly for the higher order terms in {\tt ZFITTER}.
Around the $Z$ boson resonance, however, hard photon emission is 
strongly suppressed due to the resonant Born term, resulting 
there in the better than per mil agreement with other numerical codes. 
Since the acollinearity cut is not so effective in preventing the 
radiative return to the $Z$ as the $s'$-cut, these deviations also  
survive more profoundly for the acollinearity cut at higher energies
than for the $s'$-cut. 
%Further studies hereof, also for 
%the angular distribution ${d{\sigma}}/{d{\cos\vartheta}}$, 
%and a similar update of the code for the Bhabha scattering 
%case for LEP~1 and 2 energies are planned for the future.

With this work being done in the finishing phase of LEP, it is of course
interesting to ask how the predictions of the codes look like 
when going to higher energies at a future $e^+e^-$ Linear Collider (LC)
as e.g.~the TESLA project \cite{Accomando:1997wt}. 
The overall outcome of our analysis for the LC
\cite{Christova:2000zu}
is that the codes {\tt ZFITTER}, {\tt TOPAZ0}, and
{\tt KK2f} deviate with initial state bremsstrahlung 
not worse than 5 per mil, with a better than per mil agreement 
exactly at the $Z$ peak as reference point. 
This was achieved for strong hard photon cuts using the
default settings of the code. The agreement deteriorated 
slightly with increasing $s'$-cut value, but stayed within 
a $\pm 5$ per mil margin. The comparison included different 
resummed higher order corrections, up to three-loop order
\cite{Berends:1988ab,Bardin:1991fu,Montagna:1997jv}. 
Final state bremsstrahlung was also taken into account, 
but cuts had no considerable effects here. The same is true
for QED initial state pair 
creation \cite{Arbuzov:1999uq,Passarino:1999kv} contained in 
{\tt ZFITTER} \cite{Bardin:1999yd-orig} and {\tt TOPAZ0}
\cite{Montagna:1998kp} but not in {\tt KK2f} \cite{Jadach:1999kkkz}.
The effects always stayed below the 1 to 2 per mil level.

Including the QED initial-final state interference did not 
change these observations for the {\tt ZFITTER} -- {\tt TOPAZ0} 
comparison but led to an increase of the discrepancies with 
{\tt KK2f} up to 1 per cent. This is due to an additional 
exponentiation of the soft interference terms which is 
available for {\tt KK2f} \cite{Jadach:1998jb,Jadach:1999gz}, 
but not for the other two codes. 
This is a clear indication that codes like 
{\tt ZFITTER} and {\tt TOPAZ0} may have to include higher order 
corrections from the QED interference considering its increasing 
importance at higher energies. An exponentiation
of the soft photon terms 
could be done straightforwardly \cite{Bardin:1991fu}.
A recent comparison at the LEP~2 Monte-Carlo Workshop
\cite{Jadach:200000} using the full possibilities 
of {\tt KK2f} as Monte-Carlo event generator
even showed e.g.~a better than 2 per mil agreement 
at $\sqrt{s}=189\,\mbox{GeV}$ for {\tt KK2f} and a 
quickly modified version of {\tt ZFITTER} in which 
the soft QED interference terms were resummed. All 
results were combined in Fig.~\ref{top_zf_kk_codes_a} 
and Fig.~\ref{top_zf_kk_codes_b} in Chapter \ref{ch_linac}.

Computing time is also an important issue which 
one has to reconcile when the programs shall be 
utilized for quick data-fitting routines:
While {\tt ZFITTER} needs few seconds for 30 
cross section values on a typical HP-UX workstation 
or PC, {\tt TOPAZ0} calculates 
roughly 30 minutes for the same amount of numbers,
and {\tt KKf} needs 1 or 2 days \cite{Christova:2000zu}.
The advantages of the slower numerical programs 
on the other hand can be clearly seen in the possibility 
of calculating multi-differential observables, 
including more complex kinematical cuts, helicities,
or resummed QED interference corrections.
These issues can only be treated in a very limited way 
by {\tt ZFITTER} due to its semi-analytical 
approach, so the code should be considered complementary 
to other numerical programs \cite{Boudjema:1996qg}.
%Monte-Carlo integration routines 
%or event generators \cite{Boudjema:1996qg}. 

%--------------------------------------------------------------------------
\section*{Outlook 
\label{sec_outlook}
}
%--------------------------------------------------------------------------
\addcontentsline{toc}{section}{Outlook}
%
%==========================================================================
\subsection*{Updates to the QED part of {\it ZFITTER}  
\label{sub_out_qed}
}
%--------------------------------------------------------------------------
%
Thinking of an upgrade of the {\tt ZFITTER} code for later 
applications, we discussed in Chapter \ref{ch_linac} 
two different options for which the {\tt ZFITTER} code be used 
at an $e^+e^-$ Linear Collider (LC): First, precision physics 
could again be performed on the $Z$ boson resonance, but this time
in a very high-luminosity mode at the LC, to indirectly search
e.g.~for Higgs and supersymmetric particles. Secondly, particle 
searches at TeV scale energies will also need a precise and 
fast numerical evaluation of cross sections with all theoretically 
available radiative corrections. 
 
On the $Z$ boson resonance, we have shown that the present level of 
agreement between codes like {\tt ZFITTER}, {\tt TOPAZ0}, and {\tt KK2f}
for cross section predictions is now better than $10^{-4}$  
itself and better than $0.3\times 10^{-3}$ for 
$\sqrt{s}=M_Z\pm 3\,\mbox{\GeV}$ and therefore quite satisfactory 
in comparison with the present experimental accuracies there.
With a possible later Giga $Z$ option at a LC, experimental accuracies
could increase by a factor of 100 or more and then a re-analysis 
of higher order predictions with kinematical cuts 
by the codes would become necessary.

%For applications at LEP~2 energies, around $\sqrt{s} = 160$ to 
%$200\,\mbox{GeV}$, the situation for the codes is satisfactory 
%for an $s'$-cut: the overall agreement for cross sections 
%predictions is better than roughly 2 per mil for different cut values. 
%For the acollinearity cut, however, deviations are at the 1 
%to several per cent level and troublesome compared with an 
%estimated final experimental precision at LEP~2 of at 
%least $3\%$ \cite{Grunewald:1999wn}. A detailed analysis of 
%the used prescriptions for a resumming of soft and virtual 
%photon effects including higher order hard photon corrections 
%for this cut option would be necessary.

At energies up to roughly 800 GeV like for the {\it Tesla project} 
\cite{Brinkmann:1997nb}, of course, issues like experimental and 
theoretical precisions are still quite vague, but it is clear that the demands
on 2-fermion codes will be quite challenging due to higher energies,
higher luminosities, and more refined analysis techniques compared 
to the LEP/SLC situation. On the one hand, for example, electroweak 
and QED corrections become equally important which may demand a critical 
look at the numerical validity of the effective Born approach at 
higher energies.
On the other hand, higher order QED corrections 
will also grow in importance with increasing energies.
Just to give an impression what might lie ahead 
in the near future on updates for the semi-analytical program 
{\tt ZFITTER} concerning its QED branch, 
some examples have been listed below:
\begin{itemize}
\item For the forward-backward asymmetries $A_{FB}$ still 
      the leading logarithmic $O(\alpha^2)$ corrections for 
      initial state pair creation have to be determined and 
      included in the code.

\item Already at LEP~2 energies 
      an exponentiation of the soft and virtual interference
      part in {\tt ZFITTER} appears necessary 
      in case of no or only loose cuts \cite{Jadach:200000}.
      The corresponding formulae for the one-loop case with 
      cuts on $s'$ and $\cos\vartheta$ are 
      already available \cite{Bardin:1991fu}. 
%%      An analysis of such an upgrade together with the initial state 
%%      exponentiation in {\tt ZFITTER} will of course be mandatory.

\item In order to account for incoherent higher order corrections
      from final state leptonic or hadronic pair emission 
      \cite{Passarino:1999kv}, 
      a merger between programs {\tt ZFITTER} and {\tt GENTLE} 
      \cite{Bardin:1996zz} would be a convenient first approach.

\item The Bhabha scattering case 
      is has to be correctly dealt with, analytically 
      and numerically in {\tt ZFITTER}, for an acollinearity cut.
      This is even more important due to the high statistics
      of this channel or considering applications like small-angle 
      Bhabha scattering for high-precision luminosity measurements. 
      The formulae presented in this dissertation 
      can be directly applied to the $s$-channel part, 
      while the $t$-channel and $s$-$t$-interference terms with 
      their different angular dependence still have to be calculated.
      For this, especially the treatment of mass singularities 
      and the phase-space splitting procedure presented here 
      for the $s$-channel case would be a helpful guideline.

\item Further options like the inclusion of beamstrahlung effects,
      already contained in some Monte-Carlo programs \cite{Boudjema:1996qg} 
      are also thinkable for {\tt ZFITTER} in order to obtain results  
      for more realistic, experimental set-ups.  
\end{itemize}

All of this, of course, will have to go hand-in-hand with 
further comparisons between codes {\tt ZFITTER}, {\tt TOPAZ0}, 
{\tt KK2f}, {\tt ALIBABA}, and other programs. Only then can one hope 
to guarantee the correctness and accuracy of the 
analytical and numerical results and meet 
the experimental precision demands.
 
%==========================================================================
\subsection*{Top pair production at LC energies
\label{sub_out_ttbar}
}
%--------------------------------------------------------------------------
%
One very interesting topic for the LC 
is top pair production above $\sqrt{s}\approx 350\,\mbox{GeV}$ 
\cite{Hoang:2000yr}, with real and virtual radiative corrections including 
final state mass effects and will have to be treated accurately
by the codes, at least sufficiently above the threshold region.
The correct inclusion of mass effects is crucial here. 
{\tt SM} and {\tt MSSM} calculations to $e^+e^-\to t\bar{t}$ with 
final state masses and virtual and real QED \cite{Akhundov:1991qa}, 
EW \cite{Beenakker:1991khx}, and QCD \cite{Ravindran:1998jw} 
corrections are already available. But work still has to be done 
concerning a description in the context of EW form factors or 
the inclusion of hard QED and QCD bremsstrahlung corrections 
with final state masses \cite{BKNRL:1999up}. 
First preliminary comparisons show some several per cent corrections 
from hard final state bremsstrahlung close to the threshold region
\cite{BKNRL:1999up}.

%--------------------------------------------------------------------------
\subsection*{Possible {\it beyond-the-SM} applications 
\label{sub_out_beyond}
}
%--------------------------------------------------------------------------
%
Finally, I would like to end our discussion on fermion pair 
production by giving some interesting possibilities 
of how one could extend the {\tt ZFITTER} code for 
applications beyond a {\tt SM} description:  
\begin{itemize}
\item {\it Flavor number violation in $Z$ decays}: 
      Numerical branches for the calculation
      of branching ratios of {\tt SM} flavor changing 
      neutral currents through virtual
      one-loop corrections with massive quarks and {\it CKM} mixing  
      in hadronic Z decays like $Z\to d s$ etc.~are feasible.
      The corresponding extension to the $\nu${\tt SM} case 
      for non-diagonal leptonic $Z$ decays due to massive neutrinos 
      with neutrino-lepton mixing \cite{Mann:1984dv}, 
      strongly suggested by recent experimental
      results \cite{Fukuda:1998mi}, would be straightforward. For this, a 
      recent discussion of branching ratios for different scenarios 
      has been done \cite{Illana:1999aa}. 

\item {\it Supersymmetry}:
      The inclusion of virtual corrections in the 
      {\it minimal supersymmetric model} ({\tt MSSM})
      would be the natural next step in expanding the code 
      also to non-{\tt SM} physics. The complete 
      one-loop results in the {\tt MSSM} together with masses and 
      some two-loop effects are already available \cite{Heinemeyer:1999aa}
      and could be implemented straightforwardly with the 
      effective Born approach.
 
\item {\it $Z'$, $W'$ physics}: Subroutines like 
      {\tt ZEFIT} \cite{Riemann:1997aa} calculating cross sections 
      with the exchange of extra heavy gauge bosons, $Z'$ and $W'$, 
      predicted in certain {\it GUTs}, already exist.
      This could be refined having different models, 
      higher center-of-mass energies, etc.~in mind.        

\item {\it Quantum gravity effects above the electroweak scale}:
      Recently, possible effects by the 
      exchange of spin-2 bosons in pair-production processes, 
      especially on angular cross section distributions have 
      been examined in \cite{Hewett:1998sn}.
      This is motivated by string-inspired models which may allow 
      effects of {\it extra dimensions} at \mbox{TeV} energy scales.
      For this, only the Born level part of {\tt ZFITTER}
      would have to be extended. 
\end{itemize}

These are just a few issues which could be addressed in the 
not so far future for the existing {\tt ZFITTER} code. It also illustrates 
the still rich predictive power on {\it New Physics} scenarios 
when looking at the classical fermion pair production channel.

At the beginning of LEP~1 or SLC data taking,
the precision of the codes had been sufficient, 
due to the lower experimental statistics then
and the hard QED corrections being strongly suppressed 
at the $Z$ peak. The increased experimental accuracy 
now demands a precise, better than per-mil  
determination of QED bremsstrahlung in the resonance region.
This is now guaranteed for the semi-analytical code {\tt ZFITTER} 
with the new results. The new formulae pose a realistic alternative 
to a kinematically simpler $s'$-cut for leptonic final 
states and are a direct generalization of the up-to-now 
known analytical results including cuts.

Also for LEP~2 and especially for later LC energies and luminosities 
one is for example interested in removing a disturbing 
radiative return to the $Z$ boson through hard photon emission
when searching for {\it New Physics} effects. For this,
hard photon effects have to be known accurately with 
kinematical cuts. The new results derived in this dissertation
finally form an important contribution in the 
larger framework of precision tests to the {\tt SM} or for 
{\it New Physics} searches performed at present or future 
$e^+e^-$ collider experiments.

%
%---------------------------------------------------------------
\begin{appendix}
\pagestyle{headings}
\def\thechapter{\Alph{chapter}}
\setcounter{chapter}{0}
\setcounter{section}{0}
\setcounter{subsection}{0}
\def\theequation{\Alph{chapter}.\arabic{equation}}
\def\thetable{\Alph{chapter}.\arabic{table}}
\def\thefigure{\Alph{chapter}.\arabic{figure}}
\setcounter{equation}{0}
\setcounter{table}{0}
\setcounter{figure}{0}
%
%\newpage

%%%%%%%%%%%%%%%%%%%%%%%%%%%%%%%%%%%%%%%%%%%%%%%%%%%%%%%%%%%%%%%%%%%%%%%%
\chapter{Cross Sections and Phase Space 
\label{crossphase}
}
%%%%%%%%%%%%%%%%%%%%%%%%%%%%%%%%%%%%%%%%%%%%%%%%%%%%%%%%%%%%%%%%%%%%%%%%

%%%%%%%%%%%%%%%%%%%%%%%%%%%%%%%%%%%%%%%%%%%%%%%%%%%%%%%%%%%%%%%%%%%%%%%
\section{Feynman diagrams and matrix element 
\label{feynmat}
}
\begin{figure}[htp]
\begin{center}
%%%%%%%%%%%%%%%%%%%%%%%%%%%%%%%%%%%
% Initial state radiation diagram zini1.tex%
%%%%%%%%%%%%%%%%%%%%%%%%%%%%%%%%%%%
\vfill
\setlength{\unitlength}{1pt}
%\SetScale{0.8}
\SetWidth{0.8}
%----------------------------------------
\begin{picture}(180,120)(0,0)
\thicklines
\ArrowLine(10,110)(40,80)
\Vertex(30,90){1.8}
\Photon(30,90)(70,110){2}{7}
\ArrowLine(30,90)(60,60)
\Vertex(60,60){1.8}
\ArrowLine(60,60)(10,10)
\Photon(60,60)(120,60){3}{8}
\Vertex(120,60){1.8}
\ArrowLine(120,60)(170,110)
\ArrowLine(170,10)(120,60)
\Text(5,117)[]{$e^-$}
\Text(5,5)[]{$e^+$}
\Text(90,50)[]{$\gamma,\,Z$}
\Text(178,8)[]{$\bar{f}$}
\Text(178,113)[]{$f$}
\Text(77,114)[]{$\gamma$}
\end{picture}
%============================================
%%%%%%%%%%%%%%%%%%%%%%%%%%%%%%%%%
% Final state radiation diagram %
%%%%%%%%%%%%%%%%%%%%%%%%%%%%%%%%%
\setlength{\unitlength}{1pt}
%\SetScale{0.8}
\SetWidth{0.8}
\begin{picture}(200,120)(0,0)
\thicklines
\ArrowLine(10,110)(60,60)
\Vertex(60,60){1.8}
\ArrowLine(60,60)(10,10)
\Photon(60,60)(120,60){3}{8}
\Vertex(120,60){1.8}
\ArrowLine(170,10)(150,30)
\Vertex(150,30){1.8}
\Photon(150,30)(190,50){2}{7}
\ArrowLine(150,30)(120,60)
\ArrowLine(120,60)(170,110)
\Text(5,117)[]{$e^-$}
\Text(5,5)[]{$e^+$}
\Text(90,50)[]{$\gamma\,,\,Z$} 
\Text(178,8)[]{$\bar{f}$}
\Text(178,113)[]{$f$}
\Text(197,53)[]{$\gamma$}
\end{picture}
%============================================
\vspace*{0.5cm}
\caption
[Feynman diagrams for real photon emission]
{\sf\label{phfig1}  
Examples of Feynman diagrams for real photon initial and final state radiation.
}
\end{center}
\end{figure}
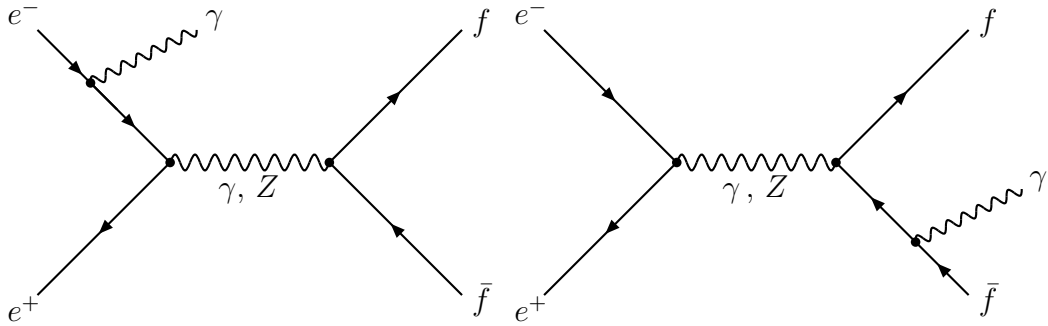

%%%%%%%%%%%%%%%%%%%%%%%%%%%%%%%%%%%%%%%%%%%%%%%%%%%%%%%%%%%%%%%%%%%%%%%
%
The real photon emission from the initial and final state 
of $s$-channel processes $e^+e^-\to \bar{f}f$ is described by 
the two Feynman diagrams depicted in the Fig.~\ref{phfig1}.

The four-momenta of the 5 particles are denoted 
as $k_{1,2}$ for $e^{-,+}$, $p_{1,2}$ 
for ${f},\bar{f}$ and $p$ for the photon $\gamma$. The c.m. energy is 
designated as $s:=(k_1+k_2)^2$ and the electron and fermion masses as 
$m_e,m_f$.
The corresponding $S$-matrix element ${\Large{\cal M}}$  
for the scattering processes $e^+e^-\to \bar{f}f$ with $O(\alpha)$ 
initial and final state real {\it bremsstrahlung} is:

%-----------------------------------------------------------------
%\vfill\eject
%-----------------------------------------------------------------
\ba
{\Large{\cal M}} &=& 
      (2 \pi)^4 \delta^{(4)}(k_1 +k_2 - p_1 - p_2 - p)
\; \frac{{M}_{ini} + {M}_{fin}} {(2 \pi)^6 (2 \pi)^{3/2} 
   (2 k^0_1\, 2 k^0_2\, 2 p^0_1\, 2 p^0_2\, 2 p^0 )^{1/2} }.
\nonumber\\
\label{br01}
%\\
%\mbox{where}&&
%%\qquad 
%{M} = {M}_{ini} + {M}_{fin}
%\label{br02}
\ea
%
%is the amplitude with initial and final state bremsstrahlung. 
The Born $S$-matrix element is:
\ba
{\Large{\cal M}}^{Born} = (2 \pi)^4 \delta^{(4)}(k_1 +k_2 - p_1 - p_2)
\; \frac{{M}^{Born}}
{(2 \pi)^6 (2 k^0_1\, 2 k^0_2\, 2 p^0_1\, 2 p^0_2)^{1/2} },
\label{b01}
\ea
The current structure for these amplitudes 
${M}_{ini}$, ${M}_{fin}$, and ${M}^{Born}$
is given below:

\ba
{M}^{Born} &=& i\, e^2\;\frac{1}{s} \, \left(\, Q_e Q_f
  {J_e}^{\mu}(\gamma)\;{ J_{f}}_{\mu}(\gamma)
  + \chiz (s)\;
 {J_e}^{\mu}(Z) \; {J_{f}}_{\mu}(Z)  \right),
\label{b02}
\\
\nonumber\\
{M}_{ini} &=& i\, e^3\; 
\frac{Q_e {\epsilon}_{\alpha}}{s'}\, \left(\, Q_e Q_f
  {I}^{\mu, \alpha}(\gamma)\; { J_f}_{\mu}(\gamma)
  + \chiz (s')\;
 {I}^{\mu, \alpha}(Z) \; {J_f}_{\mu}(Z)  \right), 
\label{br02a} 
\\
\nonumber\\
{M}_{fin} &=& i\, e^3\; 
\frac{Q_f {\epsilon}_{\alpha}}{s}\, \left(\, Q_e Q_f
  {J_e}^{\mu }(\gamma)\; { F}^{\alpha}_{\mu}(\gamma)
  + \chiz (s)\;
 {J_e}^{\mu}(Z)\;  {F}^{\alpha}_{\mu}(Z)  \right).
\label{br02b}
\ea
The vector ${\epsilon}_{\alpha}$ denotes the 
photon polarization vector. $Q_e$ and $Q_f$ are the
charges of the initial electron, $Q_e=-1$, and 
final state fermion $f$. The $Z$ boson propagator 
$\chiz(s)$ and the weak neutral current couplings 
$v_{e}$, $v_{f}$, $a_{e}$, and $a_{f}$ are defined as:
\ba
    \chiz(s) &=& \chi(s)\; \equiv\; \kappa \; \frac{s}{\iprop}
 \label{b04},        
\\
     \kappa &=&\frac{g^2}{4 e^2 \cow}
    = \frac{1}{4 \siw \cow}
    = \frac{\Gmu}{\sqrt 2 }\;\frac{\MzS}{2 \pi \alpha},
\label{b05}
\\ 
\nonumber\\   
v_{e} &=& -\frac{1}{2} + 2 \siw,\qquad\qquad a_{e} = -\frac{1}{2},
\label{b05a}
\\
v_{f} &=& I_3^f-2 Q_f \sin^2{\theta}_W,\qquad\qquad a_f = I_3^f. 
\label{b05b}
\ea
The values $M_Z$ and $\Gamma_Z$ are the $Z$ boson mass and 
width. 
The currents for real photon emission 
from the initial state, ${I}^{\mu, \alpha}(\gamma)$ and 
${ I}^{\mu, \alpha}(Z)$, and from the final state 
${F}^{\alpha}_{\mu}(\gamma)$  and ${F}^{\alpha}_{\mu}(Z)$ 
are given below together with the Born terms,
${J_e}^{\mu }(\gamma)$, ${J_e}^{\mu}(Z)$, ${J_f}_{\mu}(\gamma)$,
and ${J_f}_{\mu}(Z)$:
\ba
 { I}^{\mu, \alpha}(\gamma)&=& {\bar u}(-k_2) 
\left[\frac{2 k_2^{\alpha} - \gamma^{\alpha}\ps}{Z_2} 
  \gimu \,
- \, \gimu 
\frac{2 k_1^{\alpha}   -\ps \gamma^{\alpha}}{Z_1} \right]
 u(k_1),
\label{br03a}
         \\
 { I}^{\mu, \alpha}(Z) &=& {\bar u}(-k_2) 
\left[\frac{2 k_2^{\alpha} - \gamma^{\alpha}\ps}{Z_2} 
\gimu (v_e+a_e\gamma_5) \,  
\right.
\nonumber\\
&&
\hspace*{3.2cm}
\left.
- \, \gimu (v_e+a_e\gamma_5) 
\frac{2 k_1^{\alpha}  -\ps \gamma^{\alpha}}{Z_1} \right]
 u(k_1),  
\label{br03b}
            \\
 {F}^{\alpha}_{\mu}(\gamma)&=& {\bar u}(p_1)  
 \left[\frac{2 p_1^{\alpha} + \gamma^{\alpha}\ps}{V_1}  
\gamu \,
- \, \gamu  
\frac{2 p_2^{\alpha}   +\ps \gamma^{\alpha}}{V_2} \right]
 u(-p_2),  
\label{br03c}
\\ 
 {F}^{\alpha}_{\mu}(Z) &=& {\bar u}(p_1)
 \left[\frac{2 p_1^{\alpha} + \gamma^{\alpha}\ps}{V_1}
\gamu (v_f+a_f\gamma_5) \, 
\right.
\nonumber\\
&&
\hspace*{3.2cm}
\left.
- \, \gamu (v_f+a_f\gamma_5)
\frac{2 p_2^{\alpha}   +\ps \gamma^{\alpha}}{V_2} \right]
 u(-p_2).
\label{br03}
\\
{ J_e}^{\mu}(\gamma)&=& {\bar u}(-k_2)\; \gimu \; u(k_1),
\label{b03a}    
     \\
{ J_e}^{\mu}(Z) &=& {\bar u}(-k_2)\; \gimu (v_e+a_e\gamma_5) \; u(k_1),
\label{b03b}           
 \\
{J_{f}}_{\mu}(\gamma)&=& {\bar u}(p_1) \; \gamu \;  u(-p_2), 
  \\
{J_{f}}_{\mu}(Z) &=& {\bar u}(p_1)\; \gamu (v_f+a_f\gamma_5) \;  u(-p_2).
\label{b03}  
\ea
For brevity, in order to have a compact illustration of the 
amplitudes, the initial and final state masses $m_e$ 
and $m_f$ have been neglected in (\ref{br02a}) to (\ref{b02}). 
In the later calculation they will, however, be correctly
considered. 

For the calculation of the matrix element there are the 
following important invariants for the QED corrected scattering
process given in (\ref{br03a}) to (\ref{b03}):

\ba
\label{invs} 
s  &:=& (k_1+k_2)^2 = 2 k_1 k_2+2 m_e^2, 
\\
\label{invspr} 
s' &:=& (p_1+p_2)^2 = 2 p_1 p_2+2 m_f^2, 
\\
\label{invz1} 
Z_1 &:=& 2 p k_1 = -[(k_1-p)^2-m_e^2], 
\\
\label{invz2} 
Z_2 &:=& 2 p k_2 = -[(k_2-p)^2-m_e^2], 
\\
\label{invv1} 
V_1 &:=& 2 p p_1 = [(p_1+p)^2-m_f^2], 
\\
\label{invv2} 
V_2 &:=& 2 p p_2 = [(p_2+p)^2-m_f^2].
\ea
Using total four-momentum conservation one arrives at 
the following important relations between these invariants:

\ba
\label{invrel}
\left. 
\begin{array}{lllll}
s&=&(p_1+p_2+p)^2&=&s'+V_1+V_2
\\
&&&&
\\
s'&=&(k_1+k_2-p)^2&=&s-Z_1-Z_2
\end{array}
\right\}\,\rightarrow\, Z_1+Z_2=V_1+V_2=s-s'.
\nonumber\\ 
\ea 
We define the invariants $\lambda_s$,
$\lambda_1$, $\lambda_2$, and $\lambda_p$ with $k\equiv k_1 + k_2$:
\ba
\label{lambdas}
\lambda_s &=& \lambda \left( k^2, k_1^2, k_2^2 \right)
   = s^2 - 4 s m_e^2 = s^2 \beta_0^2,
\\
\nonumber\\
\label{lambda1}
\lambda_1 &=& \lambda \left( [k - p_1]^2, k^2, p_1^2 \right)
   = (s-V_2)^2 - 4 s m_f^2,
\\
\nonumber\\
\label{lambda2}
\lambda_2 &=& \lambda \left( [k - p_2]^2, k^2, p_2^2 \right)
   = (s'+V_2)^2 - 4 s m_f^2,
\\
\nonumber\\
\label{lambdap}
\lambda_p &=& \lambda \left( [k - p]^2, k^2, p^2 \right)
   = (s-s')^2,
\\
\nonumber\\
\label{lambda}
\mbox{with} && \lambda = 
\lambda(x,y,z) = x^2 + y^2 + z^2 - 2 x y - 2 x z - 2 y z.
\ea
We will also need the invariants:
\ba
\label{T}
 T & =& 2\,k_1 \cdot p_2
    = \frac{1}{2}\left( \,s'\,+\,V_2\, 
   +\, \beta_{0}\;\sqrt{\lambda_2} \, \cos{\vartheta} \right),
\\
\label{U}
 U & =& 2\,k_2 \cdot p_2
    = \frac{1}{2}\left(\, s'\,+\,V_2\, 
         - \,\beta_{0}\;\sqrt{\lambda_2} \, \cos{\vartheta}\right).
\ea
The differential cross sections to the bremsstrahlung 
and Born matrix elements have the form:
\ba
   d \sigma^{Born} &=& \frac{1}{\tt j}
 \; (2 \pi)^4 \delta^{(4)}(k_1 +k_2 - p_1 - p_2)
\; \frac{1}{4}\,\frac{\sum_{spin} \mid {M}^{Born} \mid^2 }
{(2 \pi)^{12} \,2 k^0_1\, 2 k^0_2\, 2 p^0_1\, 2 p^0_2  }
\nonumber\\
&&
   \;\cdot\, d^{3}\vec p_1 \; d^{3}\vec p_2,
\label{b07}
\\
\nonumber\\
   d \sigma &=& \frac{1}{\tt j}
 \; (2 \pi)^4 \delta^{(4)}(k_1 +k_2 - p_1 - p_2 - p)
\; \frac{1}{4}\,\frac{\sum_{spin} \mid {M} \mid^2 }
{(2 \pi)^{15} \,2 k^0_1\, 2 k^0_2\, 2 p^0_1\, 2 p^0_2\, 2 p^0  }
\nonumber\\
&&
   \;\cdot\, d^{3}\vec p_1 \; d^{3}\vec p_2 \; d^{3}\vec p ,
\label{br07}
\ea
where the flux of the initial particles is:

\ba
  {\tt j} = \frac{\sqrt{ (k_1\cdot k_2)^2 - \meQ } }
{2 (2 \pi)^6 \, k^0_1\,  k^0_2 }
  =\frac{s \beta_0}{(2 \pi)^6 \,2 k^0_1\, k^0_2 },
 \hspace{2cm} \beta_0 = \sqrt{1 - \frac{4 \meS}{s} }.
\label{b08}
\ea
With (\ref{b07}), (\ref{br07}), and (\ref{b08}) the differential cross-sections can 
therefore finally be written in the form:

\ba
   d \sigma^{Born} &=&
 \frac{1}{4}\,\frac{\sum_{spin}\;\mid {M}^{Born} \mid^2 }{2\, s\, \beta_0}
\; d{{\Gamma}^{(2)}}, 
\label{b09}
\\
   d \sigma &=&
 \frac{1}{4}\,\frac{\sum_{spin}\;\mid {M} \mid^2 }{2\, s\, \beta_0}
\; d{{\Gamma}^{(3)}}, 
\label{br09}
\ea
where $d{{\Gamma}^{(2)}}$ and $d{{\Gamma}^{(3)}}$
are the two- and three-particle differential 
phase space volumes of the outgoing particles
respectively.
\ba
d{{\Gamma}^{(2)}} &=& (2 \pi)^4 \delta^{(4)}(k_1 +k_2 - p_1 - p_2)
 \; \frac{d^{3}\vec p_1}{(2 \pi)^3 \, 2 p^0_1}
 \; \frac{d^{3}\vec p_2}{(2 \pi)^3 \, 2 p^0_2},
\label{b10}
\nonumber\\
\\
d{{\Gamma}^{(3)}} &=& (2 \pi)^4 \delta^{(4)}(k_1 +k_2 - p_1 - p_2 - p)
 \; \frac{d^{3}\vec p_1}{(2 \pi)^3 \, 2 p^0_1}
 \; \frac{d^{3}\vec p_2}{(2 \pi)^3 \, 2 p^0_2}
 \; \frac{d^{3}\vec p}{(2 \pi)^3 \, 2 p^0}.
\nonumber\\
\label{br10}
\ea

%%%%%%%%%%%%%%%%%%%%%%%%%%%%%%%%%%%%%%%%%%%%%%%%%%%%%%%%%%%%%%%%%%%%%%%
\section{Kinematics
\label{kinematics}
}
%%%%%%%%%%%%%%%%%%%%%%%%%%%%%%%%%%%%%%%%%%%%%%%%%%%%%%%%%%%%%%%%%%%%%%%
%
%%%%%%%%%%%%%%%%%%%%%%%%%%%%%%%%%%%%%%%%%%%%%%%%%%%%%%%%%%%%%%%%%%%%%%%
\subsection*{Angles of phase space
\label{angles}
}
%%%%%%%%%%%%%%%%%%%%%%%%%%%%%%%%%%%%%%%%%%%%%%%%%%%%%%%%%%%%%%%%%%%%%%%
%
The main objective of this discussion is to calculate 
the contributions of the $O(\alpha)$ hard bremsstrahlung
corrections to cross section observables for fermion pair 
production processes $e^+e^-\to \bar{f}{f}$, $f\ne e,\nu_e$. 
The first order soft and virtual photonic corrections are 
then added to derive finite and physical results for the 
inclusive processes $e^+e^-\to \bar{f}{f}(\gamma)$.
Leading effects from multiple soft and virtual photon emission 
can be included straightforwardly in our flux function 
description; see also (\ref{crossini}) and (\ref{generic_zf}).

\begin{figure}[htb]  
\vspace{7.0cm}
\hspace{0.0cm}
\begin{minipage}[bht]{8.5cm}
{\begin{center}
%-------------------
  \vspace{0.0cm}
  \hspace{0.0cm}
  \mbox{
  \epsfysize=16cm
  \epsffile[0 0 500 500]{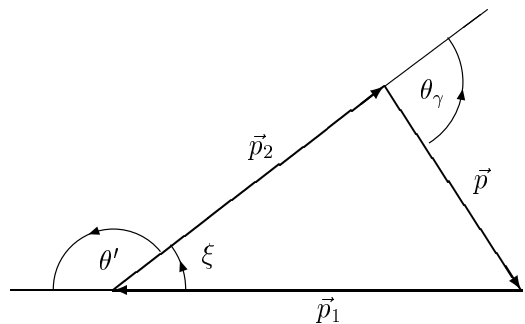}
  }
%----------------
\end{center}}
\end{minipage}
\vspace{-18cm}
\caption[Acollinearity angle]
{\sf 
\label{phfig3} 
Acollinearity angle $\xi\equiv\theta_{\rm acol}$.}
\end{figure}

We are interested in different kinematical cuts, which 
includes a cut on the maximal {\it acollinearity angle} 
of the two final state fermions. The acollinearity 
angle $\xi\equiv\theta_{\rm acol}$ of the two final 
state fermions is depicted in Fig.~\ref{phfig3}. The
directions of motion of the initial state electron and 
positron in the center-of-mass system (c.m.s.) 
define the beam axis. 
In case of the emission of an energetic photon from the initial 
or final state, the final state fermions will not move back-to-back 
anymore due to four-momentum conservation, but will be acollinear. 
The acollinearity angle can be reexpressed by the fermionic 
polar angle $\theta'$ defined in the plane 
of the two fermions' three-momenta: $\xi:=\pi-\theta'$. 
For the description of the hard photon phase space
with acollimearity cut one can also refer to \cite{Passarino:1982}
and \cite{Bilenkii:1989zg}.

\begin{figure}[htp]
\hspace{-4.0cm}
\begin{minipage}[bht]{7.9cm}
{\begin{center}
%-------------------
  \vspace{-3.0cm}
  \hspace{-2.0cm}
  \mbox{
  \epsfysize=16cm
  \epsffile[0 0 500 500]{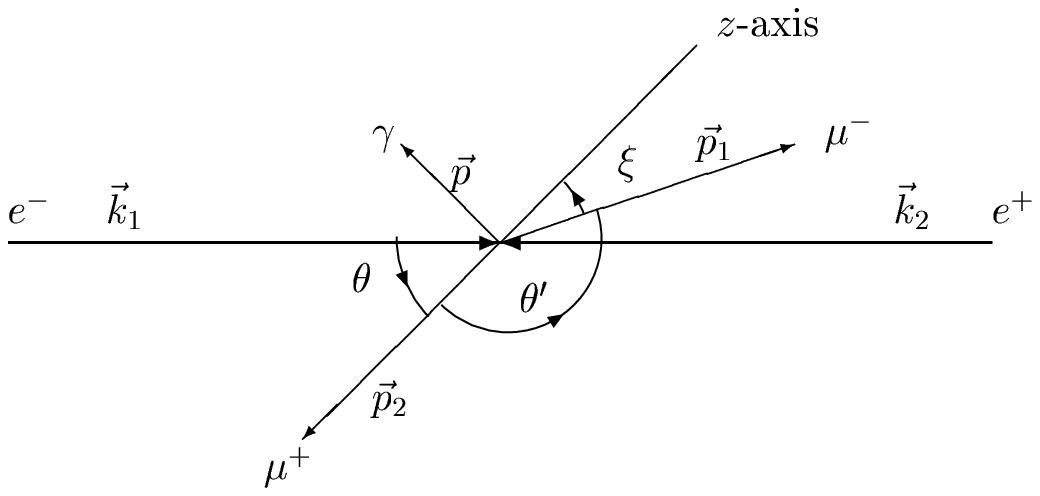}
  }
%----------------
\end{center}}
\end{minipage}
\hspace{2.0cm}
\begin{minipage}[bht]{7.9cm}
{\begin{center}
%-------------------
  \vspace{8.0cm}
  \hspace{0.0cm}
  \mbox{
  \epsfysize=16cm
  \epsffile[0 0 500 500]{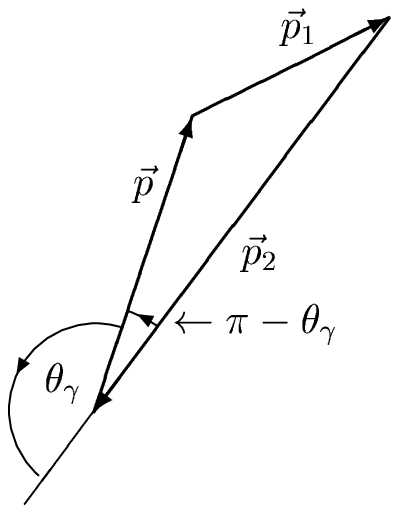}
  }
%----------------
\end{center}} 
\end{minipage}
\vspace{-18.0cm}
\caption[Angles of phase space]
{\label{phfig2}
\sf Angles of phase space and photon angle
  $\theta_\gamma$.} 
\end{figure}

The number of independent of degrees of freedom for a 
$2\to 3$ scattering process amounts to four taking into 
account four-momentum conservation, on-shell conditions,
and a rotational symmetry of the scattering process 
around the beam axis in the c.m.s. 
Having in mind a cut on the maximal acollinearity angle, 
one can use the following independent 
kinematic variables to parameterize the total phase space:

\begin{itemize}
\label{indvar}
\item $s':=(p_1+p_2)^2$ as invariant mass squared of the fermion pair,

\item $\cos\vartheta$ as cosine of the scattering angle of $\bar{f}$ 
      with respect to the $e^-$ beam axis in the c.m.s., 

\item $\cos\theta_\gamma$ as cosine of the polar angle 
      between the three-momenta of $\bar{f}$ and 
      the photon in the c.m.s.,

\item $\varphi_\gamma$ as azimuthal angle of the photon in the rest frame of 
      $({f},\gamma)$\\ 
      ($z$-axis defined by $\vec p_2$ of $\bar{f}$ in the
      c.m.s.). 
\end{itemize}
The different angles of phase space are shown in Fig.~\ref{phfig2}.

For brevity, we will from now on use the notation $\xi$ during
the calculations for the acollinearity angle.

%%%%%%%%%%%%%%%%%%%%%%%%%%%%%%%%%%%%%%%%%%%%%%%%%%%%%%%%%%%%%%%%%%%%%%%
\subsection*{Energies and three-momenta
\label{energies}
}
%%%%%%%%%%%%%%%%%%%%%%%%%%%%%%%%%%%%%%%%%%%%%%%%%%%%%%%%%%%%%%%%%%%%%%%
%
The energies and three-momenta in the c.m.s.~can now be extracted 
easily from the invariants in (\ref{lambdas}) to (\ref{lambdap})
using four-momenta conservation and the on-shell conditions.
For the energy component $q^0$ and the three-momentum $\vec q$ of 
an arbitrary 4-momemtum $q$ in the c.m.s. with 
$\vec k = \vec k_1 + \vec k_2 = 0$, it holds:
\ba
2 k^0 q^0 &=& 2 k q 
\quad\longrightarrow\quad 
q^0 = \frac{2 k q}{2 k^0} = 
\pm\frac{(k\pm q)^2 - k^2 - q^2}{2 k^0}, 
\\
\nonumber\\
|\vec q| &=& \sqrt{(q^0)^2-q^2} = 
\frac{\sqrt{\lambda([k\pm q]^2,k^2,q^2)}}{2k_0}.
\ea

We can therefore derive with (\ref{invs}) to (\ref{invz2})
and $\lambda_s$, $\lambda_1$, $\lambda_2$, and $\lambda_p$
defined in (\ref{lambdas}) to (\ref{lambdap}):
\ba
\label{momentak}
k_1^0 &=& k_2^0 = \frac{\sqrt{s}}{2},\qquad\, 
|\vec{k}_1| = |\vec{k}_2| = \frac{1}{2}\sqrt{s-4m_e^2},
\\
\label{momentap}
p^0 &=& \frac{s-s'}{2\sqrt{s}},\qquad\quad\,\, 
|\vec{p}| = \frac{\sqrt{\lambda_p}}{2\sqrt{s}}, 
\\
\label{momentap1}
p_1^0 &=& \frac{s-V_2}{2\sqrt{s}},\qquad\quad\,  
|\vec{p_1}| = \frac{\sqrt{\lambda_1}}{2\sqrt{s}},
\\
\label{momentap2}
p_2^0 &=& \frac{s'+V_2}{2\sqrt{s}},\qquad\quad  
|\vec{p_2}| = \frac{\sqrt{\lambda_2}}{2\sqrt{s}}.
\ea

%%%%%%%%%%%%%%%%%%%%%%%%%%%%%%%%%%%%%%%%%%%%%%%%%%%%%%%%%%%%%%%%%%%%%%%%%%%%
\section{The three-particle phase space with acollinearity cut
\label{threephase}
}
%%%%%%%%%%%%%%%%%%%%%%%%%%%%%%%%%%%%%%%%%%%%%%%%%%%%%%%%%%%%%%%%%%%%%%%%%%%%
%
%%%%%%%%%%%%%%%%%%%%%%%%%%%%%%%%%%%%%%%%%%%%%%%%%%%%%%%%%%%%%%%%%%%%%%%
\subsection*{Kinematical constraints
\label{kincon}
}
%%%%%%%%%%%%%%%%%%%%%%%%%%%%%%%%%%%%%%%%%%%%%%%%%%%%%%%%%%%%%%%%%%%%%%%
%
For a seperation of the two phase space regions for 
soft and hard photon emission one generally introduces 
an arbitrary parameter $\varepsilon$ in order to 
integrate over vanishing, $p^0<\varepsilon$, or 
sufficiently energetic photon momenta, $p^0\geq \varepsilon$.
The parameter $\varepsilon$ is unphysical 
and has to cancel in physical quantitites when 
combining the soft and virtual corrections with the 
hard photonic contributions. Together with the kinematically 
minimally allowed $s'$ value this gives the following 
bounds on $s'$ for the hard photon phase space:
\ba
\label{sprmin}
4 m_f^2 \leq s' \leq s(1-\varepsilon) .
\ea    

Furthermore, with (\ref{momentak}) to (\ref{momentap}) 
it is straightforward to express $\cos\theta_\gamma$ as 
function of the invariants $s$, $s'$, and $V_2$:
\ba
\label{costhg}
\cos\theta_\gamma = \frac{|\vec{p}_1|^2-|\vec{p}_2|^2-|\vec{p}|^2}
{2|\vec{p}||\vec{p}_2|} =\frac{\lambda_1-\lambda_2-\lambda_p}
{2\sqrt{\lambda_p\lambda_2}}. 
\ea 
{From} (\ref{costhg}) we gain a limiting condition on the kinematically 
maximally accessible phase space:
\ba
\label{sinthg2}
\sin^2\theta_\gamma
&=&\frac{\lambda(\lambda_1,\lambda_2,\lambda_p)}{4\lambda_2\lambda_p} \geq 0 
\quad\longrightarrow\quad
\lambda(\lambda_1,\lambda_2,\lambda_p) \geq 0,   
\\
&&4\lambda_2\lambda_p-(\lambda_1-\lambda_2-\lambda_p)^2 \geq 0.
\ea 
Thus we have the following combined kinematical 
constraint for the variables $s'$ and $V_2$:
\ba
\label{limV2exact}   
V_2^{\min}(s')\quad\leq\quad V_2\quad \leq\quad V_2^{\max}(s')
\ea
with
\ba
\label{limV2exact2}
V_2^{\max,\min}(s') = \frac{s-s'}{2}\, (1\pm\beta),
\quad \beta = \beta(s'):=\sqrt{1-\frac{4 m_f^2}{s'}}.
\ea

%%%%%%%%%%%%%%%%%%%%%%%%%%%%%%%%%%%%%%%%%%%%%%%%%%%%%%%%%%%%%%%%%%%%%%%%%%%%
\subsection*{The differential phase space volume
\label{phvol}
}
%%%%%%%%%%%%%%%%%%%%%%%%%%%%%%%%%%%%%%%%%%%%%%%%%%%%%%%%%%%%%%%%%%%%%%%%%%%%
%
The differential phase space volume $d{{\Gamma}^{(3)}}$ 
of a three-particle final state was given in (\ref{br10}):

\ba
\label{dgam}
d{\Gamma}&=&(2\pi)^4\,\frac{d^3\vec{p_1}}{(2\pi)^3 2p^0_1}\,
\frac{d^3\vec{p_2}}{(2\pi)^3 2p^0_2} 
\, \frac{d^3\vec{p}}{(2\pi)^3 2p^0} 
\, \delta^4(\underbrace{k_1+k_2}_{=k_{12}}
\underbrace{-p_1-p_2}_{=-p_{12}}-p)
\\
\nonumber\\
 &=&(2\pi)^{-5}\, d^4{p_1}\,\delta(p_1^2-m^2) \, d^4{p_2}\,\delta(p_2^2-m^2)
\, d^4{p}\,\delta(p^2)\, \delta^4(k_{12}-p_{12}-p).
\nonumber\\
\ea
When transforming the differentials of the four-momenta into those
of the independent variables $s'$, $\cos\vartheta$, $\cos\theta_\gamma$, 
and $\varphi_\gamma$, it is convenient to separate the three-particle phase
space into the two-particle phase spaces of two subsystems:
We chose one fermion and the photon, $f$ and $\gamma$, 
in their rest frame as one subsystem and boost it
in the center-of-mass system, with the anti-fermion $\bar{f}$  
and this subsystem creating the second subsystem.
This choice of subsystems is given by the definition of the 
photon angle $\theta_\gamma$ with respect to the direction of 
motion of the anti-fermion $\bar{f}$ (Fig.~\ref{phfig2}).

We therefore have the following separation of the three-particle phase space
\ba
\label{dgamsep}
d{\Gamma^{(3)}} = 
\frac{d{{M^2_{f\gamma}}}}{2\pi}
\, d{\Gamma_I^{(2)}}\, d{\Gamma_{II}^{(2)}},
\ea
into the two-particle phase spaces
\ba
\label{dgam12}
\mbox{I. } &&
d{\Gamma_I^{(2)}} = \frac{1}{(2\pi)^2}\, d^4 P\,\delta
(P^2-{M^2_{f\gamma}})\, d^4 p_2\,\delta (p_2^2-m^2)\,\delta^4 (k_1+k_2-P-p_2),
\label{dgam12a}
\nonumber\\
\\
\mbox{II.} &&
d{\Gamma_{II}^{(2)}} = \frac{1}{(2\pi)^2}\, d^4 p_1\,
\delta (p_1^2-m^2)\, d^4p\, \delta(p^2)\,\delta^4 (P-p-p_1),
\label{dgam12b}
\ea
after inserting 
\ba
\label{one}
1 = d^4 P\,\delta^4 (P-p_1-p)\, d{{M^2_{f\gamma}}}\,\delta (P^2-{M^2_{f\gamma}}) 
\ea
into (\ref{dgam}) and using
\ba
\label{deltrans} 
&& \delta^4 (k_1+k_2-p-p_1-p_2)\,\delta^4 (P-p-p_1)
\nonumber\\
&=& \delta^4 (k_1+k_2-P-p_2)\,\delta^4 (P-p-p_1).
\ea
The value ${M^2_{f\gamma}}:=P^2$ is the invariant mass squared of 
the fermion-photon subsystem $(f,\gamma)$ with $P:=p+p_1=k_1+k_2-p_2$. 

The integration of $d{\Gamma_I}$ and $d{\Gamma_{II}}$ can now be done
by inserting the invariants (\ref{invv1}) to (\ref{invz2}) into (\ref{dgam12a})
and (\ref{dgam12b}) 
and using the on-shell conditions and four-momentum conservation defined 
by the $\delta$-distributions. Starting with $d{\Gamma_I}$, we have:   
\ba
\label{dgam1a}
d{\Gamma_I^{(2)}} &=& 
\frac{1}{(2\pi)^2}\, d^4 p_2\,\delta (p_2^2-m_f^2)
\,\delta ((k_1+k_2-p_2)^2-{M^2_{f\gamma}}) 
\nonumber\\
&=&
\frac{1}{(2\pi)^2}\,\frac{|\vec{p}_2|\, d{p^0_2}}{2}
\,\delta (s'+V_2-2\sqrt{s}\, p^0_2)\, d{\Omega_{\vec{p}_2}},
\quad d{\Omega_{\vec{p}_2}} = 2\pi d {\cos\vartheta},
\nonumber\\
\\
\label{dgam1b}
\rightarrow && d{\Gamma_I^{(2)}} 
= \frac{1}{2\pi}\,\frac{\sqrt{\lambda_2}}{8s}
\, d {\cos\vartheta}\quad\mbox{ with }\quad\lambda_2 
= (s'+V_2)^2-4\,m\,s^2.
\ea
The relation (\ref{dgam1a}) was obtained using the rotational symmetry  
around the beam axis and the on-shell condition
${p_2^0}^2=|\vec{p}_2|^2+m_f^2$, which leads to $2\, p_2^0\, d{p_2^0} 
= 2\,|\vec{p}_2|\, d|\vec{p}_2|$. Similarly, we can treat $d{\Gamma_{II}}$:

\ba
\label{dgam2a}
d{\Gamma_{II}^{(2)}} &=& 
\frac{1}{(2\pi)^2}\, d^4{p}\,\delta(p^2)
\,\delta((P-p)^2-m_f^2)
\nonumber\\
&=&\frac{1}{(2\pi)^2}\,\frac{|\vec{p}|\, d{p^0}}{2}
\,\delta(s-s'-2\sqrt{s}\, p^0)\, d{\Omega_{\vec{p}}},
\quad d{\Omega_{\vec{p}}} = d {\varphi_\gamma}\, 
d {\cos\theta_\gamma},
\nonumber\\
\\
\label{dgam2b}
\rightarrow&& d{\Gamma_{II}^{(2)}} 
= \frac{1}{(2\pi)^2}\,\frac{\sqrt{\lambda_p}}{8s}
\, d {\varphi_\gamma}\, d {\cos\theta_\gamma}.
\ea
Combining the results of (\ref{dgam1b}) 
and (\ref{dgam2b}) in (\ref{dgamsep}) delivers 
the complete differential phase space volume 
in the independent variables $\varphi_\gamma$,
$\cos\theta_\gamma$, ${M^2_{f\gamma}}$, and $\cos\vartheta$: 

\ba
\label{dgamtot}
d{\Gamma^{(3)}} = \frac{1}{(2\pi)^4}
\, \frac{\sqrt{\lambda_2\lambda_p}}{(8s)^2}
\, d {\varphi_\gamma} \, d {\cos\theta_\gamma}
\, d{{M^2_{f\gamma}}}\, d {\cos\vartheta}.  
\ea
We have in mind a description of the QED bremsstrahlung 
effects to cross sections by flux functions which are
convoluted over the final state fermions' invariant 
mass squared $s'$ with effective Born observables.
For this, one can make the substitution:
\ba
\label{substitution}
{M^2_{f\gamma}},\quad \cos\theta_\gamma
\quad\longrightarrow\quad s',\quad V_2,
\ea
with the two independent variables $s'$ and $V_2$ which depends
linearly on $\cos\theta_\gamma$. Introducing $V_2$ as a new
variable of integration will also prove to be appropriate 
when applying a cut on the maximal acollinearity of the final 
state fermions. Using the relations
\ba 
\label{vartrans}
{M^2_{f\gamma}} &=& (p_1+p)^2 = -s'-V_2+s+m_f^2,
\\
V_2 &=& 2 p p_2 = 2\, p^0\, 
(p_2^0-\sqrt{{p_2^0}^2-m_f^2}\,\cos\theta_\gamma),
\ea
and expressing $p^0$ and $p_2^0$ by their invariant expressions
according to (\ref{momentap}) and (\ref{momentap2}), this 
finally delivers as differential phase volume in the
new variables:
\ba
\label{dgamtot2}
\rightarrow\quad 
d{\Gamma^{(3)}} &=& \frac{1}{(2\pi)^5}\,\frac{\pi}{16s}
\, d {\varphi_\gamma}\, d{V_2}\, d{s'}\, d {\cos\vartheta}
\\
&=&
\label{dgamtot3}
\frac{1}{2}\frac{s}{(4\pi)^4}
\, d {\varphi_\gamma}
\, d\left(\frac{V_2}{s}\right)
\, d\left(\frac{s'}{s}\right)
\, d {\cos\vartheta} 
\\
&=&
\label{dgamtot4}
\frac{1}{2}\frac{s}{(4\pi)^4}
\, d {\varphi_\gamma}\, d{x}\, d{R}\, d {\cos\vartheta}.
\ea 
In (\ref{dgamtot2}) to (\ref{dgamtot4}) 
we have introduced for the later calculation 
the more suitable dimensionless variables 
\ba
\label{Rxvar}
R\equiv \frac{s'}{s}\qquad\mbox{and}\qquad x\equiv \frac{V_2}{s}.
\ea

Before calculating cross section observables with (\ref{br09})
and (\ref{dgamtot2}), the constraints onto the integration 
variables $R$, $x$, $\cos\vartheta$, and $\varphi_\gamma$,
introduced by the kinematical cuts considered, have to be determined.

%%%%%%%%%%%%%%%%%%%%%%%%%%%%%%%%%%%%%%%%%%%%%%%%%%%%
\subsection*{Kinematical cuts and limits of integration 
\label{phcuts}
}
%%%%%%%%%%%%%%%%%%%%%%%%%%%%%%%%%%%%%%%%%%%%%%%%%%%%
%
The different kinematical cuts which shall be treated are 
stated below. They will not affect the azimuthal photon angle
$\varphi_\gamma$. It will be integrated over the complete 
angular range:
\ba
\label{phigamma}
\varphi_\gamma\, \epsilon\, [0;2\pi].
\ea
The kinematical boundaries defined by the cuts are also shown
in Fig.~\ref{dalitz} and Fig.~\ref{phasecuts} below.

\begin{itemize}
\item[1.] Cuts on the minimal and maximal scattering 
angle $\cos\vartheta$ (acceptance cut):
\ba
\label{ctcut}
-c \leq \cos\vartheta\leq c.
\ea  

\item[2.] A cut on the minimal fermion energies:
\\ 
For simplicity, equal cuts are applied on the minimal 
fermion energies $E_{\min} = E_{\min}^{f,\bar{f}}$.
With (\ref{momentap1}) and (\ref{momentap2}) 
we obtain for the variable $x$ 
kinematical limits depending on $R$:
\ba
\label{fermcut}
\left.
\begin{array}{lllll}
\mbox{For }f &:& p_1^0\geq  E_{\min} &,& p_1^0 = 
\frac{{\DS s-V_2}}{{\DS 2\sqrt{s}}}\\
\mbox{For }\bar{f} &:& p_2^0\geq E_{\min} &,& p_2^0 = 
\frac{{\DS s'+V_2}}{{\DS 2\sqrt{s}}}
\end{array}
\right\}\,
\longrightarrow
\\ 
R_E - R \leq {x} \leq 1 - R_E,
\quad R_E = \frac{2 E_{\min} }{\sqrt{s}}.
\ea
The upper and lower bounds on $x$ defined in (\ref{fermcut})
intersect with the kinematical boundaries $x^{\min,\max}(R)$, 
derived in (\ref{limV2exact}) ($x\equiv V_2/s$) in a point 
with the $R$-value $\bar{R}_E$:
\ba
\label{phcutcr}
\mbox{I. }&& 1 - R_E = {x}^{\max}(R) =
\frac{1}{2}(1-R)(1+\beta(R))
\nonumber\\
\mbox{II.}&& R_E - R = {x}^{\min}(R)  =
\frac{1}{2}(1-R)(1-\beta(R)),
\nonumber\\
\rightarrow&&
\bar{R}_E =  \frac{2\frac{m_f^2}{s}+(1-R_E)
\left(R_E+\sqrt{R_E^2-4\frac{m_f^2}{s}}\right)}
{2(1-R_E+\frac{m_f^2}{s})}.
\ea
Neglecting final state masses for $m_f^2\ll s$, 
we simply have: $\bar{R}_E = R_E$.
This automatically also defines a minimally allowed $R$-value 
$R_{cut}$ from the minimal energy cut:
\ba
\label{Rcut}
\max(4 m_f^2, R_{cut}) \leq R \leq 1 - \varepsilon
\quad\mbox{ with }\quad R_{cut}:= 2 R_E - 1.
\ea

\item[3.] A cut on the maximal acollinearity angle $\xi$:
\\
With $\bar{\xi}$ being a cut on the maximal acollinearity 
angle $\xi$ of the final state fermions, one can derive in 
analogy to (\ref{costhg}), (\ref{sinthg2}), and (\ref{limV2exact}) 
a further kinematical boundary for the phase space 
given by the variables $R$ and $x$:
\ba
\label{acol2}  
&& 
\cxi = \frac{|\vec{p}_1|^2+|\vec{p}_2|^2-|\vec{p}|^2}
{2|\vec{p}_1||\vec{p}_2|} =
\frac{\lambda_1+\lambda_2-\lambda_p}
{2\sqrt{\lambda_1\lambda_2}}\nonumber\\
\rightarrow & & \stxib \geq \stxi = \frac{1}{2}(1-\cxi) = 
\frac{\lambda_p-(\sqrt{\lambda_1}-\sqrt{\lambda_2})^2}
{4\sqrt{\lambda_1\lambda_2}}. 
\ea

This yields in terms of $x$, $R$, and $\bar{\xi}$ 
the relation ($m_f^2\ll s$):
\ba
\label{acol3}
\ctxihb\, x^2-\ctxihb\, (1-R)\, x+\stxib\,
 R\geq 0 .  
\ea   
Solving relation (\ref{acol3}) for $R$, we receive the following 
curve $R^{\min}(x)$ for minimal values of $R$ depending on the 
second variable $x$:

\ba
\label{acol4}
R\geq
{R}^{\min}(x):=\frac{x(1-x)(1-\stxib)}{x+(1-x)\stxib}.
\ea
This can be translated equivalently into a minimal 
and a maximal bound on $x$ depending on $R$. That we have either

\ba
\label{acol6}
{x}\leq x_{\xi}^{\min}(R)\qquad\mbox{or}\qquad {x}\geq
x_{\xi}^{\max}(R)\quad\mbox{for}\quad R\leq R_\xi ,
\ea
where 
\ba
\label{acol7}
x_{\xi}^{\max,\min}(R) = \frac{1-R}{2}\,\left[1\pm
\sqrt{1-\frac{R}{R_\xi}\,\frac{(1-R_\xi)^2}{(1-R)^2}}\right],
\ea
if $R\leq R_\xi$ where $R_\xi$ defines the $R$-value of
the turning point $P_t$ of the acollinearity bound, 
depicted in Fig.~\ref{dalitz}.
\ba
R_\xi &=& \frac{1-\sin\left(\bar{\xi}/2\right)}
{1+\sin\left(\bar{\xi}/2\right)},
\label{acol8}
\\
P_t &\equiv&
[R_t,x_{t}] =
\left[ R_\xi, \frac{1}{2}(1-R_\xi) \right].
\label{pt}
\ea
For $R > R_\xi$, the variable $x$ is limited by 
the relation in (\ref{limV2exact})
with $x\equiv V_2/s$.

\item[4.] A cut on the minimal invariant mass squared $s'$ 
of the fermions:
\\
This cut can be trivially applied in addition to the above 
mentioned cuts and introduces a new minimum $R$-value $\bar{R}$:
\ba
\label{sprcut}
\max(4 m_f^2, R_{cut}, \bar{R}) \leq R \leq 1 - \varepsilon.
\ea
It is equivalent to a cut on the maximal energy $\bar{E}_\gamma$
of the emitted photon.
\ba
\label{Ephcut}
E_\gamma\leq\bar{E}_\gamma\quad\leftrightarrow\quad
\bar{R} =  1 - \frac{2\bar{E}_\gamma}{\sqrt{s}}.
\ea
\end{itemize}
For the allowed complete phase space region, 
depending on the numerical values of the cuts, two different
cases may arise: 

\ba
\label{accases}
R_{cut} < R_\xi\quad\mbox{or}\quad R_{cut} \geq R_\xi,
\ea
where $R_{cut}$ was defined in (\ref{Rcut}).
The two corresponding Dalitz plots are shown in
Fig.~\ref{phasecuts}.
%
%--------------------------------------------------------------------------------
\begin{figure}[bhtp]
\begin{flushleft}
\begin{tabular}{ll}
\hspace*{-1.5cm}
\mbox{
 \epsfig{file=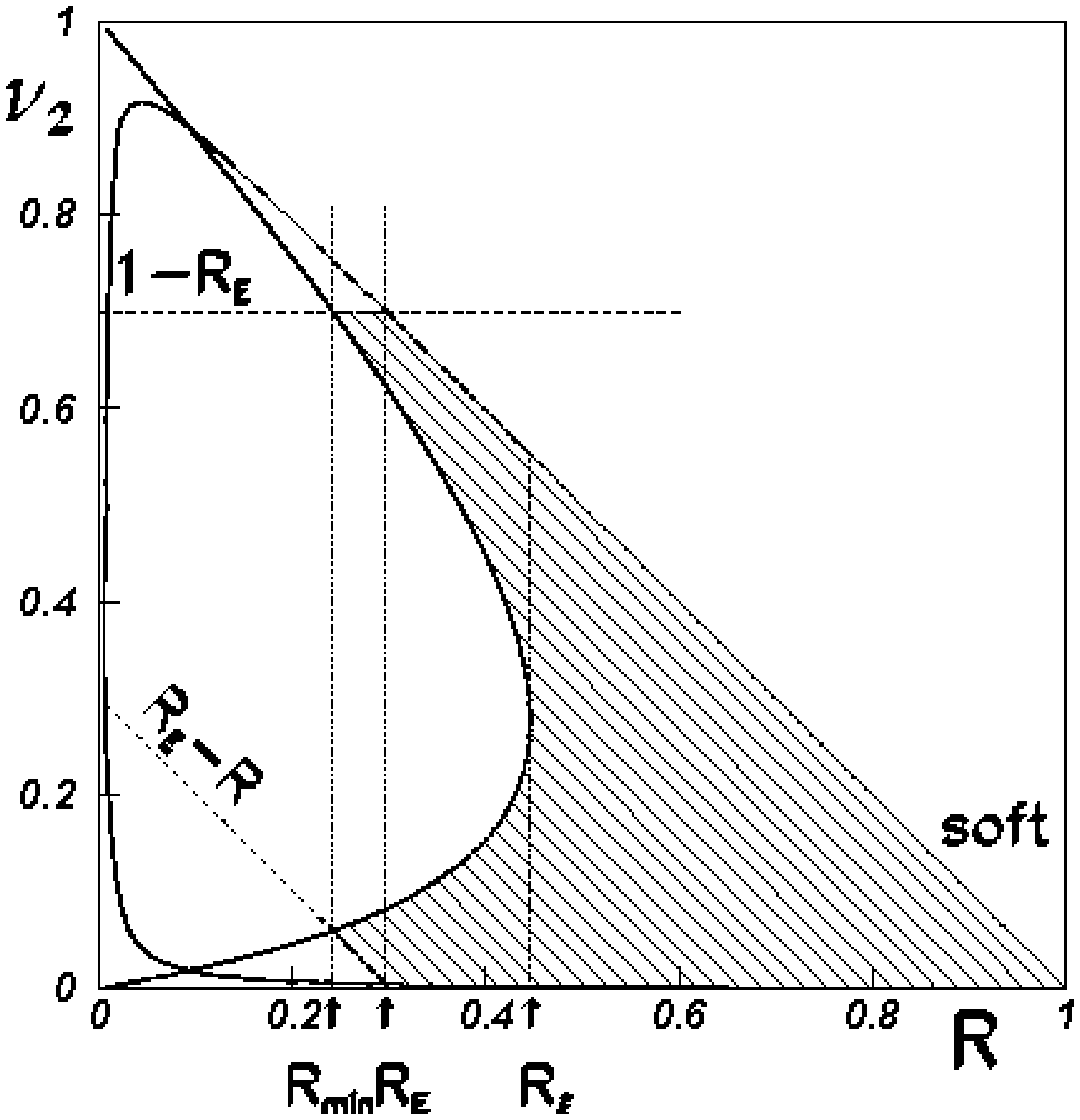,width=8.0cm}}
&
\hspace*{-1cm}
\mbox{
 \epsfig{file=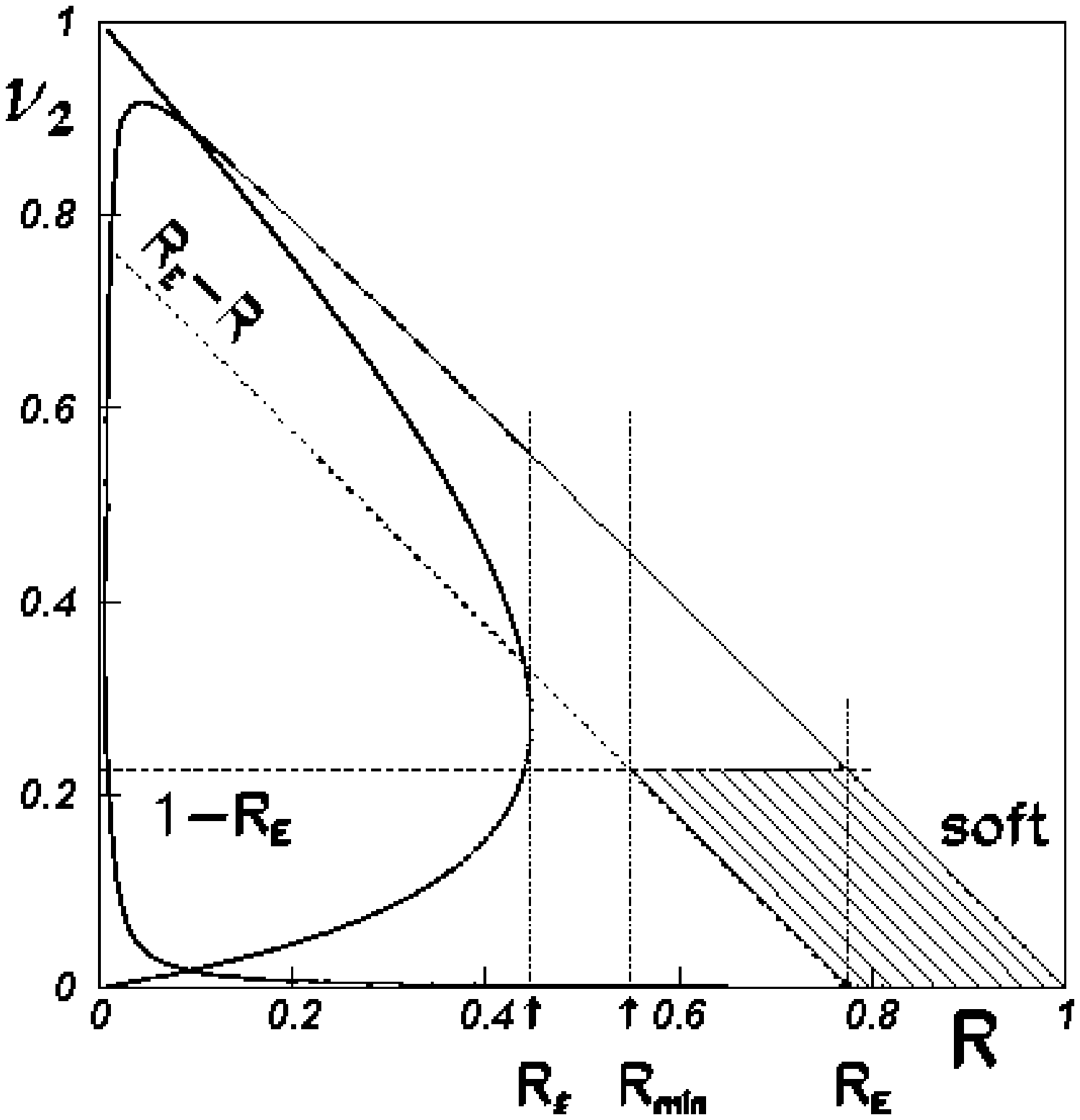,width=8.0cm}}
\end{tabular}
\end{flushleft}
\caption[Different cases in phase space with acollinearity cut]
{\sf
Phase space with acollinearity cut $R_\xi$ and equal muon 
energy cuts $R_E$: a. $R_{cut}<R_\xi$, b. $R_{cut}\geq R_\xi$;
$\nu_2:=V_2/s\equiv x$.
\label{phasecuts}
}
\end{figure}
%--------------------------------------------------------------------------------
%
In the first case, $R_{cut} < R_\xi$, the acollinearity cut has an effect
and the absolute minimum of $R$ is given by the $R$-value of the 
intersection of the kinematical bounds inflicted by the
maximal acollinearity and minimal energy cut.
\ba
R_{\min} = R_{E}\left(1-\frac{\sin^2 (\bar{\xi}/2)}
{1-R_E\cos^2 (\bar{\xi}/2)}\right).
\label{Rmin}
\ea
The complete phase space $\Gamma=\int d{\Gamma^{(3)}}$ can then
be divided into three parts where each one depends separately
on one of the applied cuts:
\ba
\label{threereg}
\Gamma &=& \Gamma_{I}\quad +\quad\Gamma_{II}\quad -\quad\Gamma_{III}
\\
\label{regions1}
&=&\int_{\bar{R}_E}^{1} d{R} \int_{{x}^{\min}(R)}^{{x}^{\max}(R)} d{{x}}
+\int_{R_{\min}}^{\bar{R}_E} d{R} \int_{R_E-R}^{1-R_E} d{{x}}
-\int_{R_{\min}}^{R_\xi} d{R} \int_{x_{\xi}^{\min}(R)}
^{x_{\xi}^{\max}(R)} d{{x}}.
\nonumber\\
\ea
In the other case, $R_{cut}\geq R_\xi$, the energy cuts are so stringent
that the acollinearity cut has no influence. The minimum value of $R$ is
\ba
R_{\min} = R_{cut},
\label{rmin2} 
\ea
and the integration region is simplified to a trapezoid: 
\ba
\label{tworeg}
\Gamma\quad =\quad\Gamma_{I}\,+\,\Gamma_{II}
\quad =\quad\int_{\bar{R}_E}^{1} d{R}
\int_{{x}^{\min}(R)}^{{x}^{\max}(R)} d{{x}} 
+\int_{R_{\min}}^{\bar{R}_E} d{R} \int_{R_E-R}^{1-R_E} d{{x}}.
\ea

These findings have been summarized in Table \ref{nulimits}, 
introducing the parameter $A=A_a(R)$, $a=m,E,\xi$,
as defined in (\ref{A1}), (\ref{A2}), and (\ref{A3}):
\begin{center}
$A_m = [1- 4 m_f^2 / (s R)]^{\frac{1}{2}}$
\hspace{1cm},\hspace{1cm}
$A_E = (1+R-2\,R_E)/(1-R)$,\\
and
$\qquad A_\xi = [ 1 - R (1-R_{\xi})^2 / (R_{\xi}(1-R)^2) ]^{\frac{1}{2}}$.
\end{center}
Neglecting the final state mass for the integration limits
of regions II and III, $m_f^2\ll s$, allows to set 
${\bar{R}_E}\approx R_E$ and is justified 
as they will only play a role for integrands proportional to
$1/(1-R)$ for $R\to 1$, i.e.~in region I.

The above relations are independent of the scattering angle and thus,
as mentioned, compatible with the angular acceptance cut.
The acollinearity cut has an indirect influence on the acceptance cut.
It is easy to see that the maximal scattering angle of 
the other final state fermion becomes limited by an acollinearity 
cut, i.e. the scattering angle of the second fermion is limited to
$[-({\xi}+\vartheta^{\max}),({\xi}+\vartheta^{\max})]$.
\begin{table}[htp]
\begin{center}
\begin{tabular}{|lccc|}
\hline
&&&\\
Region I:
&$\bar{R}_E\le R\le 1-\varepsilon$,&${x}^{\min}(R)\le{x}\le{x}^{\max}(R)$,&\\
&&&\\
&${x}^{\min}(R)=(1-R) d(R)$,&${x}^{\max}(R)=(1-R) (1-d(R))$,&\\
&&&\\
&$d(R) = (1-A_m(R))/2$,&$1-d(R) = (1+A_m(R))/2$,
&\\
&&&\\
\hline\hline
&&&\\
Region II:
&$R_{\min}\le R\le\bar{R}_E$,&${x}_E^{\min}(R)\le{x}\le{x}_E^{\max}(R)$,&\\
&&&\\
&${x}_E^{\min}(R)=(1-R) d_E(R)$,&${x}_E^{\max}(R)=(1-R) (1-d_E(R))$,&\\
&&&\\
&$d_E(R) = (1-A_E(R))/2$,&$1-d_E(R) = (1+A_E(R))/2$,
&\\
&&&\\
\hline\hline
&&&\\
Region III:
&$R_{\min}\le R\le R_{\xi}$,&${x}_{\xi}^{\min}(R)\le{x}\le{x}_{\xi}^{\max}(R)$,&\\
&&&\\ 
&${x}_{\xi}^{\min}(R)=(1-R) d_{\xi}(R)$,
&${x}_{\xi}^{\max}(R)=(1-R) (1-d_{\xi}(R))$,&\\
&&&\\
&$d_{\xi}(R) = (1-A_\xi(R))/2$,&$1-d_{\xi}(R) = (1+A_\xi(R))/2$.
&\\
&&&\\
\hline
\end{tabular}
\caption[Regions of phase space for an acollinearity cut]
{\sf
Regions of phase space with cuts on maximal acollinearity 
and minimal energy of the final state fermions.
\label{nulimits}
}
\end{center}
\end{table}

\newpage

%%%%%%%%%%%%%%%%%%%%%%%%%%%%%%%%%%%%%%%%%%%%%%%%%%%%%%%%%%%%%%%%%%%%%%%%
\chapter{Hard bremsstrahlung corrections 
\label{hardbrem}
}
%%%%%%%%%%%%%%%%%%%%%%%%%%%%%%%%%%%%%%%%%%%%%%%%%%%%%%%%%%%%%%%%%%%%%%%%

%%%%%%%%%%%%%%%%%%%%%%%%%%%%%%%%%%%%%%
\section{Initial state radiation
\label{hardini}} 
%%%%%%%%%%%%%%%%%%%%%%%%%%%%%%%%%%%%%%
%
%%%%%%%%%%%%%%%%%%%%%%%%%%%%%%%%%%%%%%%%%%%%%%%%%%
\subsection*{Matrix element and differential cross section
\label{hardmatini}} 
%%%%%%%%%%%%%%%%%%%%%%%%%%%%%%%%%%%%%%%%%%%%%%%%%%
%
The matrix element for real initial state bremsstrahlung 
is given below:
\ba
\label{dhardmatini}
{\cal M}_{ini} &=& (2 \pi)^4 
\delta^{(4)}(k_1 + k_2 - p_1 - p_2 - p)
\nonumber\\
&&\Bigl[(2 \pi)^{15} \, 2 k_1^0 \, 2 k_2^0 \, 2 p_1^0 \, p_2^0 \, 
 2 p^0 \Bigr]^{-\frac{1}{2}} \, {\tt{M}}_{ini} .
\ea
The initial state differential cross section with
its contributions from the three phase space 
regions I, II, and III created by the cuts 
(see Appendix \ref{threephase}) is: 

\ba
\label{dhardini1a}
\frac{d \sigma^{hard}_{ini}}{d \cos{\vartheta} } =
\frac{1}{2}\frac{s}{(4 \pi)^4}
\Biggl[\int\limits_{I}+\int\limits_{II} - \int\limits_{III} \Biggr]
\frac{1}{4}\,\frac{\sum_{spin} |M_{ini}|^2}{2\,s\beta_0}
\, d {\varphi}_{\gamma}\, {d\,x}\, {d\,R}.
\ea
Introducing the coupling constant 
combinations ${\cal V}(s')$ and ${\cal A}(s')$,
with the neutral current couplings 
$v_e$, $a_e$, $v_f$, $a_f$ and $Z$ propagator 
$\chi(s')$ defined in (\ref{b04}) to (\ref{b05b}),
\ba
\label{V}
{\cal V}(s) &=& {Q}^2_e {Q}^2_f 
  + 2 Q_e Q_f \, v_e\, v_f\,{\Re}e\, \chi(s') 
  + [v_e^2+a_e^2]\, [v_f^2+a_f^2]\, {\mid \chi(s') \mid}^2,      
\\
\nonumber\\
\label{A}
{\cal A}(s) &=&  
    2 Q_e Q_f\, a_e\, a_f\,{\Re}e\, \chi(s') 
  + 4\, v_e\, a_e\, v_f\, a_f\, {\mid\chi(s')\mid}^2,
\ea 
the squared amplitude $|M_{ini}|^2$,
including all mass terms reads ($e^2 = 4\pi\alpha$):

\vfill\eject
%---------------------------------------------------------------------
\ba
\label{dhardini1b}
|M_{ini}|^2 &=& 
(4\pi\alpha)^3
\Biggl\{
\frac{{\cal V}(s')}{s {s'}} 
\Biggl[
-\frac{2\, m_e^2}{Z_1^2} 
\left( 2\, T^2 - 2\, T\, s' + {s'}^2 + 2\,{m_f^2}\,s' \right)
 \nonumber\\
&&-
\frac{2\, m_e^2}{Z_2^2} 
\left( 2\, U^2 - 2\, U\, s' + {s'}^2 + 2\,{m_f^2}\,s' \right)
\nonumber\\
&&+
\frac{s'}{Z_1 Z_2} 
\left( 2\, T^2 + 2\, U^2 + 2\, {s'}^2 - 2\, (T + U)\, s' 
+ 4\,{m_f^2}\,s' \right)
\nonumber\\
&&+
\frac{1}{Z_1} 
\left( s\,s' - 2\, U\, s' + {s'}^2 + 2\,{m_f^2}\,( s + s' ) \right)
 \nonumber\\
&&+
\frac{1}{Z_2} 
\left( s\,s' - 2\, T\, s' + {s'}^2 + 2\,{m_f^2}\,( s + s' ) \right)
- 2\,s' - 4\,{m_f}^2
\Biggr]
\nonumber\\
&+&
\frac{{\cal A}(s)}{s {s'}}  
\Biggl[
-\frac{2\, m_e^2\,s'\, (s' - 2\,T) }{Z_1^2}
+\frac{2\, m_e^2\,s'\, (s' - 2\,U) }{Z_2^2}
\nonumber\\
&&
-\frac{2\, (T - U)\, {s'}^2}{Z_1 Z_2} 
- \frac{s' ( s + s' - 2\, U)}{Z_1}
+ \frac{s' ( s + s' - 2\, T)}{Z_2} 
\Biggr]
\nonumber\\
&+&
\frac{2\,m_f^2}{s {s'}}\,a_f^2\, ( v_e^2 + a_e^2 )\,\mid\chi(s')^2\mid
\nonumber\\
&&
\cdot\Biggl[
\frac{4\, m_e^2\, s'}{Z_1^2} + \frac{4\, m_e^2\, s'}{Z_2^2} 
- \frac{4\,{s'}^2}{Z_1 Z_2} - \frac{2\,(s + s')}{Z_1} 
- \frac{2\,(s + s')}{Z_2} 
\Biggr]
\Biggr\}.
\ea
The bremsstrahlung kinematic invariants 
$Z_1$, $Z_2$, $V_1$, $V_2$, $T$, and $U$ 
are taken from (\ref{invs}) to (\ref{invv2}) 
and (\ref{T}) and (\ref{U}). 
It is important to note that the coupling constant 
functions ${\cal V}(s')$ and ${\cal A}(s')$
in (\ref{dhardini1b}) depend on the final state invariant
mass squared $s'$. This is immediately clear due to
the emission of the hard photon from the initial state,
with the propagators of the effective Born terms
then depending on $s'$, and not on $s$.

The integration shall be performed analytically over 
the variables $\varphi_{\gamma}$, $x\equiv V_2/s$,
and $\cos\vartheta$ to deliver QED flux functions
for hard photon emission. A convolution integral of 
these flux functions with effective Born terms 
over $R\equiv s'/s$ is kept for numerical integration 
(see (\ref{crossini}) and (\ref{generic_zf})).

The hard photon radiator functions $H^{ini}_{T,FB}(R)$
together with the soft and virtual correction terms $S_{ini}$ for the 
totally integrated results $\sigma_T^{ini}$ and $A_{FB}^{ini}$
have been presented in the main part of this dissertation
in Section \ref{sub_lep1slc_formacol}. 
Here, we just want to present the basic
steps of the analytical integration and describe especially
the treatment of mass singularities arising during the calculation:
The denominators $1/Z_i$, $i=1,2$ appearing
in (\ref{dhardini1b}) are the propagators of the electron 
and positron and will determine the logarithmic 
structure of the final integrated results for 
$d{\sigma^{hard}_{ini}}/{d \cos{\vartheta} }$ 
and $\sigma^{hard}_{ini}$. These propagator terms 
deliver artificial singularities, i.e. mass 
singularities, when neglecting the initial state 
mass $m_e$ for the analytical integration over
$\cos\vartheta$ 
(see Section \ref{sub_lep1slc_mass_sing}). 
These unphysical 
singularities will however cancel each other when 
combining all integrated terms in the completely
integrated radiator $H^{ini}_{T,FB}(R)$.
This shall be presented and discussed below. 

%%%%%%%%%%%%%%%%%%%%%%%%%%%%%%%%%%%%%%%%%%%%%%%%%%
\subsection*{Integration over $\varphi_{\gamma}$
\label{phigtab}} 
%%%%%%%%%%%%%%%%%%%%%%%%%%%%%%%%%%%%%%%%%%%%%%%%%%
%
Performing the first integration over $\varphi_{\gamma}$
in (\ref{dhardini1a}), only the denominators $1/Z_i$, $i=1,2$ 
are relevant. They depend linearly on $\cos\varphi_{\gamma}$.
\vspace*{0.2cm}
\begin{center}
{\large  Table of integrals:}
\end{center}
\ba
\label{intphig1}
\frac{1}{2\pi}\int\limits^{2\pi}_0 d{{\varphi_{\gamma}}} &=& 1,
\\
\label{intphig2}
\frac{1}{2\pi}\int\limits^{2\pi}_0\frac{d{{\varphi_{\gamma}}}}{Z_i} &=& 
\frac{1}{2\pi}\int\limits^{2\pi}_0\frac{d{{\varphi_{\gamma}}}}
{A_i\pm b\cos{\varphi_{\gamma}}} =
\frac{1}{\sqrt{A^2_i-B^2}}, 
\\
\label{intphig3}
\frac{1}{2\pi}\int\limits^{2\pi}_0 \frac{d{{\varphi_{\gamma}}}}{Z_i^2} &=&
\frac{1}{2\pi}\int\limits^{2\pi}_0\frac{d{{\varphi_{\gamma}}}}
{(A_i\pm B\cos{\varphi_{\gamma}})^2} =
\frac{A_i}{(A_i^2-B^2)^{\frac{3}{2}}}, 
\ea
with
\ba
\label{intphigex}
\frac{1}{2\pi}\int\limits^{2\pi}_0\frac{d{{\varphi_{\gamma}}}}
{A_i\pm B\cos{\varphi_{\gamma}}} 
&=&
\left.\frac{1}{2\pi}\frac{2}{\sqrt{A_i^2-B^2}}\arctan\frac{\sqrt{A_i^2-B^2}
\tan(\varphi_{\gamma}/2)}{A\pm B}\right|^{2\pi}_0 
\nonumber\\
&=& \frac{1}{\sqrt{A_i^2-B^2}}, 
\\
\frac{1}{2\pi}\int\limits^{2\pi}_0\frac{d{{\varphi_{\gamma}}}}
{(A_i\pm B\cos{\varphi_{\gamma}})^2} 
&=&
\frac{1}{2\pi}\frac{A_i}{A_i^2-B^2}
\int\limits^{2\pi}_0\frac{d{{\varphi_{\gamma}}}}
{A_i\pm B\cos{\varphi_{\gamma}}} 
= \frac{A_i}{(A_i^2-B^2)^{\frac{3}{2}}}.
\nonumber\\
\ea
The coefficients $A_i$, $i=1,2$ and $B$ depend on $\cos\vartheta_{\gamma}$, or
equivalently $V_2$, and $s'$:
\ba
\label{coefthgai}
A_i &=& \frac{s}{2}(1-R) [1\pm\beta_0\cos\vartheta\cos\vartheta_{\gamma}]
\quad (i=1\to +\,,\,2\to -),
\\
\label{coefthgb}
B &=& \frac{s}{2}(1-R)\beta_0\sin\vartheta\sin\vartheta_{\gamma},
\ea
with
\ba
\label{rvbeta}
R=\frac{s'}{s},\quad v= 1-R,
\quad \beta_0 = \sqrt{1-\frac{4\,m_e^2}{s}},
\quad \beta=\sqrt{1-\frac{4\,m_f^2}{s\,R}}.
\ea

%%%%%%%%%%%%%%%%%%%%%%%%%%%%%%%%%%%%%%%%%%%%%%%%%%
\subsection*{Integration over $V_2$
\label{vtab}} 
%%%%%%%%%%%%%%%%%%%%%%%%%%%%%%%%%%%%%%%%%%%%%%%%%%
%
The dependence of the integrals in (\ref{intphig1}) to 
(\ref{intphig3}) on $\cos\vartheta_{\gamma}$ can be transformed 
into a dependence on $V_2$ and $s'$
using the identities from (\ref{costhg}) and (\ref{sinthg2})
with the definitions (\ref{lambdas}) to (\ref{lambdap}):
\ba
\label{tgv2}
\cos\vartheta_{\gamma} = \frac{\lambda_1-\lambda_2-\lambda_p}
{2\sqrt{\lambda_2}\sqrt{\lambda_p}},\qquad
\sin\vartheta_{\gamma} = \frac{\sqrt{-\lambda(\lambda_1,\lambda_2,\lambda_p)}}
{2\sqrt{\lambda_2}\sqrt{\lambda_p}}.
\ea
With (\ref{coefthgai}) 
and (\ref{coefthgb}) and (\ref{tgv2}) one gets the following results 
depending on $V_2$ and $s'$:
\ba
\label{intphigv2}
\frac{1}{2\pi}\int\limits^{2\pi}_0\frac{d{{\varphi_{\gamma}}}}{Z_i} = 
\frac{\sqrt{\lambda_2}}{C_i^{\frac{1}{2}}}\quad,\quad
\frac{1}{2\pi}\int\limits^{2\pi}_0\frac{d{{\varphi_{\gamma}}}}{Z_i^2} = 
\frac{\sqrt{\lambda_2} B_i}{C_i^{\frac{3}{2}}},
\ea
with 
\ba
\label{coefv2}
B_i &=& \frac{1}{2}\left[(s-s')\lambda_2\pm(2 s s'-(V_2+s')(s+s'))
\sqrt{\lambda_2}\beta_0\cos\vartheta\right],
\label{coefv2Bi}
\\
C_i &=& \frac{1}{4}[2 s s'-(V_2+s')(s+s')\pm (s-s')
\sqrt{\lambda_2}\beta_0\cos\vartheta]^2
\nonumber\\
&+&4 m_e^2 [s'V_2(s-s'-V_2)-(s-s')^2 m_f^2],
\label{coefv2Ci}
\\
\label{coefv2lambda}
\lambda_2 &=& (s' + V_2)^2-4\,m_f^2\,s 
\approx (s' + V_2)^2
\quad \mbox{for}\quad m_f \approx 0.
\ea

Rewriting (\ref{coefv2Bi}) and (\ref{coefv2Ci}) in a more suitable 
form for the integration over $V_2$ and later over $\cos\vartheta$,
we get:

\ba
\label{coefv22}
C_i(x, R, \cos\vartheta) 
&=& s^2 a_i x^2 - 2 s b_i x + c_i,
\label{coefv2Ci2}
\\
B_i(x, R, \cos\vartheta) 
&=& s^3 \left[ (x + R)^2 y_i\pm (x + R ) R\beta_0\cos\vartheta\right],
\label{coefv2Bi2}
\\
a_i(R,\cos\vartheta) &=& s^2 \left( z_i^2-(1-\be^2) R \right),
\label{coefv2ai}
\\
b_i(R,\cos\vartheta) &=& s^3 \left[ R z_i (1 - z_i)-\frac{1}{2}(1-\be^2) {R}(1 - R) \right],
\label{coefv2bi}
\\
c_i(R,\cos\vartheta) &=& s^4 R^2 (1 - z_i)^2,
\label{coefv2c0}
\\
\nonumber\\
{z_1}(R,\cos\vartheta) &=& \frac{1-\beta_0\cos\vartheta}{2}+R
\frac{1+\beta_0\cos\vartheta}{2},
\label{coefv2z1}
\\
{z_2}(R,\cos\vartheta) &=& \frac{1+\beta_0\cos\vartheta}{2}+R
\frac{1-\beta_0\cos\vartheta}{2},
\label{coefv2z2}
\\
y_1(R,\cos\vartheta) &=& \frac{1-\beta_0\cos\vartheta}{2}-R
\frac{1+\beta_0\cos\vartheta}{2},
\label{coefv2y1}
\\
y_2(R,\cos\vartheta) &=& \frac{1+\beta_0\cos\vartheta}{2}-R
\frac{1-\beta_0\cos\vartheta}{2}.
\label{coefv2y2}
\ea
Alternatively, one can also write:
\ba
\label{coefv2g}
y_1 &=& \frac{1}{2}\left[(1-R)-(1+R)\beta_0\cos\vartheta\right]
,\, {y_2}(R,\cos\vartheta) = y_1(R,-\cos\vartheta),
\\
{z_1} &=& \frac{1}{2}\left[(1+R)-(1-R)\beta_0\cos\vartheta\right]
,\, {z_2}(R,\cos\vartheta) = {z_1}(R,-\cos\vartheta),
\\
\eta_0^2&:=& 1-\beta_0^2 =\frac{4\,m_e^2}{s}.
\ea
{From} (\ref{coefv2Ci2}) to (\ref{coefv2y2}), it is immediately obvious 
that the possible integrands are symmetric to each other with respect
to a change of sign of $\cos\vartheta$.

In (\ref{coefv2lambda}), the final state mass $m_f$ is neglected.
This is justified as for initial state bremsstrahlung only 
final state mass terms proportional to $m_f^2/s$ may arise
which can be neglected having in mind applications at energies 
around or above the $Z$ boson resonance, $m_f^2 \ll M_Z^2 \le s$. 
This approximation also greatly simplifies the analytical 
integrations over $x$ and $\cos\vartheta$. 

We will perform an integration over $x$ in the three different 
regions of phase space I, II, and III , suggested by the cuts 
and depending on the variables $x$ and $R$. This was summarized
in Table (\ref{nulimits}). 
\ba
\label{phspints}
\frac{d \sigma^{hard}_{ini_I}}{d \cos{\vartheta}}
&=&\int\limits_{\bar{R}_E}^{1} d{R} 
\,\int\limits_{x_m^{\min}(R)}^{x_m^{\max}(R)}\, d{{x}}\,
\frac{d \sigma^{hard}_{ini}}{d R\, d x\, d \cos{\vartheta}},
\\
\frac{d \sigma^{hard}_{ini_{II}}}{d \cos{\vartheta}}
&=&\int\limits_{R_{\min}}^{\bar{R}_E} d{R} 
\,\int\limits_{x_E^{\max}(R)}^{x_E^{\min}(R)}\, d{{x}}\,
\frac{d \sigma^{hard}_{ini}}{d R\, d x\, d \cos{\vartheta}},
\\
\frac{d \sigma^{hard}_{ini_{III}}}{d \cos{\vartheta}}
&=&\int\limits_{R_{\min}}^{R_\xi} d{R} 
\,\int\limits_{x_{\xi}^{\min}(R)}^{x_{\xi}^{\max}(R)}\, d{{x}}\,
\frac{d \sigma^{hard}_{ini}}{d R\, d x\, d \cos{\vartheta}}.
\ea
All three regions of phase space can be described 
equivalently by introducing a general, cut dependent 
parameter $0\leq A_a(R)\leq 1$, $a=m,E,\xi$ 
defined in (\ref{A1}), (\ref{A2}), and (\ref{A3}). 
The limits of integration $x_a^{\max}(R)$
and $x_a^{\min}$ can then be treated according to (\ref{nulimits}) 
in the general form:
\ba
\label{xlimits}
{x}_a^{\max,\min}(R) = 
\frac{1}{2}\,(1-R)\,\left[1 \pm A_a(R)\right], a = m,E,\xi.
\ea

In order to obtain short analytical formulae which can later 
be integrated analytically over $\cos\vartheta$, we will have  
to make different approximations for small initial state masses, 
$m_e^2\ll s$. This will essentially define three intervals for 
$\cos\vartheta$ with different analytical expressions for 
the hard photon radiators. The boundaries of these intervals  
depend on $R$ and the cut parameter $A_a=A_a(R)$ defined 
in (\ref{A1}), (\ref{A2}), and (\ref{A3}). 
For brevity, we will use from now on $A\equiv A_a$ and only 
consider the integrands (\ref{intphigv2}) with index `1', as 
the results with index `2' can trivially be obtained by the 
substitution `$\cos\vartheta \to -\cos\vartheta$'
(see (\ref{coefv2Ci2}) to (\ref{coefv2y2})).

%%%%%%%%%%%%%%%%%%%%%%%%%%%%%%%%%%%%%%%%%%%%%%%%%%%%%%%%%%%%%%%%%%
\subsubsection*{Integrands proportional to $C_i^{-\frac{1}{2}}$}
\label{vtabZ1g} 
%%%%%%%%%%%%%%%%%%%%%%%%%%%%%%%%%%%%%%%%%%%%%%%%%%%%%%%%%%%%%%%%%%
%
We have as basic set of integrals arising from 
integrands proportional to $C_i^{-\frac{1}{2}}$:
\ba
\label{I01def} 
I^0_1 &=& \int\limits^{V_2^{\max}}_{V_2^{\min}}\frac{d{V_2}}
{\sqrt{C_1(V_2,s',\cos\vartheta)}} 
= s\,\int\limits^{{x}^{\max}}_{{x}^{\min}}\frac{d{{x}}}
{\sqrt{C_1(x,R,\cos\vartheta)}},
\\
\nonumber\\
\label{I1def} 
{I}^i_1 &=& \int\limits^{V_2^{\max}}_{V_2^{\min}}\frac{d{V_2}}
{\sqrt{C_1(V_2,s',\cos\vartheta)}} (V_2+{s'})^i
=s^{i+1}\,\int\limits^{{x}^{\max}}_{{x}^{\min}}\frac{d{{x}}}
{\sqrt{C_1(x,R,\cos\vartheta)}}
({x}+R)^i,
\nonumber\\
&&\hspace*{9cm} i=1,2,3.
\ea
We will start with the basic integral $I^0_1$:

\ba
\label{I01g} 
I^0_1 &=&
\frac{1}{\sqrt{a_1}}\,\left.
\ln\left[\sqrt{a_1}\, C_1^{\frac{1}{2}}(x,R,\cos\vartheta)
+s\, a_1\,{x}-b_1\right]\right|^{{x}^{\max}}_{{x}^{\min}}.
\ea
With (\ref{coefv2g}) and (\ref{coefv2}) we can express 
e.g.~$C_1({x}^{\max},R,\cos\vartheta)$ and $C_1({x}^{\min},R,\cos\vartheta)$ as: 
\ba
\label{c1max}
C_1({x}^{\max},R,\cos\vartheta) &=& \frac{1}{4} s^4 (1-R)^2\,
\left[(y_1+A\,{z_1})^2+R\, (1-A^2)\,\eta_0^2\right],
\\
\label{c1min}
C_1({x}^{\min},R,\cos\vartheta) &=& \frac{1}{4} s^4 (1-R)^2\,
\left[(y_1-A\,{z_1})^2+R\, (1-A^2)\,\eta_0^2\right]. 
\ea
The function $(y_1\pm A\,{z_1})$ 
can change sign, which means that 
$C_1({x}^{\max,\min},R,\cos\vartheta)$ can be of the order 
$O(\eta_0^2=\frac{4\,m_e^2}{s})$, if $y_1\pm A\,{z_1}$ disappears. As we
want to integrate all angles of phase space analytically and 
fermion masses do not play a role at the center-of-mass
energies in mind, we will neglect all non-logarithmic 
contributions of $m_e^2$.
So, the exact analytical result for $I_1^0$ from (\ref{I01g})
can first be summarized as follows:
\ba
\label{I01gex}
I^0_1 &=& \frac{1}{s}\,\frac{1}{\sqrt{{z_1}^2-R\eta_0^2}}
\,\ln\left(\frac{Z}{N}\right).
\ea
Depending on the signs of $(y_1\pm A\,{z_1})$, it is useful 
to apply the following expressions for $\ln{Z}$ and 
$\ln{N}$ because otherwise the nominator $Z$ or denominator $N$ 
will lead to an exact zero under the logarithm, when neglecting
small mass terms of $O(\eta_0^2)$. This occurs for (\ref{yaz}): 
\ba
\label{zeros}
R\, (1-A(R)^2)\eta_0^2 = 
\left[{y_1}(R,\cos\vartheta)
\pm A(R){z_1}(R,\cos\vartheta)\right]^2,
\ea
which defines the following limiting parameters:
\ba
\label{c1p}
c_1^{+}(R) &=& -\frac{1-R+A(R) (1+R)}{1+R+A(R)(1-R)},
\\
\label{c1m}
c_1^{-}(R) &=& -\frac{1-R-A(R) (1+R)}{1+R-A(R) (1-R)},
\\
\label{c2p}
c_2^{+}(R) &=& -c_1^{+}(R), 
\\
\label{c2m}
c_2^{-}(R) &=& -c_1^{-}(R), 
\ea
with $c_1^{+} \le c_1^{-}$ and $c_2^{-} \le c_2^{+}$.
These values were already given in (\ref{c1+}) to (\ref{c2+})
but for convenience are presented here again.
Therefore we use:
\\
\noindent a. for $y_1+A{z_1}>0$:
\ba
\label{I01gex1}
\ln{Z}\hspace*{-0.2cm} &=&\hspace*{-0.2cm}\left\{
\begin{array}{ll}
&\hspace*{-0.2cm}\ln\left[A\sqrt{{z_1}^2-R\,\eta_0^2}
+\sqrt{(y_1+A\,{z_1})^2+R\, (1-A^2)\,\eta_0^2}
+\sqrt{y_1^2+R\,\eta_0^2}\right]
\\&
\\
+&\hspace*{-0.2cm}\ln\left[A\sqrt{{z_1}^2-R\,\eta_0^2}
+\sqrt{(y_1+A\,{z_1})^2+R\, (1-A^2)\,\eta_0^2}
-\sqrt{y_1^2+R\,\eta_0^2}\right]
\\&
\\
-&\hspace*{-0.2cm}\ln(2A);
\end{array}
\right.
\nonumber\\
\ea
b. for $y_1+A{z_1}\leq 0$:
\ba
\label{I01gex2}
\ln{Z}\hspace*{-0.2cm} &=&\hspace*{-0.2cm}\left\{
\begin{array}{ll}
-&\hspace*{-0.2cm}\ln\left[\sqrt{y_1^2+R\,\eta_0^2}
-\sqrt{(y_1+A\,{z_1})^2+R\, (1-A^2)\,\eta_0^2}
+A\sqrt{{z_1}^2-R\,\eta_0^2}\right]
\\&
\\
-&\hspace*{-0.2cm}\ln\left[\sqrt{y_1^2+R\,\eta_0^2}
+\sqrt{(y_1+A\,{z_1})^2+R\, (1-A^2)\,\eta_0^2}
-A\sqrt{{z_1}^2-R\,\eta_0^2}\right]
\\&
\\
+&\hspace*{-0.2cm}\ln(2A)+\ln\left[R^2\eta_0^2
\beta_0^2(1-\cos^2\vartheta)\right];
\end{array}
\right.
\nonumber\\
\ea
c. for $y_1-A{z_1} \geq 0$:
\ba
\label{I01gex3}
\ln{N}\hspace*{-0.2cm} &=&\hspace*{-0.2cm}\left\{
\begin{array}{ll}
&\hspace*{-0.2cm}\ln\left[-A\sqrt{{z_1}^2-R\,\eta_0^2}
+\sqrt{(y_1-A\,{z_1})^2+R\, (1-A^2)\,\eta_0^2}
+\sqrt{y_1^2+R\,\eta_0^2}\right]
\\&
\\
+&\hspace*{-0.2cm}\ln\left[A\sqrt{{z_1}^2-R\,\eta_0^2}
-\sqrt{(y_1-A\,{z_1})^2+R\, (1-A^2)\,\eta_0^2}
+\sqrt{y_1^2+R\,\eta_0^2}\right]
\\&
\\
-&\hspace*{-0.2cm}\ln(2A);
\end{array}
\right.
\nonumber\\
\ea
d. for $y_1-A{z_1} < 0$:
\ba
\label{I01gex4}
\ln{N}\hspace*{-0.3cm} &=&\hspace*{-0.3cm}\left\{
\begin{array}{ll}
-&\hspace*{-0.3cm}\ln\left[\sqrt{y_1^2+R\,\eta_0^2}
+\sqrt{(y_1-A\,{z_1})^2+R\, (1-A^2)\,\eta_0^2}
-A\sqrt{{z_1}^2-R\,\eta_0^2}\right]
\\&
\\
-&\hspace*{-0.3cm}\ln\left[-\sqrt{y_1^2+R\,\eta_0^2}
+\sqrt{(y_1-A\,{z_1})^2+R\, (1-A^2)\,\eta_0^2}
+A\sqrt{{z_1}^2-R\,\eta_0^2}\right]
\\&
\\
+&\hspace*{-0.3cm}\ln(2A)+\ln\left[R^2\eta_0^2
\beta_0^2(1-\cos^2\vartheta)\right].
\end{array}
\right.
\nonumber\\
\ea
Now we neglect terms of $O(\eta_0^2)$ in (\ref{I01gex1}) to (\ref{I01gex4}).
The parameters $c^{+}_{1}$ and $c^{-}_{1}$ from (\ref{c1p}) and 
(\ref{c1m}) thus distinguish three 
different intervals for the variable  $\cos\vartheta$ 
with different analytical expressions for $I_1^0$ in each interval:

\ba
\label{I01gapp}
1.\quad && \cos\vartheta<c^{-}_{1}\quad (y_1\pm A{z_1} > 0):
\nonumber\\
\nonumber\\
I_1^0&=&\frac{1}{s{z_1}}\,\ln\left(\frac{y_1+A{z_1}}
{y_1-A{z_1}}\right)
\\
&=&\frac{1}{s{z_1}}\,\ln\left\{\frac{[(1-R)+A(1+R)]
-[(1+R)+A(1-R)]\cos\vartheta}{[(1-R)-A(1+R)]
-[(1+R)-A(1-R)]\cos\vartheta}\right\},
\label{I01gapp1}
\nonumber\\
\\
2.\quad && c^{-}_{1}<\cos\vartheta<c^{+}_{1}\quad (y_1+A{z_1} > 0\wedge y_1-A{z_1}< 0):
\nonumber\\
\nonumber\\
I_1^0&=&\frac{1}{s{z_1}}\,\left\{\ln\left[(y_1+A{z_1})
(A{z_1}-y_1)\right]+2\ln{z_1}\right.
\nonumber\\
&&
\left.
-\ln(1-\cos^2\vartheta)
+\ln\left(\frac{s}{m_e^2}\right)-2\ln{R}\right\},
\label{I01gapp2}
\\
\nonumber\\
3.\quad && \cos\vartheta>c^{+}_{1}\quad (y_1\pm A{z_1}< 0):
\nonumber\\
\nonumber\\
I_1^0&=& -\frac{1}{s{z_1}}\,\ln\left(\frac{y_1+A{z_1}}
{y_1-A{z_1}}\right)
\\
&=& 
-\frac{1}{s{z_1}}\,\ln\left\{\frac{[(1-R)+A(1+R)]
-[(1+R)+A(1-R)]\cos\vartheta}{[(1-R)-A(1+R)]
-[(1+R)-A(1-R)]\cos\vartheta}\right\}.
\label{I01gapp3}
\nonumber\\
\ea
In case 2., the term $\ln(1-\cos^2\vartheta)$
in (\ref{I01gapp2}) 
does not pose a problem for $\cos\vartheta\to \pm 1$ 
because it cancels with the logarithm 
$\ln\left[(y_1+A{z_1})(A{z_1}-y_1)\right]$.
To see this one has to take into account that 
the condition $\cos\vartheta\to \pm 1$ 
demands $c_1^{\pm}\pm 1$ which only occurs 
for $A\to 1$ (see (\ref{c1p}) and (\ref{c1m}))
This finally implies: 
\ba
\label{criticallog}
\ln\left[(y_1 + A {z_1})(A {z_1} - y_1)\right]
\to\ln\left[z_1^2 - y_1^2\right] = \ln(1-\cos^2\vartheta) + O(\eta_0^2).
\ea
In the limit $\cos\vartheta\to c^{-}_{1}$ or 
$\cos\vartheta\to c^{+}_{1}$, $I_1^0$ simplifies to:
\ba
\label{I01gapps}
I_1^0\to \frac{1}{s{z_1}}\,\left[
\frac{1}{2}\,\ln\left(\frac{s}{m_e^2}\right)
+\ln{z_1}-\frac{1}{2}\ln{R}
-\frac{1}{2}\ln\left(\frac{1-A^2}{4\,A^2}\right)\right].
\ea
In order to determine the remaining integrals 
\ba
\label{I1deftypg}
{I}^i_1 &=& \int\limits^{V_2^{\max}}_{V_2^{\min}}\frac{d{V_2}}
{\sqrt{C_1(V_2,s',\cos\vartheta}} (V_2+{s'})^i
= s^{i+1}\,\int\limits^{{x}^{\max}}_{{x}^{\min}}\frac{d{{x}}}
{\sqrt{C_1({x,R,\cos\vartheta})}}
({x}+R)^i,
\nonumber\\
\ea
we can use the recurrence relations:
\ba
\label{I1defg}
\hat I^1_1 &=&
\frac{{\DS\sqrt{C_1}}}{{\DS a_1}}\Biggr|
^{x^{\max}}_{x^{\min}}
+\frac{{\DS b_1}}{{\DS a_1}} \hat I^0_1
\hspace*{2.8cm}
\rightarrow
{I}^1_1 = \hat I^1_1+{s'} \hat I^1_0,
\label{I1defg1} 
\\
\nonumber\\
\hat I^2_1 &=&
\frac{{\DS s\,x\sqrt{C_1}}}{{\DS 2\,a_1}}\Biggr|
^{x^{\max}}_{x^{\min}}
+\frac{{\DS 3}}{{\DS 2}}\frac{{\DS b_1}}{{\DS a_1}} \hat I^1_1
-\frac{{\DS 1}}{{\DS 2}}\frac{{\DS c_1}}{{\DS a_1}} \hat I^0_1
\hspace*{0.3cm}
\rightarrow
{I}^2_1 = \hat I^2_1+2\,{s'}\hat I^1_1+\hat I^1_0,
\label{I1defg2}
\\ 
\nonumber\\
\hat I^3_1 &=&
\frac{{\DS s^2\,x^2\sqrt{C_1}}}{{\DS 3 a_1}}\Biggr|
^{x^{\max}}_{x^{\min}}
+\frac{{\DS 5}}{{\DS 3}}\frac{{\DS b_1}}{{\DS a_1}} \hat I^2_1
-\frac{{\DS 2}}{{\DS 3}}\frac{{\DS c_1}}{{\DS a_1}} \hat I^1_1
\rightarrow 
{I}^3_1 = \hat I^3_1+3\,{s'}\hat I^2_1
\\
&&
\hspace*{7cm} 
+ 3\,{s'}^2 \hat I^1_1+{s'}^3 \hat I^1_0,
\label{I1defg3}
\nonumber
\ea
with the integrals 
\ba
\label{I1defhelp}
\hat I^i_1 &=& \int\limits^{V_2^{\max}}_{V_2^{\min}}\frac{d{V_2}}
{\sqrt{C_1(V_2,s',\cos\vartheta)}} V_2^i
=s^{i+1}\,\int\limits^{{x}^{\max}}_{{x}^{\min}}\frac{d{{x}}}
{\sqrt{C_1({x,R,\cos\vartheta})}}
{x}^i
\ea
which can be calculated via recurrence relations 
from the basic integral $I_1^0$.
In (\ref{I1defg}), the approximations for $m_e^2\ll s$ were used:
\ba
\label{coefhelp}
\frac{{\DS c_1}}{{\DS a_1}}\approx\left(\frac{{\DS b_1}}{{\DS a_1}}\right)^2,
\quad
\frac{{\DS b_1}}{{\DS a_1}}\approx\, s R\,\frac{{\DS 1-{z_1}}}{{\DS {z_1}}},
\quad
\frac{{\DS 1}}{{\DS a_1}}\approx\frac{{\DS 1}}{{\DS s^2{z_1}^2}}.
\ea
So, for the three intervals in $\cos\vartheta$, given 
e.g.~through (\ref{I01gapp1}), (\ref{I01gapp2}), and (\ref{I01gapp3}), 
we obtain for ${I}^i_1$: 

\vfill\eject
%---------------------------------------------------------------------
\ba
\label{I1gapp}
1.&& \cos\vartheta<c^{-}_{1}\quad (y_1\pm A{z_1}> 0):
\nonumber\\
&&{I}^1_1 = \frac{A(1-R)}{{z_1}}+\left(\frac{sR}{{z_1}}\right)\, I^0_1,
\label{I1gapp1a}
\\
&&{I}^2_1 = s\,\frac{A(1-R)}{{z_1}}\left(
\frac{1+R}{2}+\frac{R}{{z_1}}\right)
+\left(\frac{sR}{{z_1}}\right)^2\, I^0_1,
\label{I1gapp1b}
\\
&&{I}^3_1 = s^2\,\frac{A(1-R)}{{z_1}}
\left\{
\left[\frac{A^2(1-R)^2}{12}+\frac{(1+R)^2}{4}\right]
+\frac{R(1+R)}{2{z_1}}+\frac{R^2}{{z_1}^2}
\right\}
\nonumber\\
&&
\hspace*{1cm}+\left(\frac{sR}{{z_1}}\right)^3\, I^0_1,
\label{I1gapp1c}
\\
\nonumber\\
2.&& c^{-}_{1}<\cos\vartheta<c^{+}_{1}\quad (y_1+A{z_1}> 0\wedge y_1-A{z_1}< 0):
\nonumber\\
&&{I}^1_1 =\frac{1}{{z_1}}\left[(1+R)-\frac{2R}{{z_1}}\right]
+\left(\frac{sR}{{z_1}}\right)\, I^0_1,
\label{I1gapp2a}
\\
&&{I}^2_1 = s\,\frac{1}{{z_1}}
\left\{
\frac{1}{4}\left[(1+R)^2+A^2(1-R)^2\right]
+\frac{R(1+R)}{{z_1}}-\frac{3R^2}{{z_1}^2}
\right\}
+\left(\frac{sR}{{z_1}}\right)^2\, I^0_1,
\label{I1gapp2b}
\nonumber\\
\\
&&{I}^3_1 = s^2\,\frac{1}{{z_1}}
\left\{
\frac{1+R}{12}\left[(1+R)^2+3A^2(1-R)^2\right]\right.
\nonumber\\
&&\left.
\hspace*{2cm}
+\frac{R}{4{z_1}}\left[(1+R)^2+A^2(1-R)^2\right]
+\frac{R^2(1+R)}{{z_1}^2}-\frac{11R^3}{3{z_1}^3}
\right\}
\nonumber\\
&&
\hspace*{2cm}
+\left(\frac{sR}{{z_1}}\right)^3\, I^0_1,
\label{I1gapp2c}
\\
\nonumber\\
3.&& \cos\vartheta>c^{+}_{1}\quad (y_1\pm A{z_1}< 0):
\nonumber\\
&&{I}^1_1 =-\frac{A(1-R)}{{z_1}}
+\left(\frac{sR}{{z_1}}\right)\, I^0_1,
\label{I1gapp3a}
\\
&&{I}^2_1 =-s\,\frac{A(1-R)}{{z_1}}\left(
\frac{1+R}{2}+\frac{R}{{z_1}}\right)
+\left(\frac{sR}{{z_1}}\right)^2\, I^0_1,
\label{I1gapp3b}
\\
&&{I}^3_1 = -s^2\,\frac{A(1-R)}{{z_1}}
\left\{
\frac{1}{12}\,\left[3(1+R)^2+A^2(1-R)^2\right]
+\frac{R(1+R)}{2{z_1}}+\frac{R^2}{{z_1}^2}
\right\}
\nonumber\\
&&
\hspace*{1cm}+\left(\frac{sR}{{z_1}}\right)^3\, I^0_1.
\label{I1gapp3c}
\ea
This type of integrals ${I}^i_1$
can generally be transformed into a sum of {\it rational functions} in 
powers of $1/{z_1}(\cos\vartheta)$ and the logarithmic function 
$I^0_1(\cos\vartheta)$ with powers of $1/{z_1}$ as coefficient.

%%%%%%%%%%%%%%%%%%%%%%%%%%%%%%%%%%%%%%%%%%%%%%%%%%%%%%%%%%%%%%%%%%%%%%%%%%
\subsubsection*{Integrands proportional to $m_e^2\,C_i^{-\frac{3}{2}}$}
\label{vtabZ3g} 
%%%%%%%%%%%%%%%%%%%%%%%%%%%%%%%%%%%%%%%%%%%%%%%%%%%%%%%%%%%%%%%%%%%%%%%%%%
%
For the second type of integrands proportional to 
$m_e^2\,C_i^{-\frac{3}{2}}$ is stated below:
\ba
\label{J01def} 
J^0_1 &=& m_e^2\,\int\limits^{V_2^{\max}}_{V_2^{\min}}
\frac{d{V_2}}{C_1^{\frac{3}{2}}} 
= s\,\int\limits^{{x}^{\max}}_{{x}^{\min}}
\frac{d{{x}}\,m_e^2}{C_1(x,R,\cos\vartheta)^{\frac{3}{2}}},
\\
\nonumber\\
\label{J1def}
{J}^i_1&=&\int\limits^{V_2^{\max}}_{V_2^{\min}}\frac{d{V_2}
\,m_e^2\, B_1\, (V_2+{s'})^i}
{C_1^{\frac{3}{2}}} 
=s^{i+1}\,\int\limits^{{x}^{\max}}_{{x}^{\min}}\frac{d{{x}}
\,m_e^2\, B_1(x,R,\cos\vartheta)\, ({x}+R)^i}
{C_1(x,R,\cos\vartheta)^{\frac{3}{2}}},
\nonumber\\
&&\hspace*{9cm} i=1,2,3, 
\ea
with $C_1$ and $B_1$ defined in (\ref{coefv2Ci2}) and (\ref{coefv2Bi2}).
We again start with the basic integral $J^0_1$. Its integration
yields:

\ba
\label{J01g} 
J^0_1&=&\frac{1}{4\,s^4}\,\frac{{z_1}}{R^2(1-{z_1})({z_1}-R)}
\\
&&\,\left\{\frac{y_1+A{z_1}}{\sqrt{(y_1+A{z_1})^2+R(1-A^2)\eta_0^2}}
-\frac{y_1-A{z_1}}{\sqrt{(y_1-A{z_1})^2+R(1-A^2)\eta_0^2}}
\right\}.
\nonumber
\ea
It can be shown that this integral $J^0_1$ cancels in the  
integrated matrix element, but all other integrals of the type 
${J}^i_1$ will be related to $J^0_1$ by recurrence relations.
The denominator in (\ref{J01g}) vanishes 
if $R(1-A^2)\eta_0^2\ll (y_1\pm A{z_1})^2$. Thus, an approximation 
for small $m_e^2$ again delivers three different analytical 
expressions for $J_1^0$ depending on the relative values
of $\cos\vartheta$ and $c^{\pm}_{1}$ defined in (\ref{c1p}) and (\ref{c1m}):
\ba
\label{J01gapp}
1.\quad 
&& \cos\vartheta<c^{-}_{1}\quad (y_1\pm A{z_1}> 0):
\nonumber\\
J_1^0&\approx&\eta_0^2\,\frac{1}{4\,s^4}\,
\frac{2 A(1-A^2){z_1}^2 y_1}{R(1-{z_1})({z_1}-R)(y_1^2-A^2{z_1}^2)^2}
\label{J01gapp1b}
\\
&=&\frac{\eta_0^2}{s^4}\,
\frac{2 A(1-A^2){z_1}^2 y_1}{R(1-R)^2(1-\beta^2_0\cos^2\vartheta)},
\label{J01gapp1}
\\
\nonumber\\
\nonumber\\
2.\quad 
&& c^{-}_{1}<\cos\vartheta<c^{+}_{1}\quad (y_1+A{z_1}> 0\wedge y_1-A{z_1}< 0):
\nonumber\\
J_1^0&\approx&\frac{1}{4\,s^4}\,\frac{2{z_1}}{R^2(1-{z_1})({z_1}-R)}
=\frac{1}{s^4}\,
\frac{2{z_1}}{R^2(1-R)^2(1-\beta^2_0\cos^2\vartheta)},
\label{J01gapp2}
\ea

\vfill\eject
%------------------------------------------------------------------------
\ba
3.\quad 
&& \cos\vartheta>c^{+}_{1}\quad (y_1\pm A{z_1}< 0):
\nonumber\\
J_1^0&\approx& -\eta_0^2\,\frac{1}{4\,s^4}\,
\frac{2 A(1-A^2){z_1}^2 y_1}{R(1-{z_1})({z_1}-R)(y_1^2-A^2{z_1}^2)^2}
\label{J01gapp3b}
\\
&=& -\frac{\eta_0^2}{s^4}\,
\frac{2 A(1-A^2){z_1}^2 y_1}{R(1-R)^2(1-\beta^2_0\cos^2\vartheta)}.
\label{J01gapp3}
\ea

So, for $J_1^0$ we have the interesting situation
that after integration over $\cos\vartheta$ the expressions
in (\ref{J01gapp1}) and (\ref{J01gapp3})
are suppressed by an additional factor 
\ba
\label{suppression}
\eta_0^2=\frac{4\,m_e^2}{s}
\ea
and can be neglected in the integrated matrix element. 
This `drop' of $J_1^0$ by a factor $\eta_0^2$ occurs in a 
very narrow region with a width of the order $O(\eta_0^2)$, 
i.e.~for $|\cos\vartheta-c^{-}_{1}| < \eta_0^2$
and $|\cos\vartheta-c^{+}_{1}|<\eta_0^2$.
In the limit $\cos\vartheta\to c^{-}_{1}$ 
or $\cos\vartheta\to c^{+}_{1}$, we have:
\ba
\label{J01gapps}
J_1^0\to\frac{1}{4\,s^4}\,\frac{{z_1}}{R^2(1-{z_1})({z_1}-R)}
=\frac{1}{s^4}\,
\frac{{z_1}}{R^2(1-R)^2(1-\beta^2_0\cos^2\vartheta)}.
\ea
The remaining integrals ${J}^i_1$ from (\ref{J1def}) 
can be written as
\ba
\label{J1grel}
{J}^i_1=\frac{1}{1-R}\left\{
\left[(1+R){z_1}-2R\right]\,\hat{J}^{i+2}_1
+R\left[(1+R)-2{z_1}\right]\,\hat{J}^{i+1}_1\right\},
\ea
with $z_1$ from (\ref{coefv2z1}) and where we introduce the integrals
\ba
\label{J1gdef2}
\hat{J}^i_1=\int\limits^{V_2^{\max}}_{V_2^{\min}}\frac{d{V_2}
\,m_e^2\, (V_2+{s'})^i}{C_1(V_2)^{\frac{3}{2}}} 
=s^{i+1}\,\int\limits^{{x}^{\max}}_{{x}^{\min}}\frac{d{{x}}
\,m_e^2\, ({x}+R)^i}{C_1({x})^{\frac{3}{2}}}. 
\ea
If we neglect terms of $O(\eta_0^2)$, (\ref{J1grel})
simplifies with (\ref{coefhelp}) to:
\ba
\label{J1gall}
\hat{J}^i_1\approx \left(\frac{b_1}{a_1}+{s'}\right)^i\cdot J^0_1
\approx \left(\frac{s\, R}{{z_1}}\right)^i\cdot J^0_1,
\ea
which produces with (\ref{J01gapp2}) for only the interesting case, 
$c^{-}_{1}<\cos\vartheta<c^{+}_{1}$:
\ba
\label{J1gall2}
{J}^i_1 &=& \frac{2s(1-{z_1})({z_1}-R)}{1-R}
\,\left(\frac{s R}{{z_1}}\right)^{i+2}\!\!\!\!\, J^0_1
\\
&=&\frac{1}{2}\, s(1-R)(1-\beta^2_0\cos^2\vartheta)
\,\left(\frac{s R}{{z_1}}\right)^{i+2}\!\!\!\!\, J^0_1,
\\
\label{J1gall3}
{J}^i_1 &=& \frac{1}{s(1-R){z_1}}\,\left(\frac{s R}{{z_1}}\right)^i.
\ea
At the end we discover the remarkable result that all integrands 
proportional to $m_e^2\,C_i^{-\frac{3}{2}}$ yield integrals
proportional to the basic integral $J^0_1$. 
They will therefore, just like $J^0_1$, only lead to non-neglectable 
contributions in the defined region $c^{-}_{1}<\cos\vartheta<c^{+}_{1}$.
So, the dependence on the parameter $A(R)$, which contains the 
cut dependence of the different phase space regions I, II, and III, 
is only introduced through the basic integral $J^0_1$. 

The only terms, however, which arise in integrated matrix
element are proportional to ${J}^i_1$ with $i=1,2,3$.
All other terms proportional to $J^0_1$ have to cancel 
for $c^{-}_{1}<\cos\vartheta<c^{+}_{1}$ due to its 
mass singular behaviour there. {From} (\ref{J01gapp2}) we see:
\ba
J_1^0&\sim&\frac{1}{1-\beta^2_0\cos^2\vartheta}
\quad\approx\quad \frac{1}{\eta_0^2}
\label{J01gapp2b}
\ea
for $\cos\vartheta\approx \pm 1$, 
and $c^{\pm}_{1}\approx 1$ respectively which 
can happen for $A(R)\approx 1$ in phase space
region I (see (\ref{c1p}) and (\ref{c1m})). Due to 
\cite{Kinoshita:1962ur,Lee:1964is} these 
poles have to cancel.
In ${J}^i_1$, $i\geq 1$, this problem does not occur 
because the factor $1/(1-\beta^2_0\cos^2\vartheta)$  
is cancelled.

All the results obtained in this section 
can now be used straightforwardly for the corresponding integrands 
proportional to $C_2^{-\frac{1}{2}}$ and 
$m_e^2\,C_2^{-\frac{3}{2}}$, just by replacing formally
in the expressions the term `${z_1}$' by `${z_2}$', 
or, more explicitly, substituting `$\cos\vartheta$' 
by `$-\cos\vartheta$'.

%%%%%%%%%%%%%%%%%%%%%%%%%%%%%%%%%%%%%%%%%%%%%%%%%%
\subsection*{Integration over $\cos\vartheta$}
\label{cttab} 
%%%%%%%%%%%%%%%%%%%%%%%%%%%%%%%%%%%%%%%%%%%%%%%%%%
%
Following the discussion, above the zero conditions
defined in (\ref{yaz}) and (\ref{zeros}) fix 
different values 
$c_1^+(R)$, $c_1^-(R)$, $c_2^+(R)$, and $c_2^-(R)$ 
(see (\ref{c1p}) to (\ref{c2m}))
which separate the symmetric integration interval 
$[-c;c]$ into several different regions with different
analytical expressions for the hard radiator functions.
With $c\ge 0$, we define below the different 
intervals for an integration over $\cos\vartheta$.
$H^{ini}_{T,FB}(R,c,A)$ and $h^{ini}_{T,FB}(\cos\vartheta,R,A)$
are the total or differential hard flux functions
which factorize from the (improved) Born cross sections
or asymmetries, not demonstrated here.

\vfill\eject
%------------------------------------------------------------------------
\ba
\label{intervalcbp}
1.&& 0\le c < c^{-}_2 \quad ( c^{-}_2 > 0 ):
%\quad\rightarrow\quad I=0\quad,\quad J=0\quad (\Rightarrow K = 0)
\nonumber\\
&&H^{ini}_{T,FB}(R,c,A)= 
\int\limits^{c}_{c^{-}_2} d{\cos\vartheta}
\,\, h^{ini}_{T,FB}(\cos\vartheta,R,A),
\\
\nonumber\\
%---------------------------------------------------------------------
2.&& 0\le c < c^{-}_1 \quad ( c^{-}_1\ge 0 ):
%\quad\rightarrow\quad I=1\quad,\quad J=1\quad (\Rightarrow K = 1)
\nonumber\\
&&H^{ini}_{T,FB}(R,c,A)= 
\int\limits^{c}_{c^{-}_1} d{\cos\vartheta}
\,\, h^{ini}_{T,FB}(\cos\vartheta,R,A),
\\
\nonumber\\
%---------------------------------------------------------------------
3.&& c^{-}_1,c^{-}_2\le c < c^{+}_1: 
%\quad\rightarrow\quad I=1\quad,\quad J=0
\nonumber\\
&a.&  c^{-}_1\ge 0:  
%\quad\rightarrow\quad K = 1
\nonumber\\
&&H^{ini}_{T,FB}(R,c,A)= \left\{
\int\limits^{c^{-}_1}_0 d{\cos\vartheta}
+\int\limits^{c}_{c^{-}_1} d{\cos\vartheta}
\right\}\, h^{ini}_{T,FB}(\cos\vartheta,R,A),
\\
&b.&  c^{-}_2 > 0:  
%\quad\rightarrow\quad K = 0
\nonumber\\
&&H^{ini}_{T,FB}(R,c,A)= \left\{
\int\limits^{c^{-}_2}_0 d{\cos\vartheta}
+\int\limits^{c}_{c^{-}_2} d{\cos\vartheta}
\right\}\, h^{ini}_{T,FB}(\cos\vartheta,R,A),
\\
\nonumber\\
%---------------------------------------------------------------------
4.&& c^{+}_1\le c\le 1: 
%\quad\rightarrow\quad I=1\quad,\quad J=-1
\nonumber\\
&a.&  c^{-}_1\ge 0: 
%\quad\rightarrow\quad K = 1
\nonumber\\
&& H^{ini}_{T,FB}(R,c,A)= \left\{
\int\limits^{c^{-}_1}_0 d{\cos\vartheta}
+\int\limits^{c^{+}_1}_{c^{-}_1} d{\cos\vartheta}
+\int\limits^{c}_{c^{+}_1} d{\cos\vartheta}
\right\}\, h^{ini}_{T,FB}(\cos\vartheta,R,A),
\nonumber\\
\\
&b.&  c^{-}_2 > 0: 
%\quad\rightarrow\quad K = 0
\nonumber\\
&& H^{ini}_{T,FB}(R,c,A)= \left\{
\int\limits^{c^{-}_2}_0 d{\cos\vartheta}
+\int\limits^{c^{+}_1}_{c^{-}_2} d{\cos\vartheta}
+\int\limits^{c}_{c^{+}_1} d{\cos\vartheta}
\right\}\, h^{ini}_{T,FB}(\cos\vartheta,R,A).
\nonumber\\
\ea
For $H^{ini}_{T}(R,c,A)$ the results 
for $c^{-}_1\ge 0$ and $c^{-}_2\ge 0$  
have to reproduce which serves as a check of the
analytical integration.

\vfill\eject
%------------------------------------------------------------------------
%%%%%%%%%%%%%%%%%%%%%%%%%%%%%%%%%%%%%%%%%%%%%%%%%%%%%%%%%%%%%%%%
\subsubsection*{Integral types of the integration over $\cos\vartheta$
\label{ctinttypes}} 
%%%%%%%%%%%%%%%%%%%%%%%%%%%%%%%%%%%%%%%%%%%%%%%%%%%%%%%%%%%%%%%%
%
Introducing the the following abbreviations,
\ba
\label{abbrev}
R_c^{+} &=& (1+R)+{c}\,(1-R) = 2\,{z_2}(c),
\label{abbrev1}
\\
R_c^{-} &=& (1+R)-{c}\,(1-R) =: R_c = 2\,{z_1}(c),
\label{abbrev2}
\\
R_A^{+} &=& (1+R)+A\,(1-R) = 2\,{z_2}(A),
\label{abbrev3}
\\
R_A^{-} &=& (1+R)-A\,(1-R) = 2\,{z_1}(A),
\label{abbrev4}
\\
\bar{R}_A^{+} &=& (1-R)+A\,(1+R) = 2\, y_2(A),
\label{abbrev5}
\\
\bar{R}_A^{-} &=& (1-R)-A\,(1+R) = 2\, y_1(A),
\label{abbrev6}
\\
R_{Ac}^{+} &=& (1-R)(1-{c}A)-(1+R)({c}-A) = 2 ( y_1(c)+A{z_1}(c) ),
\label{abbrev7}
\\
R_{Ac}^{-} &=& (1-R)(1+{c}A)-(1+R)({c}+A) = 2 ( y_2(c)+A{z_2}(c) ).
\label{abbrev8}
\ea
It is obvious that all 
integrals with $ z_2$ in the integrand can uniformly be obtained 
from those with index `1' by introducing an extra overall minus sign 
and substituting $c$ by $-c$.~\footnote{To see this,
just substitute $\cos\vartheta$ in the integrand 
by $-\cos\vartheta$ with ${z_1}(R,-\cos\vartheta)
={z_2}(R,\cos\vartheta)$ and the new upper limit $-c$}.

We first have the following rational integrals over $\cos\vartheta$:
%
%%%%%%%%%%%%%%%%%%%%%%%%%%%%%%%%%%%%%%%%%%%%%%%%%%%%%%%%%%%%%%%%%%%%%%%%%%%%%%%%%
%
\begin{center}
{\large  Table of integrals}
\end{center}
{\it Type 0:}
\ba
\label{intctrat}
\left[f_0(\cos\vartheta)\right]_{(c)}  
&=&\int\limits^{c}_{0} d(\cos\vartheta) \,f_0(\cos\vartheta),
\\
\nonumber\\
\left[\frac{1}{{z_1}}\right]_{(c)}   
&=&
\int\limits^{{c}}_{0} d{\xi}\,\frac{2}{1-R}
\frac{1}{\frac{1+R}{1-R}-\xi} = 
-\frac{2}{1-R}\left[\ln{R_c^{-}}-\ln(1+R)\right],  
\\
\nonumber\\
\left[\frac{1}{{z_1}^2}\right]_{(c)}   
&=&
-\frac{4}{1-R}\,\left[\frac{1}{1+R}-\frac{1}{R_c^{-}}\right]=
\frac{4 {c}}{ R_c^{-}(1+R)},
\\
\left[\frac{1}{{z_1}^3}\right]_{(c)}   
&=&
-\frac{4}{1-R}\,\left[\frac{1}{(1+R)^2}-\frac{1}{{R_c^{-}}^2}\right]=
\frac{4 {c}}{ R_c^{-}(1+R)}\,\left[\frac{1}{1+R}+\frac{1}{R_c^{-}}\right],
\nonumber\\ 
\\
\left[\frac{1}{{z_1}^4}\right]_{(c)}   
&=&
-\frac{16}{3(1-R)}\,\left[\frac{1}{(1+R)^3}-\frac{1}{{R_c^{-}}^3}\right]
\nonumber\\
&=&\frac{16 {c}}{3 R_c^{-}(1+R)}\,\left[\frac{1}{(1+R)^2}
+\frac{1}{R_c^{-}(1+R)}+\frac{1}{{R_c^{-}}^2}\right], 
\\
\nonumber\\
\left[\frac{1}{{z_2}}\right]_{(c)}   
&=&
\frac{2}{1-R}\,\left[\ln{R_c^{+}}-\ln(1+R)\right],  
\\
\left[\frac{1}{{z_2}^2}\right]_{(c)}   
&=&
\frac{4}{1-R}\,\left[\frac{1}{1+R}-\frac{1}{R_c^{+}}\right]=
\frac{4 {c}}{ R_c^{+}(1+R)}, 
\\
\left[\frac{1}{{z_2}^3}\right]_{(c)}   
&=&
\frac{4}{1-R}\,\left[\frac{1}{(1+R)^2}-\frac{1}{{R_c^{+}}^2}\right]=
\frac{4 {c}}{ R_c^{+}(1+R)}\,\left[\frac{1}{1+R}+\frac{1}{R_c^{+}}\right], 
\nonumber\\
\\
\left[\frac{1}{{z_2}^4}\right]_{(c)}   
&=&
\frac{16}{3(1-R)}\,\left[\frac{1}{(1+R)^3}-\frac{1}{{R_c^{+}}^3}\right]
\nonumber\\
&=&\frac{16 {c}}{3 R_c^{+}(1+R)}\,\left[\frac{1}{(1+R)^2}
+\frac{1}{R_c^{+}(1+R)}+\frac{1}{{R_c^{+}}^2}\right].
\ea
And secondly, there are the following logarithmic integrals, just showing the 
integrals with index `1' and $R_c:=R_c^{-}$ and keeping the 
above said in mind:

%%%%%%%%%%%%%%%%%%%%%%%%%%%%%%%%%%%%%%%%%%%%%%%%%%%%%%%%%%%%%%%%%%%%%%%%%%%%%%%%%
\noindent {\it Type 1:}
\ba
\label{intctln1} 
\left[\frac{1}{ z_i^k}\ln\left(\frac{s}{m_e^2}
\frac{ z_i^2}{R}\right)\right]_{(c)}  
&=&
\ln\left(\frac{s}{m_e^2}\frac{1}{R}\right)\,\int\limits^{c}_{0} d(\cos\vartheta) 
\frac{1}{ z_i^k}+2\,\int\limits^{c}_{0} d(\cos\vartheta)\frac{\ln{ z_i}}{ z_i^k}
\\
&=&\pm\frac{2}{1-R}\,\left\{\ln\left(\frac{s}{m_e^2}\frac{1}{R}\right) 
\,\!\!\int\limits^{\frac{1}{2}R_c^{\pm}}_{\frac{1}{2}(1+R)}\!\!d{ z_i}
\,\,\frac{1}{ z_i^k}+
2\,\!\!\int\limits^{\frac{1}{2}R_c^{\pm}}_{\frac{1}{2}(1+R)}\!\!d{ z_i}
\,\,\frac{\ln{ z_i}}{ z_i^k}
\right\},
\nonumber\\
\\
\mbox{with} && i=1\mbox{ (`-') or } 2 \mbox{ (`+')},\quad k=2,3,4,
\nonumber
\ea

\ba
\label{intctln1b}
%--------------------------------------------------------------------------
\frac{1}{2}\,\left[\frac{\ln{{z_1}}}{{z_1}^2}\right]_{(c)}   
&=&
2\,\left\{
\frac{1}{v}\,\frac{1}{1+R}\,\ln\left(\frac{R_c}{1+R}\right)
+\frac{{c}}{R_c(1+R)}\,\left[\ln\left(\frac{R_c}{2}
\right)+1\right]\right\},
\\
\nonumber\\
\nonumber\\
%-------------------------------------------------------------------------- 
\frac{1}{2}\,\left[\frac{\ln{{z_1}}}{{z_1}^3}\right]_{(c)}    
&=&
2\,\left\{
\frac{1}{v}\,\left(\frac{1}{1+R}\right)^2
\,\ln\left(\frac{R_c}{1+R}\right)
\right.
\nonumber\\
&&
\left.
+\frac{{c}}{R_c(1+R)}\left(\frac{1}{R_c}+\frac{1}{1+R}\right)
\,\left[\ln\left(\frac{R_c}{2}\right)+\frac{1}{2}\right]\right\},
\\
\nonumber\\
\nonumber\\
%-------------------------------------------------------------------------- 
\frac{1}{2}
\left[\frac{\ln{{z_1}}}{{z_1}^4}\right]_{(c)}    
&=&
\frac{8}{3}\,\left\{
\frac{1}{v}\,\left(\frac{1}{1+R}\right)^3
\,\ln\left(\frac{R_c}{1+R}\right)
\right.
\\
&&\left.+\frac{{c}}{R_c(1+R)}\left[\frac{1}{R_c^2}+\frac{1}{R_c(1+R)}
+\frac{1}{(1+R)^2}\right]
\,\left[\ln\left(\frac{R_c}{2}\right)+\frac{1}{3}\right]\right\}.
\nonumber
\ea

\bigskip

%%%%%%%%%%%%%%%%%%%%%%%%%%%%%%%%%%%%%%%%%%%%%%%%%%%%%%%%%%%%%%%%%%%%%%%%%%%%%%%%%
\noindent {\it Type 2:}
\ba
\label{intctln2} 
\left[\frac{\ln(1-\be^2\cos^2\vartheta)}{ z_i^k}\right]_{(c)}  
&=&
\int\limits^{c}_{0} d(\cos\vartheta)
\,\frac{\ln[( z_i-R)(1- z_i)]-2\ln\left(\frac{1-R}{2}\right)}{ z_i^k}
\\
&=&
\pm\frac{2}{1-R}\,\!\!\int\limits^{\frac{1}{2}R_c^{\pm}}
_{\frac{1}{2}(1+R)}\!\!d{ z_i}
\,\,\frac{\ln[( z_i-R)(1- z_i)]-2\ln\left(\frac{1-R}{2}\right)}{ z_i^k},
\nonumber\\
\nonumber\\
\mbox{with} && i=1\mbox{ (`-') or } 2 \mbox{ (`+')},\quad k=2,3,4,
\ea

\ba
\label{intctln2b}
%--------------------------------------------------------------------------
&&
\frac{1}{2}\,
\left[\frac{\ln(1-\be^2\cos^2\vartheta)}{{z_1}^2}\right]_{(c)}  
\nonumber\\
&=&
\frac{1}{v}\,
\frac{1+R}{R}\,\ln\left(\frac{R_c}{1+R}\right)
+\frac{1+{c}}{R_c}\,\ln(1+{c})
-\frac{1-{c}}{R R_c}\,\ln(1-{c}),
\\
\nonumber\\
\nonumber\\
%--------------------------------------------------------------------------
&&
\frac{1}{2}\,
\left[\frac{\ln(1-\be^2\cos^2\vartheta)}{{z_1}^3}\right]_{(c)}  
\nonumber\\
&=&
\frac{1}{2}\,\left\{\frac{1}{v}\,
\frac{1+R^2}{R^2}\,\ln\left(\frac{R_c}{1+R}\right)
+
\left(\frac{2}{R_c}+1\right)\,\frac{1+{c}}{R_c}\,\ln(1+{c})\right.
\nonumber\\
&&-\left.\left(\frac{2}{R_c}+\frac{1}{R}\right)\,\frac{1-{c}}{RR_c}\,\ln(1-{c})
-\frac{2{c}}{R R_c}\right\},
\\
\nonumber\\
\nonumber\\
%--------------------------------------------------------------------------
&&
\frac{1}{2}\,
\left[\frac{\ln(1-\be^2\cos^2\vartheta)}{{z_1}^4}\right]_{(c)}  
\\
&=&
\frac{1}{3}\,\left\{\frac{1}{v}\,
\frac{1+R^3}{R^3}\,\ln\left(\frac{R_c}{1+R}\right)
+
\left(\frac{4}{R_c^2}+\frac{2}{R_c}+1\right)\,\frac{1+{c}}{R_c}\,\ln(1+{c})
\right.
\nonumber\\
&&\left.
-\left(\frac{4}{R_c^2}+\frac{2}{R R_c}+\frac{1}{R^2}\right)\,\frac{1-{c}}{RR_c}
\,\ln(1-{c})
-\frac{2{c}}{R R_c}\left[\frac{1}{R_c}
+\frac{1+R+R^2}{R(1+R)}\right]\right\}.\nonumber
\ea

\bigskip

%%%%%%%%%%%%%%%%%%%%%%%%%%%%%%%%%%%%%%%%%%%%%%%%%%%%%%%%%%%%%%%%%%%%%%%%%%%%%%%%%
\noindent {\it Type 3:}
\ba
\label{intctln3} 
\left[\frac{1}{ z_i^k}\,\ln\left|\frac{y_i+A z_i}
{y_i-A z_i}\right|\right]_{(c)}  
&=&\pm\frac{2}{1-R}\,\!\!\int\limits^{\frac{1}{2}R_c^{\pm}}
_{\frac{1}{2}(1+R)}\!\!d{ z_i}
\,\,\frac{1}{ z_i^k}\,\ln\left|\frac{y_i+A z_i}
{y_i-A z_i}\right|
\nonumber\\
\nonumber\\
\mbox{with} && i=1\mbox{ (`-') or } 2 \mbox{ (`+')},\quad k=2,3,4,
\ea

\ba
\label{intctln3b}
%--------------------------------------------------------------------------
&&
\frac{1}{2}\,
\left[\frac{1}{{z_1}^2}\,
\ln\left|\frac{y_1+A{z_1}}{y_1-A{z_1}}\right|\right]_{(c)}  
\nonumber\\
&=&
\quad\frac{A}{R}\,\ln\left(\frac{R_c}{1+R}\right)
+\frac{1}{2R(1+R)}\,\left[\RApb\ln(\RApb)-\RAmb\ln|\RAmb|\right]
\nonumber\\
&&
+\frac{1}{2R R_c}\,\left[\RAcp\ln|\RAcp|-\RAcm\ln|\RAcm|\right],
\\
\nonumber\\
\nonumber\\
%--------------------------------------------------------------------------
&&
\frac{1}{2}\,
\left[\frac{1}{{z_1}^3}\,
\ln\left|\frac{y_1+A{z_1}}{y_1-A{z_1}}\right|\right]_{(c)}  
\nonumber\\
&=&
\frac{1}{2}\,\left\{
\frac{A(1+R)}{R^2}\,\ln\left(\frac{R_c}{1+R}\right)
+\frac{1}{2R(1+R)}
\,\left[\left(\frac{2}{1+R}+\frac{\RAp}{2R}\right)\,\RApb\ln(\RApb)
\right.\right.
\nonumber\\
&&
\quad\left.\left.
-\left(\frac{2}{1+R}+\frac{\RAm}{2R}\right)\,\RAmb\ln|\RAmb|\right]
\right.
\nonumber\\
&&\quad+\left.\frac{1}{2R R_c}
\,\left[\left(\frac{2}{R_c}+\frac{\RAp}{2R}\right)\,\RAcp\ln|\RAcp|
-\left(\frac{2}{R_c}+\frac{\RAm}{2R}\right)\,\RAcm\ln|\RAcm|\right]
\right.
\nonumber\\
&&\quad-\left.\frac{2A{c}}{R R_c}\frac{1-R}{1+R}\right\},
\\
\nonumber\\
\nonumber\\
%--------------------------------------------------------------------------
&&
\frac{1}{2}
\,\frac{1}{{z_1}^4}\,
\left[\ln\left|\frac{y_1+A{z_1}}{y_1-A{z_1}}\right|\right]_{(c)}  
\nonumber\\
&=&
\frac{1}{3}\,\left\{\frac{A}{4R^3}\left[A^2(1-R)^2+3(1+R)^2\right]
\,\ln\left(\frac{R_c}{1+R}\right)
\right.
\nonumber\\
&&\quad+\left.\frac{1}{2R(1+R)}
\,\left[
\left(\frac{4}{(1+R)^2}+\frac{\RAp}{R(1+R)}
+\frac{{\RAp}^2}{4R^2}\right)\,\RApb\ln(\RApb)
\right.\right.
\nonumber\\
&&\quad\qquad\qquad\qquad
-\left.\left.\left(\frac{4}{(1+R)^2}+\frac{\RAm}{R(1+R)}
+\frac{{\RAm}^2}{4R^2}\right)\,\RAmb\ln|\RAmb|
\right]
\right.
\nonumber\\
&&\quad+\left.\frac{1}{2R R_c}
\,\left[
\left(\frac{4}{R_c^2}+\frac{\RAp}{R R_c}+\frac{{\RAp}^2}{4R^2}\right)
\,\RAcp\ln|\RAcp|
\right.\right.
\nonumber\\
&&
\quad\left.\left.
-\left(\frac{4}{R_c^2}+\frac{\RAm}{R R_c}
+\frac{{\RAm}^2}{4R^2}\right)\,\RAcm\ln|\RAcm|\right]
\right.
\nonumber\\
&&\quad-\left.\frac{2A{c}(1-R)}{R R_c(1+R)}\,\left(
\frac{1}{R_c}+\frac{1}{1+R}+\frac{1+R}{R}\right)\right\}.
\ea

\bigskip

%%%%%%%%%%%%%%%%%%%%%%%%%%%%%%%%%%%%%%%%%%%%%%%%%%%%%%%%%%%%%%%%%%%%%%%%%%%%%%%%%
\noindent {\it Type 4:}
\ba
\label{intctln4}
\left[\frac{1}{ z_i^k}\,\ln\left|(y_i+A z_i)
(A z_i-y_i)\right|\right]_{(c)}  
&=&
\pm\frac{2}{1-R}\,\!\!\int\limits^{\frac{1}{2}R_c^{\pm}}_{\frac{1}{2}(1+R)}
\!\!d{ z_i}\,\,\frac{1}{ z_i^k}\,\ln\left|(y_i+A z_i)
(A z_i-y_i)\right|,
\nonumber\\
\nonumber\\
\mbox{with} && i=1\mbox{ (`-') or } 2 \mbox{ (`+')},\quad k=2,3,4,
\ea 

\ba
\label{intctln4b}
%--------------------------------------------------------------------------
&&\frac{1}{2}
\,\left[\frac{1}{{z_1}^2}\,
\ln\left|(y_1+A{z_1})(A{z_1}-y_1)\right|\right]_{(c)}  
\nonumber\\
&=&
\quad\frac{1}{v}\,\frac{1+R}{R}\,\ln\left(\frac{R_c}{1+R}\right)
+\frac{1}{2R(1+R)}\,\left[\RApb\ln(\RApb)+\RAmb\ln|\RAmb|\right]
\nonumber\\
&&
\hspace*{4.5cm}
-\frac{1}{2R R_c}\,\left[\RAcp\ln|\RAcp|+\RAcm\ln|\RAcm|\right],
\\
\nonumber\\
\nonumber\\
%--------------------------------------------------------------------------
&&\frac{1}{2}
\,\left[\frac{1}{{z_1}^3}\,
\ln\left|(y_1+A{z_1})(A{z_1}-y_1)\right|\right]_{(c)}  
\nonumber\\
&=&
\frac{1}{2}\,\left\{
\frac{1}{2R^2}
\left[\frac{1}{v}\,(1+R)^2+A^2(1-R)\right]
\,\ln\left(\frac{R_c}{1+R}\right)\right.
\nonumber\\
&&\quad+\left.\frac{1}{2R(1+R)}
\,\left[\left(\frac{2}{1+R}+\frac{\RAp}{2R}\right)\,\RApb\ln(\RApb)
\right.\right.
\nonumber\\
&&\hspace*{3cm}
\left.\left.
+\left(\frac{2}{1+R}
+\frac{\RAm}{2R}\right)\,\RAmb\ln|\RAmb|\right]
\right.
\nonumber\\
&&\quad-\left.\frac{1}{2R R_c}
\,\left[\left(\frac{2}{R_c}+\frac{\RAp}{2R}\right)\,\RAcp\ln|\RAcp|
+\left(\frac{2}{R_c}+\frac{\RAm}{2R}\right)\,\RAcm\ln|\RAcm|\right]
\right.
\nonumber\\
&&\quad-\left.\frac{2{c}}{R R_c}\right\},
\\
\nonumber\\
\nonumber\\
%--------------------------------------------------------------------------
&&\frac{1}{2}
\,\left[\frac{1}{{z_1}^4}\,
\ln\left|(y_1+A{z_1})(A{z_1}-y_1)\right|\right]_{(c)}  
\nonumber\\
&=&
\frac{1}{3}\,\left\{
\frac{(1+R)}{4R^3}
\left[\frac{1}{v}\,(1+R)^2+3A^2(1-R)\right]
\,\ln\left(\frac{R_c}{1+R}\right)
\right.
\nonumber\\
&&\quad+\left.\frac{1}{2R(1+R)}
\,\left[
\left(\frac{4}{(1+R)^2}+\frac{\RAp}{R(1+R)}
+\frac{{\RAp}^2}{4R^2}\right)\,\RApb\ln(\RApb)
\right.\right.
\nonumber\\
&&\quad\qquad\qquad\qquad
+\left.\left.\left(\frac{4}{(1+R)^2}+\frac{\RAm}{R(1+R)}
+\frac{{\RAm}^2}{4R^2}\right)\,\RAmb\ln|\RAmb|
\right]
\right.
\nonumber\\
&&\quad-\left.\frac{1}{2R R_c}
\,\left[
\left(\frac{4}{R_c^2}+\frac{\RAp}{R R_c}+\frac{{\RAp}^2}{4R^2}\right)
\,\RAcp\ln|\RAcp|
\right.\right.
\nonumber\\
&&\qquad\quad+\left.\left.
\left(\frac{4}{R_c^2}+\frac{\RAm}{R R_c}+\frac{{\RAm}^2}{4R^2}\right)
\,\RAcm\ln|\RAcm|
\right]
\right.
\nonumber\\
&&\quad-\left.\frac{{c}}{R R_c}\,\left[\frac{(1+R)^2+A^2(1-R)^2}{R(1+R)}
+2\,\left(\frac{1}{R_c}+\frac{1}{1+R}\right)\right]\right\}.
\ea

As the integrals of  (\ref{intctln4}) always appear in combination 
with the integrals of (\ref{intctln2}) as differences of the form 
\ba  
\label{intctln24}
\int\limits^{c}_{0} d(\cos\vartheta)
\frac{1}{ z_i^k}\,\left\{\ln\left|(y_i+A z_i)
(A z_i-y_i)\right|-\ln(1-\be^2\cos^2\vartheta)\right\},
\ea
%
%inserting to \ref{I01gapp},
we can show that powers of the order $\frac{1}{v^k}$, $k\geq 2$, 
with $v:=1-R$ being proportional to the photon energy,
%(as factor
%of the logarithms $\ln\left(\frac{R_c}{1+R}\right)$ in
% (\ref{intctln2b}) and (\ref{intctln4b}) 
completely cancel out, as they should.
So, the unregularized hard photon radiators 
will only contain physical poles in the photon energy of the 
order $\frac{1}{v}$. After adding the soft terms and integrating 
over $R$ also the remaining singularities proportional 
to $\ln\varepsilon$ will disappear ($\varepsilon$: cut-off 
between soft and hard photon phase space).   

%%%%%%%%%%%%%%%%%%%%%%%%%%%%%%%%%%%%%%%%%%%%%%%%%%%%%%%%%%%%%%%%
\subsubsection*{Some coefficient functions 
\label{coeffun}} 
%%%%%%%%%%%%%%%%%%%%%%%%%%%%%%%%%%%%%%%%%%%%%%%%%%%%%%%%%%%%%%%%
%
The hard photon radiator functions 
${H}^{ini}_{T,FB} = {H}^{ini}_{T,FB}(c,{\theta_{{\rm acol}}},E_{\min})$
for total cross sections 
and asymmetries were presented in Section \ref{sub_lep1slc_formacol}. 
For completeness we just give some additional coefficient functions
here appearing in ${H}^{ini}_{T,FB}$ which were not shown there
for brevity (functions $y(R,c)$ and $z(R,c)$ defined 
in (\ref{coef_fungy2})).

\ba
\label{coef_fun2a}
f_{01}(R,c)
&=&
-\frac{4}{3 z^3}\,\left(7+12c+8c^2+4c^3+c^4+8R\right)
\nonumber\\
&&\,+\,\frac{2}{3 z^2}\,\left(24+25c+11c^2+c^3-c^4+12R\right)
\nonumber\\
&&\,-\,\frac{1}{3 z}\,\left(26+19c+3c^2-c^3+c^4+12R\right)
\nonumber\\
&&\,+\,\frac{2}{3}\left(4+R\right),
\ea

\vfill\eject
%---------------------------------------------------------------
\ba
\label{coef_fun2b}
f_{02}(R,c)
&=&\frac{4}{3 z^3}\,\left(15+28c+22c^2+12c^3+3c^4+16R\right)
\nonumber\\
&&\,-\,\frac{2}{3 z^2}\,\left(52+63c+37c^2+9c^3-c^4+24R\right)
\nonumber\\
&&\,+\,\frac{1}{3 z}\, (52+2R+41c+10c^2-3c^3)
\nonumber\\
&&\,-\,\frac{1}{12}\left[23-9R+2(11-R)c-(5+R)c^2\right],
\\
\nonumber\\
\label{coef_fun2c}
f_{03}(R,A,c)
&=& \frac{A^2}{12}\left\{-\frac{4}{ z}\,\left[(1+c)^2+2 R\right]
\right.
\nonumber\\
&&\,+\,\left. \left[3-R+12c+3(1+R)c^2-2(1-R)c^3\right]\right\}
\nonumber\\
&&\,+\, \frac{2}{3}\frac{c(1-A^2)}{v}
+ \frac{c}{2}\left(1+\frac{1}{3} A^2 c^2\right)\, (1-R),
\\
\nonumber\\
\label{coef_fun3}
f_{11}(R,A,c)
&=&A\left\{ -\frac{2}{3} g_{L_z}( R,c)
+\frac{1}{4v}\,\left[y^2-\frac{1}{3}A^2( z^2+8 R)\right]\right\},
\\
\nonumber\\
f_{12}(R,A,c)
&=&\frac{2}{3}\frac{A(1-c^2)}{v}
+ \frac{A}{2}\left(1+\frac{1}{3} A^2 c^2\right)\, (1-R),
\\
\nonumber\\
\label{coef_fun4}
g_{01}( R,c) &=&
\frac{1-c^2}{ z}\,\left[c+\frac{y+2R(1+R)}{ z}\right],
\\
\nonumber\\
g_{02}(R,A,c)
&=&\frac{1}{2}c(1+R)\,\left\{A^2+\left[\frac{y}{ z}\right]^2
\right\},
\\
\nonumber\\
g_{11}(R,A,c)
&=& A\,\left[\frac{2R}{v}+(1+R)c\right]\,\frac{y}{ z}.
\ea

\newpage

%%%%%%%%%%%%%%%%%%%%%%%%%%%%%%%%%%%%%%%%%%%%%%%%%%
\section{Initial-final state interference 
\label{int}}
%%%%%%%%%%%%%%%%%%%%%%%%%%%%%%%%%%%%%%%%%%%%%%%%%%
%
%%%%%%%%%%%%%%%%%%%%%%%%%%%%%%%%%%%%%%%%%%%%%%%%%%
\subsection*{Matrix element and differential cross section
\label{hardmatint}} 
%%%%%%%%%%%%%%%%%%%%%%%%%%%%%%%%%%%%%%%%%%%%%%%%%%
%
The matrix element for the initial-final state 
interference contribution to real bremsstrahlung 
is given by:
\ba
\label{dhardmatint}
{\cal M}_{int} &=& (2 \pi)^4 
\delta^{(4)}(k_1 + k_2 - p_1 - p_2 - p)
\nonumber\\
&\cdot&\Bigl[(2 \pi)^{15} \, 2 k_1^0 \, 2 k_2^0 \, 2 p_1^0 \, p_2^0 \, 
 2 p^0 \Bigr]^{-\frac{1}{2}} \, 
\frac{1}{4}\,
\frac{
\sum_{spin} 2\,\Re{e}
\left\{ 
M_{ini} M_{fin}^*
\right\}
}
{2\,s\,\beta_0} .
\ea
The hard photon cross section part
\ba
\label{dhardint1a}
\hspace*{-0.2cm}
\frac{d \sigma^{hard}_{int}}{d \cos{\vartheta} } =
\frac{1}{2}\,\frac{s}{(4 \pi)^4}
\Biggl[\int_{I}+\int_{II} - \int_{III} \Biggr]
\frac{1}{4}\,
\frac{
\sum_{spin} 2\,\Re{e}
\left\{ 
M_{ini} M_{fin}^*
\right\}
}
{2\,s\,\beta_0}
{d {\varphi}_{\gamma}}\, {d\,x}\, {d\,R},
\ea
then contains the following squared amplitude $|M_{ini}|^2$
including all mass terms:
\ba
\label{dhardint1b}
&&2\,\Re{e}\left\{ M_{ini} M_{fin}^* \right\} = 
\nonumber\\
\nonumber\\ 
&=&(4\pi\alpha)^3
\Biggl\{
\frac{{\cal V}(s,s')}{s\, s'} 
\Biggl[
\frac{1}{V_1}
\left( s\, (Z_1 - Z_2) + (U - T)(3 s - s' - 4\, m_f^2) \right)
 \nonumber\\
&&\,+\,
\frac{1}{V_2}
\left( - (Z_1 - Z_2) (s' + 4\,m_f^2) + (U - T)( - 3 s' + s - 4\, m_f^2) \right)
\nonumber\\
&&\,+\,
\frac{s-T}{Z_1\, V_1} 
\left( T^2 + (s' - T)^2 + 2\, m_f^2 s' - 2\,T U + s^2 + 2\,{m_f^2}\,s \right)
\nonumber\\
&&\,-\,
\frac{s'-U}{Z_1\, V_2}
\left( U^2 + (s - U)^2 + 2\, m_f^2 s - 2\,T U + {s'}^2 + 2\,{m_f^2}\,s' \right)  
 \nonumber\\
&&\,+\,
\frac{s'-T}{Z_2\, V_2} 
\left( T^2 + (s - T)^2 + 2\, m_f^2 s - 2\,T U + {s'}^2 + 2\,{m_f^2}\,s' \right)
\nonumber\\
&&\,-\,
\frac{s-U}{Z_2\, V_1}
\left( U^2 + (s' - U)^2 + 2\, m_f^2 s' - 2\,T U + s^2 + 2\,{m_f^2}\,s \right)
\nonumber\\
&&\,+\,
\left(\frac{1}{Z_2}-\frac{1}{Z_1}\right) 
\left( s^2 + {s'}^2 + 2\, m_f^2 (s + s') \right)
\Biggr]
\nonumber\\
&&\,+\,
\frac{{\cal A}(s,s')}{s\, s'}   
\Biggl[
 2\,( s+s') + 4\, m_f^2 
+ 2 s\, s'\left(\frac{1}{V_1}-\frac{1}{V_2}\right) 
\nonumber\\
&&
+ (-s\,U + s'\,T)\frac{1}{Z_1} + (s'\,U - s\,T)\frac{1}{Z_2} 
\nonumber\\
&&
+ (V_2 - 2\,s)\frac{2\,m_f^2}{V_1} + (V_1 - 2\,s')\frac{2\,m_f^2}{V_2} 
\nonumber\\
&&\,+\,
\frac{s - U}{Z_1\, V_1}
\left( {s'}^2 - 2\,s' T - s\,(T - U) \right)   
\nonumber\\
&&\,+\,
\frac{s - T}{Z_2\, V_1}
\left( {s'}^2 - 2\,s' U + s\,(T - U) \right) 
 \nonumber\\
&&\,+\,
\frac{s' - T}{Z_1\, V_2}
\left( s^2 - 2\,s\, U + {s'}\,(T - U) \right) 
\nonumber\\
&&\,+\,
\frac{s' - U}{Z_2\, V_2}
\left( s^2 - 2\,s\, T - {s'}\,(T - U) \right) 
\Biggr]
\nonumber\\
&&\,+\,
\frac{m_f^2}{s\, s'}\,{\cal C}_1(s,s')
\,
\Biggl[
(Z_1 - Z_2)\left(\frac{1}{V_2}-\frac{1}{V_1}\right)
\nonumber\\ 
&&
+ (s + s') \left(-\frac{s - U}{Z_1\, V_1} + \frac{s - T}{Z_2\, V_1}
+\frac{U}{Z_1\, V_2} - \frac{T}{Z_2\, V_2}\right)
\Biggr]
\nonumber\\
&&\,+\,
\frac{m_f^2}{s\, s'}\,{\cal C}_2(s,s')
\,
(s + s')\left(\frac{1}{V_1}+\frac{1}{V_2}\right) 
\Biggr\}.
\ea

The coupling functions ${\cal V}(s,s')$ and ${\cal A}(s,s')$ 
are generalizations of (\ref{V}) and (\ref{A}) with the 
kinematic invariants $Z_1$, $Z_2$, $V_1$, $V_2$, $T$, and $U$ 
from (\ref{invs}) to (\ref{invv2}) and (\ref{T}) and (\ref{U}), 
while the extra coupling factors $C_{1,2}(s,s')$
are new for the interference part:

\ba
\label{Vsspr}
{\cal V}(s,s') &=& 
Q^2_e Q^2_f 
  +  Q_e Q_f \, v_e\,v_f\, 
\left(|\chi(s)|\eta(s) + |\chi(s')|\eta(s')\right)
\\
&&\,+\,
(v_e^2+a_e^2)\,(v_f^2+a_f^2)
\, |\chi(s)\chi(s')|\,\left(\eta(s) \eta(s') + \zeta(s)\zeta(s')\right),
\nonumber\\
\nonumber\\
\label{Asspr}
{\cal A}(s,s') &=& 
 Q_e Q_f \, a_e\, a_f\, 
\left(|\chi(s)|\eta(s) + |\chi(s')|\eta(s')\right)
\nonumber\\
&&\,+\,
4\, v_e\,v_f\,a_e\,a_f\,
|\chi(s)\chi(s')|\,\left(\eta(s) \eta(s') + \zeta(s)\zeta(s')\right),
\\
\nonumber\\
\label{C1}
{\cal C}_1(s,s') &=& 
4\, ( v_e^2 + a_e^2 )\, a_f^2\,
|\chi(s)\chi(s')|\,\left(\eta(s) \eta(s') + \zeta(s)\zeta(s')\right),
\\
\nonumber\\
\label{C2}
{\cal C}_2(s,s') &=&  
2\, a_e\, a_f\, Q_e Q_f  
\left(|\chi(s)|\eta(s) - |\chi(s')|\eta(s')\right),
\\
\nonumber\\
\mbox{with}&& \eta(s) = \kappa  \;\frac{s-M_Z^2}
{\sqrt{(s-M_Z^2)^2 + \left(\frac{s}{M_Z}\Gamma_Z\right)^2}},
\\
&& \zeta(s) = \kappa \;\frac{s \Gamma_Z / M_Z}
{\sqrt{(s-M_Z^2)^2 + \left(\frac{s}{M_Z}\Gamma_Z\right)^2}},
\\
&&
\chiz(s) = \chi(s)\; \equiv\; \kappa \; \frac{s}{\iprop},        
\\
&&     
\kappa = \frac{g^2}{4 e^2 \cow}
       = \frac{1}{4 \siw \cow}
       = \frac{\Gmu}{\sqrt 2 }\;\frac{\MzS}{2 \pi \alpha}.
\ea 
All other neutral current properties are 
summarized in Appendix \ref{feynmat}.
Interesting to note is that the coupling constant 
functions ${\cal V}(s,s')$, ${\cal A}(s,s')$, 
and ${\cal C}_{1,2}(s,s')$ in the interference term 
naturally depend on both $s'$ and $s$. 
It holds: ${\cal V}(s')= {\cal V}(s',s')$,
${\cal A}(s')= {\cal A}(s',s')$,
which reproduces the initial state factors
${\cal V}(s')$ and ${\cal A}(s')$.
In (\ref{dhardint1b}) we again will neglect final state 
mass terms proportional to $m_f^2/s$ as for the initial
state bremsstrahlung.  

Furthermore, (\ref{dhardint1b}) contains as denominators 

\ba
\label{denomint}
\frac{1}{V_i}, \quad \frac{1}{Z_i}, 
\quad\mbox{and}\quad \frac{1}{Z_i\,V_j}, \quad i,j=1,2. 
\ea
In comparison, the matrix elements for initial state 
or final state bremsstrahlung terms are proprortional 
to $1/Z_i$, $1/(Z_i Z_j)$, and $1/Z_i^2$, or  
to $1/V_i$, $1/(V_i V_j)$, and $1/V_i^2$ respectively.
For the integration of terms proportional to $\frac{1}{Z_i}$ 
in the matrix element for the QED interference
the same separation of phase space into 
different regions for the variables $\cos\vartheta$ occurs 
as in Appendix \ref{hardini}. This is done 
in order to treat the occurring mass singularities after
neglecting initial mass terms $m_e^2/s$.
The invariants $\frac{1}{V_i}$, whether appearing by themselves
or as factors of $1/(Z_j)$ are regular and do not possess
this singular behaviour and therefore do not affect the 
phase space splitting for the variable $\cos\vartheta$. 

Finally, there is a symmetric description of the initial
state and initial-final state interference results possible, 
for the angular cross section distributions, as well as for 
the totally integrated results.
This fact also went into the implementation 
of these new results in the updated program version 
${\tt ZFITTER}$ \cite{Bardin:1999yd-orig}.

%------------------------------------------------------------
\subsection*{The integrated hard radiators
${H}^{int}_{T,FB}$
\label{hardintfun}
}
%------------------------------------------------------------
%
The integrated results ${H}^{int}_{T,FB}(R,A(R),c)$, 
for the interference of initial and final state hard photon
emission are presented below. They can be factorized 
in one-loop approximation from an improved Born cross 
section or asymmetry $\sigma_{T,FB}^0(s,s')$ for the 
differential cross section contribution $d{\sigma^{int}}$. 
From $d{\sigma^{int}}$ then the QED interference contributions
$\sigma_{T,FB}^{int}(s,\bar{\xi},E_{\min},c)$ can be
derived.

Again, the value $c$ is defined 
as symmetric cut on the minimal and maximal scattering angle of one 
final-state fermion and $\bar{\xi}$ and $E_{\min}$ 
as cuts on the maximal acollinearity and minimal energy
of the fermions. In the hard photon radiators 
${H}^{int}_{T,FB}(R,A(R),c)$ the dependence on the 
acollinearity and energy cut is introduced through 
the value $A=A(R;\bar{\xi},E_{\min})$. It parameterizes
the three different contributions 
from different regions of phase space for the variables 
$x$ and $R$, each region depending on only one of the cuts 
(see (\ref{A1}), (\ref{A2}), and (\ref{A3})).

The distinction of analytical expressions according to the 
splitting of phase space due to mass singularities, 
which was discussed in Section \ref{sub_lep1slc_mass_sing},
follows the same lines as Section \ref{sub_lep1slc_formacol} 
and Appendix \ref{hardini}.
See for this also the analytical results on 
${H}^{ini}_{T,FB}(R,A(R),c)$ 
in (\ref{hard_fg_1}) to (\ref{hard_fg_6}).

%-----------------------------------------------------------------------
\ba
\label{sighard}
\sigma_{{int}_{A}}^{hard}
(s,\bar{\xi},E_{\min},c,\varepsilon) 
&=&\int^{1 - \varepsilon}_{R_{min}} 
d{R}\,\left[ 
{H}^{int}_{A}(R,A(R),c)\, \sigma_{A}^0(s,s')\right], 
\quad A = T,FB,
\nonumber\\
\ea
\ba
\label{hardinttot1}
\mbox{I. case}:\qquad
&&(1)^{+\,+}
\leftrightarrow~ (2)^{+\,+}\quad (\Rightarrow A\ge A_0(R))  
%1.\qquad I=0\quad &,& J=0\quad (\Rightarrow K =0) : 
\nonumber\\
{H}^{int}_{T}(R,A,c)& = & 
\frac{\alpha}{\pi}\,{Q_e Q_f}\,
\Biggl\{
{\cal F}_{00}(R,A,c) - {\cal F}_{00}(R,-c,A) 
+ {\cal C}_0(R,A,c)
\Biggr\},
\nonumber\\
\\
{H}^{int}_{FB}(R,A,c)& = & 
\frac{\alpha}{\pi}\,{Q_e Q_f}\,
\Biggl\{
{\cal G}_{00}(R,A,c) + {\cal G}_{00}(R,-c,A)
\Biggr\},
\\
\nonumber\\
\mbox{II. case}:\qquad
&&(1)^{+\,-}
\leftrightarrow~ (2)^{+\,-}\quad (\Rightarrow A< A_0(R)) 
%2.\qquad I=1\quad &,& J=1\quad (\Rightarrow K = 1) : 
\nonumber\\
{H}^{int}_{T}(R,A,c)& = & 
\frac{\alpha}{\pi}\,{Q_e Q_f}\,
\Biggl\{
{\cal F}_{11}(R,A,c) - {\cal F}_{11}(R,-c,A) 
+ {\cal C}_0(R,A,c)
\Biggr\},
\nonumber\\
\\
{H}^{int}_{FB}(R,A,c)& = & 
\frac{\alpha}{\pi}\,{Q_e Q_f}\,
\Biggl\{
{\cal G}_{11}(R,A,c) + {\cal G}_{11}(R,-c,A)
+ {\cal G}_1(R,A,c) 
\nonumber\\
&& 
+\, 2\,{\cal G}_{10}(R,A,c)
\Biggr\},
\\
\nonumber\\
\mbox{III. case}:\qquad
&&(1)^{+\,+}
\leftrightarrow~ (2)^{+\,-}  
%3.\qquad I=1\quad &,& J=0\quad (c\ge 0) : 
\nonumber\\
{H}^{int}_{T}(R,A,c)& = &  
\frac{\alpha}{\pi}\,{Q_e Q_f}\,
\Biggl\{
{\cal F}_{11}(R,A,c) - {\cal F}_{00}(R,-c,A) 
+ {\cal F}_{10}(R,c)
\nonumber\\
&&
+\, {\cal C}_0(R,A,c)
\Biggr\},
\\
a.~  A< A_0(R) &:&
\nonumber\\
{H}^{int}_{FB}(R,A,c)& = & 
\frac{\alpha}{\pi}\,{Q_e Q_f}\,
\Biggl\{
{\cal G}_{11}(R,A,c) + {\cal G}_{00}(R,-c,A) 
+ {\cal G}_1(R,A,c)
\nonumber\\
&&
-\, {\cal G}_0(R,A,c) + {\cal G}_{10}(R,A,c)
\Biggr\},
\\
b.~ A\ge A_0(R) &:&
\nonumber\\
{H}^{int}_{FB}(R,A,c)& = & 
\frac{\alpha}{\pi}\,{Q_e Q_f}\,
\Biggl\{
{\cal G}_{11}(R,A,c) + {\cal G}_{00}(R,-c,A) 
+ {\cal G}_0(R,A,c)
\nonumber\\
&&
+\, {\cal G}_{10}(R,A,c)
\Biggr\},
\ea

\vfill\eject
%------------------------------------------------------
\ba
\mbox{IV. case}:\qquad
&&(1)^{-\,-}
\leftrightarrow~ (2)^{+\,+} 
%4.\qquad I=1\quad &,& J=-1\quad (c\ge 0) : 
\nonumber\\
{H}^{int}_{T}(R,A,c)& = & 
\frac{\alpha}{\pi}\,{Q_e Q_f}\,
\Biggl\{
{\cal F}_{11}(R,A,c) + {\cal F}_{11}(R,-c,A) 
+ {\cal C}_0(R,A,c)
\Biggr\},
\nonumber\\
\\
a.~  A< A_0(R) &:&
\nonumber\\
{H}^{int}_{FB}(R,A,c)& = & 
\frac{\alpha}{\pi}\,{Q_e Q_f}\,
\Biggl\{
{\cal G}_{11}(R,A,c) - {\cal G}_{11}(R,-c,A) 
+ {\cal G}_1(R,A,c)
\nonumber\\
&&
+\, 2\,{\cal G}_{10}(R,A,c)
\Biggr\},
\\
b.~  A\ge A_0(R) &:&
\nonumber\\
{H}^{int}_{FB}(R,A,c)& = & 
\frac{\alpha}{\pi}\,{Q_e Q_f}\,
\Biggl\{
{\cal G}_{11}(R,A,c) - {\cal G}_{11}(R,-c,A) 
+ 2\,{\cal G}_0(R,A,c)
\nonumber\\
&&
+\, 2\,{\cal G}_{10}(R,A,c)
\Biggr\}.
\ea
%-----------------------------------------------------------------------
%
The functions ${\cal F}_{i(j)}$, ${\cal G}_{i(j)}$, 
and ${\cal C}_{0}$ appearing above depend on the following logarithms,
linear functions, variables, and cut parameters ($A=A(R)$)
($i,j=0,1$):
\ba
\label{hardinttot2}
{\cal F}_{ii}(R,A,c) &=&
 {\cal F}_{ii}\left(L_z(R,c);L_z(R,\pm A);
 L^{\pm}(c);L^{\pm}(A);z(R,c);R,A,c\right),
\nonumber\\
\\
{\cal G}_{ii}(R,A,c) &=&
 {\cal G}_{ii}\left(L_z(R,c);L_z(R,\pm A);
 L^{\pm}(c);L^{\pm}(A);z(R,c);R,A,c\right),
\nonumber\\
\\
{\cal G}_{0,1,10}(R,A,c) &=& 
{\cal G}_{0,1,10}\left(L_z(R,\pm A);L^{\pm}(A);R,A,c\right),
\\
\nonumber\\
{\cal C}_{0} &=& {\cal C}_{0}(R,A,c),
\\
\nonumber\\
{\cal F}_{10} &=& {\cal F}_{10}(R,c).
\ea
%-----------------------------------------------------------------------
%
${\cal F}_{00}$, ${\cal G}_{00}$, ${\cal F}_{11}$, and
${\cal G}_{11}$ together with ${\cal C}_{0}$, ${\cal F}_{10}$,
and ${\cal G}_{0,1,10}$ are now illustrated below ($v\equiv 1-R$):

%-----------------------------------------------------------------------
\ba
\label{hardinttot3}
%------------------------------------------------
{\cal F}_{00}(R,A,c) &=& 
\frac{2}{v} \left[(1-c^2)\,\ln\left(\frac{1+c}{1-c}\right)- 2 c \right]
\\
&&\,-\, ( 1 + 2 R - c^2 + 2 c R )\,
\left[\ln\left(\frac{1+c}{1-c}\right)-\ln{R}\right]
\nonumber\\
&&\,-\, \frac{1}{2}\left[(1 - R)^2 - c^2 (1 + R)^2 \right]
\nonumber\\
&&\,
\left\{\ln\left[\frac{z^2(R,c)}{4 R^2(1-c)^2}\right]+ \ln(1-A^2) + 1\right\}
\nonumber\\
&&\,+\, c (3 + 2 R + R^2)- (1+R^2), 
\nonumber\\
\nonumber\\
%------------------------------------------------
{\cal F}_{11}(R,A,c) &=& 
\frac{2}{v} \left[(1-c^2)\,\ln\left(\frac{1+A}{1-A}\right) - 2 A \right]
\\
&&\,-\, \frac{1}{2} \left[(1-c^2) ( 3 + 2 R + R^2 ) + 4 c R\right]\,
\ln\left(\frac{1+A}{1-A}\right)
\nonumber\\
&&\,+\, A \left[ 2 (1+R+R^2) + \frac{1}{2} (1+c^2) (1-R^2) - c (1-R)^2\right],
\nonumber\\
\nonumber\\
{\cal C}_{0}(R,c) &=& 2 A c (1-R^2),
\\
\nonumber\\
{\cal F}_{10}(R,c) &=& \frac{2}{v} (1-c^2) \ln{R},
\\
\nonumber\\
\label{hardinttot4}
{\cal G}_{00}(R,A,c) &=& 
-\frac{2}{v} \left\{ \left(c + \frac{c^3}{3}\right)\, 
\left[\ln\left(\frac{1+c}{1-c}\right)-\ln{R}\right]\right.
\\
&&\,+\,\left. \frac{1}{3}\left[ 
4\ln(1-c^2)-8\ln\left[\frac{z(R,c)}{1+R}\right] 
+ c^2\right] \right\}
\nonumber\\
&&\,+\, ( 1 + 2 R - c^2 + 2 c R )\, \ln\left(\frac{1+c}{1-c}\right)
\nonumber\\
&&\,+\, \frac{1}{3} (1 + 8 R + 5 R^2 )\, 
\left\{\ln(1-c)-\ln\left[\frac{z(R,c)}{1+R}\right]\right\} 
\nonumber\\
&&\,+\,\left[  \frac{c^3}{3} (1+R+R^2) + \frac{c^2}{2} (1-R^2)+ c (1-R+R^2) \right]
\nonumber\\
&&\, \left\{\ln\left[\frac{z^2(R,c)}{4 R^2(1-c)^2}\right] + \ln(1-A^2) \right\}
\nonumber\\
&&\,+\, \frac{1}{2}c^2\left\{ (1-R^2)\, \ln\left(\frac{1+A}{1-A}\right)
+2 R\,\ln\left[\frac{z(R,A)}{z(R,-A)}\right] \right\}
\nonumber\\
&&\,+\, c (c-2 R)\, \ln{R}
\nonumber\\
&&\,-\,  \frac{4}{z(R,c)}\,\left[ 2 (1+R) + 4 c + \frac{c}{1+R} + \frac{5 c^2}{1+R} 
+ \frac{2 c^3}{(1+R)^2} \right]
\nonumber\\
&&\,+\, \frac{c^3}{12} \left[ A^2 (1-R)^2 + 5 (1+R)^2 \right]
\nonumber\\
&&\,+\, c^2\left[- \frac{A}{2} (1-R^2)
- \frac{1}{2} R^2 + \frac{4}{3} R + \frac{19}{6}
+ \frac{2}{1+R} + \frac{4}{(1+R)^2} \right]
\nonumber\\
&&\,+\, c\left[ \frac{A^2}{4} (1-R)^2 +\frac{1}{4} R^2
- \frac{1}{6} R + \frac{17}{4} + \frac{8}{1+R} \right] + 8 ,
\nonumber
\ea

\vfill\eject
%------------------------------------------------
\ba
{\cal G}_{11}(R,A,c) &=& 
- \frac{2}{v}\left( c + \frac{c^3}{3}\right)
\,\ln\left(\frac{1+A}{1-A}\right)
\nonumber\\
&&\,+\,\frac{1}{2}\left\{ 2\left[c + \frac{c^3}{3}\right] (1 + R + R^2) 
- c^2 (1+R^2)- \frac{1}{3}\left(5 - 4 R + 5
R^2\right)\right\}
\nonumber\\
&&\,\ln\left(\frac{1+A}{1-A}\right)
\nonumber\\
&&\,+\, \frac{1}{2} A ( 1 + c^2 ) (1-R) \left[(1-R) + c (1+R) \right],
\\
\nonumber\\
%------------------------------------------------
{\cal G}_{0}(R,A,c) &=& 
- \frac{2}{3v}
\left\{ 4 \left[ \ln(1-A^2) + \ln\left[\frac{(1+R)^2}{4 R}\right]\right] 
+ A^2 \right\}
\nonumber\\
&&\,+\,\frac{1}{6} ( 11 + 8 R + 5 R^2 ) \,
\left\{\ln(1-A^2)+\ln\left[\frac{(1+R)^2}{4 R}\right]-\ln{R}\right\}
\nonumber\\
&&\,+\, ( 1 + 2 R )\, \ln{R}
\nonumber\\
&&\,-\, \frac{1}{2} A c^2 (1-R^2)
\nonumber\\
&&\,+\, \frac{1}{3} A^2 (1 + R + R^2)
\nonumber\\
&&\,-\,  \frac{2}{1+R}+ \frac{11}{6} - R + \frac{1}{6} R^2,
\\
\nonumber\\
%------------------------------------------------
{\cal G}_{1}(R,A,c) &=& 
\frac{1}{3} ( 5- 4 R + 5 R^2 )\,
\ln\left(\frac{1+A}{1-A}\right) 
\nonumber\\
&&\,-\, A (1-R)\left[(1-R) + c^2 (1+R)\right], 
\\
\nonumber\\
%------------------------------------------------
{\cal G}_{10}(R,A,c) &=& 
 \frac{1}{2} c^2\left\{(1-R^2)\, \ln\left(\frac{1+A}{1-A}\right) 
+  2 R\, \ln\left[\frac{z(R,A)}{z(R,-A)}\right]\right\}.
\ea
%-----------------------------------------------------------------------

\newpage

%####################################
\section{Final state radiation
\label{fin}
}
%####################################
%
%%%%%%%%%%%%%%%%%%%%%%%%%%%%%%%%%%%%%%%%%%%%%%%%%%
\subsection*{Matrix element and differential cross section
\label{hardmatfin}} 
\begin{figure}[htp]
\vspace*{-0.5cm}
\begin{flushleft}
\begin{center}
\hspace{0.0cm}  
%--- 
\mbox{%
  \epsfig{file=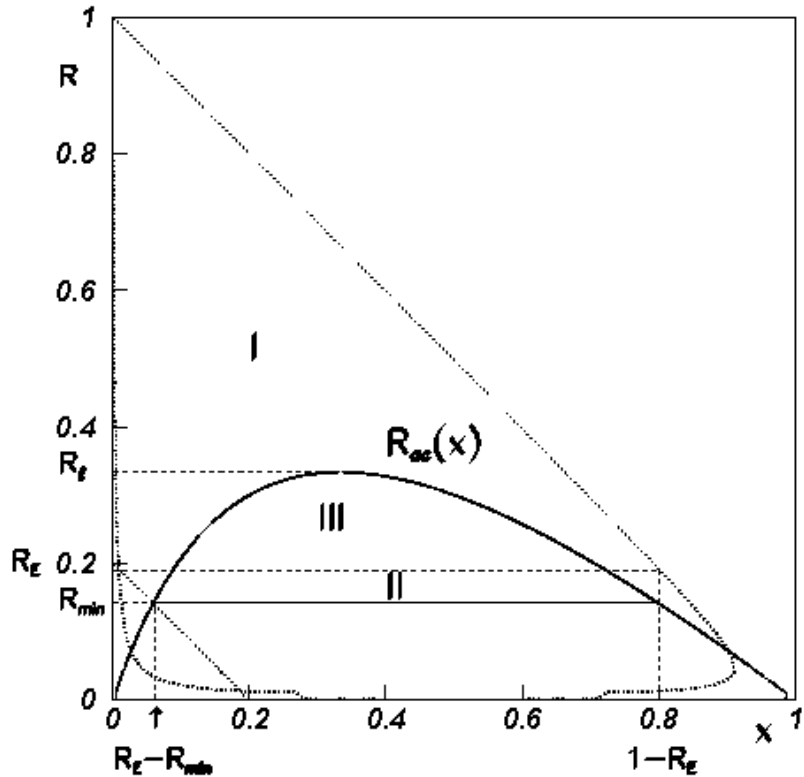
          ,height=12cm  % this is the height of the figure (optional)
          ,width=12cm   % this is the width of the figure (optional)
         }%
  } 
\end{center}
\end{flushleft}
\vspace*{-1cm}
\caption[Phase space with acollinearity cut (axis rotated)]{\sf 
 Phase space with cuts on maximal acollinearity and minimal 
fermion energy.
\label{dalitz_fin}
}
\end{figure}

%%%%%%%%%%%%%%%%%%%%%%%%%%%%%%%%%%%%%%%%%%%%%%%%%%
%
The kinematically allowed phase space region is given in
Fig.~\ref{dalitz_fin}. For the third region of phase space
III, it will be necessary to exchange the order
of the last two integrations over $x$ and $R$
in order to integrate the final state 
terms analytically completely for total 
or differential cross sections.
%
%%%%%%%%%%%%%%%
%---------------------
%---------------------
%%%%%%%%%%%%%%%%%%%%%%%%%%%%%%

The kinematics for the hard photon phase space has been
extensively discussed in Appendix \ref{crossphase} with
all kinematical variables, invariants, and boundaries 
defined by the cuts on maximal acollinearity and minimal
energy of the final state. A cut on one scattering
angle can be treated in addition.

The equation of the phase space boundary defined by the 
acollinearity cut, shown in Fig.~\ref{dalitz_fin},
can be written as 
\ba
\label{Racol}
R= R_{ac}(x)&=&\frac{x\,(\,1\,-\,x\,)}
                       {x\,+\, \rho\, - \,1},
\\
\label{Rbar}
 R_{\min}&=& \frac{R_E \,(1\, - \,R_E)}{\rho \,- \, R_E},
\\
\label{rho}
  \rho &=& 1\,+\, \tan^2{\frac{\bar{\xi}}{2}}\, = 
\,\frac{(1+R_{\xi})^2}{4 R_{\xi}}\geq 1.
\ea

Here we have introduced $R_{\min}$ as starting value of 
the variable $R$ which can be substituted
by a sufficiently large see (\ref{sprcut})
and the parameter $\rho$ following the notation 
of \cite{Montagna:1993mf}.  
Results on the complete final state corrections to 
the total cross section $\sigma_T$ were derived there
first. The calculation is repeated here and 
some small misprints in the original work 
can be corrected. Moreover, the complete 
differential cross section ${d{\sigma}}/{d{\cos\vartheta}}$
is presented here from which $\sigma_T$ and also 
the forward-backward asymmetry $A_{FB}$, not 
given in \cite{Montagna:1993mf}, trivially can be
extracted. 

Using the separation of phase space into three 
different regions I, II, and III, discussed 
in Section \ref{sec_lep1slc_corracol} and Appendix \ref{threephase} 
(see also Fig.~\ref{dalitz} and Fig.~\ref{dalitz_fin}),
hard photonic final state contribution to the differential 
cross section can be written as:
\ba
\label{dhard1}
 \frac{d {\sigma}^{hard}_{fin}}{d \cos{\vartheta} }  \,=\,
\Biggl[\int_{I}+\int_{II} - \int_{III} \Biggr]
\,d{\varphi_{\gamma}}\,d\,x\,d\,R
\frac{d {\sigma}^{hard}_{fin}}
{d{\varphi_{\gamma}}\, d x\, d R\, d \cos{\vartheta} }. 
\ea
Starting from the final state bremsstrahlung matrix element
\ba
\label{dhardmat}
{\cal M}_{fin} &=& (2 \pi)^4 
\delta^{(4)}(k_1 + k_2 - p_1 - p_2 - p)
\nonumber\\
&&\Bigl[(2 \pi)^{15} \, 2 k_1^0 \, 2 k_2^0 \, 2 p_1^0 \, p_2^0 \, 
 2 p^0 \Bigr]^{-\frac{1}{2}} \, {M}_{fin} ,
\ea
with ${M}_{fin}$ given in (\ref{br02b}),
one can write the contribution of the final state hard 
bremsstrahlung in terms of the four independent kinematic 
variables $\varphi_\gamma$, $x$, $R$, and $\cos{\vartheta}$:
\ba
\label{dhard1a}
\frac{d \sigma^{hard}_{fin}}{d \cos{\vartheta} } 
=
\frac{s}{(4 \pi)^4}\, 
\Biggl[\int_{I}+\int_{II} - \int_{III} \Biggr]
\frac{\sum_{spin}\,|M_{fin}|^2}{2\,s\,\beta_0}
\,{d{\varphi}_{\gamma}}\, {d\,x}\, {d\,R}.
\ea

\vfill\eject
%-----------------------------------------------------------
\noindent The squared amplitude $|M_{fin}|^2$ is given as:
\ba
\label{dhard1b}
|M_{fin}|^2 &=& 
(4\pi\alpha)^3
\Biggl\{
\frac{{\cal V}(s)}{s^2} \Biggl[
\frac{2}{V_1 V_2} 
\left( s^2 - T\, U - (U+Z_1)\,(T+Z_2) + 2\,s'\,{m_f^2} \right)
 \nonumber\\
&&+\frac{V_2}{V_1} +\frac{V_1 - 2 s}{V_2} 
+ 
4 \frac{{m_f^2}}{s} \left(\frac{U}{V_1} \,
                           +\,\frac{U+Z_1}{V_2} \right)
\left(\frac{T}{V_1} \,+\,\frac{T+Z_2}{V_2} \right)
 \nonumber\\
&&-  2 {m_f^2} \,(s + 2 {m_f^2})\,
    \left(\frac{1}{V_1}+\frac{1}{V_2} \right)^2  \Biggr] 
\\
&&+
\frac{{\cal A}(s)}{s^2} \Biggl[
\left(- \frac{2 {m_f^2}}{V_1}+\frac{s'-2 {m_f^2}}{V_2} +1 \right)
             \frac{U-T}{V_1} \nonumber\\
&&+ 
\left(- \frac{2 {m_f^2}}{V_2}+\frac{s'-2 {m_f^2}}{V_1} +1 \right)
             \frac{U+Z_1-T-Z_2}{V_2}   \Biggr]  \nonumber\\
&&+ 
\frac{8 {m_f^2}}{s}\,(v_e^2+a_e^2)\,a_f^2 
\,\frac{1}{s^2} \, {\mid \chi(s) \mid}^2
 \left[\frac{Z_1 \,Z_2 - s\,s'}{V_1 V_2}
+ s\,{m_f^2} \left(\frac{1}{V_1}+\frac{1}{V_2} \right)^2 \right] 
\Biggr\},
\nonumber
\ea
where ${\cal V}(s)$ and ${\cal A}(s)$ are functions 
of the center-of-mass energy squared $s$ in the final state
case. They contain the neutral current couplings
and $Z$ boson propagator and are given in (\ref{V}) and (\ref{A}).
%
%\ba
%\label{V}
%{\cal V}(s) &=& {Q}^2_e {Q}^2_f 
%  + 2 \mid {Q}_e \mid \mid {Q}_f \mid\, v_e\, v_f\,{\Re}e \chi(s) 
%  + [v_e^2+a_e^2]\, [v_f^2+a_f^2]\, {\mid \chi(s) \mid}^2,      
%\nonumber\\
%\\
%\label{A}
%{\cal A}(s) &=&  
%  + 2 \mid {Q}_e \mid \mid {Q}_f \mid\, a_e\, a_f\,{\Re}e \chi(s) 
%  + 4\, v_e\, a_e\, v_f\, a_f\, {\mid\chi(s)\mid}^2.
%\ea 
%
The bremsstrahlung kinematic invariants 
$Z_1$, $Z_2$, $V_1$, $V_2$, $T$, and $U$ 
are taken from (\ref{invs}) to (\ref{invv2}) 
and (\ref{T}) and (\ref{U}). It is useful
to also introduce the dimensionless quantities:
\ba
\label{invv1tuv}
v_1\equiv \frac{V_1}{s},\quad t\equiv\frac{T}{s}, 
\quad u\equiv\frac{U}{s},
\quad\mbox{and}\quad v\equiv 1-R = 1-\frac{s'}{s}.
\ea

%---------------------------------------------------------
\subsection*{Integration over $\varphi_{\gamma}$}
%--------------------------------------------------------- 
%
The only invariants depending on $\varphi_{\gamma}$ are: 
\ba
\label{1Z1,Z2}
Z_{1,2} &=& a_{1,2} \pm b\, \cos{\varphi_{\gamma} }\,, 
\ea
where 
\ba
\label{3Z1,Z2}
a_{1,2} &=& \frac{1}{2} \, s\, v \, 
\left(1 \pm \beta_{0}\,\cos{\vartheta}\,
\cos{\theta_{\gamma}} \right) \,, \\
%\label{4Z1,Z2}
%a2 &=& \frac{1}{2} \, s\, {v} \, \left(1 - \beta_{0}\,\cos{\vartheta}\,
%\cos{\theta_{\gamma}} \right) \, , \\
\label{1cosgamma}
b^2\,&=& \, \frac{1}{4} \,s^2\,{{v}}^2\,\beta^2_{0}
\left( \,1\,-\,\cos^2{\theta}\,\right)
\, \left(\,1\,-\,\cos^2{\theta_{\gamma}}\,\right) \, ,\\  
\label{2cosgamma}
 \cos{\theta_{\gamma}} &=& \quad \frac{R\,v_1\, -\, x }
    {\,{v} \,\sqrt{\lambda_2} }.
\ea

\vfill\eject
%-----------------------------------------------------------
\vspace*{0.2cm}
\begin{center}
{\large  Table of integrals:}
\end{center}
\ba
\label{1intPHI}
 \frac{1}{ 2\pi} \int\limits_{0}^{ 2\pi} d\,\varphi_{\gamma}
 \,&=& 1,\quad 
\label{2intPHI}
 \frac{1}{ 2\pi} \int\limits_{0}^{ 2\pi} d\,\varphi_{\gamma}
 \,\cos{\varphi_{\gamma}}  = 0, \quad   
\label{3intPHI}
 \frac{1}{ 2\pi} \int\limits_{0}^{ 2\pi} d\,\varphi_{\gamma}
 \,\cos^2{\varphi_{\gamma}}  =  \frac{1}{2}.
\nonumber\\
%\label{4intPHI}
% \frac{1}{ \pi} \int\limits_{0}^{ \pi} d\,\varphi_{\gamma}
% \, Z_1\, &=& a_1  \\
%\label{5intPHI}
% \frac{1}{ \pi} \int\limits_{0}^{ \pi} d\,\varphi_{\gamma}
% \, Z_2\,&=& a_2 \\
%\label{6intPHI}
% \frac{1}{ \pi} \int\limits_{0}^{ \pi} d\,\varphi_{\gamma}
% \,Z_1\, Z_2 &=& a_1\,a_2 - \frac{1}{2}\,b^2 \,,
\ea 
The complete integration over $\varphi_{\gamma}$ yields:

\ba
\label{dhard1c}
\frac{1}{ 2\pi} \int\limits_{0}^{ 2\pi} d\,\varphi_{\gamma}
 \,|M_{fin}|^2 &=&   
%      \nonumber\\  &&\hspace{-25mm}
(4\pi\alpha)^3
\Biggl\{  
\frac{{\cal V}(s)}{s} \Biggl[
  \frac{1\,+\,\cos^2{\vartheta}}{2} \biggl(
\, -\,\frac{2 {m_f^2}}{s} \,\Bigl(\frac{1}{v_1^2} \,
     +\,\frac{1}{x^2} \Bigr)  +\frac{2}{v_1\,x}\nonumber\\
& & \qquad - \,\frac{1+R}{v_1} 
                            +\,\frac{1+R}{x}  \,-\,2
                          \biggr)   
%   \nonumber\\&& \hspace{-25mm}  \qquad \qquad \qquad 
+\,\left(\,1\,-\,3\,\cos^2{\vartheta} \right)
                     \frac{R}{(R+x)^2}\, \Biggr] 
\nonumber\\
\nonumber\\
&+& \frac{{\cal A}(s)}{s}\, \cos{\vartheta}\,  
\Biggl[
\, -\,\frac{2 {m_f^2}}{s} \,\Bigl(\frac{1}{v_1^2} \,
                           +\,\frac{1}{x^2} \Bigr)
\, +\,\frac{2}{v_1\,x}\, 
\nonumber\\
& & \qquad - \,\frac{1+R}{v_1} +\,\frac{1+R}{x}  \,
  -\,2 \,\frac{1+R}{R+x}\, 
\Biggr]\Biggr\} .
\nonumber\\
\ea
We only keep terms proportional to ${m_f^2}/v_{1,2}^2$
because their contributions could be non-negligible.
Elsewhere we set $m_f=0$.

%---------------------------------------------------------
\subsection*{Integration over $x$}
%---------------------------------------------------------
%
The integration over $x$ can be performed 
using the parameterization with $A(R)$ in (\ref{A1}),
(\ref{A2}), and (\ref{A3}). So we integrate over $x$ in the limits:
\ba
\label{limitsA}
    (1-R)\,\frac{1-A(R)}{2} \quad  \leq \quad x(R) 
  \quad  \leq  \quad   (1-R)\,\frac{1+A(R)}{2}.
\ea
The variable $R$ then remains as last variable of integration:

\vfill\eject
%-----------------------------------------------------------
\begin{center}
{\large Table of integrals }
\end{center}
\ba
\left[ f(x) \right]_{(x)} &=&
 \int\limits_{{v} \frac{1-A(R)}{2}}^{{v} \frac{1+A(R)}{2}} 
          d\,x  f(x),
\\
\nonumber\\
\label{1intV2}
\left[1 \right]_{(x)}  &=&    {v} \, A(R), \\
\label{2intV2}
\left[\frac{1}{v_{1,2}} \right]_{(x)} & =& \, \ln{\frac{1+A(R)}{1-A(R)}},\\
\label{3intV2}
\left[\frac{1}{v_1^2,x^2} \right]_{(x)}  &=&
         \, \frac{1}{{v}}\frac{4\,A(R)}{1-A(R)^2},    \\
%\label{4intV2}
%\Bigl[\frac{1}{x} \Bigr] &=& \, \ln{\frac{1+A}{1-A}} \\
%\label{5intV2}
%\left[\frac{1}{x^2} \right] &=&
%         \, \frac{1}{{v}}\frac{4\,A}{1-A^2}    \\
\label{6intV2}
\left[\frac{1}{x +R} \right]_{(x)}  &=& 
\, \ln{\frac{1+R\,+\,A(R)(1-R)}{1+R\, -\,A(R)(1-R)}},   \\
\label{7intV2}
\left[\frac{1}{(x+R)^2} \right]_{(x)}  &=&
         \, \frac{4\,A(R) \,{v}}{(1+R)^2-A(R)^2\,{{v}}^2}. 
\ea   
The result of the integration of (\ref{dhard1a}) with (\ref{dhard1c}) 
over $x$ is:

\ba
\label{dhard2}
\hspace{-0.7cm}  \frac{d {\sigma}^{hard}_{fin}}{d \cos{\vartheta} } &=&
 \frac{\pi \alpha^2}{s}  \frac{\alpha}{\pi} \, {Q_f^2} \,
    \Biggl[\int_{I}+\int_{II} - \int_{III} \Biggr] \, d\, R \,
\\
&&-
     \Biggl\{                                               
 {\cal V}(s)   \Biggl[
           \frac{1 + \cos^2{\vartheta}}{2}   
     \biggl( \frac{1+R^2}{{v}} \ln{\frac{1+A}{1-A}}      
    - A {v}  \nonumber\\
&& 
\hspace*{1cm}-\,\frac{8 {m_f^2} A}{s {v} (1-A^2)}  \biggr)       
   + \frac{1 - 3 \cos^2{\vartheta}}{2}
           \frac{4\,A\,{v}\,R}{(1+R)^2-A^2{v}^2}\,
                                                   \Biggr] 
\nonumber\\
&&+
  {\cal A}(s) \cos{\vartheta}  \Bigl[
    \frac{1+R^2}{{v}} \ln{\frac{1+A}{1-A}} \nonumber\\
&&\hspace*{1cm} -\, (1+R) \ln{\frac{1+R + A(1-R)}{1+R - A(1-R)}} 
  -\, \frac{8 {m_f^2} A}{s {v} (1-A^2)}  \Bigr] 
      \Biggr\}.
\nonumber
\ea

\vfill\eject
%---------------------------------------------------------
\subsection*{Integration over $R$ 
\label{intR}
}
%---------------------------------------------------------
%
%---------------------------------------------------------
\subsubsection*{Region I
\label{intR-I}
}
%---------------------------------------------------------
%
The value $R$ varies in the interval 
\ba
\label{limR_I}
 R_E \, \leq \, R \, \leq \, 1 - \varepsilon,
\ea
where $\varepsilon/2$ is the normalized minimal photon 
energy. The value $\varepsilon$ separates soft and 
hard photon phase space and cancels when adding the 
soft contributions (see Appendix \ref{kincon}). 

Only region I has to take into account 
the soft-photon corner of phase space, i.e. only
the expressions derived there have to be regularized.
For this, the soft and virtual photonic corrections from 
(\ref{dsigsvini}) with (\ref{deltasvini}) have to be added, 
with the necessary modifications for the final state case.
The hard-photon results for regions II and III are 
finite.

For region I, logarithmic mass terms 
$L_f=\ln(s'/m_f^2)$ have to be taken into account.
The parameter $A(R)$ for the final state expressions 
of region I therefore has to be taken correctly with mass:

\ba
\label{1-A_I}
 1 - A^2(R) \,=\, 4 {\displaystyle \frac{{m_f^2}}{s R} }.
\ea
For the initial state bremsstrahlung and initial-final state 
interference expressions this be could set to $A \,\cong \,1$ 
(see Sections \ref{ssub_lep1slc_ini} and \ref{ssub_lep1slc_int}).
Inserting (\ref{1-A_I}) into (\ref{dhard2}), we get: 
\ba
\label{dhard2_I}
\frac{d {\sigma}^{hard}_{finI}}{d \cos{\vartheta} } &=&
  \frac{\pi\, \alpha^2}{s} \, \frac{\alpha}{\pi}\,{Q_f^2} \,
               \int_{I} d \, R \,
    \Biggl\{ \,   {\cal V}(s)      \,\biggl[
          \frac{1 + \cos^2{\vartheta}}{2}   \,
 \biggl(\ln{\frac{s}{{m_f^2}}} - 1 + \ln{R} \biggr)  
  \frac{1+R^2}{{v}} \nonumber\\ 
&& \qquad \qquad \qquad \qquad  \qquad
+ \frac{1 - 3 \cos^2{\vartheta}}{2} {v} 
                                   \biggr]   \nonumber\\
&& \qquad +\,  {\cal A}(s) \cos{\vartheta}\,  
   \left[ \biggl( \ln{\frac{s}{{m_f^2}}} - 1 \biggr)
         \frac{1+R^2}{{v}} \,+\, \frac{2}{{v}} \ln{R} + {v} 
     \right]       \Biggr\}.
\ea

The over $\cos\vartheta$ in region I integrated flux 
functions of (\ref{dhard2_I}) are given for the total cross section 
$\sigma_T$ in \cite{Bardin:1991fu} and for the 
forward-backward cross section $\sigma_{FB}$ 
in \cite{Bardin:1989cw}. 
The final integration of (\ref{dhard2_I}) over $R$ uses
the following

\vfill\eject
%--------------------------------------------------------------------
\begin{center}
{\large Table of integrals }
\end{center} 
\ba
 \left[ f(R) \right]_{(R)} &=&
 \int\limits_{R_E}^{1-\varepsilon}
          d\,R \, f(R), 
\\
\nonumber\\
\label{1intR_I}
\left[1\, \right]_{(R)}  &=&  1 \,- \,R_E, \\
\nonumber\\
\label{2intR_I}
\left[\ln{R} \right]_{(R)}  &=& - 1  + R_E  - R_E \ln{R_E},   \\
\label{3intR_I}
\left[R\, \right]_{(R)}  &=&  \frac{1}{2}\  \Bigl(1 - R_E^2\Bigr), \\ 
\label{4intR_I}
\left[R \ln{R} \right]_{(R)}  &=&  \frac{1}{4}\  \Bigl(- 1 + R_E^2\Bigr)
                      -  \frac{1}{2} R_E^2 \ln{R_E},          \\
\label{5intR_I}
\left[\frac{1}{1-R} \right]_{(R)} &=& 
         - \ln{\frac{2\varepsilon}{\sqrt{s}}} + \ln(1-R_E), \\
\label{6intR_I}
\left[\frac{\ln{R}}{1-R} \right]_{(R)}  &=& 
   - {\rm Li}_2(1) + {\rm Li}_2(R_E) + \ln{R_E} \ln{(1-R_E)}.      
\ea
After integration over  $R$ we have obtained the hard 
bremsstrahlung contribution to the differential cross-section 
for region I:

\ba
\label{dhard3_I}
 \frac{d {\sigma}^{hard}_{finI}}{d \cos{\vartheta} }  \,&=&\,
     \frac{\pi\, \alpha^2}{s}  \frac{\alpha}{\pi}\,{Q_f^2} 
\Biggl\{   {\cal V}(s) \, \biggl[
           \frac{1 + \cos^2{\vartheta}}{2}  
  \Biggl(      \Bigl(\ln{\frac{s}{{m_f^2}} } - 1 \Bigr) 
  \Bigl(- 2\ln{\varepsilon} - 2\ln{2}\,
\nonumber\\
&&
+\, 2 \ln(1-R_E) 
                - \frac{3}{2} 
+\, R_E \,+\, \frac{R^2_E}{2} \Bigr) 
\,+\,  \frac{5}{4} \,-\, R_E  \,-\, \frac{R^2_E}{4} 
\nonumber\\
&&  
\,+\, R_E \bigl(1 + \frac{R_E}{2} \bigr) \ln{R_E} 
-\, \frac{\pi^2}{3} \,+\, 2 {\rm Li}_2(R_E)
+\, 2 \ln{R_E} \ln(1-R_E)
  \Biggr)                                                      
\nonumber\\
&&
+\, \frac{1 - 3 \cos^2{\vartheta}}{4} (1-R_E)^2     \biggr]        
\nonumber\\
&&
\,+\,  
{\cal A}(s)  \cos{\vartheta}
  \biggl[     \Bigl(\ln{\frac{s}{{m_f^2}} } - 1 \Bigr) 
 \Bigl(- \ln{\frac{(2\varepsilon)^2}{s}} \,+\, 2 \ln(1-R_E)     
       \,-\, \frac{3}{2} 
 \nonumber\\   
&&  
\,+\, R_E \,+\, \frac{R^2_E}{2} \Bigr)
\,+\,  \frac{1}{2}\bigl(1 - R_E \bigr)^2 
  \,-\, \frac{\pi^2}{3} 
  +\, 2 {\rm Li}_2(R_E) 
\nonumber\\ 
&&
\,+\, 2 \ln{R_E} \ln(1-R_E)
   \biggr]      \Biggr\}. 
\ea
Adding the soft and virtual contributions 
(see e.g.~(\ref{dsigsvini}) with (\ref{deltasvini})), 
the regularized result for phase space 
region I can be written as:

\ba
\label{dregul_I}
 \frac{d \sigma_{finI}}{d \cos{\vartheta}} 
& = &
\frac{d \sigma^{soft+virtual}_{fin}}{d \cos{\vartheta}}
+ \frac{d \sigma^{hard}_{finI}}{d \cos{\vartheta}}
\\
&=&
  \frac{\pi\, \alpha^2}{s}  \frac{\alpha}{\pi}\,{Q_f^2} \, 
 \nonumber\\ 
&&
\cdot\,
      \Biggl\{      {\cal V}(s) \biggl[
          \frac{1 + \cos^2{\vartheta}}{2}  
  \Biggl( \Bigl(\ln{\frac{s}{{m_f^2}} } - 1 \Bigr) 
      \Bigl( 2 \ln(1-R_E) + R_E + \frac{R^2_E}{2} \Bigr) 
\,+\, \frac{3}{4} \nonumber\\
&& \,-\, R_E  - \frac{1}{4} R^2_E 
  + R_E \Bigl(1 + \frac{1}{2} R_E \Bigr) \ln{R_E} 
\, + \, 2 {\rm Li}_2(R_E)
 \nonumber\\        
&&\, + \, 2 \ln{R_E} \ln{(1-R_E)}    \Biggr) 
 \,+\, \frac{1 - 3 \cos^2{\vartheta}}{4} (1-R_E)^2    \biggr] 
\\ 
&&
\,+\, 
 {\cal A}(s) \, \cos{\vartheta}  \biggl[
       \Bigl(\ln{\frac{s}{{m_f^2}} } - 1 \Bigr) 
     \Bigl( 2 \ln(1-R_E)  
\nonumber\\
&& \, + \, R_E \, + \, \frac{R^2_E}{2} \Bigr)    
 - R_E + \frac{R^2_E}{2} \, + \, 2 {\rm Li}_2(R_E) 
           \,  +\, 2 \ln{R_E} \ln(1-R_E)   \biggr]        
\Biggr\}.
\nonumber
\ea
One can see that this expression does not contain divergences nor
the parameter $\varepsilon$ used to distinguish the soft photons 
from hard photons. 

%---------------------------------------------------------
\subsubsection*{Region II
\label{intR-II}
}
%---------------------------------------------------------
%
The value for $A(R)$ here is given in (\ref{A2}). 
The value $R$ varies in the interval 
\ba
\label{limR_II}
 R_{\min} \, \leq \, R \, \leq \, \bar{R}_E,
\ea
where $\bar{R}_E \cong R_E$ (see Fig.~\ref{dalitz_fin}). 
The final state mass $m_f^2$ has now been neglected because
only mass terms proportional to $m_f^2$ or $m_f^2 L_f$ 
can appear which vanish for $m_f^2\to 0$.  
Inserting $A(R)$ from (\ref{A2}) into (\ref{dhard2}) we get
after partial fraction decomposition:
\ba
\label{dhard2_II}
\frac{d {\sigma}^{hard}_{finII}}{d \cos{\vartheta} }  &=&
  \frac{\pi\, \alpha^2}{s}  \frac{\alpha}{\pi}\,{Q_f^2} \,
                   \int_{II} \,d \, R \,
\\
&&
\cdot\,   
\Biggl\{     {\cal V}(s) \, \Biggl[
         \frac{1 + \cos^2{\vartheta}}{2}  \,
     \biggl( \frac{1+R^2}{{v}} \ln{\frac{1 - R_E}{R_E - R}}
                  - 1 - R + 2 R_E  \biggr)                     
\nonumber\\  
& & \qquad \qquad \qquad \qquad \qquad \quad \quad
 +   \frac{1 - 3 \cos^2{\vartheta}}{2}
   \biggl(\frac{R}{R_E} - \frac{R}{1 + R - R_E} \biggr) \Biggr]
\nonumber\\ 
&&+  
{\cal A}(s)\, \cos{\vartheta}  \Biggl[
         \frac{1+R^2}{{v}} \ln{\frac{1 - R_E}{R_E - R}}
          - (1 + R) \ln{\frac{1 + R - R_E}{R_E} }   \Biggr] 
  \Biggr\}.
\nonumber  
\ea
Introducing the notation from \cite{Montagna:1993mf},
\ba
\label{x_rho}
 x_{\rho} &=& 1 - R_E + R_{\min}  = 
   \frac{\rho (1 - R_E)}{\rho - R_E} ,
\\
\label{1-x_rho}
R_{\rm emin} &=& R_{cut} = 2 R_E-1,
\ea
we have with (\ref{Rbar}) and (\ref{rho}) 
the general inequalities: 
\ba
\label{Remin}
R_{\rm emin}  \leq     R_{\min},
&&
1 + R_{\min} - 2 R_E  \geq  0,
\\
\nonumber\\
\label{unequal}
 R_{\min}  \leq  R_E,  
&&
 0  \leq  R_E -  R_{\min}  \leq  1 - R_E, 
\\
\nonumber\\
\label{unequal2}
R_E \leq  x_{\rho} \leq 1,
&&
0 \leq 1 -  x_{\rho} \leq 1 - R_E. 
\ea
Now we can integrate over $R$ using the following
\vspace*{.2cm}
\begin{center}
{\large Table of integrals }
\end{center}
\ba
 \left[ f(R) \right]_{(R)} &=&
 \int\limits_{R_{\min}}^{R_E }
          d\,R \, f(R),
\\
\nonumber\\
\label{1intR_II}
\left[1 \right]_{(R)}  
&=&  
R_E - R_{\min},                       \\  
\label{2intR_II}
\left[R \right]_{(R)}  
&=&
      \frac{1}{2} \Bigl(R^2_E \,- R^2_{\min} \Bigr),       \\ 
\label{3intR_II}
\left[\frac{1}{1 + R - R_E} \right]_{(R)}  
&=& 
                        - \ln(1 +  R_{\min} - R_E),        \\
\label{4intR_II}
\left[\ln{\frac{1 - R_E}{R_E - R}} \right]_{(R)} 
&=& 
      \Bigl(R_E - R_{\min}\Bigr)  \left( 1 +
           \ln{\frac{1-R_E}{R_E - R_{\min}}} \right),     \\
\label{5intR_II}
\left[R \ln{\frac{1 - R_E}{R_E - R}} \right]_{(R)} 
&=& 
  \frac{1}{2} \Bigl(R^2_E - R^2_{\min}\Bigr)  \left( 
  \frac{1}{2} + \ln{\frac{1-R_E}{R_E - R_{\min}}} \right) \\
&& + \frac{1}{2} R_E \Bigl(R_E - R_{\min}\Bigr),    
\nonumber\\
\label{6intR_II}
\left[\frac{1}{1-R} \ln{\frac{1 - R_E}{R_E - R}} \right]_{(R)} 
&=& 
   - \frac{1}{2} \ln^2{\frac{1-R_E}{1 - R_{\min}}} + {\rm Li}_2(1)
   -  {\rm Li}_2\Bigl(\frac{1 - R_E}{1 - R_{\min}} \Bigr), 
\nonumber\\
       \\
\label{7intR_II}
\left[\ln{\frac{1 + R - R_E}{R_E}} \right]_{(R)} 
&=& 
    - \Bigl(R_E - R_{\min}\Bigr) \left( 1 + \ln{R_E} \right)
\nonumber\\
&&    - \Bigl(1 +  R_{\min} - R_E \Bigr)
                    \ln(1 + R_{\min} - R_E),                \\
\label{8intR_II}
\left[R \ln{\frac{1 + R - R_E}{R_E}} \right]_{(R)} 
&=& 
  - \frac{1}{2} \Bigl(R^2_E - R^2_{\min}\Bigr)  \left( 
  \frac{1}{2} + \ln{R_E} \right)    
\nonumber\\
&&   + \frac{1}{2} \Bigl( (1 - R_E)^2 - R^2_{\min} \Bigr) 
                    \ln(1 + R_{\min} - R_E)             
\nonumber\\ 
&& 
   + \frac{1}{2} \Bigl(1 - R_E \Bigr) \Bigl(R_E - R_{\min}\Bigr).          
\ea
So, performing the intagration over $R$ in (\ref{dhard2_II})
we obtain the hard bremsstrahlung contribution to the differential
cross-section for region II:

\ba
\label{dhard3_II}
\frac{d {\sigma}^{hard}_{finII}}{d \cos{\vartheta} }  &=&
  \frac{\pi\, \alpha^2}{s}  \frac{\alpha}{\pi}\,{Q_f^2} \,
     \Biggl\{     {\cal V}(s) \, \Biggl[
         \frac{1 + \cos^2{\vartheta}}{2}  \,
     \biggl( - 2 (R_E - R_{\min})   
  + \frac{3}{4} \Bigl(R_E -  R_{\min} \Bigr)^2 
\nonumber\\
&& \qquad \qquad \qquad \qquad \qquad
 - \ln^2{\frac{1 - R_E}{1 -  R_{\min}} } + \frac{\pi^2}{3}  
- 2 {\rm Li}_2{\left(\frac{1 - R_E}{1 -  R_{\min}} \right)}
 \nonumber\\
&&+ \Bigl[R_E \bigl(1 + \frac{R_E}{2} \bigr) 
 -  R_{\min} \bigl(1 + \frac{ R_{\min}}{2} \bigr) \Bigr]
  \Bigl[\ln(R_E -  R_{\min}) - \ln(1 - R_E) \Bigr]  \biggr) 
\nonumber\\ 
&  &  \qquad \qquad \qquad \qquad  \frac{1 - 3 \cos^2{\vartheta}}{2}
   \biggl( - \frac{\bigl(R_E -  R_{\min}\bigr)^2 }{2 R_E}  
   - (1 - R_E) \ln{x_{\rho}}  \biggr) \Biggr] 
\nonumber\\
&&+  
{\cal A}(s)\, \cos{\vartheta}  \Biggl[
  - \frac{R_E}{2} + \frac{ R_{\min}}{2} 
  + x_{\rho} \Bigl(R_E + \frac{x_{\rho}}{2} \Bigr) \ln{x_{\rho}}
\\  && 
 - \ln^2{\frac{1 - R_E}{1 -  R_{\min}} } + \frac{\pi^2}{3} 
 - 2 {\rm Li}_2{\left(\frac{1 - R_E}{1 -  R_{\min}} \right)}  
\nonumber\\
&&+ \Bigl[R_E \bigl(1 + \frac{R_E}{2} \bigr) 
 -  R_{\min} \bigl(1 + \frac{ R_{\min}}{2} \bigr) \Bigr]
\nonumber\\
&&
\,\cdot \Bigl[\ln{R_E} + \ln(R_E -  R_{\min}) - \ln(1 - R_E) \Bigr] 
  \Biggr] 
  \Biggr\}.   \nonumber
\ea

%---------------------------------------------------------
\subsubsection*{Region III 
\label{intR-III}
}
%---------------------------------------------------------
%
Region III is parameterized by:
\ba
\label{1intervalsIII}
III: \quad R_E -  R_{\min}  \leq  x
 \leq 1 - R_E,     \\
\label{2intervalsIII}
 R_{\min}  \leq  R   \leq  R_{ac}(x) ,
\ea  
where the acollinearity bound $R_{ac}(x)$ was 
defined in (\ref{Racol}).
Starting from the expression (\ref{dhard1a}) with (\ref{dhard1c}),
one sees that for a complete analytical integration 
over $x$ and $R$ one should integrate over $R$ first, and then 
over $x$:
\vspace*{.2cm}
\begin{center}
{\large Table of integrals }
\end{center} 
\ba
 \left[ f(R) \right]_{(R)} &=&
 \int\limits_{R_{\min}}^{R_{ac}(x) }
          d\,R \, f(R),
\\
\nonumber\\
\label{1intR_III}
\left[1\, \right]_{(R)}  &=&  R_{ac}(x) \, - \, R_{\min} ,      \\
\label{2intR_III}
 \left[R\, \right]_{(R)}  &=&  \frac{1}{2} \Bigl(R^2_{ac} - 
          R^2_{\min} \Bigr),   \\
\label{3intR_III}
\left[\frac{1}{v_1} \right]_{(R)}  &=&  
-\ln{\frac{1 - R_{ac}(x) - x}{1 - R_{\min} - x} } , \\
\label{4intR_III}
\left[\frac{1}{R+x} \right]_{(R)}  &=& 
\ln{\frac{ R_{ac}(x) + x}{R_{\min} + x} },  \\
\label{5intR_III}
\left[\frac{1}{(R+x)^2} \right]_{(R)}  &=&
   -\frac{1}{R_{ac}(x) + x} + \frac{1}{R_{\min} + x}.
\ea
The result of the integration over $R$ is:  

\ba
\label{dhard2_III}
\frac{d {\sigma}^{hard}_{finIII}}{d \cos{\vartheta} }  &=&
             \frac{\alpha}{\pi}\,{Q_f^2} \,  \int_{III}
         d \, x  
\nonumber\\
&&
\Biggl\{     
\frac{d {\sigma}^{born}}{d \cos{\vartheta} }
  \Biggl[
           \,  \frac{1}{2} \Bigl(1 + R_{\min}\Bigr) +
             \frac{1}{4} x + \frac{ R_{\min}}{2} 
  \Bigl(1 + \frac{ R_{\min}}{2} \Bigr) \frac{1}{x} 
  \nonumber\\      
&&
-  \frac{\rho}{2} \Bigl( 1 + \frac{\rho}{2}\Bigr) 
                                \frac{1}{x + \rho -1}
+ \frac{1}{4} \rho^2 (\rho-1) \frac{1}{(x + \rho - 1)^2} 
 \nonumber\\
&&
   + \Bigl(1 - \frac{x}{2} - \frac{1}{x} \Bigr) 
\ln{\frac{1 - R_{ac}(x) - x}{1 - R_{\min} - x} }   \Biggr] 
\nonumber\\      
&&
+   \frac{\pi\, \alpha^2}{s} 
   \Biggl[    {\cal V}(s) \, \frac{1 - 3 \cos^2{\vartheta}}{2}  
  \Biggl(            \frac{x - 1}{\rho}                    
            +  \frac{ R_{\min}}{x +  R_{\min} } 
+ \ln{\frac{ R_{ac}(x) + x}{R_{\min} + x} }       \Biggr)  \nonumber\\
 &&  \qquad \quad {\cal A}(s)  \cos{\vartheta}  \bigl( x - 1 \bigr) 
                  \ln{\frac{ R_{ac}(x) + x}{R_{\min} + x} }   
          \Biggr]          \Biggr\}.    
\ea
Now one has to integrate over $x$ in the interval 
  $ R_E -  R_{\min}  \leq  x  \leq 1 - R_E $ .
It is suitable to also introduce the notation 
\cite{Montagna:1993mf}:
\ba
\label{ROX}
  \rho_x = \rho - 1 + R_E - R_{\min}  =
                  \frac{\rho (\rho - 1)}{\rho - R_E}.
\ea

{From} Figure (\ref{dalitz_fin}) one observes one very nice
property of the function $R_{ac}(x)$: It has the same value
$R_{\min}$ in two different points where $x = 1 - R_E$ 
and  $x = R_E  - R_{\min}$, which removes some of the 
logarithmic terms arising during the integration:
\ba
\label{relationsIII}
     \ln{\frac{1 - R_{ac}(x) - x}{1 - R_{\min} - x} }
 \bigg|^{1-R_E} & = & 
   \ln{\frac{1 - R_{ac}(x) - x}{1 - R_{\min} - x} }
 \bigg|_{R_E- R_{\min}} = \, 0 ,
\\
    \ln{\frac{R_{ac}(x) + x}{R_{\min} + x} }
 \bigg|^{1-R_E} & = & 
   \ln{\frac{R_{ac}(x) + x}{R_{\min} + x} }   
 \bigg|_{R_E- R_{\min}} = \, 0.
\ea

\vfill\eject
%--------------------------------------------------------------------
\vspace*{0.2cm}
\begin{center}
{\large Table of integrals }
\end{center} 
\ba
  \left[ f(x) \right]_{(x)} &=&
 \int\limits_{R_E - R_{\min}}^{1-R_E} d\,x  f(x),            
\\
\nonumber\\
\label{1intV2_III}
 \left[1 \right]_{(x)}  &=&    x_{\rho}-R_E, 
\\
\label{2intV2_III}
    \left[x \right]_{(x)}  &=&  \frac{1}{2} 
    \Bigl(x_{\rho}-R_E\Bigr)  
      \Bigl(1 - R_{\min} \Bigr) ,      
\\
\label{3intV2_III}          
\left[\frac{1}{x} \right]_{(x)}  &=&   \ln(1 -  R_E)
             - \ln(R_E - R_{\min}) ,     
\\
\label{4intV2_III}
 \left[\frac{1}{x +  R_{\min}}\right]_{(x)}  
&=&  \ln{x_{\rho}} - \ln{R_E},                     
\\
\label{5intV2_III}
\left[\frac{1}{x+\rho-1} \right]_{(x)}  &=&   \ln(\rho -  R_E)
             - \ln(\rho - x_{\rho}),      
\\ 
\label{6intV2_III}
\left[\frac{1}{(x + \rho - 1)^2}\right]_{(x)}    &=&     
     \frac{1 + R_{\min} - 2 R_E}{\rho (\rho - 1)},    
\\
\label{7intV2_III}
\left[\ln{\frac{1 - R_{ac}(x) - x}{1 - R_{\min} - x}} \right]_{(x)} 
&=&  x_{\rho}-R_E + \ln\frac{x_{\rho}}{R_E}
- (\rho -1) \ln\frac{\rho-R_E}{\rho-x_{\rho}}   
\nonumber\\
&& 
- (1- R_{\min})\ln\frac{1 -  R_E}{R_E - R_{\min}},
  \\  
\label{8intV2_III}
\left[\frac{1}{x} 
    \ln{\frac{1 - R_{ac}(x) - x}{1 - R_{\min} - x}} \right]_{(x)} 
&=& -\ln(1- R_{\min}) \ln\frac{1-R_E}{1-x_{\rho}}
  + {\rm Li}_2\left(\frac{1-R_E}{1- R_{\min} }\right)  
\nonumber\\
&& - 
{\rm Li}_2\left(\frac{1-x_{\rho}}{1- R_{\min} }\right)       
 -   {\rm Li}_2(1-R_E) + {\rm Li}_2(1-x_{\rho})    
\nonumber\\
&& 
   + {\rm Li}_2\left(\frac{1-R_E}{1- \rho }\right)   -
  {\rm Li}_2\left(\frac{1-x_{\rho}}{1- \rho }\right),         
\\
\label{9intV2_III}
\left[\ln{\frac{R_{ac}(x) + x}{R_{\min} + x}} \right]_{(x)} 
&=&  x_{\rho}-R_E 
    -  R_{\min}\ln\frac{x_{\rho}}{R_E}
\nonumber\\
&&
- (\rho - 1) \ln\frac{\rho-R_E}{\rho-x_{\rho}},  
\\
\label{10intV2_III}
\left[x \ln{\frac{1 - R_{ac}(x) - x}
{1 - R_{\min} - x}} \right]_{(x)} 
&=&  \frac{1}{2} \bigl(x_{\rho}-R_E\bigr)
      \Bigl(1-\rho- R_{\min} \Bigr)
       \nonumber\\
&&
+ \frac{1}{4} (1-R_E)^2 
-  \frac{1}{4} (R_E -  R_{\min})^2
+ \frac{1}{2} \ln\frac{x_{\rho}}{R_E}
       \nonumber\\
&&
+ \frac{1}{2} (\rho -1)^2 
    \ln\frac{\rho-R_E}{\rho-x_{\rho}}
\nonumber\\
&&   
-    \frac{1}{2} (1- R_{\min})^2  
     \ln\frac{1 -  R_E}{R_E - R_{\min}},  
\\ 
\label{11intV2_III}
\left[x \ln{\frac{R_{ac}(x) + x}{R_{\min} + x}} \right]_{(x)} 
&=&  \frac{1}{2} \bigl(x_{\rho}-R_E\bigr)
          \Bigl(1-\rho- R_{\min} \Bigr)
      \nonumber\\
&&+ \frac{1}{4} (1-R_E)^2
-  \frac{1}{4} (R_E -  R_{\min})^2
 \nonumber\\
&&+ 
\frac{1}{2}  R^2_{\min}   
     \ln\frac{x_{\rho}}{R_E}  + \frac{1}{2} (\rho -1)^2 
        \ln\frac{\rho-R_E}{\rho-x_{\rho}}. 
\ea 
This provides as hard photonic contribution of the third
region to ${d {\sigma}^{hard}_{finIII}}/{d\cos{\vartheta}}$:

\ba
\label{dhard3_III}
\frac{d {\sigma}^{hard}_{finIII}}{d \cos{\vartheta} }  &=&
  \frac{\pi\, \alpha^2}{s}  \frac{\alpha}{\pi}\,{Q_f^2} 
\Biggl\{     {\cal V}(s) \, \Biggl(
         \frac{1 + \cos^2{\vartheta}}{2}   
 \Biggl[\frac{1}{2} \bigl(x_{\rho}-R_E \bigr)
    \bigl(\rho +\frac{3  R_{\min}}{2} + \frac{5}{2}\bigr)
\nonumber\\
&&
+\rho (1+\frac{\rho}{2}) \ln\frac{\rho-x_{\rho}}{\rho-R_E}
\nonumber\\
&&
+\Bigl[ R_{\min} \bigl(
1 + \frac{ R_{\min}}{2}\bigr)
+\ln(1- R_{\min}) \Bigr]
\ln\frac{1-R_E}{1-x_{\rho}}
\nonumber\\
&&
+{\rm Li}_2\left(\frac{1-x_{\rho}}{1- R_{\min} }\right)
- {\rm Li}_2\left(\frac{1-R_E}{1- R_{\min} }\right)       
+{\rm Li}_2\left(\frac{1-x_{\rho}}{1- \rho }\right)
\nonumber\\
&& 
- {\rm Li}_2\left(\frac{1-R_E}{1- \rho }\right)  
+{\rm Li}_2\left(1-x_{\rho}\right) 
-{\rm Li}_2\left(1-R_E\right)
 \Biggr]
\nonumber\\
&&+\frac{1 - 3\cos^2{\vartheta}}{2}  
\Biggl[
\bigl(x_{\rho}-R_E \bigr)
\bigl(1 - \frac{R_{\min}}{2 \rho} - \frac{1}{2 \rho}\bigr)
- (1-\rho) \ln\frac{\rho-x_{\rho}}{\rho-R_E}
\Biggr]
\Biggr)
\nonumber\\
&&+
{\cal A}(s) \,\cos\vartheta \,\Biggl(
x_{\rho}-R_E 
+\Bigl[\frac{1}{2} + \rho + R_{\min} \bigl(
1 + \frac{ R_{\min}}{2}\bigr)\Bigr]
\ln\frac{\rho-x_{\rho}}{\rho-R_E}
\nonumber\\
&&+\Bigl[ 2 R_{\min} \bigl(
1 + \frac{ R_{\min}}{2}\bigr)
+\ln(1- R_{\min}) \Bigr]
\ln\frac{1-R_E}{1-x_{\rho}}
\nonumber\\
&&
+{\rm Li}_2\left(\frac{1-x_{\rho}}{1- R_{\min} }\right)       
-{\rm Li}_2\left(\frac{1-R_E}{1- R_{\min} }\right)
+{\rm Li}_2\left(\frac{1-x_{\rho}}{1- \rho }\right) 
\nonumber\\
&&
-{\rm Li}_2\left(\frac{1-R_E}{1- \rho }\right)
+{\rm Li}_2\left(1-x_{\rho}\right) 
-{\rm Li}_2\left(1-R_E\right) 
\Biggr)
\Biggr\}.  
%\nonumber\\
\ea

\vfill\eject
%---------------------------------------------------------
\subsection*{Integration over $\cos{\vartheta}$ }
%---------------------------------------------------------
%
It is clear looking at the final results for 
${d {\sigma}^{hard}_{fin(a)}}/{d\cos{\vartheta}}$,
$a=I,II,III$ that it factorizes in each region 
into a term proportional to ${\cal V}(s) $ and 
a term proportional to ${\cal A}(s)$. The term with factor ${\cal V}(s)$ 
contributes solely to the total cross section, 
while the terms with factor ${\cal A}(s)$ 
only contribute to the integrated forward-backward result. 

\vspace*{0.2cm}
\begin{center}
{\large Table of integrals }
\end{center} 
\ba
\label{0intCOS}
  \left[ f(c) \right]_T&=&
 \int\limits_{-c}^{c}
          d \cos{\vartheta}\, f(\cos{\vartheta}),
\qquad    \left[ f(c) \right]_{FB}=
 \left[\int\limits_{0}^{+c}-\int\limits_{-c}^{0}\right]
          d \cos{\vartheta}\, f(\cos{\vartheta}),          
\nonumber\\
\\
\label{1intCOS}
\left[ 1 \right]_T
 \,&=& 2\, c, \quad    
\label{12intCOS}
\left[ \cos{\vartheta}\right]_T
 \,= 0, \quad       
\label{13intCOS}
\left[ \cos^2{\vartheta}\right]_T
 \,= \frac{2}{3}  c^3, \\
\label{14intCOS}
\left[ 1 \right]_{FB} &=& 0, \quad         
\label{15intCOS}
\left[ \cos{\vartheta} \right]_{FB} = c^2, \quad         
\label{16intCOS}
\left[ \cos^2{\vartheta} \right]_{FB} = 0.
\ea

One can trivially obtain the total cross section terms
${\sigma}_{T,fin(a)}$, $a=I,II,III$, from (\ref{dregul_I}),
(\ref{dhard2_III}), and (\ref{dhard3_III})
taking the terms proportional to ${\cal V}(s)$ 
and replacing the angular factors
$(1 + \cos^2{\vartheta})/2$   
and $(1 - 3\cos^2{\vartheta})/2$ by $c + c^3/3$   
and $c - c^3$ respectively.
Similarly the forward-backward cross section 
${\sigma}_{FB,fin(a)}$ is given by the 
terms proportional to ${\cal A}(s)$ in (\ref{dhard3_III})
with the factor $\cos\vartheta$ replaced by a factor $c^2$.
This observation holds at any stage of the integration,
i.e. the final state cross section distributions
which appeared during the successive steps of integration 
all factorize into terms proportional to $(1 + \cos^2{\vartheta})/2$, 
$(1 - 3\cos^2{\vartheta})/2$, and $\cos\vartheta$.

The comparison of the integrated results with results
given in \cite{Montagna:1993mf} shows that they 
coincide, except for some small misprints there. 
In eq. (5) of \cite{Montagna:1993mf} 
a term $\ln^2(y^{-}_b)$ is missing 
and in eq. (6) a minus sign is missing, i.e.
one has to make the following replacement:
\ba
\frac{1}{\rho_e}+\frac{1}{\rho_x}
\to -\frac{1}{\rho_e}+\frac{1}{\rho_x}.
\ea
The term $\rho^{-}$ has to be corrected there 
with an overall minus sign.
Our $R_E$  and $R_{\min}$ are denoted there by $e$ 
and $y_b$ respectively and our results are more compact 
and simplified. The expressions for $d{\sigma}{d{\cos\vartheta}}$
and $\sigma_{FB}$, however, were not presented there and are 
therefore completely new.

%\newpage

%%%%%%%%%%%%%%%%%%%%%%%%%%%%%%%%%%%%%%
\chapter{Soft and virtual photonic corrections
\label{softvirtual}}
%%%%%%%%%%%%%%%%%%%%%%%%%%%%%%%%%%%%%%

%%%%%%%%%%%%%%%%%%%%%%%%%%%%%%%%%%%%%%
\section{Soft photonic corrections
\label{soft}}
%%%%%%%%%%%%%%%%%%%%%%%%%%%%%%%%%%%%%%
%
For the general case with all soft photonic corrections
and the real photon momentum $k\to 0$, one can show
that the soft amplitudes ${M}^{soft}_{ini}$ 
and ${M}^{soft}_{fin}$ factorize from the Born 
amplitude \cite{Bloch:1937}:
\ba
{M}^{soft}_{ini} &=& e\, Q_e {\epsilon}_{\alpha}
\left(\frac{2 k_2^{\alpha} }{Z_2} - 
\frac{2 k_1^{\alpha} }{Z_1} \right)\cdot {M}^{Born}(s'),
\label{msoftini}     \\ 
{M}^{soft}_{fin} &=& e\, Q_f {\epsilon}_{\alpha}
 \left(\frac{2 p_1^{\alpha}}{V_1} -
\frac{2 p_2^{\alpha}   }{V_2} \right)\cdot {M}^{Born}(s).
\label{msoftfin}
\ea
So, having in mind that $s'$ becomes $s$, the soft photon 
contribution to the differential cross section takes the form:
\ba
\hspace{-1.cm}  
d \sigma^{soft} &=& -  d \sigma^{Born} e^2 \left[ 
Q_e \left(\frac{2 k_2 }{Z_2} - 
\frac{2 k_1 }{Z_1} \right) + 
Q_f \left(\frac{2 p_1}{V_1} -
\frac{2 p_2 }{V_2} \right)  \right]^2\theta(\bar{\varepsilon} - p^0)  
\nonumber\\
&&\qquad \cdot\, 
         \frac{d^3 \vec{p  }}{(2\pi)^3 2 p^0}.
\label{dsoft}
\ea
Here we have applied the sum rule for photon polarization vectors:
\ba
\label{spinsum}
\sum_{spin} {\epsilon}_{\alpha} {\epsilon}_{\beta} = 
    - g_{\alpha\, \beta}.
\ea
The soft photon cut-off parameter $\varepsilon$ 
from (\ref{sprmin}) is related to $\bar{\varepsilon}$ through 
$\varepsilon = 2 \bar{\varepsilon} / \sqrt{s}$.
The expression in (\ref{dsoft}) can be written in the 
following way ($ e^2 = 4 \pi \alpha $):
\ba
 d \sigma^{soft} = 
d \sigma^{soft}_{ini} 
+ d \sigma^{soft}_{fin} 
+ d \sigma^{soft}_{int} = 
\frac{\alpha}{\pi} \,\delta^{soft}\, d \sigma^{Born}.
\label{desoft}
\ea
The correction $\delta^{soft}$, which we have to calculate, 
has the form:
\ba
\label{delsoft}
  \delta^{soft} = -\, 4 \pi^2 \int 
\frac{d^3 \vec p}{(2\pi)^3 2p^0} \left[ 
Q_e \left(\frac{2 k_2 }{Z_2} - 
\frac{2 k_1 }{Z_1} \right) + 
Q_f \left(\frac{2 p_1}{V_1} -
\frac{2 p_2 }{V_2} \right)  \right]^2 
\theta(\bar{\varepsilon} -p^0).
\nonumber\\
\ea
Taking the expression squared explicitly, we obtain:

\ba
\label{soft1}
    \delta^{soft} &=&  16 \pi^2 \int 
\frac{d^3 \vec p}{(2\pi)^3 2p^0} \biggl[
    Q_e^2 \left( -\frac{ m_e^2}{Z_1^2} -\frac{ m_e^2}{Z_2^2} 
+ \frac{2 k_1 \cdot k_2}{Z_1 Z_2} \right) \nonumber\\
&+& Q_e Q_f \left(  \frac{2 k_1 \cdot p_1}{Z_1 V_1}
+ \frac{2 k_2 \cdot p_2}{Z_2 V_2} - \frac{2 k_1 \cdot p_2}{Z_1 V_2}
- \frac{2 k_2 \cdot p_1}{Z_2 V_1} \right) 
\nonumber\\
&+& Q_f^2 \left( -\frac{ m_f^2}{V_1^2} -\frac{ m_f^2}{V_2^2} 
+ \frac{2 p_1 \cdot p_2}{V_1 V_2} \right)
 \biggr] \theta(\bar{\varepsilon} - p^0)
\\
\nonumber\\
\label{soft2}
&=&  16 \pi^2 \int 
\frac{d^3 \vec p}{(2\pi)^3 2p^0} \biggl[
    Q_e^2 \left( -\frac{ m_e^2}{Z_1^2} -\frac{ m_e^2}{Z_2^2} 
+ \frac{s-2 m_e^2}{Z_1 Z_2} \right) 
\nonumber\\
&+& Q_e Q_f \left(  \frac{T}{Z_1 V_1}
+ \frac{T}{Z_2 V_2} - \frac{U}{Z_1 V_2}
- \frac{U}{Z_2 V_1} \right) 
\nonumber\\
&+& Q_f^2 \left( -\frac{ m_f^2}{V_1^2} -\frac{ m_f^2}{V_2^2} 
+ \frac{s-2 m_f^2}{V_1 V_2} \right)
 \biggr] \theta(\bar{\varepsilon} - p^0).
\ea
The kinematic invariants are defined in (\ref{invs}) to (\ref{invz2})
and in (\ref{T}) and (\ref{U}).

%-------------------------------------------------------
\subsection{Initial state radiation
\label{softini}}
%-------------------------------------------------------
%
As an example, the soft photon factor
$\delta^{soft}_{ini}$ for initial state bremsstrahlung 
to the differential cross section contribution $d{\sigma_{ini}}$
shall be explicitly evaluated.
The differential cross section $d{\sigma_{ini}}$ can be separated into a
hard and soft with virtual photon part, 
$d{\sigma^{hard}}$ and $d{\sigma^{soft+virtual}}$.

\ba
\label{intgam}
d{\sigma_{ini}}&=&d{\sigma^{hard}_{ini}}+d{\sigma^{soft}_{ini}}
+d{\sigma^{virtual}_{ini}}.
\ea
The derivation of the virtual correction term $d{\sigma^{virtual}_{ini}}$ 
we will keep to the next Section \ref{virtual}.
The basic idea for the calculation of the soft photonic part 
is to introduce an arbitrary small 
cut-off $\bar{\varepsilon}$ for the photon energy $p^0$, thus distinguishing 
the hard photon region of phase space, $p^0\geq\bar{\varepsilon}$, from 
the soft photon region with $p^0<\bar{\varepsilon}$. 
\ba
\label{intgam2}
d{\sigma^{hard}}+d{\sigma^{soft}}
&=& 
-16\pi\alpha Q_e^2\,\left(\frac{k_1}{Z_1}-\frac{k_2}{Z_2}\right)^2
\, d{\sigma^{Born}}\,
d{\Gamma^{\gamma}}\,\Biggl[
\theta(\bar{\varepsilon}-p^0)
+
\theta(p^0-\bar{\varepsilon})
\Biggr],
\nonumber\\
\ea
\ba
\label{intgam3}
&=& d{\sigma^{Born}}\cdot 
\left(\delta^{soft}(\bar{\varepsilon})
+\delta^{hard}(\bar{\varepsilon})\right),
\\
\nonumber\\
\mbox{with}&&d{\sigma^{Born}} =
\frac{1}{2\sqrt{\lambda_s}}
\,\sum_{spin}\, \overline{|M_{Born}|^2}\, d{\Gamma^{(2)}},
\\
\nonumber\\
\mbox{and}&&
\delta^{soft}(\bar{\varepsilon}) = 
16\pi^2 Q_e^2\,\int\,\left[\frac{s-2m_e^2}
{Z_1 Z_2}-\frac{m_e^2}{Z^2_1}-\frac{m_e^2}{Z^2_2}\right]
d{\Gamma^\gamma}\,\Theta(\bar{\varepsilon}-p^0),
\nonumber\\
\\
\mbox{and}&&
\delta^{hard}(\bar{\varepsilon}) = 
16\pi^2 Q_e^2\,\int\,\left[\frac{s-2m_e^2}
{Z_1 Z_2}-\frac{m_e^2}{Z^2_1}-\frac{m_e^2}{Z^2_2}\right] 
d{\Gamma^\gamma}\,\Theta(p^0-\bar{\varepsilon}),
\nonumber\\
\\
&&
d{\Gamma^{\gamma}}= 
\frac{d^3\vec{p}}{(2\pi)^3 2 p^0}\,,
\quad d{\Gamma^{(2)}}= \frac{\beta}{16\pi}\, d(\cos\vartheta),
\\
&& 
\lambda_S = s^2\beta_0^2,\quad \beta_0=\sqrt{1-\frac{4 m_e^2}{s}}, 
\quad \beta=\sqrt{1-\frac{4 m_f^2}{s'}}.
\ea
The Born squared amplitude $\overline{|M_{Born}|^2}$ for 
$e^+e^-\to \bar{f}f$ was given in Section \ref{feynmat}.
The hard photon contribution 
$d{\sigma^{hard}_{ini}}(\bar{\varepsilon})$ has 
been calculated straightforwardly using the 
parameterization and the kinematical cuts
presented in Appendix \ref{hardini}. The corresponding 
phase space volume for hard photon emission is
from (\ref{dgamtot2}) to (\ref{dgamtot4}) :  

\ba
\label{gamhard}
d{\Gamma^{(3)}_{hard}} 
= 
d{\Gamma^{(2)}_{hard}}\cdot d{\Gamma^{\gamma}}
= 
\frac{1}{(2\pi)^5}\,\frac{\pi}{16s}
\, d {\varphi_\gamma}\, d{V_2}\, d{s'}\, d {\cos\vartheta}
\,\Theta(p^0-\bar{\varepsilon}).
\ea
%
%For the integration of $\delta^{soft}$ we will follow~[(\ref{lehner})]. 
%
In the soft photon case, if we introduce the Feynman 
parameter $\alpha$ and define 
$k_{\alpha}:=k_1\alpha+k_2(1-\alpha)$, we can transform $\delta^{soft}$
with the definitions of $Z_{1,2}$ into:
\ba
\label{delsoft1}
\delta^{soft}(\bar{\varepsilon}) =  
\frac{Q_e^2}{2\pi}\int^1_0 d{\alpha}
\int\,\frac{d^3\vec{p}}{2 p^0}\,\Theta(\bar{\varepsilon}-p^0)
\,\left\{\frac{s-2m_e^2}{(k_\alpha\, p)^2}
-\frac{m_e^2}{(k_1\, p)^2}-\frac{m_e^2}{(k_2\, p)^2} 
\right\}.
\ea
Before we can proceed, we have to be cautious as the underlying integrals
 
\ba
\label{delint}
\int\,\frac{d^3{\vec{p}}}{2 p^0}
\,\frac{1}{(k_a\, p)^2}\, 
\Theta(\bar{\varepsilon}-p^0)
=\int d^4{p}\,\delta(p^2)
\,\frac{1}{(k_a\, p)^2}\, 
\Theta(\bar{\varepsilon}-p^0),\quad a=1,2,\alpha,
\nonumber\\
\ea
are infrared divergent, i.e. they become singular for vanishing photon 
momenta $p^0\to 0$. According to \cite{Bloch:1937}, the infrared divergence 
has to cancel in the full scattering amplitude with soft and virtual 
corrections, i.e. when adding the term $d \sigma^{virtual}_{ini}$ 
(see Section \ref{virtual}).
 
In order to calculate the integrals (\ref{delint}) over the photon 
momentum $p$, it is necessary to use a regularization technique 
to circumvent the infrared singular behaviour and to deal with
regularized, i.e finite expressions. We use {\it dimensional regularization}
\cite{'tHooft:1973mm,'tHooft:1979xw,Bardin:1980fe}. The four-dimensional 
integration over $p$ is replaced by an $n$-dimensional integration,
using now the appropriately generalized on-shell condition
${p^0}^2:=p_1^2+p_2^2+\ldots +p_{n-1}^2$. 
For $n\neq 4$ the integrals (\ref{delint}) 
exist, and we can use $(n-1)$-dimensional 
spherical coordinates to determine $\delta^{soft}$.
As side remark, dimensional regularization breaks the Lorentz structure,
but preserves the symmetries and the gauge invariance of the theory
and is therefore one strongly advocated regularization technique 
among other possible procedures.

In $(n-1)$ dimensional spherical coordinates (\ref{delsoft1}) reads:
\ba
\label{delreg}
&&\int\,\frac{d^3{\vec{p}}}{2 p^0}\Theta(\bar{\varepsilon}-p^0)
\longrightarrow
\int\,\frac{d^{n-1}{\vec{p}}}{2 p^0}\Theta(\bar{\varepsilon}-p^0)=
\nonumber\\
&=&\int^{\bar{\varepsilon}}_0\,
\frac{d{p^0}}{2 p^0}(p^0)^{n-2}\,\int^\pi_0 d{\theta_{n-2}}
(\sin\theta_{n-2})^{n-3}\ldots\int^\pi_0 d{\theta_2}
\sin\theta_2\int^{2\pi}_0 d{\theta_1}\,.
\nonumber\\
\ea
We can evaluate the integrals over $p^0$ and the first $(n-3)$ angles 
$\theta_i$ for each of the three integrals in (\ref{delsoft1}), but 
have to keep the remaining angle $\theta_{n-2}=:\theta_p$ to parameterize  
each one of the propagators $k_a p=k_a^0 p^0(1-\beta_a\cos\theta_p)$ 
separately with $\beta_a=|\vec{k}_a|/k_a^0$, $a:=1,2,\alpha,$ for the last
integration (isotropy of soft photon radiation). 
The integration of the first $(n-3)$ angular coordinates yields:
% %
\ba 
\label{delang}
\int^\pi_0 d{\theta_{n-4}}
(\sin\theta_{n-3})^{n-4}\ldots\int^\pi_0 d{\theta_2}
\sin\theta_2\int^{2\pi}_0 d{\theta_1}
=2\pi\,\pi^{\frac{n-4}{2}}\,\frac{\Gamma(1)}
{\Gamma\left(\frac{n-2}{2}\right)}\,.
\ea
So, for the propagators $1/(k_a\, p)^2$ we obtain: 
\ba
\label{delprop1}
&&
\frac{1}{2\pi\,\mu^{n-4}}
\int^1_0 
d{\alpha}
\int\,\frac{d^{n-1}{p}}{2 p^0}
\frac{1}{(k_a p)^2}\,
\Theta(\bar{\varepsilon}-p^0)=
\\
&=&
\frac{1}{2}\pi^{\frac{n-4}{2}}
\Gamma\left(\frac{n}{2}-1\right)
\int^1_0
\frac{d{\alpha}}{\mu^{n-4}}
\int^{\bar{\varepsilon}}_0\,d{p^0}(p^0)^{n-5}
\int^\pi_0 d{\theta_p}
\frac{(\sin\theta_p)^{n-3}}
{(k_a^0)^2(1-\beta_a\cos\theta_p)^2}.
\nonumber
\ea
We have used in (\ref{delprop1}) the additional parameter 
$\mu$ as regulator mass term for vanishing photon 
mass. It will eventually drop out as unphysical quantity
when adding $\delta^{virtual}_{ini}$. We obtain further 
for (\ref{delprop1}):
\ba
\label{delprop2}
1.&&\frac{1}{\mu^{n-4}}
\int^{\bar{\varepsilon}}_0\,d{p^0}(p^0)^{n-5}
=\frac{1}{n-4}\,\left(\frac{\bar{\varepsilon}}{\mu}\right)^{n-4}
\nonumber\\
&=&\frac{1}{n-4}\,\left[1+(n-4)
\ln\left(\frac{\bar{\varepsilon}}{\mu}\right)
+o\left((n-4)^2\right)\right],
\\
2.&&\int^\pi_0 d{\theta_p}\,\frac{(\sin\theta_p)^{n-3}}
{(1-\beta_{a}\cos\theta_p)^2}=
\int^{1}_{-1} d{\xi}\,\frac{(1-\xi^2)^{\frac{n-4}{2}}}
{(1-\beta_{a}\xi)^2}\nonumber\\
&=&\int^{1}_{-1} d{\xi}\,\frac{1}{(1-\beta_a\xi)^2}
\left[1+\frac{n-4}{2}\ln(1-\xi^2)+o((n-4)^2)\right].
\ea

After isolating the infrared pole proportional to 
$1/(n-4)$, all terms of $o(n-4)$ and higher can be neglected,
as the final results have to be derived with $n\to 4$ 
in the physical 4-dimensional Lorentz space.
Inserting the results of (\ref{delprop2}) into 
(\ref{delprop1}), and with the definition (C: Euler constant)
% %
\ba
\label{coldiv}
P_{IR} := \frac{1}{n-4}+\frac{1}{2}C+\frac{1}{2}\ln{\pi} 
= \frac{1}{2\varepsilon_{IR}},
\ea%
we have ($\beta_{1,2}=\beta_0$):
\ba
\label{delsoft2}
\delta^{soft}(\bar{\varepsilon},\varepsilon_{IR},\mu)&=& 
\frac{Q_e^2}{2 (k_a^0)^2}\int^1_0 d{\alpha} 
\int^{+1}_{-1}d{\xi}
\left[P_{IR}(\mu)+\ln\left(\frac{\bar{\varepsilon}}{\mu}\right)
+\frac{1}{2}\ln(1-\xi^2)\right]
\nonumber\\
&&
\hspace*{1cm}
\cdot\left[
\frac{s-2m_e^2}
{(1-\beta_{\alpha}\xi)^2}
-\frac{m_e^2}{(1+\beta_0\xi)^2}
-\frac{m_e^2}{(1-\beta_0\xi)^2}\right],
\ea
or eliminating the Feynman parameter $\alpha$ again
for this simple case:
\ba
\label{delsoft3}
\delta^{soft}(\bar{\varepsilon},\varepsilon_{IR},\mu)&=&
\frac{2 Q_e^2}{s}\int^{+1}_{-1}d{\xi}
\,\left[
P_{IR}(\mu) 
+ \ln\left(\frac{\bar{\varepsilon}}{\mu}\right)
+ \frac{1}{2}\ln(1-\xi^2)
\right]
\nonumber\\
&&
\hspace*{1cm}
\cdot\left[
\frac{s-2m_e^2}{1-\be^2\xi^2}
- \frac{m_e^2}{(1+\be\xi)^2}
- \frac{m_e^2}{(1-\be\xi)^2}
\right].
\ea
In (\ref{delsoft2}) and (\ref{delsoft3}) terms of $o(n-4)$ have been neglected.
The integrals over $\xi$ can be determined relatively easily, leading
to logarithmic and dilogarithmic mass terms.

\begin{center}
{\large  Table of integrals:}
\end{center}%
\ba
\label{intxi}
[f(\xi)]_{(\xi)} &=& \frac{1}{2}\int\limits^{+1}_{-1}\, d{\xi}\, f(\xi),
\\
\nonumber\\
\left[1\right]_{(\xi)} &=& 1,
\\
\nonumber\\
\left[\frac{1}{1\pm\be\xi}\right]_{(\xi)} &=& 
\frac{1}{2\be}\ln\left(\frac{1+\be}{1-\be}\right) 
\approx 
\frac{1}{2}\ln\left(\frac{s}{m_e^2}\right),
\\
\left[\frac{1}{(1\pm\be\xi)^2}\right]_{(\xi)} &=& 
\frac{1}{1-\be^2} 
\approx 
\frac{s}{4 m_e^2},
\\
\left[\frac{1}{2}\ln(1-\xi^2)\right]_{(\xi)} &=& \ln{2}-1,
\\
\left[\frac{1}{2}\frac{\ln(1-\xi^2)}{1\pm\be\xi}\right]_{(\xi)} &=&
\frac{1}{2\be}\left\{\ln{2}\,\ln\left(\frac{1+\be}{1-\be}\right)+\frac{1}{2}
\left[\Phi\left(\frac{2\be}{\be-1}\right)
-\Phi\left(\frac{2\be}{\be+1}\right)\right]\right\}
\nonumber\\
\\
&\approx&
\frac{1}{2}\,\ln{2}\,
\ln\left(\frac{s}{m_e^2}\right)
-\frac{1}{4}
\left\{
\frac{\pi^2}{3}
+\frac{1}{2}\left[\ln\left(\frac{s}{m_e^2}\right)\right]^2
\right\},
\\
\left[\frac{1}{2}\frac{\ln(1-\xi^2)}{(1\pm\be\xi)^2}\right]_{(\xi)} &=&
\frac{1}{1-\be^2}\left\{\ln{2}-\frac{1}{2\be}
\ln\left(\frac{1+\be}{1-\be}\right)\right\}
\\
&\approx&
\frac{s}{4m_e^2}\left\{\ln{2}
-\frac{1}{2}\ln\left(\frac{s}{m_e^2}\right)
\right\},
\\
\nonumber\\
\mbox{with}\quad
\Phi(y)&:=&\mbox{Li}_2(y)=-{\int}^1_0
\frac{{\ln(1-xy)}}{{ x}} d{x}
=-{\int}^y_0\frac{{\ln(1-x)}}{{ x}} d{x}.
\ea
If we neglect mass terms for $4 m_e^2\ll s$ except for 
the logarithms $\ln(s/m_e^2)$, we obtain the right hand side 
of the above Table. Due to \cite{Kinoshita:1962ur,Lee:1964is} 
these are also the only mass singularities allowed to arise 
if one puts $m_e\to 0$. They are due to 
collinear photon emission from one of the initial state fermions.  
The final result for the initial state soft photon contribution 
$\delta^{soft}$ is:
\ba
&&\delta^{soft} =
\delta^{soft}(\bar{\varepsilon},\varepsilon_{IR},\mu)
\label{deltasoftini}
\\
&=&
Q_e^2
\left[
(L_e-1) 
\left(
2\ln{\frac{\bar{\varepsilon}}{\mu}} + 2\ln{2} 
+ \ln\frac{m_e^2}{s} 
+ \frac{1}{\varepsilon_{IR}}
\right)
+ \frac{1}{2} L_e^2 
- \frac{\pi^2}{3}
\right],
\nonumber\\
\mbox{with} && \quad L_e := \ln\frac{s}{m_e^2}.
\label{deltasoftinile}
\ea

%%%%%%%%%%%%%%%%%%%%%%%%%%%%%%%%%%%%%%%%%%%%%%%%%%
\subsection{The complete soft photonic corrections
\label{softintcom}}
%%%%%%%%%%%%%%%%%%%%%%%%%%%%%%%%%%%%%%%%%%%%%%%%%% 
%
Again using for the integration of (\ref{soft1}), 
(\ref{soft2}) dimensional regularization and the 
transformation into spherical coordinations as
in (\ref{delreg}), we arrive at:
\ba
\label{dsigsoft}
d{\sigma^{soft}} &=& 
\frac{\alpha}{\pi}\,
\left(
\delta^{soft}_{ini}
+ \delta^{soft}_{fin} + 
\delta^{soft}_{int}
\right)
\,d{\sigma^{Born}}
\\
&=&
\frac{1}{4}
\int^{+1}_{-1}d{\xi}\,
\left[ 2 P_{IR}
+\ln\frac{\bar{\varepsilon}^2}{\mu^2}+\ln\left(1-\xi^2\right)
\right] 4 {p^0}^2 
\nonumber\\
&&\cdot\left[
 Q_e^2 \left( -\frac{ m_e^2}{Z_1^2} -\frac{ m_e^2}{Z_2^2} 
+ \frac{s - 2 m_e^2}{Z_1 Z_2} \right)
+ Q_f^2 \left( -\frac{ m_f^2}{V_1^2} -\frac{ m_f^2}{V_2^2} 
+ \frac{s - 2 m_f^2}{V_1 V_2} \right)\right. 
\nonumber\\
&&+\left. Q_e Q_f \left(  \frac{U}{Z_1 V_1}
+ \frac{U}{Z_2 V_2} - \frac{T}{Z_1 V_2}
- \frac{T}{Z_2 V_1} \right) 
\right].
\ea

We can repeat the calculation of the initial state 
case completely analogously for the final state terms, 
while for the interference contribution
a suitable Feynman parameterization is necessary. 
Here, just the results from
\cite{Bardin:1989cw,Bardin:1991de,Bardin:1991fu}
are presented:
\ba
\label{dsigsoft3}
%------------------------------------------------------
\delta^{soft}_{ini}(\bar{\varepsilon},\varepsilon_{IR},\mu) &=& 
Q_e^2
\left[
2\left(  P_{IR}
+\ln\frac{\bar{\varepsilon}}{\mu}+\ln{2}
\right)
\left(
\ln\frac{s}{m_e^2}-1
\right)
\right.
\nonumber\\
&&
\left.
+\,\ln\frac{s}{m_e^2}-\frac{1}{2}\ln^2\frac{s}{m_e^2}-2 \mbox{Li}_2(1)
\right],
\\
\nonumber\\
%------------------------------------------------------
\delta^{soft}_{fin}(\bar{\varepsilon},\varepsilon_{IR},\mu) &=&
Q_f^2
\left[
2\left( P_{IR}+\ln\frac{\bar{\varepsilon}}{\mu}+\ln{2}
\right)
\left(
\ln\frac{s'}{m_f^2}-1
\right)
\right.
\nonumber\\
&&
\left.
+\,\ln\frac{s'}{m_f^2}-\frac{1}{2}\ln^2\frac{s'}{m_f^2}-2 \mbox{Li}_2(1)
\right],
\\
\nonumber\\
%------------------------------------------------------
\delta^{soft}_{int}(\bar{\varepsilon},\varepsilon_{IR},\mu) &=&
2 Q_e Q_f
\left[
2\left( P_{IR}
+\ln\frac{\bar{\varepsilon}}{\mu}+\ln{2}
\right)
\,
\ln\left(\frac{c_{-}}{c_{+}}\right)
\right.
\nonumber\\
&&
\left.
+\,\mbox{Li}_2(c_{+})-\mbox{Li}_2(c_{-})
-\frac{1}{2}\left(\ln^2(c_{+})-\ln^2(c_{-})\right)
\right],
\\
\mbox{with} && 
c_{\pm} = \frac{1}{2}(1 \pm \beta\beta_0\cos\vartheta).
\ea
The unregularized soft photon contributions $s_{a}$ and 
$S_{a}$, $a = ini$, $fin$, $int$ (without virtual corrections) 
to the total cross section $\sigma_T$ and forward-backward 
asymmetry $A_{FB}$ can be summarized as follows:
\ba
\label{dsigsoft5}
%------------------------------------------------------
S_{A}^{ini}(c,\bar{\varepsilon},\varepsilon_{IR},\mu) 
&=&
\frac{\alpha}{\pi} Q_e^2 D_A(c)\,
\delta^{soft}_{ini}(\bar{\varepsilon},\varepsilon_{IR},\mu),
\\
\nonumber\\
%------------------------------------------------------
S_{A}^{fin}(c,\bar{\varepsilon},\varepsilon_{IR},\mu) 
&=&
\frac{\alpha}{\pi} Q_f^2 D_A(c)\,
\delta^{soft}_{fin}(\bar{\varepsilon},\varepsilon_{IR},\mu),
\\
\nonumber\\
%------------------------------------------------------
{S}_{T}^{int}(c,\bar{\varepsilon},\varepsilon_{IR},\mu) 
&=&
\frac{\alpha}{\pi}
\int_{-c}^{c} d{\cos\vartheta}\, d_{FB}(\cos\vartheta)
\,\delta^{soft}_{int}
(\cos\vartheta,\bar{\varepsilon},\varepsilon_{IR},\mu)
\nonumber\\
&=&
\frac{\alpha}{\pi} Q_e Q_f
\Biggl\{
-4\left( P_{IR}
+\ln\frac{\bar{\varepsilon}}{\mu}+\ln{2}
\right)
\,
\left[
(c^2-1)\ln\frac{c_{+}}{c_{-}}+2c
\right]
\nonumber\\
&&
+2(c^2-1)\left[\mbox{Li}_2(c_{+})-\mbox{Li}_2(c_{-})\right]
-(c^2-1)\ln(c_{+}c_{-})\ln\frac{c_{+}}{c_{-}}
\nonumber\\
&&
-4c\ln(c_{+}c_{-})
-4 \ln\frac{c_{+}}{c_{-}}+8c,
\Biggr\}
\\
\nonumber\\
%......................................................
\frac{3}{4}{S}_{FB}^{int}(c,\bar{\varepsilon},\varepsilon_{IR},\mu) 
&=&
\frac{3}{4}
\frac{\alpha}{\pi}
\left\{\int_{0}^{c} - \int_{-c}^{0}\right\} 
d{\cos\vartheta}\, d_{T}(\cos\vartheta) 
\,\delta^{soft}_{int}
(\cos\vartheta,\bar{\varepsilon},\varepsilon_{IR},\mu)
\nonumber\\
&=&
\frac{\alpha}{\pi} Q_e Q_f
\Biggl\{
-\left( P_{IR}
+\ln\frac{\bar{\varepsilon}}{\mu}+\ln{2}
\right)
\nonumber\\
&&
\cdot\,
\left[
3(\frac{1}{3}c^3+c)\ln\frac{c_{+}}{c_{-}}
+4\ln(c_{+}c_{-})+c^2+8\ln{2}
\right]
\nonumber\\
&&
+\frac{3}{2}(\frac{1}{3}c^3+c)
\left[\mbox{Li}_2(c_{+})-\mbox{Li}_2(c_{-})\right]
+2\left[\mbox{Li}_2(c_{+})+\mbox{Li}_2(c_{-})\right]
\nonumber\\
&&
-\frac{3}{4}(\frac{1}{3}c^3+c)
\ln(c_{+}c_{-})\ln\frac{c_{+}}{c_{-}}
-(\ln^2{c_{+}}+\ln^2{c_{-}})
\nonumber\\
&&
+\frac{1}{2}(1-c^2)\ln(c_{+}c_{-})+\frac{1}{2}c^2
+4\ln^2{2}+\ln{2}-2\mbox{Li}_2(1)
\Biggr\},
\nonumber\\
\\
%......................................................
\mbox{with}&& D_{T}(c) = \frac{3}{4}\left(c+\frac{c^3}{3}\right),
\qquad D_{FB}(c) = c^2,
\\
&& d_{T}(\cos\vartheta) = 1+\cos^2\vartheta,
\qquad d_{FB}(\cos\vartheta) = 2\cos\vartheta,
\\
&& 
c_{\pm} = \frac{1}{2}(1 \pm c),\quad (\beta=\beta_0=1).
\ea
The soft photon flux functions $S_{A}^{a}$, $a=ini,fin,int$, $A=T,FB$
are inserted into convolution integrals for total cross sections
and asymmetries as for example given in (\ref{generic_zf})
in Section \ref{sec_lep1slc_radcuts}.

The infrared pole term $P_{IR}$ will be cancelled together with 
the regulator photon mass term $\mu^2$ when combining the soft 
and virtual corrections. In the initial and final state case
it is the initial and final state vertex corrections,  
while the soft interference term $\delta^{soft}_{int}$  
has to be combined with the virtual corrections from the 
interference of the Born graphs with the $\gamma\gamma$ 
and $\gamma Z$ exchange box diagrams. This will be shown
in the next Section \ref{virtual}.
The logarithmic soft-photon cut-off $2\ln(\varepsilon)$ 
cancels together with the corresponding term in the 
hard photon results derived in Appendix \ref{hardbrem}.

%%%%%%%%%%%%%%%%%%%%%%%%%%%%%%%%%%%%%%
\section{Virtual corrections
\label{virtual}}
%%%%%%%%%%%%%%%%%%%%%%%%%%%%%%%%%%%%%%

%%%%%%%%%%%%%%%%%%%%%%%%%%%%%%%%%%%%%%
\subsection{Initial state and final state virtual corrections
\label{virtinifin}}
%%%%%%%%%%%%%%%%%%%%%%%%%%%%%%%%%%%%%%

%-------------------------------------
\subsubsection*{The vertex function
\label{vertex}}
\begin{figure}[tbhp]
\begin{center}
% new vertices:
%%%%%%%%%%%%%%%%%%%%%%%%%%%%%
% Vertex correction diagram zvert1.tex%
%%%%%%%%%%%%%%%%%%%%%%%%%%%%%
\vfill
\setlength{\unitlength}{1pt}
%\SetScale{0.8}
\SetWidth{0.8}
\begin{picture}(180,120)(0,0)
\thicklines
\ArrowLine(10,110)(40,80)
\Vertex(40,80){1.8}
\ArrowLine(40,80)(60,60)
\Photon(40,80)(40,40){2}{5}
\ArrowLine(60,60)(40,40)
\Vertex(40,40){1.8}
\ArrowLine(40,40)(10,10)
\Vertex(60,60){1.8}
\Photon(60,60)(120,60){3}{8}
\Vertex(120,60){1.8}
\ArrowLine(120,60)(170,110)
\ArrowLine(170,10)(120,60)
\Text(5,117)[]{$e^-$}
\Text(5,5)[]{$e^+$}
\Text(90,50)[]{$\gamma\,,\,Z$}
\Text(178,8)[]{$\bar{f}$}
\Text(178,113)[]{$f$}
\Text(32,45)[]{$\gamma$}
\Text(32,70)[]{$p$}
\Text(16,33)[]{$-k_2$}
\Text(16,95)[]{$k_1$}
\Text(60,35)[]{$p-k_2$}
\Text(60,85)[]{$p+k_1$}
\end{picture}
%%%%%%%%%%%%%%%%%%%%%%%%%%%%%
% Vertex correction diagram zvert2.tex%
%%%%%%%%%%%%%%%%%%%%%%%%%%%%%
\setlength{\unitlength}{1pt}
%\SetScale{0.8}
\SetWidth{0.8}
\begin{picture}(180,120)(0,0)
\thicklines
\ArrowLine(10,110)(60,60)
\Vertex(60,60){1.8}
\ArrowLine(60,60)(10,10)
\Photon(60,60)(120,60){3}{8}
\Vertex(120,60){1.8}
\ArrowLine(170,10)(140,40)
\Vertex(140,40){1.8}
\ArrowLine(140,40)(120,60)
\Photon(140,40)(140,80){2}{5}
\ArrowLine(120,60)(140,80)
\Vertex(140,80){1.8}
\ArrowLine(140,80)(170,110)
\Text(5,117)[]{$e^-$}
\Text(5,5)[]{$e^+$}
\Text(90,50)[]{$\gamma\,,\,Z$}
\Text(178,8)[]{$\bar{f}$}
\Text(178,113)[]{$f$}
\Text(149,60)[]{$\gamma$}
\end{picture}

\vspace*{0.5cm}
\caption[Photonic vertex corrections]
{\label{fig2paw}
{\sf
The photonic vertex corrections. % from initial (a) and final (b) states
}}
\label{Fig.virtph}
\end{center}
\end{figure}
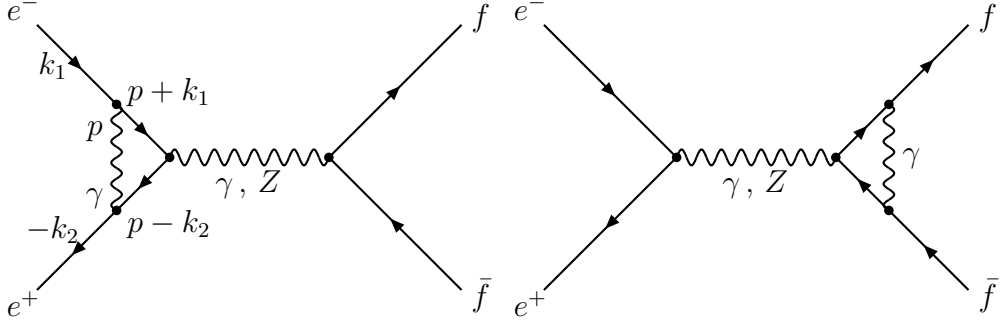

%-------------------------------------
%
In Fig.~\ref{Fig.virtph} we depict the one-loop
vertex corrections, i.e. the initial state and 
final state vertex diagrams. In order to completely 
cancel the infrared divergences from the initial and final
state soft photon contributions, their contributions 
have to be added.   

%------------------------------------------------------------

The integration is again done in dimensional regularization 
over the following matrix element, easily derived as real part 
of the interference between the Born diagram and the one-loop 
vertex diagrams illustrated above. 
The amplitude for the initial state vertex correction 
in Fig.~\ref{Fig.virtph},
\ba
{\Large{\cal M}}^{virtual} &=& 
      (2 \pi)^4 \delta^{(4)}(k_1 +k_2 - p_1 - p_2)
\, \frac{{M}^{vertex}} {(2 \pi)^6  
   (2 k^0_1\, 2 k^0_2\, 2 p^0_1\, 2 p^0_2)^{1/2} },
\label{vert01}
\ea
leads to the following cross section contribution:
\ba
d{\sigma^{virtual}} &=& 
\frac{1}{\tt j}
 \, (2 \pi)^4 \delta^{(4)}(k_1 +k_2 - p_1 - p_2)
\,
\frac{1}{4} \,
\sum_{spin} 
\,2\, {\Re}e
\left( 
{{M}^{Born}}^{*} \cdot {M}^{vertex}
\right)
\nonumber\\
&&
\cdot\,
\frac{d^{3}\vec p_1 \, d^{3}\vec p_2}
{(2 \pi)^{12} \,2 k^0_1\, 2 k^0_2\, 2 p^0_1\, 2 p^0_2  }
\label{dsigvirt1}
\\
\nonumber\\
&=&  
(2 \pi)^4 \delta^{(4)}(k_1 + k_2 - p_1 - p_2)
\,
\frac{1}{4} 
\frac{\sum_{spin}
2\, {\Re}e\left( 
{{M}^{Born}}^{*} \cdot {M}^{vertex}
\right)
}
{2 s\, \beta_0}
\nonumber\\
&&
\cdot
\, \frac{d^{3}\vec p_1}{(2 \pi)^3 \, 2 p^0_1}
 \,\frac{ d^{3}\vec p_2}{(2 \pi)^3 \, 2 p^0_2}
\label{dsigvirt2}.
\ea
In (\ref{dsigvirt1}) and (\ref{dsigvirt2}) it is summed
over the final state helicities and averaged over the 
initial state helicities.
Here, we only sketch the initial state calculation as 
the final state calculation can trivially be deduced 
from the initial state one. For the initial state vertex 
we have to distinguish the two different contributions from the 
$\gamma e^{+}e^{-}$ and $Z e^{+}e^{-}$ vertices,
\footnote{In the final state case, it is the $\gamma \bar{f}f$
and $Z \bar{f}f$ vertices correspondingly.} 
$\Gamma^{\gamma}_\mu$
and $\Gamma^{Z}_\mu$, with their respective couplings:
\ba
{M}^{vertex} &=& i Q_e Q_f e^2 %
\bar{v}(k_2)\,{\Gamma^{\gamma}_\mu}\,{u}(k_1)
\,\frac{g^{\mu\nu}}{s}\,%
\bar{u}(p_1)\,{\gamma_\nu}\,{v}(p_2)
\nonumber\\
&+&
i\frac{g^2}{4\cos^2\theta_W} %
\bar{v}(k_2)\,{\Gamma^{Z}_\mu}\,{u}(k_1)
\,\frac{g^{\mu\nu}-\frac{k_\mu k_\nu}{M_Z^2}}{s-m_Z^2+i\bar{\varepsilon}}\,
\,%
\bar{u}(p_1)\,{\gamma_\nu}\,(v_f+a_f\gamma_5)\,{v}(p_2),
\nonumber\\
&&
\hspace*{3cm} 
m_z^2 = M_Z^2 - i\,\Gamma_Z\frac{s}{M_Z},\quad k = k_1 + k_2,
\label{vert02}
\ea
with the following vertex functions $\Gamma^{\gamma}_\mu$
and $\Gamma^{Z}_\mu$ as loop integrals over the virtual 
photon momentum $p$:
\ba
\Gamma^{\gamma}_\mu(k) &=&
-i Q_e^2 e^2
\int\frac{d^4{p}}{(2\pi)^4}\, 
\,\gamma^{\alpha}\,
\frac{(\ps-\kst)+m_e}{(p-k_2)^2-m_e^2+i\bar{\varepsilon}}
\,\gamma_{\mu}\,
\frac{(\kso+\ps)+m_e}{(k_1+p)^2-m_e^2+i\bar{\varepsilon}}
\,\gamma_{\alpha}\,
\frac{1}{p^2},
\label{vert03a}
\nonumber\\
\\
\Gamma^{Z}_\mu(k) &=&
-i Q_e^2 e^2
\int\frac{d^4{p}}{(2\pi)^4}\,  
\,\gamma^{\alpha}\,
\frac{(\ps-\kst)+m_e}{(p-k_2)^2-m_e^2+i\bar{\varepsilon}}
\,\gamma_{\mu}\,
(v_e+a_e\gamma_5)
\nonumber\\
&&
\cdot\frac{(\kso+\ps)+m_e}{(k_1+p)^2-m_e^2+i\bar{\varepsilon}}
\,\gamma_{\alpha}\,
\frac{1}{p^2},
\label{vert03}
\ea
with the final dependence on the momentum $k=k_1-k_2$.
There are the following three Feynman integrals to compute:
\ba
I_0^F &=& 
\mu^{4-n}\,
\int{d^4{p}}\, 
\frac{1}{[p^2 + 2 p k_1]\, [p^2 - 2 p k_2]\, p^2},
\\
\label{vert07a}
\nonumber\\
I_{\rho}^F &=& 
\mu^{4-n}\,
\int{d^4{p}}\, 
\frac{p_{\rho}}
{[p^2 + 2 p k_1]\, [p^2 - 2 p k_2]\, p^2}, 
\\
\label{vert07b}
\nonumber\\
I_{\rho\,\sigma}^F &=& 
\mu^{4-n}\,
\int{d^4{p}}\, 
\frac{p_{\rho} p_{\sigma}}
{[p^2 + 2 p k_1]\, [p^2 - 2 p k_2]\, p^2}. 
\label{vert07c}
\ea
Calculating these integrals using suitable Feynman
parameterizations and isolating the infrared singularities
finally delivers for the photonic vertex function:  

\ba 
\Gamma^{\gamma}_{\mu}(k)
&=&
-\frac{e^2 Q_e^2}{16\pi^2}
\left[
\gamma_\mu\, 
V^{\gamma}(-s,m_e,m_e)
+ m_e\,\sigma_{\mu\nu}\,k^\nu\,
{\cal J}(-s,m_e,m_e)
\right],
\label{vert21}
\nonumber\\
\\
\mbox{with}&&
V^{\gamma}(-s,m_e,m_e) = 2 (s-2 m_e^2) C_0(-s,m_e,0,m_e)
\nonumber\\
&&
\qquad + B_0(-s,m_e,m_e) - 2 (s - 3 m_e^2){\cal J}(-s,m_e,m_e) - 2.
\label{vert22}
\nonumber\\
\ea
$B_0$ and $C_0$ denote the well-known {\it Passarino-Veltman functions}
as two- and three-point scalar one-loop integrals
given in \cite{Passarino:1979jh}. They deliver two singularities 
$1/\varepsilon_{UV}$ and $1/\varepsilon_{IR}$
which can be associated with an infrared and an ultraviolet 
divergence for vanishing or large photon momenta:  
\ba
\label{vert22a}
B_0(-s,m_e,m_e) 
&=& 
\frac{1}{\varepsilon_{UV}}-\ln\frac{m_e^2}{\mu^2}
+ 2 - \beta_0\ln\frac{\beta_0+1}{\beta_0-1},
\\
\label{vert22b}
C_0(-s,m_e,0,m_e) 
&\approx& 
\frac{1}{2\varepsilon_{IR}}
\cdot {\cal J}(-s,m_e,m_e) 
+\frac{1}{2}{\cal K}(-s,m_e,m_e),
\ea

\vfill\eject
%-------------------------------------------------------------
\ba
\label{vert22c}
{\cal J}(-s,m_e,m_e) &=&
\int^1_0{d{x}}
\frac{1}{m_e^2-x(1-x)s} 
\\
&\approx&
\label{vert22c2} 
-\frac{2}{s\beta_0}\ln\frac{1+\beta_0}{1-\beta_0} 
= -\frac{2}{s}\ln\frac{s}{m_e^2},
\\
\label{vert22d}
{\cal K}(-s,m_e,m_e)& =&
\int^1_0{d{x}}\,
\frac{1}{m_e^2+x(1-x)s}\,
\ln\left[\frac{m_e^2+x(1-x)s}{\mu^2}\right]
\\
&\approx&
\label{vert22d2}  
-\frac{1}{s}\,
\left[
2\ln\frac{s}{m_e^2}\frac{m_e^2}{\mu^2}
+\ln^2\frac{s}{m_e^2}-\frac{4}{3}\pi^2
\right].
\ea
The relations (\ref{vert22b}), (\ref{vert22c2}), and 
(\ref{vert22d2}) have been derived for small electron 
masses $m_e^2$. We can show similarly for the $Z e^{+}e^{-}$ 
vertex correction applying the Dirac equations 
for the bispinorial factors:
\ba 
\Gamma^{Z}_{\mu}(k)
&=&
\frac{e^2 Q_e^2}{16\pi^2}
\Biggl[
\gamma_\mu\, 
(v_e + a_e\gamma_5)\, V^{\gamma}(-s,m_e,m_e)
\nonumber\\
&&
+ m_e\,v_e\,\sigma_{\mu\nu}\,k^\nu\,{\cal J}(-s,m_e,m_e)
+ a_e\,A^Z_\mu(k)\Biggr],
\\
A^Z_\mu(k) &=& -2 m_e^2\gamma_\mu\gamma_5\,{\cal J}(-s,m_e,m_e)
\nonumber\\
&&
- m_e k_\mu \gamma_5 
\Biggl[
\frac{4}{s} + 
\left(3 - \frac{4 m_e^2}{s}\right){\cal J}(-s,m_e,m_e)
\Biggr].
\label{vert23}
\ea
So, neglecting masses, we can nicely correlate the 
two neutral current vertex functions simply by 
a vector- and axial-vector coupling term:
\ba
\Gamma^{Z}_{\mu}(k) 
=
\Gamma^{\gamma}_{\mu}(k) 
(v_e + a_e\gamma_5). 
\label{vert24}
\ea
The final state vertex functions can be derived completely 
analoguously, replacing $m_e, Q_e, v_e, a_e$ by 
$m_f, Q_f, v_f, a_f$ and repeating the above calculation with:
\ba
k_1 \to -p_2,\quad k_2 \to -p_1,\quad k \to -(k_1+k_2) = p_1 + p_2.
\label{findef05}
\ea

In order to finalize the calculation of the first order
vertex correction, we now have to renormalize the vertex 
by adding the {\it counter terms} to the vertex.
They are calculated from the self energy corrections to
the external fermionic lines.
They will remove the ultraviolet divergent contributions
encountered in the $B_0$ function in (\ref{vert22a}).

\vfill\eject
%-----------------------------------------------
\subsubsection*{The vertex counter terms
\label{counter}}
%-----------------------------------------------
%
The procedure of renormalization can be looked up 
in any advanced textbook on quantum field theory
or particle physics, so this will not be illustrated here 
\cite{Veltman:1968ki,'tHooft:1971rn,'tHooft:1972fi,'tHooft:1972ue}. 
We just want to give a quick illustration how to 
derive the necessary {\it counter terms} 
to the above calculated $\gamma$ and $Z$ vertex 
functions.

The counter term for the $\gamma e^{+}e^{-}$ 
vertex function is nothing else but
\ba
{\Gamma^{\gamma}_{\mu}}^{c.t.}
= (Z-1) {\gamma}_{\mu},
\label{barelagr4}
\ea
with the $Z$-factor still to be determined.
This $Z$-factor arised from the renormalization 
of the fermion fields and can be extracted
from the {\it self-energy corrections} to the external
fermion legs. To see this connection, we use 
{\it on-mass-shell renormalization} and give the
{\it renormalization conditions} which fix the 
$Z$-factor for the fermion fields' renormalization.
The two renormalization conditions for the external
fermion fields for on-mass-shell renormalization are:
\begin{itemize}
\item[1.] The renormalized fermion mass is the 
pole of the renormalized fermion propagator on-mass shell. 

\item[2.] On-mass shell, the residuum of the renormalized  
propagator is 1.
\end{itemize}
For this we need the {\it regularized} self-energy 
corrections to the fermion lines. 

\begin{figure}[th]
\[
\begin{array}{ccccccc}
\begin{picture}(75,20)(0,7.5)
\SetScale{2.}
  \ArrowLine(0,5)(13.5,5)
  \ArrowLine(25,5)(37.5,5)
\SetScale{1.}
\Text(12.5,14)[bc]{$f$}
\Text(62.5,14)[bc]{$f$}
\GCirc(37.5,10){12.5}{0.5}
\end{picture}
&=&
\begin{picture}(75,20)(0,7.5)
\Text(12.5,13)[bc]{$f$}
\Text(62.5,13)[bc]{$f$}
\ArrowLine(0,10)(75,10)
\end{picture}
&+&
\begin{picture}(75,20)(0,7.5)
\Text(12.5,13)[bc]{$f$}
\Text(62.5,13)[bc]{$f$}
\Text(37.5,26)[bc]{$f$}
\Text(37.5,-8)[tc]{$\gamma$}
  \ArrowLine(0,10)(25,10)
  \ArrowArcn(37.5,10)(12.5,180,0)
  \PhotonArc(37.5,10)(12.5,180,0){3}{15}
  \Vertex(25,10){2.5}
  \Vertex(50,10){2.5}
  \ArrowLine(50,10)(75,10)
\end{picture}
&+&
\begin{picture}(75,20)(0,7)
  \Text(12.5,13)[bc]{$f$}
  \Text(62.5,13)[bc]{$f$}
  \ArrowLine(0,10)(37.5,10)
  \ArrowLine(37.5,10)(75,10)
  \Line(27.5,0)(47.5,20)
  \Line(27.5,20)(47.5,0)
  \Vertex(37.5,10){1}
\SetScale{1.}
\end{picture}
\end{array}
\]
\vspace*{3mm}
\caption[Fermion self-energies and counter term diagrams]
{\sf
Fermion line with QED self-energy insertion and counter term diagrams.
\label{fig.fsesm}}
\end{figure}
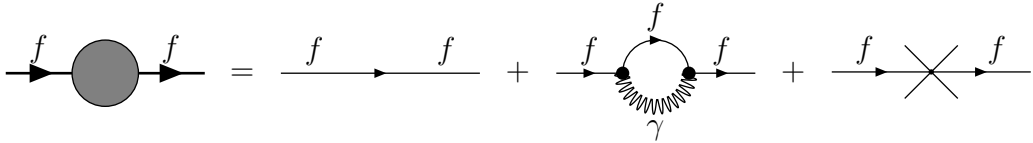
%-----------
\vspace*{3mm}
We can extract the two renormalization conditions:
\ba
{\delta}m &=& \Sigma^{\gamma}(m_e),
\\
Z - 1 &=& -\left.
\frac{\partial\Sigma^{\gamma}(\ps)}{\partial\ps}
\right|_{\ps \to m_e}.
\label{renself4}
\ea
For the fermionic QED self-energy correction we have 
the following loop integral after dimensional regularization:
\ba
\Sigma^{\gamma}(\ps)
&=&
- e^2 Q_e^2 \int\frac{d^4{q}}{(2\pi)^4}
\,\gamma_\alpha\,
\frac{ (\qs+\ps) +m_e }{ (q+p)^2 - m_e^2 + i\bar{\varepsilon}}
\gamma_\beta\,
\frac{g_{\alpha\,\beta}}{q^2},
\label{renself5}
\\
&\to&
\frac{e^2 Q_e^2}{(2\pi)^4}\mu^{4-n} 
\int{d^n{q}}\,
\frac{(2-n)(\qs+\ps)+ n\, m_e}
{\left[(q+p)^2-m_e^2+i\bar{\varepsilon}\right]q^2},
\label{renself6}
\ea
because in $n$ dimensions we have:
\ba
\gamma^\alpha (\qs+\ps)\gamma_\alpha = (2-n)(\qs+\ps),
\qquad \gamma^\alpha \gamma_\alpha = n.
\label{gammandim6}
\ea
Again with a suitable Feynman parameterization we have 
a scalar and a tensorial two-point function to compute and
finally get for $\Sigma^{\gamma}(\ps)$:

\ba 
\Sigma^{\gamma}(\ps) &=& 
\frac{e^2 Q_e^2}{16\pi^2}
\left[
-2\ps B_1(p^2,0,m_e) - 4 m_e B_0(p^2,m_e,0) - \ps +2 m_e
\right].
\label{renself12}
\nonumber\\
\ea
Calculating the derivative of  (\ref{renself12})
produces the $Z$ factor which we need 
to fix the vertex counter term:

\ba 
%------------------------------------------------------------
Z - 1 &=& - \frac{\partial \Sigma^{\gamma}(\ps)}{\partial \ps}
\biggr|_{\ps\to m_e}\qquad \longrightarrow
\\
\nonumber\\
Z-1 &=& 
-\frac{e^2 Q_e^2}{16\pi^2}
\left[
\frac{1}{\varepsilon_{UV}}-2\frac{1}{\varepsilon_{IR}}
-3\ln\frac{m_e^2}{\mu^2}+4
\right],
\label{zfact}
\ea
with 
\ba
\frac{1}{\varepsilon_{UV}} &=& -\frac{2}{n_{UV}-4}-C-\ln\pi = - 2\, P_{UV},
\label{ultrav2}
\\
\frac{1}{\varepsilon_{IR}} &=& \frac{2}{n_{IR}-4}+C+\ln\pi =  2\, P_{IR},
\label{infrar2}
\ea
thus fixing the vertex counter term to:
\ba
{\Gamma^{\gamma}}^{c.t.}_\mu &=& (Z-1) \gamma_\mu=
-\frac{e^2 Q_e^2}{16\pi^2}
\gamma_\mu
\left[
\frac{1}{\varepsilon_{UV}}-2\frac{1}{\varepsilon_{IR}}
-3\ln\frac{m^2}{\mu^2}+4
\right].
\label{vertcount2}
\ea
We can write down as regularized vertex function 
removing the ultra-violet pole of the 
bare vertex function 
${\Gamma^{\gamma}}_{\mu}(q)$:

\ba 
&&{\Gamma^{\gamma}_{\mu}}^{reg}(q)
=
\Gamma^{\gamma}_{\mu}(q)
+
{\Gamma^{\gamma}_{\mu}}^{c.t.}(q)
\nonumber\\
&=&
\frac{e^2 Q_e^2}{16\pi^2}
\left[
\gamma_\mu\, 
{V^{\gamma}}^{reg}(q^2,m_e,m_e)
+ m_e\,\sigma_{\mu\nu}
\left(
k_1^\nu - k_2^\nu
\right)
{\cal J}(q^2,m_e,m_e)
\right],
\\
\nonumber\\
\label{vert25}
\mbox{with}&&{V^{\gamma}}^{reg}(q^2,m_e,m_e)
=
V^{\gamma}(q^2,m_e,m_e)+
\left[
-\frac{1}{\varepsilon_{UV}}+2\frac{1}{\varepsilon_{IR}}
+3\ln\frac{m_e^2}{\mu^2}-4
\right].
\nonumber\\
\label{vert26}
\ea
Equivalently, we can show in absolutely the same manner 
for the corrected final state vertex function replacing
the Lorentz index, the fermion mass and the momenta by
the corresponding final state terms:

\ba 
{\Gamma^{\gamma}_{\mu}}^{reg}(q)
&=&
\Gamma^{\gamma}_{\mu}(q)
+
{\Gamma^{\gamma}_{\mu}}^{c.t.}(q)
\nonumber\\
&=&
\frac{e^2 Q_f^2}{16\pi^2}
\left[
\gamma_\mu\, 
{V^{\gamma}}^{reg}(q^2,m_f,m_f)
+ m_f\,\sigma_{\mu\nu}
\left(
p_1^\nu + p_2^\nu
\right)
\right].
\label{vert25b}
\ea
Inserting this (\ref{vert25}) together with (\ref{vert24}) 
into (\ref{dsigvirt2}) and calculating the cross section 
contribution from the 
virtual corrections one can show that the ultraviolet pole 
$1/\varepsilon_{UV}$ cancels. We obtain as virtual 
correction factor ${d{\sigma^{virtual}_{ini}}}$:

\ba
{d{\sigma^{virtual}}}
&=& 
\frac{\alpha}{\pi}\delta^{virtual}(\varepsilon_{IR},\mu)
\, {d{\sigma^{Born}}},
\label{dsigvirtini} 
\\
\nonumber\\
\delta^{virtual}(\varepsilon_{IR},\mu)
&=&
Q_e^2
\left[
(L_e-1) \left(\frac{3}{2}
-\ln\frac{m_e^2}{\mu^2} 
- \frac{1}{\varepsilon_{IR}}
\right)
-\frac{1}{2} L_e^2 
-\frac{1}{2} + \frac{2\pi^2}{3}
\right].
\nonumber\\
\label{deltavirtini}
\ea
Combining the soft  and virtual photon contributions 
from (\ref{deltasoftini}) and (\ref{deltavirtini}) to the 
fermion-pair production differential cross section 
we can see that finally also the infrared divergences 
cancel and the complete result for the initial state soft 
and virtual contribution to the cross sections can be given:
\footnote{Of course, also the regulator term $\mu$
for the vanishing photon mass finally has to cancel 
and it does.
}
\ba
{d{\sigma^{soft+virtual}}}
&=& \frac{\alpha}{\pi}
\left(\delta^{soft}(\bar{\varepsilon},\varepsilon_{IR},\mu)
+\delta^{virtual}(\varepsilon_{IR},\mu)\right)
\,{d{\sigma^{Born}}},
\label{dsigsvini} 
\\
\nonumber\\
\delta^{soft+virtual}(\bar{\varepsilon})
&=&
Q_e^2\,
\left[
(L_e-1) 
\left(
\frac{3}{2}
+ 2\ln\varepsilon
\right)
- \frac{1}{2} + \frac{\pi^2}{3}
\right].
\nonumber\\
\label{deltasvini}
\ea
The final state results for $\delta^{virtual}_{fin}$, 
$\delta^{soft}_{fin}$, and $\delta^{soft+virtual}_{fin}$
are of course easily obtained merely replacing 
$L_e=\ln(s/m_e^2)$ by $L_f=\ln(s'/m_f^2)$.

The parameter $\bar{\varepsilon}$ is again the arbitrarily chosen
soft photon cut-off to distinguish from the hard photon
region. Adding the hard photon contribution -- calculated  
for the initial state, final state, and initial-final state 
interference in Appendix \ref{hardbrem} -- then also removes 
this arbitrarily chosen and therefore unphysical parameter. 

This can be shown of course separately for the initial 
state and the final state real and virtual photon 
contributions to $e^+e^-\to \bar{f}f$.
For the initial-final state interference 
we have to combine the soft photon
terms with the virtual corrections from the interference
of Born and virtual box corrections 
for an infrared-finite result. 

%\newpage 

%####################################################################
\subsection{Virtual box corrections 
\label{boxes}}
%####################################################################
%
After determining all first order soft and hard photonic 
contributions to $e^+e^-\to \bar{f}f$ adding the initial and 
final state vertex corrections,
we also have to include the virtual QED corrections 
from the interference of the $\gamma\gamma$ and $\gamma Z$ 
box diagrams with the Born graphs in order to remove all infrared 
singularities. Due to the exchange of both photon and $Z$ vector boson
in the Born amplitudes also all box diagrams with $\gamma\gamma$
and $\gamma Z$ exchange have to be added in order to remove the divergences. 
The weak box contributions with $ZZ$ and $WW$ exchange are infrared-finite 
and are contained as weak virtual corrections in {\it improved Born observables} 
in an {\it effective Born approximation} (see e.g.~Section \ref{sec_lep1slc_precobs}).

\begin{figure}[thbp]
\begin{center}
%%%%%%%%%%%%%%%
% Box diagram zbox1.tex%
%%%%%%%%%%%%%%%
\vfill
\setlength{\unitlength}{1pt}
%\SetScale{0.8}
\SetWidth{0.8}
\begin{picture}(200,120)(0,0)
\thicklines
\ArrowLine(10,90)(70,90)
\Vertex(70,90){1.8}
\ArrowLine(70,90)(70,30)
\Vertex(70,30){1.8}
\ArrowLine(70,30)(10,30)
\Photon(70,90)(130,90){2}{8}
\Photon(70,30)(130,30){2}{8}
\ArrowLine(190,30)(130,30)
\Vertex(130,30){1.8}
\ArrowLine(130,30)(130,90)
\Vertex(130,90){1.8}
\ArrowLine(130,90)(190,90)
\Text(4,92)[]{$e^-$}
\Text(4,32)[]{$e^+$}
\Text(197,30)[]{$\bar{f}$}
\Text(197,90)[]{$f$}
\Text(100,100)[]{$\gamma\,,\,Z$}
\Text(100,20)[]{$\gamma\,,\,Z$}
\Text(102,80)[]{$q+k_1+k_2$}
\Text(100,40)[]{$q$}
\Text(40,60)[]{$-(q+k_2)$}
\Text(160,60)[]{$-(q+p_2)$}
\Text(30,100)[]{$k_1$}
\Text(30,20)[]{$k_2$}
\Text(170,100)[]{$p_1$}
\Text(170,20)[]{$p_2$}
\end{picture}
%\begin{Feynman}{45,30}{0,0}{1.0}
%\put(0,5){\fermionrighthalf}
%\put(0,20){\fermionlefthalf}
%\put(15,5){\fermionuphalf}
%\put(15,5){\gaugebosonrighthalf}
%\put(21,0){$\gamma$}
%\put(15,20){\gaugebosonrighthalf}
%\put(19,23){$\gamma,Z$}
%\put(30,5){\fermiondownhalf}
%\put(30,5){\fermionrighthalf}
%\put(30,20){\fermionlefthalf}
%\end{Feynman}
%%%%%%%%%%%%%%%
% Box diagram zbox2.tex%
%%%%%%%%%%%%%%%
\setlength{\unitlength}{1pt}
%\SetScale{0.8}
\SetWidth{0.8}
\begin{picture}(200,120)(0,0)
\thicklines
\ArrowLine(10,90)(70,90)
\Vertex(70,90){1.8}
\ArrowLine(70,90)(70,30)
\Vertex(70,30){1.8}
\ArrowLine(70,30)(10,30)
\Photon(70,90)(130,30){2}{9}
\Photon(70,30)(130,90){2}{9}
\ArrowLine(190,30)(130,30)
\Vertex(130,30){1.8}
\ArrowLine(130,30)(130,90)
\Vertex(130,90){1.8}
\ArrowLine(130,90)(190,90)
\Text(4,92)[]{$e^-$}
\Text(4,32)[]{$e^+$}
\Text(197,30)[]{$\bar{f}$}
\Text(197,90)[]{$f$}
\Text(90,88)[]{$\gamma,Z$}
\Text(90,32)[]{$\gamma,Z$}
\Text(40,60)[]{$-(q+k_2)$}
\Text(150,60)[]{$q+p_1$}
\Text(30,100)[]{$k_1$}
\Text(30,20)[]{$k_2$}
\Text(170,100)[]{$p_1$}
\Text(170,20)[]{$p_2$}
\end{picture}

\caption[Direct and crossed box diagrams]
{\label{fig3paw}
{\sf
The $\gamma\gamma$ and $\gamma Z$ direct and crossed box diagrams.
 % with virtual photons
}}
\label{fig.virtbx}
\end{center}
\end{figure}
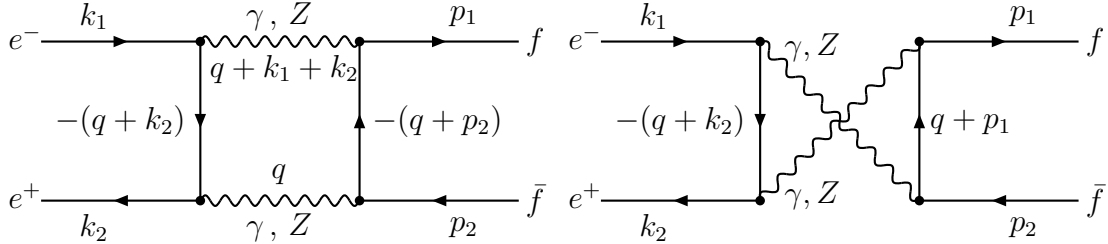

As QED box diagrams (Fig.~\ref{fig.virtbx}) we have two different topologies:
the {\sl direct} and {\sl crossed} graphs. 
%
%----------------------------------------------------------------------
%------------------------------------------------------------
%
The corresponding $S$-matrix element of the box graphs can be written as
\ba
{\Large{\cal M}}^{box} = 
      (2 \pi)^4 \delta^{(4)}(k_1 +k_2 - p_1 - p_2)
\; \frac{{M}^{box}} {(2 \pi)^6  
   (2 k^0_1\, 2 k^0_2\, 2 p^0_1\, 2 p^0_2)^{1/2} },
\label{box01}
\ea
where ${M}^{box}$ contains the direct and crossed diagrams 
with $\gamma\gamma$ and $\gamma Z$ (and $Z\gamma$) exchange. 
\ba
{M}^{box} = 
{M}^{\gamma\gamma}_{di}\, +\, {M}^{\gamma\gamma}_{cr}\,
 +\, {M}^{\gamma Z}_{di}\, +\, {M}^{\gamma Z}_{cr}\, 
 +\, {M}^{Z \gamma }_{di}\, +\, {M}^{Z \gamma }_{cr}.
\label{box02}
\ea
For the Born amplitude we of course similarly have:

\ba
{\Large{\cal M}}^{Born} = 
      (2 \pi)^4 \delta^{(4)}(k_1 +k_2 - p_1 - p_2)
\; \frac{{M}^{Born}} {(2 \pi)^6  
   (2 k^0_1\, 2 k^0_2\, 2 p^0_1\, 2 p^0_2)^{1/2} },
\label{box1}
\ea
and can straightforwardly derive the interference of the Born and 
box diagrams as contributions to the differential cross section:
\ba
d{\sigma^{box}} &=& \frac{1}{j}  
      (2 \pi)^4 \delta^{(4)}(k_1 +k_2 - p_1 - p_2)\; 
\frac{1}{4}\,
\frac{\sum_{spin} 
2\,{\Re}e\left({M}^{box}\cdot {{M}^{Born}}^{*}\right) } 
{(2 \pi)^6 (2 k^0_1\, 2 k^0_2\, 2 p^0_1\, 2 p^0_2)^{1/2} }
d^3{\vec{p}_1}d^3{\vec{p}_2},
\nonumber\\
\\
\mbox{with} && {\tt j} = \frac{\sqrt{ (k_1\cdot k_2)^2 - \meQ } }
{(2 \pi)^6 \, k^0_1\,  k^0_2 }
  =\frac{s \beta_0}{(2 \pi)^6 \,2 k^0_1\, k^0_2 }.
\label{box2}
\ea
We sum over all final-state helicities and average over
the initial-state polarizations (for the unpolarized cross section
contribution).

\ba
   d \sigma^{box} &=&
  (2 \pi)^4 \delta^{(4)}(k_1 +k_2 - p_1 - p_2)
\; \frac{1}{4}\,\frac{\sum_{spin}\; 
2\,{\Re}e \left({M}^{box}\cdot {{M}^{Born}}^{*}\right) } 
 {2 s\, \beta_0}
\nonumber\\
   &&
\cdot \frac{d^{3}\vec p_1}{(2 \pi)^3 \, 2 p^0_1}
 \;\frac{ d^{3}\vec p_2}{(2 \pi)^3 \, 2 p^0_2}.
\label{box3}
\ea
The current structure of ${M}^{Born}$ can be seen 
in Section \ref{feynmat}.
The matrix element to the box diagrams contains one 
loop integration over the free momentum $q$.  

\ba
\label{box5a}
{M}^{\gamma\gamma}_{di} &=&
e^4 Q_e^2 Q_f^2\int\frac{d^n{q}}{(2\pi)^n}\;
{\bar u}(-k_2)\; \gimu \; 
\frac{-(\qs+\kst)+m_e}{(q+k_2)^2-m_e^2 + i\bar{\varepsilon} }
\; \ginu  \; u(k_1)
\nonumber\\
&&\cdot\,\, {\bar u}(p_1)\; \gabe \; 
\frac{-(\qs+\pst)+m_f}{(q+p_2)^2-m_f^2 + i\bar{\varepsilon} }
\; \gaal  \; u(-p_2)
\cdot\,\, \frac{g_{\mu\alpha}}{q^2}
\frac{g_{\nu\beta}}{Q^2},
\\
\nonumber\\
\nonumber\\
\label{box5b}
{M}^{\gamma\gamma}_{cr} &=&
e^4 Q_e^2 Q_f^2\int\frac{d^n{q}}{(2\pi)^n}\;
{\bar u}(-k_2)\; \gimu \; 
\frac{-(\qs+\kst)+m_e}{(q+k_2)^2-m_e^2 + i\bar{\varepsilon} }
\; \ginu  \; u(k_1)
\nonumber\\
&&\cdot\,\, {\bar u}(p_1)\; \gabe \; 
\frac{\qs+\pso+m_f}{(q+p_1)^2-m_f^2 + i\bar{\varepsilon} }
\; \gaal  \; u(-p_2)
\cdot\,\, \frac{g_{\mu\beta}}{q^2}
\frac{g_{\nu\alpha}}{Q^2},
\ea

\vfill\eject
%----------------------------------------------------------------
\ba
\label{box5c}
{M}^{\gamma Z}_{di} &=&
e^2 Q_e Q_f \frac{g^2 }{4\cos^2\theta_W}
\int\frac{d^n{q}}{(2\pi)^n}\;
{\bar u}(-k_2)\; \gimu \; 
\frac{-(\qs+\kst)+m_e}{(q+k_2)^2-m_e^2 + i\bar{\varepsilon} }
\nonumber\\
&&
\cdot\,
\ginu \vae  \; u(k_1)
\; {\bar u}(p_1)\; \gabe \vaf \; 
\frac{-(\qs+\pst)+m_f}{(q+p_2)^2-m_f^2 + i\bar{\varepsilon} }
\nonumber\\
&&
\cdot\, 
\gaal  \; u(-p_2)
\frac{g_{\mu\alpha}}{q^2}
\frac{g_{\nu\beta}-\frac{1}{m_Z^2}Q_{\nu}Q_{\beta}}{Q^2-m_Z^2},
\\
\nonumber\\
\nonumber\\
\label{box5d}
{M}^{\gamma Z}_{cr} &=&
e^2 Q_e Q_f \frac{g^2 }{4\cos^2\theta_W}
\int\frac{d^n{q}}{(2\pi)^n}\;
{\bar u}(-k_2)\; \gimu \; 
\frac{-(\qs+\kst)+m_e}{(q+k_2)^2-m_e^2 + i\bar{\varepsilon} }
\nonumber\\
&&
\cdot\,
\ginu \vae  \; u(k_1)
{\bar u}(p_1)\; \gabe \; 
\frac{\qs+\pso+m_f}{(q+p_1)^2-m_f^2 + i\bar{\varepsilon} }
\nonumber\\
&&
\cdot\, 
\gaal  \vaf \; u(-p_2)
\frac{g_{\mu\beta}}{q^2}
\frac{g_{\nu\alpha}-\frac{1}{m_Z^2}Q_{\nu}Q_{\alpha}}{Q^2-m_Z^2},
\\
\nonumber\\
\nonumber\\
\label{box5e}
{M}^{Z\gamma}_{di} &=&
e^2 Q_e Q_f \frac{g^2 }{4\cos^2\theta_W}
\int\frac{d^n{q}}{(2\pi)^n}\;
{\bar u}(-k_2)\; \gimu \vae\; 
\frac{-(\qs+\kst)+m_e}{(q+k_2)^2-m_e^2 + i\bar{\varepsilon} }
\nonumber\\
&&
\cdot\,
\ginu \; u(k_1)
{\bar u}(p_1)\; \gabe \; 
\frac{-(\qs+\pst)+m_f}{(q+p_2)^2-m_f^2 + i\bar{\varepsilon} }
\; \gaal  \vaf \; u(-p_2)
\nonumber\\
&&
\cdot\,
\frac{g_{\mu\alpha}-\frac{1}{m_Z^2}q_{\mu}q_{\alpha}}{q^2-m_Z^2}
\frac{g_{\nu\beta}}{Q^2},
\\
\nonumber\\
\nonumber\\
\label{box5f}
{M}^{Z\gamma}_{cr} &=&
e^2 Q_e Q_f \frac{g^2 }{4\cos^2\theta_W}
\int\frac{d^n{q}}{(2\pi)^n}\;
{\bar u}(-k_2)\; \gimu \vae \; 
\frac{-(\qs+\kst)+m_e}{(q+k_2)^2-m_e^2 + i\bar{\varepsilon} }
\nonumber\\
&&
\cdot\,
\ginu \; u(k_1)
{\bar u}(p_1)\; \gabe \vaf \; 
\frac{\qs+\pso+m_f}{(q+p_1)^2-m_f^2 + i\bar{\varepsilon} }
\; \gaal \; u(-p_2)
\nonumber\\
&&
\cdot\,\frac{g_{\mu\beta}-\frac{1}{m_Z^2}q_{\mu}q_{\beta}}{q^2-m_Z^2}
\frac{g_{\nu\alpha}}{Q^2},
\\
\nonumber\\
\nonumber\\
\mbox{with}&&Q\equiv q+k_1+k_2,
\\
&&v_e = - \frac{1}{2} + 2 \sin^2{\theta}_W,
\qquad\qquad a_{e} = -\frac{1}{2},
\\
&&v_f = I_3^f - 2 Q_f \sin^2{\theta}_W,\qquad a_f = I_3^f,
\\
&& m_Z^2 = M_Z^2 - i M_z \Gamma_Z.
\ea

The box terms have been calculated 
in \cite{Bardin:1989cw,Bardin:1991de,Bardin:1991fu}
and they are shown below. One can show that the infrared 
singularities in the box contributions exactly cancel 
the divergences of the soft initial-final state interference 
terms, derived in Section \ref{soft}.

The integrated cross section contributions from 
the hard and soft initial-final interference terms 
${\bar H}_{{A}}^{int}(v,c)$ and ${\bar S}_{{A}}^{int}(c)$
(see also Appendix \ref{int} and \ref{softintcom})
and its virtual box corrections $\bar{B}_{{A}}(c,m,n)$ 
presented in terms of the regularized, infrared-finite 
flux functions are given below (we use $\sigma_{{A}}^0(s,s';m,n)$ 
from Section \ref{sec_lep1slc_radcuts}, see for 
example (\ref{generic_zf})):

%----%
\ba
\sigma_{{T}}^{int} &=&
\sum_{m,n} \int_{0}^{\Delta} d{v}  \, \Re e 
\Biggl[
\sigma_{{FB}}^0(s,s';m,n)  \, R_{{T}}^{int}(v,c;m,n) \Biggr],
\label{int2}
%--------------------------------------
\\
\sigma_{{FB}}^{int} &=&
\sum_{m,n} \int_{0}^{\Delta} d{v} 
 \, \Re e 
\Biggl[
\sigma_{{T}}^0(s,s';m,n)  
\, R_{{FB}}^{int}(v,c;m,n) 
\Biggr],
\label{int3}
\\
\nonumber\\
R_{{A}}^{int}(v,c,m,n)
&=&
\delta(v) 
\left[{\bar S}_{{A}}^{int}(c) 
+ \bar{B}_{{A}}(c,m,n) \right] 
+ {\bar H}_{{A}}^{int}(v,c), 
\label{Rint2}
\nonumber\\
\ea
The $\gamma Z$ exchange box contribution can be derived from the  
$\gamma\gamma$ and $ZZ$ exchange functions ($m,n=\gamma,Z$, $A=T,FB$): 

\ba 
\bar{B}_{{A}}(c;m,n) 
&=& 
\frac{1}{2}
\left[
\bar{B}_{{A}}(c;m,m)
+
\bar{B}_{{A}}(c;n,n)^{*}
\right], 
\label{RgZ}
\\
\nonumber\\
\bar{B}_{A}(c;n,n) 
&=& 
\frac{\alpha}{\pi} Q_e Q_f
\Biggl\{ 
\bar{b}_{A}(c;n,n) 
\mp \bar{b}_{A}(-c;n,n)
\Biggr\}.
\label{bpm}
\ea
%------------------
%
The corresponding box terms contained in (\ref{bpm}) finally are
($c_{\pm} = \frac{1}{2}(1 \pm \cos{c})$):

%------------------%
\ba
\bar{b}_{T}(c;\gamma,\gamma)  &=&
\frac{1}{2}\left(c^{2}-1\right)\ln^{2}c_{+} 
+ \left(- c^{2}+ 2c  -3    \right) \ln c_{+}  - 3c
\nonumber\\
&& -~i \pi (c^{2}-1) \ln c_{+},
\label{bt00}
%------------------    
\\ \nonumber\\
%------------------------%
\bar{b}_{FB}(c;\gamma,\gamma) 
 &=& -\frac{1}{2}\left(c^{2}-1\right)\ln^{2}c_{+} 
- \left(- c^{2}+ 2c  -3    \right) \ln c_{+}
\nonumber\\
&&
- \frac{1}{2}\left(\ln^{2}2 + 6\ln 2 +
      c^{2}\right) 
 \nonumber \\
&&   -~ \frac{i \pi}{3} 
\left[5 \ln 2 + \left(2c^3+3c^{2}+6c+5 \right)\ln c_{+} 
- \frac{2}{3}c^{2}\right], 
\label{bfb00}
\ea

\vfill\eject
\pagestyle{headings}
%----------------------------------------------------------------
\ba
 \bar{b}_{T}(c;Z,Z) &=&
 -2
\Biggl\{-c R_Z (R_Z+1) -2c R_Z (1-R_Z)\ln c_{+}
%\nonumber 
\\
&& -~ \left[-2R_Z^{2} + R_Z\left(c^{2}+1\right)+c^{2}-1\right]\ln c_{+}
\Biggr\} 
L_Z %\ln\left(1-\frac{1}{R_Z}\right)
 \nonumber \\
&& + 2R_Zc - 6c + 4  c R_Z (R_Z-1) l(1) %\litwo \left(1-\frac{1}{R_Z}\right) 
- 2c\ln R_Z 
 \nonumber \\
&&   + \left(c^{2}-1\right)\ln^{2}c_{+} 
+ 2\left[R_Z\left(c^{2}-1\right)-c^{2}+2c+3\right]\ln c_{+} 
 \nonumber \\  \nonumber 
&&-~ 2 \left[2R_Z^{2}+2c R_Z(R_Z-1)-R_Z\left(c^{2}+1\right)\right]
l(c_+), %\litwo\left(1-\frac{c_{+}}{R_Z}\right)
\label{btnn}
%-------------------------------
\\
\nonumber\\
%----------------------------------------
 \bar{b}_{FB}(c;Z,Z) &=& 
%\left(\frac{c^{3}}{3} + c\right)
\left[C_{{T}}(c) + \frac{4}{3} \right] \ln^{2}c_{+} 
% \nonumber 
\\
&&+~ %\ln\left(1-\frac{1}{R_Z}\right)
\Biggl\{ c^{2}\left(-R_Z^{2}+3R_Z-\frac{4}{3}\right) +
\left(4R_Z^{2} - 2R_Z + \frac{10}{3}\right) \ln 2
\nonumber\\
&&+~\left[4R_Z^{2}-2R_Z\left(c^{2}+1\right)+2c^{2}+\frac{10}{3}\right]\ln c_{+}
 \nonumber \\
&&  +~ 4 \left[ cR_Z(R_Z-1)+C_{{T}}(c) \right] \ln c_{+}\Biggr\} L_Z
 \nonumber \\
&& +~ c^{2}\left(\frac{4}{3}R_Z-\frac{5}{3}\right)\ln R_Z  
+   \left(\frac{16}{3}R_Z^{3} - 4R_Z^{2} + 2R_Z -\frac{2}{3}\right)
l \left( \frac{1}{2} \right)  %\litwo\left(1-\frac{1}{2R_Z}\right)
 \nonumber \\
&& +~ 2c^{2}(R_Z-1) l(1) %\litwo\left(1-\frac{1}{R_Z}\right)  
 - \frac{4}{3}\ln^{2}2 + \left(\frac{8}{3}R_Z^{2}+\frac{8}{3}R_Z - 6\right)
\ln 2
 \nonumber \\
&&+~\left[\frac{8}{3}R_Z^{2} + R_Z\left(-\frac{4}{3}c^{2}+\frac{8}{3}\right)
+ 2\left(c^{2}-3\right)\right] \ln c_{+} 
\nonumber\\ && +~\left(\frac{8}{3}R_Z^{2}+\frac{4}{3}R_Z-4\right)c\ln c_{+} 
 \nonumber \\
&& +~ 2\left[-\frac{8}{3}R_Z^{3} + 2R_Z^{2}-R_Z\left(c^{2}+1\right)+c^{2}
 +\frac{1}{3}\right] l(c_+) %\litwo\left(1-\frac{c_{+}}{R_Z}\right)
 \nonumber \\ \nonumber
&& +~ 2\left[
2c\left(R_Z^{2}-R_Z\right) +
C_{{T}}(c)\right]  l(c_+) %\litwo\left(1-\frac{c_{+}}{R_Z}\right) 
 + \left(\frac{2}{3}R_Z-1\right)c^{2},
\label{bfbnn}
\ea
%--------------------------------------
with the following abbreviations:
\ba
 l(a) &=& \mbox{Li}_2\left(1 - aR_{Z}^{-1}\right),
\label{la}
\\ 
   L_{Z} &=& \ln\left(1 - R_{Z}^{-1}\right) ,
\label{LZ}
\\ 
R_Z &=& \frac{m_Z^2}{s}.
\label{Rz}
\ea

\end{appendix}
%---------------------------------------------------------------
%	\nocite{*}
%	\def\href#1#2{#2}

%
\pagestyle{myheadings}
\markright{\it BIBLIOGRAPHY}
\providecommand{\href}[2]{#2}
%---------------------------------------------------------------------------
\chapter*{Acknowledgements}

First, I would very much like to thank my advisor Tord Riemann 
for giving me the opportunity to finally make a dream  
come true. I am indebted to him, his expertise,
his professionalism and his constant support on any
issue, on and off the job. If I've learnt one thing 
during this time, then that there is more to success
than just reading books and doing calculations.

I also would like to express many thanks to Fred Jegerlehner,
Rainer Sommer and the theory staff at DESY Zeuthen 
for letting me be a part of the group and for the support 
until the very end, even in hectic moments. Many thanks also 
to Sabine Riemann, Arnd Leike and Professor Fritzsch for all 
their support. I would especially like to thank Silvia Arkadova 
and Penka Christova for kindly reading through the manuscript 
and for the many helpful suggestions.

I thank my parents for their support, love and wisdom 
and I owe them the knowledge, strength and self-discipline 
to carry through the last twenty-one years, and especially the 
last three. Mick, Butz, Ayrin, and Amina -- thank
you for just being there for me.

I would like to thank you, Marta, Jenny and Jochen, Francesco, Axel, 
Ilaria and Alessandro, Andreas, Lisa and Kurt, Ravindran, Martin and Stefan,
for your friendship, the fun and the many lovely moments spent at DESY 
or during our spare time. Silvia A. and Silvia N., Avto, Chris, 
and Oleg, I thank you guys just as much for your cameraderie.
And Axel, I will never forget that late-night `burgers-and-beer-phase' \ldots

Fany and Jos\'{e}, I owe you more than you can imagine, your 
kindness and friendship I will always treasure. 

Tanja -- if it hadn't been for you I don't think that I would
have been able to finish this. You are special and this I know
from the bottom of my heart.

\newpage

\end{document}